\documentclass[twocolumn]{article}

\usepackage{amsmath}
\usepackage{amsthm}
\usepackage{amssymb}
\usepackage{array}
\usepackage{todonotes}
\usepackage{enumitem}
\usepackage{multirow}
\usepackage{rotating}
\usepackage{morefloats}
\usepackage{comment}
\usepackage{authblk}

\usepackage[ruled]{algorithm2e}

\SetAlFnt{\small}
\SetAlCapFnt{\small}
\SetAlCapNameFnt{\small}
\SetAlCapHSkip{0pt}
\IncMargin{-\parindent}

\newtheorem{theorem}{Theorem}
\newtheorem{lemma}{Lemma}
\newtheorem{definition}{Definition}

\DeclareMathOperator*{\argmin}{arg\,min}
\DeclareMathOperator*{\argmax}{arg\,max}

\newcommand{\R}{\mathbb{R}}
\newcommand{\N}{\mathbb{N}}
\newcommand{\ROC}{\mathit{AUC}}
\newcommand{\PR}{\mathit{PR}}
\newcommand{\AP}{\mathit{AP}}

\newcommand{\ER}{\bar{E}}
\newcommand{\FA}{\widehat{A}}
\newcommand{\FR}{\widehat{R}}
\newcommand{\Add}{A^*}
\newcommand{\Remove}{R^*}
\newcommand{\Hide}{H}
\newcommand{\PA}{P_{A}}

\newcommand{\sCN}{s^{\text{\normalfont CN}}}
\newcommand{\sSal}{s^{\text{\normalfont Sal}}}
\newcommand{\sJac}{s^{\text{\normalfont Jac}}}
\newcommand{\sSor}{s^{\text{\normalfont S{\o}r}}}
\newcommand{\sHPI}{s^{\text{\normalfont HPI}}}
\newcommand{\sHDI}{s^{\text{\normalfont HDI}}}
\newcommand{\sLHN}{s^{\text{\normalfont LHN}}}
\newcommand{\sAA}{s^{\text{\normalfont AA}}}
\newcommand{\sRA}{s^{\text{\normalfont RA}}}

\newcommand{\sACT}{s^{\text{\normalfont ACT}}}
\newcommand{\sCos}{s^{\text{\normalfont Cos}}}
\newcommand{\sRWR}{s^{\text{\normalfont RWR}}}
\newcommand{\sSR}{s^{\text{\normalfont SR}}}

\newcommand{\SKatz}{S^{\text{\normalfont Katz}}}
\newcommand{\SLHNG}{S^{\text{\normalfont LHNG}}}

\newcommand{\SMFI}{S^{\text{\normalfont MFI}}}

\begin{document}

\title{Attack Tolerance of Link Prediction Algorithms:\\ How to Hide Your Relations in a Social Network}

\author[a,b]{Marcin Waniek}
\author[c]{Kai Zhou}
\author[c]{Yevgeniy Vorobeychik}
\author[d,e]{\\Esteban Moro}
\author[b,*]{Tomasz P. Michalak}
\author[a,*]{Talal Rahwan}

\renewcommand*{\Affilfont}{\normalsize}

\affil[a]{Department of Computer Science, Khalifa University of Science and Technology, Abu Dhabi, UAE}
\affil[b]{Institute of Informatics, University of Warsaw, Warsaw, Poland}
\affil[c]{Computer Science and Engineering, Washington University in Saint Louis, Saint Louis, MO, USA}
\affil[d]{Department of Mathematics \& GISC, Universidad Carlos III de Madrid, Madrid, Spain}
\affil[e]{Media Lab, Massachusetts Institute of Technology, Cambridge, MA, USA}
\affil[*]{To whom correspondence should be addressed:  tpm@mimuw.edu.pl, talal.rahwan@ku.ac.ae}

\date{}

\maketitle

\begin{abstract}
Link prediction is one of the fundamental research problems in network analysis.
Intuitively, it involves identifying the edges that are most likely to be added to a given network, or the edges that appear to be missing from the network when in fact they are present. Various algorithms have been proposed to solve this problem  over the past decades. For all their benefits, such algorithms raise serious privacy concerns, as they could be used to expose a connection between two individuals who wish to keep their relationship private. With this in mind, we investigate the ability of such individuals to \emph{evade link prediction algorithms}. More precisely, we study their ability to strategically alter their connections so as to increase the probability that some of their connections remain unidentified by link prediction algorithms. We formalize this question as an optimization problem, and prove that finding an optimal solution is NP-complete. Despite this hardness, we show that the situation is not bleak in practice. In particular, we propose two heuristics that can easily be applied by members of the general public on existing social media. We demonstrate the effectiveness of those heuristics on a wide variety of networks and against a plethora of link prediction algorithms.
\end{abstract}

\section*{Introduction}
The Internet and social media have fueled enormous interest in developing new social network analysis tools~\cite{scott2012social}.
With such tools, data about our social connections, email exchanges, and even financial transactions may all be analysed to infer personal information that would otherwise remain confidential \cite{bird2006mining}. This raises both privacy and security related concerns as our data may be valueable not only to enterprises and public entities, but also to cyber criminals who are increasingly relying on network analysis tools for malicious purposes \cite{altshuler2013stealing}. 

One of the main network analysis tools is \textit{link prediction}~\cite{liben2007link,lu2011link}. Intuitively, based on the current network topology, this problem involves predicting the connections that are most likely to form in the future~\cite{liben2007link}. An alternative interpretation of this problem is to identify the connections that are \textit{hidden} from an observer, either due to data scarcity, or due to the deliberate concealment of information~\cite{brantingham2011co}. Link prediction has numerous applications, from providing recommendations to customers in e-commerce~\cite{crone2005predicting}, through discovering the interactions between proteins in biological networks~\cite{cannistraci2013link}, to finding hidden connections between terrorists~\cite{al2006link} or criminals~\cite{tayebi2011locating}. 

A plethora of different link prediction algorithms have been proposed in the literature~\cite{liben2007link,lu2011link,al2011survey}.
We focus in this article on the mainstream class of link-prediction algorithms based on \textit{similarity indices}~\cite{lu2011link} which analyse the network topology to quantify the similarity between any two disconnected nodes in that network. The underlying assumption in this class of algorithms is that the greater the similarity between two nodes, the greater the likelihood of having a link between them.

If used with malicious intent, link prediction algorithms may constitute a serious threat to both the privacy and the security of the general public.
In particular, inspired by the saying ``\emph{tell me who your friends are and I'll tell you who are}'', a network analyser may use link-prediction algorithms to perform a \textit{link reconstruction attack}~\cite{fire2013links}, which not only reveals your undisclosed ``friends'', but may also enhance the severity of the more general \textit{attribute inference attack} \cite{kumar2016improving} whereby the goal is to infer various private information about ``who you are'' \cite{Zheleva:2009:WWW,Mislove:2010}. 

Driven by these concerns, a number of studies recommended that social media users conceal some of their attributes, and especially their connections~\cite{lindamood2009inferring,heatherly2013preventing}. Nevertheless, although the literature identified a variety of reasons \emph{why} one should conceal his or her private connections, unfortunately far less attention has been paid to \emph{how} this should be done.

Driven by these observations, we study settings in which a ``\emph{seeker}'' runs link-prediction algorithms, and ``\emph{evaders}'' wish to hide some of their connections by making them harder to identify. More specifically, we focus on two questions: (i) how may individuals effectively evade such algorithms by rewiring the connections within their neighbourhood? and (ii) how do such evasion efforts influence the network structure?
Since, from a graph-theoretic perspective, the problem of evading link prediction is in essence an optimization problem, we analyse its computational complexity to illuminate the theoretical limits of evading link-prediction algorithms. We prove that an optimal solution is hard to compute given nine link-prediction algorithms that are widely studied in the literature. Based on this finding, we move our attention towards identifying effective, albeit not optimal, solutions. To this end, we propose two alternative heuristics that can easily be implemented by members of the general public on existing social media platforms. The first heuristic \textit{removes} strategically-chosen links from the network, and another that \textit{adds} new ones. We show that both heuristics are effective in practice, although the former seems more effective than the latter, suggesting that in order to hide a relationship, ``unfriending'' carefully-chosen individuals can provide a better disguise than befriending new ones. Finally, we evaluate the attack tolerance of different link-prediction algorithms, and find that their resilience tends to increase with the number of nodes, and tends to decrease with the average degree in the network.

\section*{Results}

\subsection*{Theoretical Analysis}\label{sec:theoreticalAnalysis}

Given an undirected network, $G = (V, E)$, where $V$ is the set of nodes and $E$ is the set of edges,
we will use the term ``\emph{non-edge}'' to refer to any pair of nodes that is \emph{not} in $E$, and will denote the set of all non-edges by $\ER$.
Our problem of \textit{evading link prediction} involves a \emph{seeker} who ranks all non-edges based on a \emph{similarity index} (Section~S1), and identifies the highly-ranked ones as edges that are likely to be part of the network, or likely to form in the future. An \emph{evader}, on the other hand, has a set of undeclared relationships that he or she wishes to keep private; the fact  that these relationships are undeclared means that they are \textit{non-edges} as far as the seeker is concerned, and we will model them as such. The evader's goal is then to rewire the network in order to minimize the likelihood of those non-edges being highlighted by the seeker. Note that a non-edge becomes less exposed to the seeker if it drops in the similarity-based ranking of all non-edges. To quantify the degree to which a non-edge is exposed in any such a ranking, we use two alternative measures, namely the \emph{area under the ROC curve} ($\ROC$) \cite{fawcett2006introduction} and the \emph{average precision} ($\AP$) \cite{boyd2013area} (Section~S2). Our problem is then formally defined as follows:

\begin{definition}[Evading Link Prediction]\label{def:EvadingLinkPrediction}
This problem is defined by a tuple, $(G,s_G,f,\Hide,b,\FA,\FR)$, where $G=(V,E)$ is a network, $s_G: \ER \rightarrow \R$ is a similarity index, $f \in \{\ROC, \AP\}$ is a performance evaluation metric, $\Hide \subset \ER$ is the set of non-edges to be hidden, $b\in\N$ is a budget specifying the maximum number of edges that can be modified (i.e., added or removed), $\FA \subseteq \ER \setminus \Hide$ is the set of edges that can be added, and $\FR \subseteq E$ is the set of edges that can be removed.
The goal is then to identify two sets, $\Add \subseteq \FA$ and $\Remove \subseteq \FR$, such that the resulting set, $E^*=(E \cup \Add)\setminus \Remove$, is in:
$$
\argmin_{E'\in\big\{(E \cup A)\setminus R\ :\ A \subseteq \FA,\ R \subseteq \FR,\ |A|+|R| \leq b\big\}} f(E', \Hide, s_G).
$$
\end{definition}

In this definition, we introduced the budget $b$ as well as the sets $\FA$ and $\FR$ to model scenarios in which the evader's ability to modify the network is limited.
The following theorem implies that, given a budget specifying the number of permitted network modifications, it is extremely challenging to identify an \emph{optimal} way to spend this budget in order to best hide a given set of non-edges; see the proof in Section~S3.

\begin{theorem}
\label{theorem:NPcompleteness}
The problem of Evading Link Prediction is NP-complete for each of the following similarity indices:
Common Neighbours~\cite{newman2001clustering}, Salton~\cite{salton1986introduction}, Jaccard~\cite{jaccard1901etude}, S{\o}rensen~\cite{sorensen1948method}, Hub Promoted~\cite{ravasz2002hierarchical}, Hub Depressed~\cite{ravasz2002hierarchical}, Leicht-Holme-Newman~\cite{leicht2006vertex}, Adamic-Adar~\cite{adamic2003friends} and Resource Allocation~\cite{zhou2009predicting}.
\end{theorem}

To put it differently, given any of the indices outlined in Theorem~\ref{theorem:NPcompleteness}, the theorem implies that the problem of evading link prediction is at least as hard as any of the problems in the class NP (Non-deterministic Polynomial-time), implying that no known algorithm can solve it in polynomial time. Despite this hardness, the situation is not necessarily bleak, especially in situations where a reasonable, albeit not optimal, solution would suffice. With this in mind, we will present two heuristic algorithms that run in polynomial time; the first, called \emph{CTR}, focuses on removing edges whereas the second, called \emph{OTC}, focuses on adding edges.

\subsection*{The CTR Heuristic}
\label{sec:CTRheuristic}

\begin{figure}[thb]
\centering
\includegraphics[width=1\linewidth]{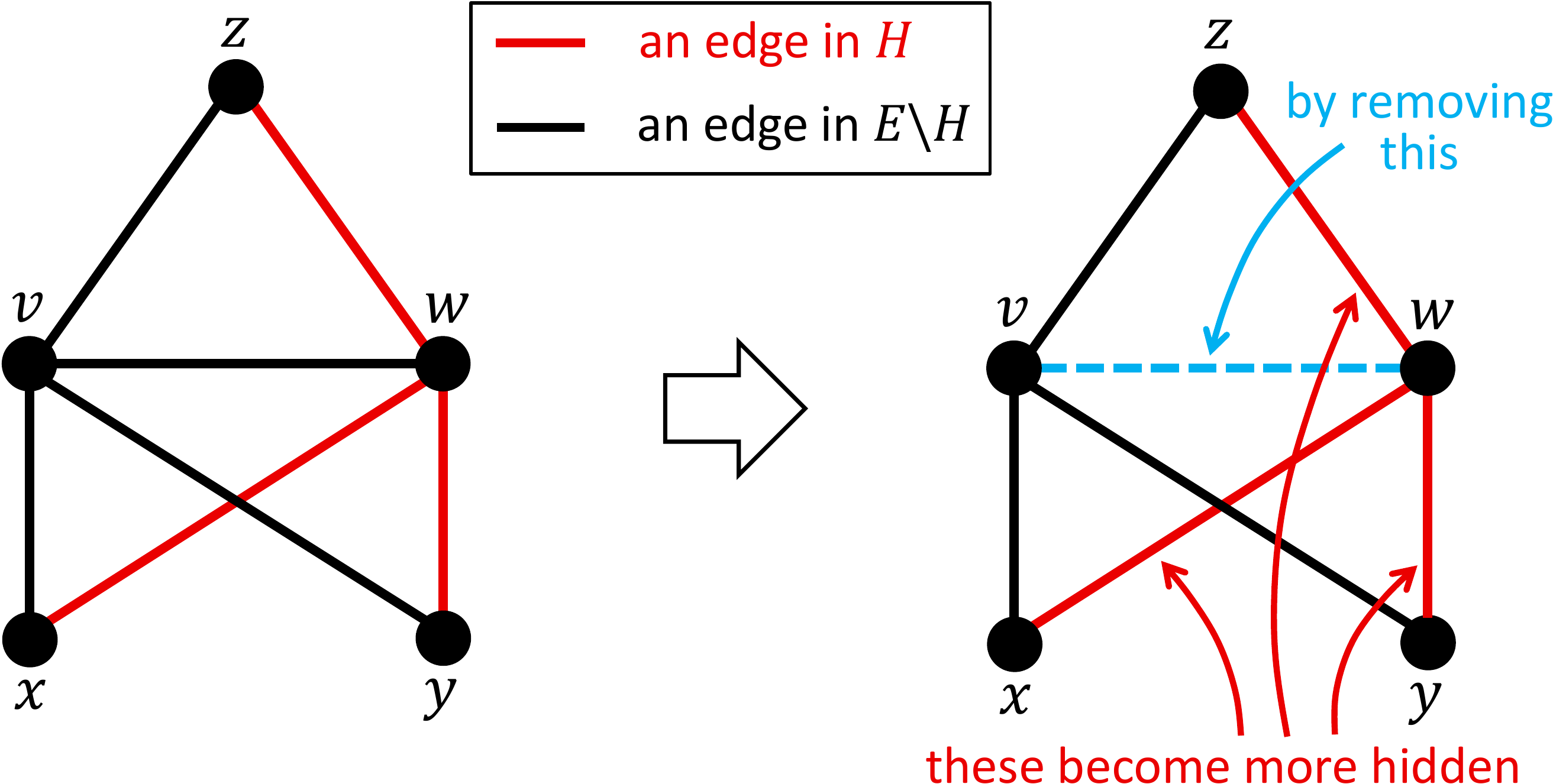}
\caption{An illustration of the main idea behind the CTR heuristic. Here, by removing $(v,w)$, we remove from the network three closed triads: one containing the nodes $v,w,x$, another containing $v,w,y$, and a third containing $v,w,z$. Consequently, the similarity scores of $(x,w)$, $(w,y)$ and $(w,z)$ can only decrease based on the analysis in \textit{Materials and Methods}.}
\label{fig:CTR-illustration}
\end{figure}

Our first heuristic, called CTR (which stands for \textit{Closed-Triad-Removal}) works by selecting an edge, $(v,w)\in E$, such that:
$$
\exists x\in V: \big((v,x)\in E\big) \wedge \big((x,w)\in\Hide\big),
$$
which implies that $(v,x)$, $(x,w)$ and $(v,w)$ form a closed triad. The algorithm then removes $(v,w)$ from the network, thereby \emph{removing the closed triad} whose nodes are $v$, $w$ and $x$; hence the name \textit{Closed-Triad-Removal (CTR)}; see the pseudo-code in Section~S5 (although $v$, $w$ and $x$ form a closed triad, this is initially unknown to the seeker since $(x,w)$ is undeclared, i.e., it is a non-edge as far as the seeker is concerned). Importantly, the removal of $(v,w)$ can only decrease the similarity score of $(x,w)$ according to any of the similarity indices outlined in Theorem~\ref{theorem:NPcompleteness}; see the analysis in \emph{Materials and Methods}. The algorithm can be even more effective if the removal of $(v,w)$ results in the removal of multiple closed triads, each containing a non-edge in $H$. In Figure~\ref{fig:CTR-illustration} for example, the removal of $(v,w)$ decreases the similarity scores of not one, but three non-edges in $\Hide$, namely $(x,w)$, $(w,y)$ and $(w,z)$. Based on this observation, the CTR heuristic is designed to maximize the number of such non-edges, by examining all possible choices of $(v,w)$ and selecting one that affects the greatest number of edges in $\Hide$. 

CTR can readily be applied by members of the general public on existing social media platforms. In Figure~\ref{fig:CTR-illustration} for example, if $w$ wishes to hide his or her relationships with $x$, $y$, and $z$, then CTR simply requires $w$ to ``unfriend'' as many people as possible who are friends of $x$, $y$ and $z$. This can easily be applied on Facebook for instance, since the mutual friends of a person and any of his or her friends are always visible.

\subsection*{The OTC Heuristic}
\label{sec:OTCheuristic}

\begin{figure}[thb]
\centering
\vspace*{-0.25cm}\includegraphics[width=1\linewidth]{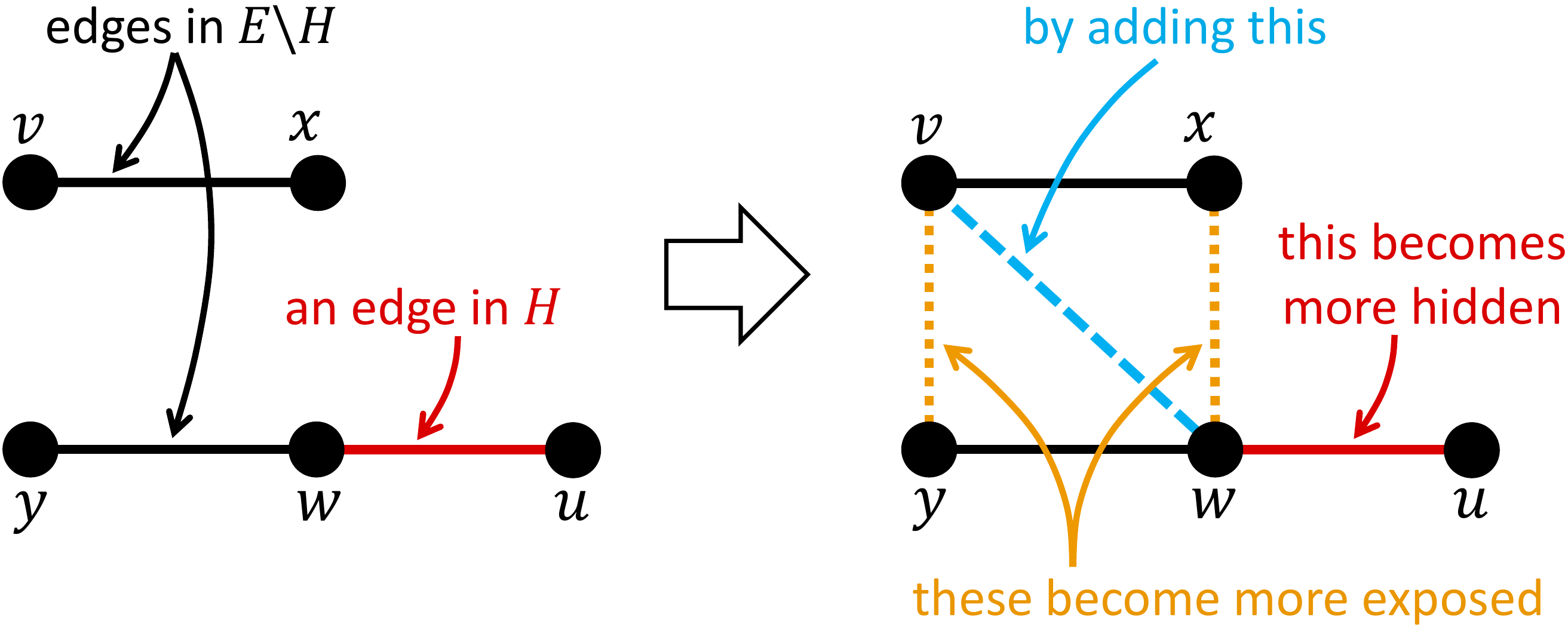}
\caption{An illustration of the main idea behind the OTC heuristic. Here, the addition of $(v,w)$ creates two open triads: one contains the nodes $x,v,w$; the other contains $v,w,y$. Consequently, the similarity scores of $(x,w)$ and $(y,v)$ increase while that of $(w,u)$ decreases; see the analysis in \textit{Materials and Methods}.}
\label{fig:OTC-illustration}
\end{figure}

Our second heuristic, called OTC (which stands for \emph{Open-Triad-Creation}) works by \emph{adding} edges to the network, unlike CTR which worked by \emph{removing} edges. Generally speaking, OTC ``hides'' a non-edge, $e\in\Hide$, by \textit{decreasing} the similarity score of $e$ while at the same time \emph{increasing} the similarity scores of (some of) the non-edges that fall within the neighbourhood of $e$. This, in turn, decreases the position of $e$ in the similarity-based ranking of all non-edges, thereby reducing the likelihood of $e$ being highlighted by a seeker armed with a link-prediction algorithm. To achieve this goal, OTC rewires the network as illustrated in Figure~\ref{fig:OTC-illustration}. More formally, it selects a non-edge $(v,w)$ to be added to the network such that:
\begin{itemize}\itemsep-0.25em
\item $\exists u\in V\!:\!(w,u)\in \Hide$;
\item $\exists x\in V\!:\!\big((x,v)\in E\big) \wedge \big((x,w)\in\ER\setminus \Hide\big)$.
\end{itemize}
As shown in Figure~\ref{fig:OTC-illustration}, the addition of $(v,w)$ \emph{creates open triads}--one containing $x,v,w$ and another containing $v,w,y$---hence the name \textit{Open-Triad-Creation (OTC)}. Importantly, given the similarity indices outlined in Theorem~\ref{theorem:NPcompleteness}, the addition of $(v,w)$ in Figure~\ref{fig:OTC-illustration} can only decrease the similarity score of $(w,u)$ and can only increase that of $(x,w)$ and $(y,v)$; see \emph{Materials and Methods} for a more formal analysis. More generally, since the creation of an open triad can only increase the similarity score of the non-edge therein, the more open triads we create by adding $(v,w)$ the better, since this may \textit{increase} the similarity scores of a greater number of non-edges, all of which contribute towards \textit{reducing} the position of $(w,u)$ in the similarity-based ranking of all non-edges. Based on this observation, OTC examines all possible choices of $(v,w)$, and selects one that results in the greatest reduction in the ranking of the non-edges in $\Hide$; see the pseudo-code in Section~S6.

OTC can be applied on popular social media platforms in a straightforward manner. For instance, if $u$ and $w$ wish to hide their relationship, then any one of them, say $w$, can send friendship requests to individuals whose list of friends contains as many people as possible who are not connected to $w$. Even if such individuals are hard to find, one can still send random friendship requests to highly-connected strangers, hoping that some of them would accept the request. This is indeed plausible, as an estimated 55\% of people accept friendship requests from complete strangers on Facebook \cite{Nagle:Singh:2009}.

\subsection*{Simulation Results}

\begin{figure}[tbhp]
\centering
\setlength\tabcolsep{1pt}
\renewcommand{\arraystretch}{0.01}
\begin{tabular}{m{.03\linewidth}m{.48\linewidth}m{.48\linewidth}}
& \multicolumn{1}{c}{\footnotesize \hspace*{0.3cm}OTC}
& \multicolumn{1}{c}{\footnotesize \hspace*{0.4cm}CTR}\\
\rotatebox{90}{\hspace*{0.3cm}\scriptsize WTC 9/11} &
\includegraphics[width=\linewidth]{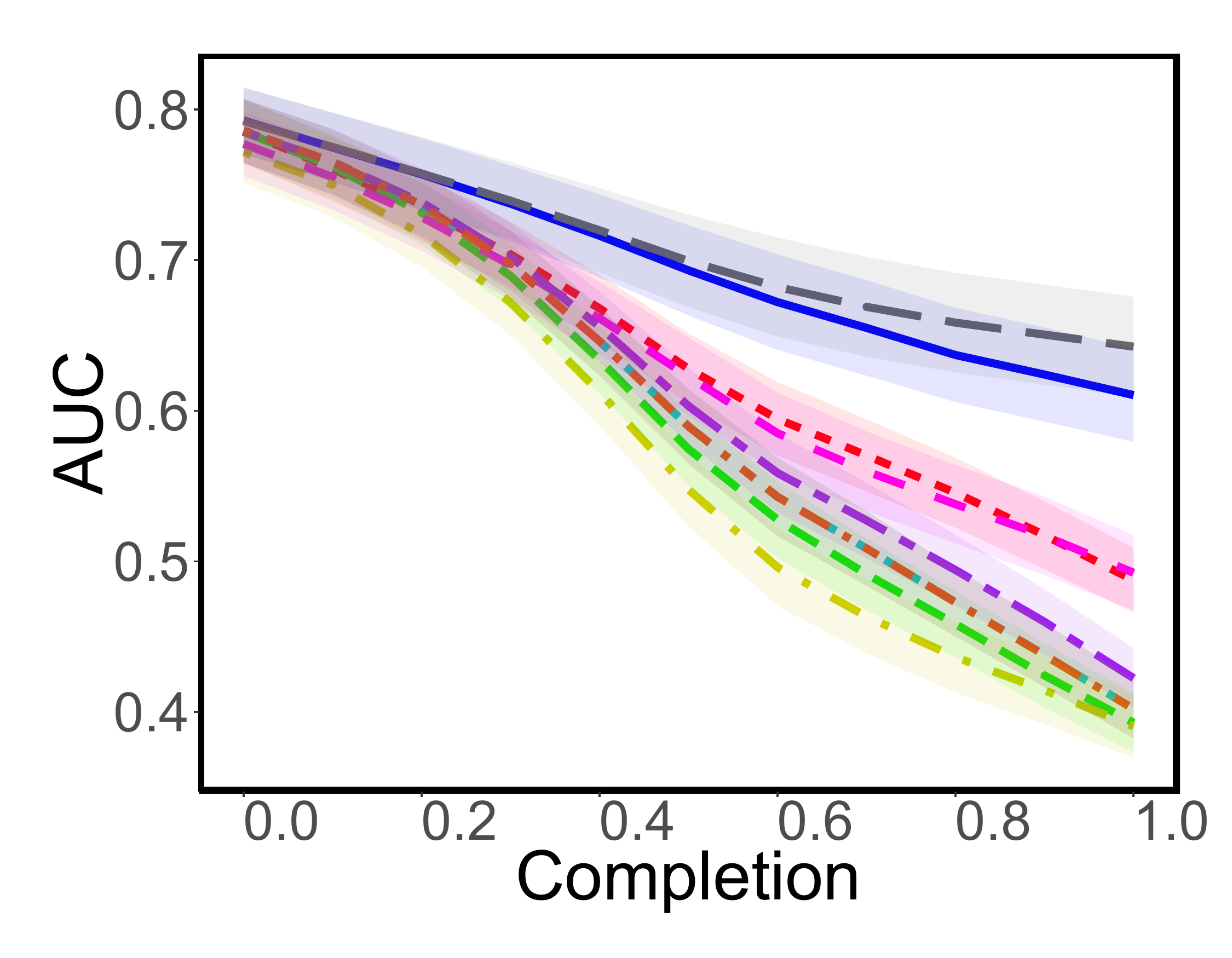}\vspace*{-0.1cm} &
\includegraphics[width=\linewidth]{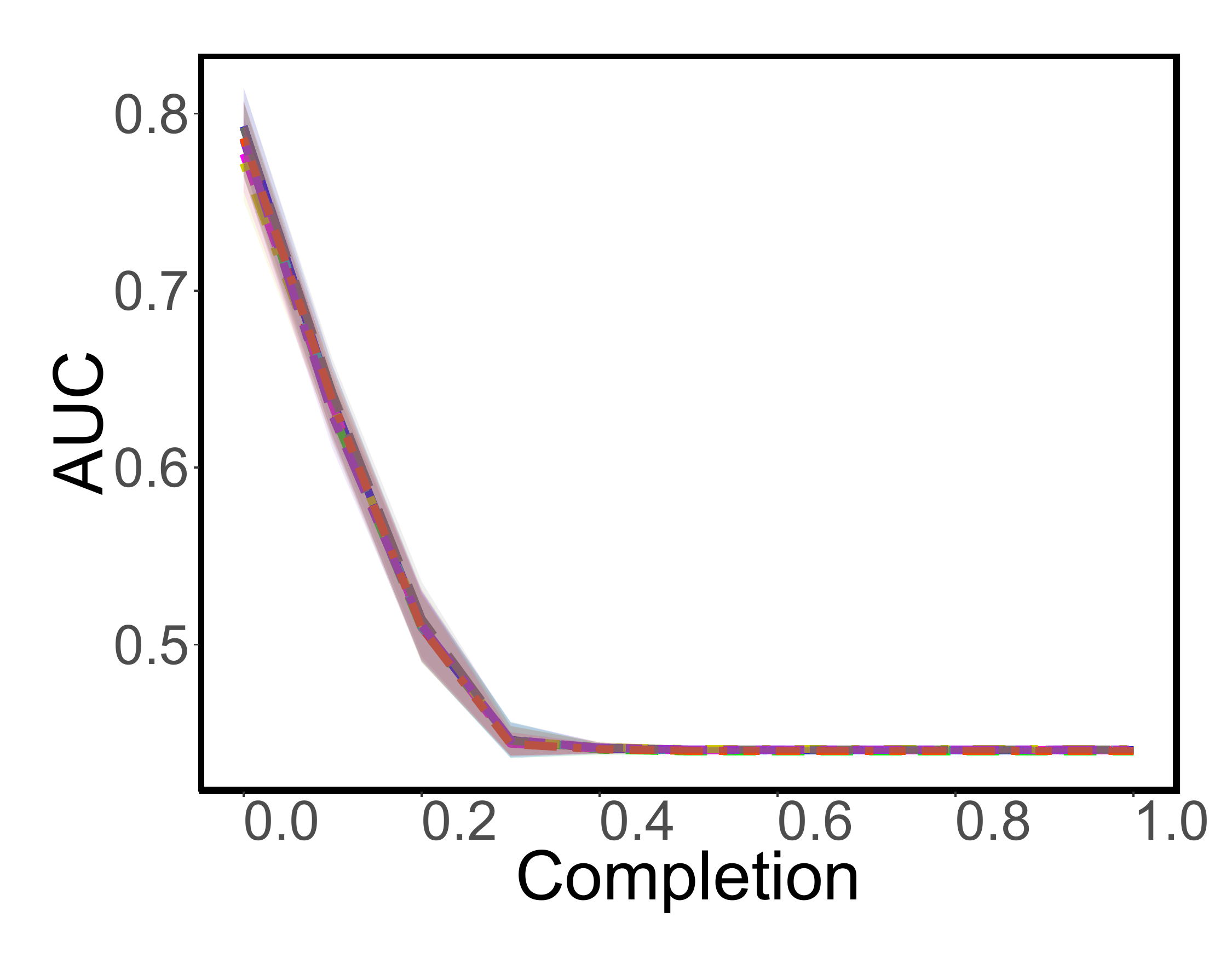}\vspace*{-0.1cm}\\
\rotatebox{90}{\hspace*{0.3cm}\scriptsize ScaleFree$(100,3)$} &
\includegraphics[width=\linewidth]{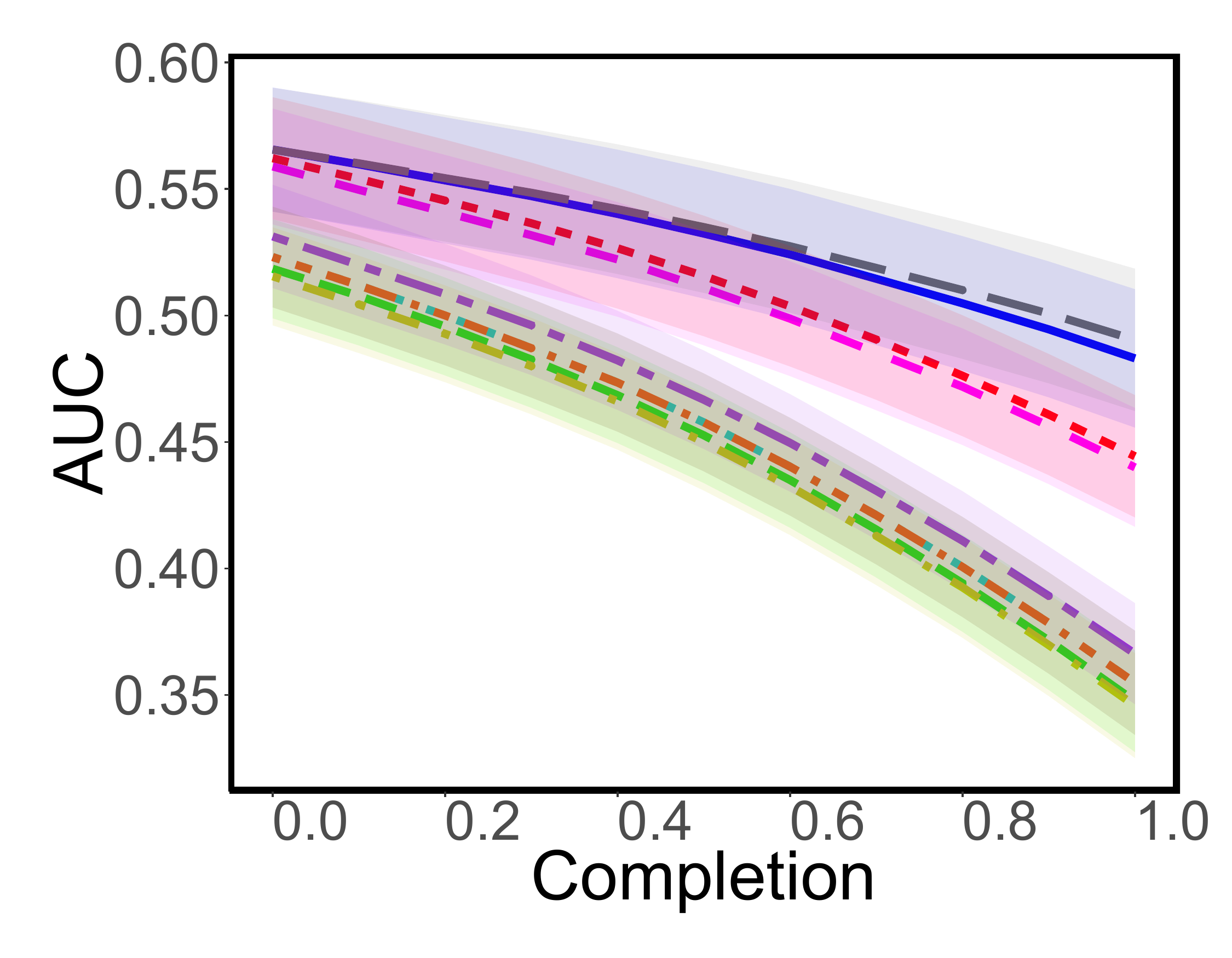}\vspace*{-0.1cm} &
\includegraphics[width=\linewidth]{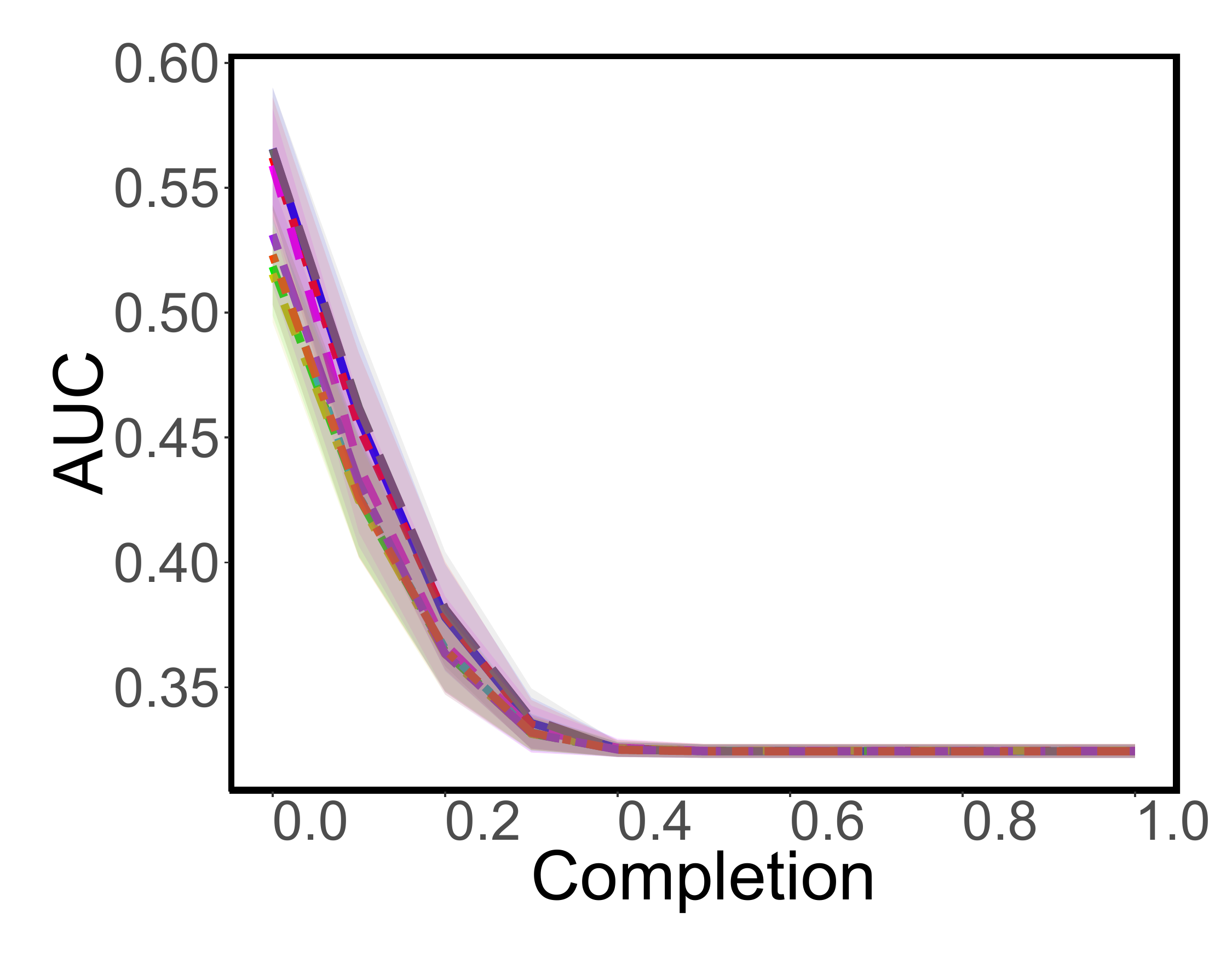}\vspace*{-0.1cm}\\
\rotatebox{90}{\hspace*{0.2cm}\scriptsize Facebook (medium)} &
\includegraphics[width=\linewidth]{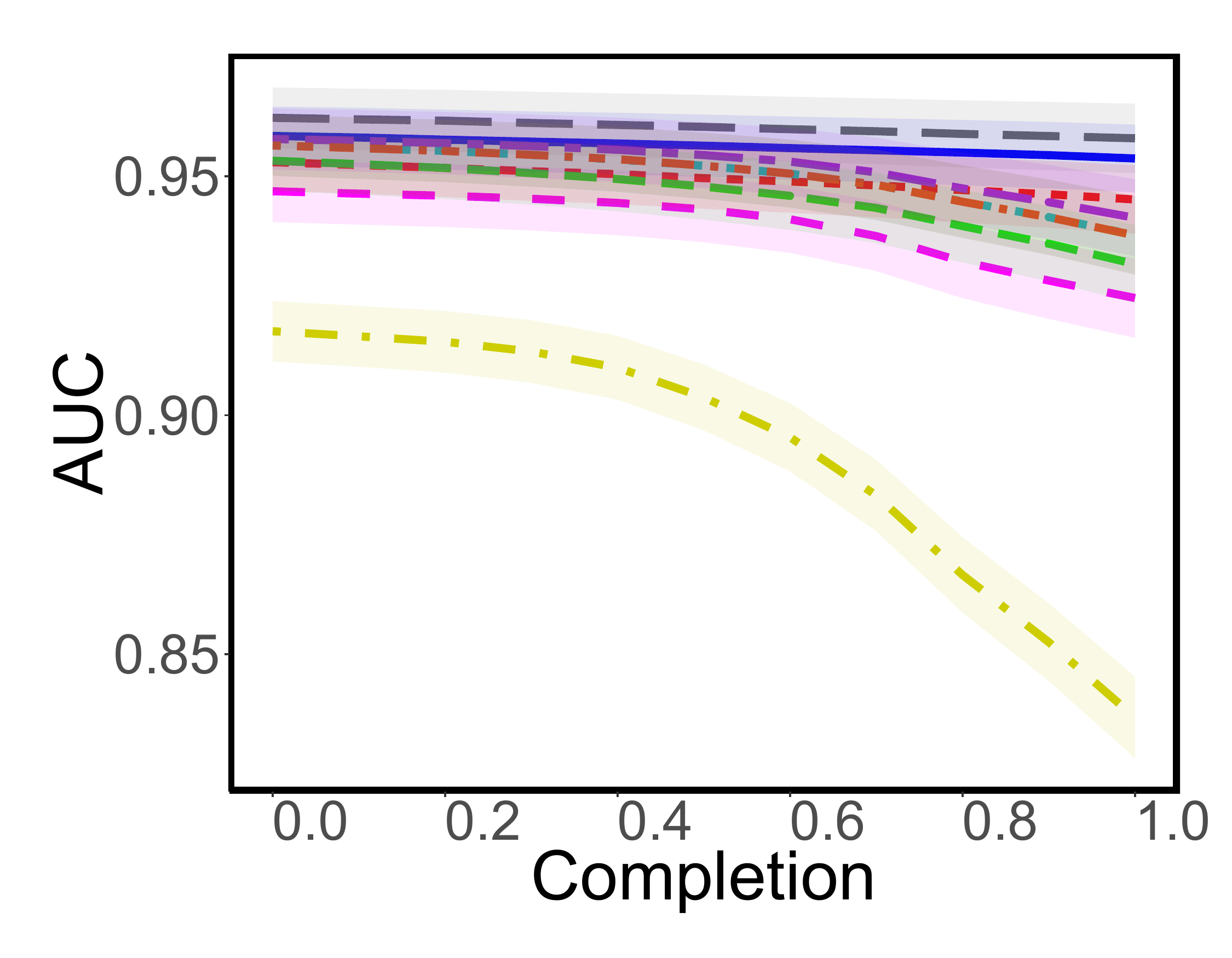} &
\includegraphics[width=\linewidth]{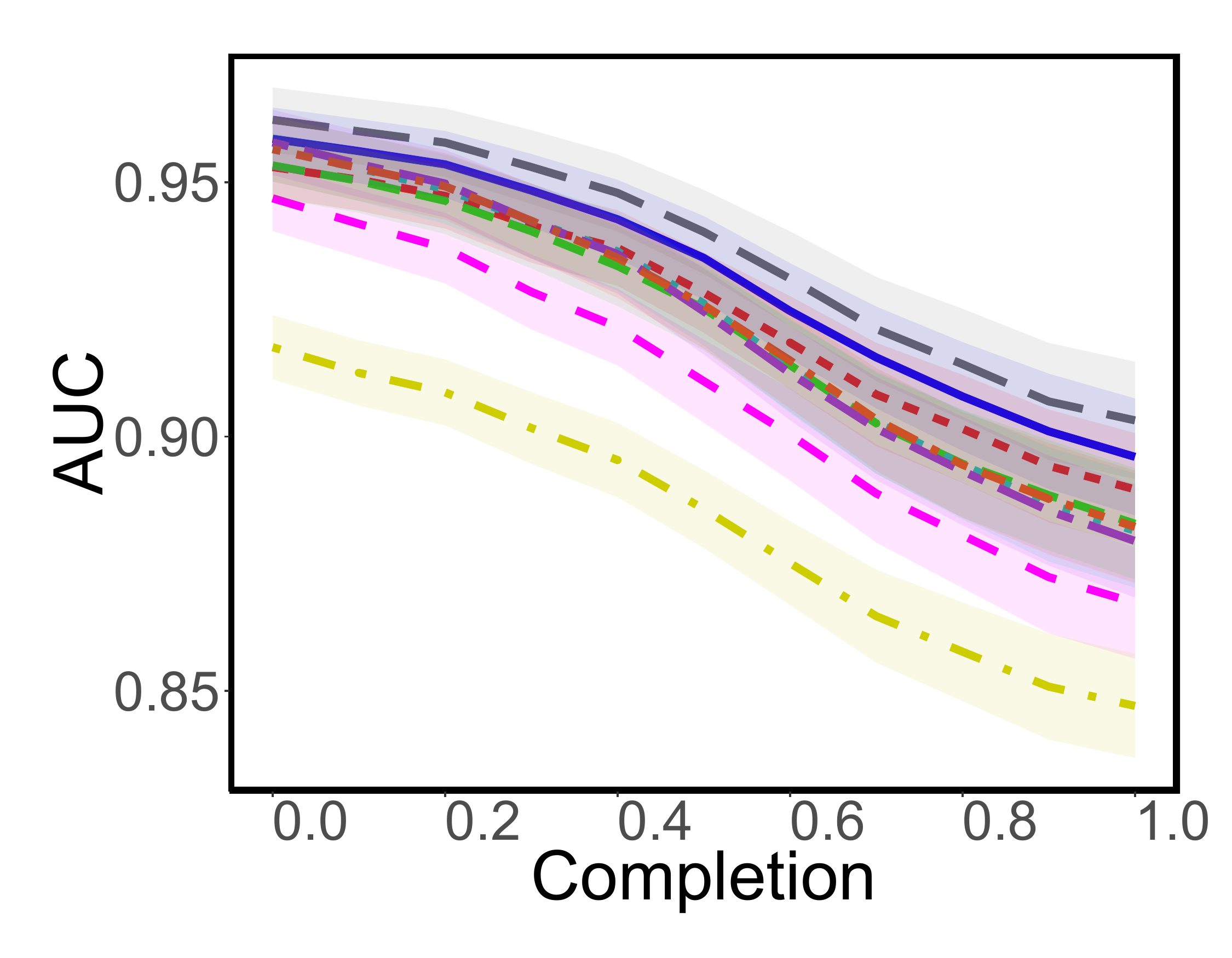}\\
\multicolumn{3}{c}{\includegraphics[width=\linewidth]{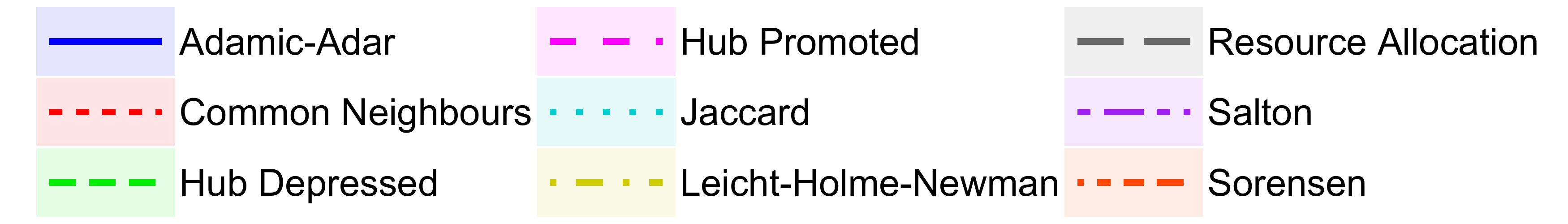}}
\end{tabular}
\caption{For each of the similarity indices outlined in Theorem~\ref{theorem:NPcompleteness}, the figure depicts $\ROC$ during the execution of OTC and CTR given different networks, where $|\Hide|=\max(10,|E|/100)$ and $b=4|\Hide|$, and the links in $\Hide$ are chosen at random. Results are taken as the average over $50$ executions, with coloured areas representing the $95\%$ confidence intervals.}
\label{fig:evaluatingOurHeuristics}
\end{figure}

A typical and intuitive way to evaluate a similarity index is as follows. First, the links of the network are divided into a training set, $T$, and a probe set, $Q$. The index trains on $T$ and assigns a similarity score to every pair of nodes accordingly. Those scores are then evaluated based on the \emph{area under the ROC curve} ($\ROC$) \cite{fawcett2006introduction}, which can be interpreted as the probability that the index assigns a greater score to a random link in $Q$ than to a random non-edge; see Section~S2 for more details. With this in mind, we evaluate the effectiveness of each heuristic against a similarity index in a given network as follows: we run the heuristic iteratively, and after each iteration, we compute $\ROC$ given a training set consisting of every link in the network and a probe set consisting of every link in $H$; this way we can assess \textit{the probability that the index assigns a greater score to a random link in $H$ than to a random non-edge}.
Figure~\ref{fig:evaluatingOurHeuristics} depicts the results in three networks given the similarity indices outlined in Theorem~\ref{theorem:NPcompleteness}. As can be seen, both heuristics are able to reduce $\ROC$ and thus hide the links in $\Hide$, although CTR seems more effective than OTC, suggesting that in order to hide a relationship, ``unfriending'' carefully-chosen individuals can provide a better disguise than befriending new ones. Similar trends where observed when replacing $\ROC$ with a different performance metric---the \emph{average precision} ($\AP$) \cite{boyd2013area}---and when experimenting with other networks and similarity indices; see Section~S8.

\begin{figure}[htbp]
\centering
\setlength\tabcolsep{1pt}
\renewcommand{\arraystretch}{0.01}
\begin{tabular}{m{.03\linewidth}m{.35\linewidth}m{.35\linewidth}}
\rotatebox{90}{\scriptsize OTC-$\ROC$} &
\includegraphics[width=\linewidth]{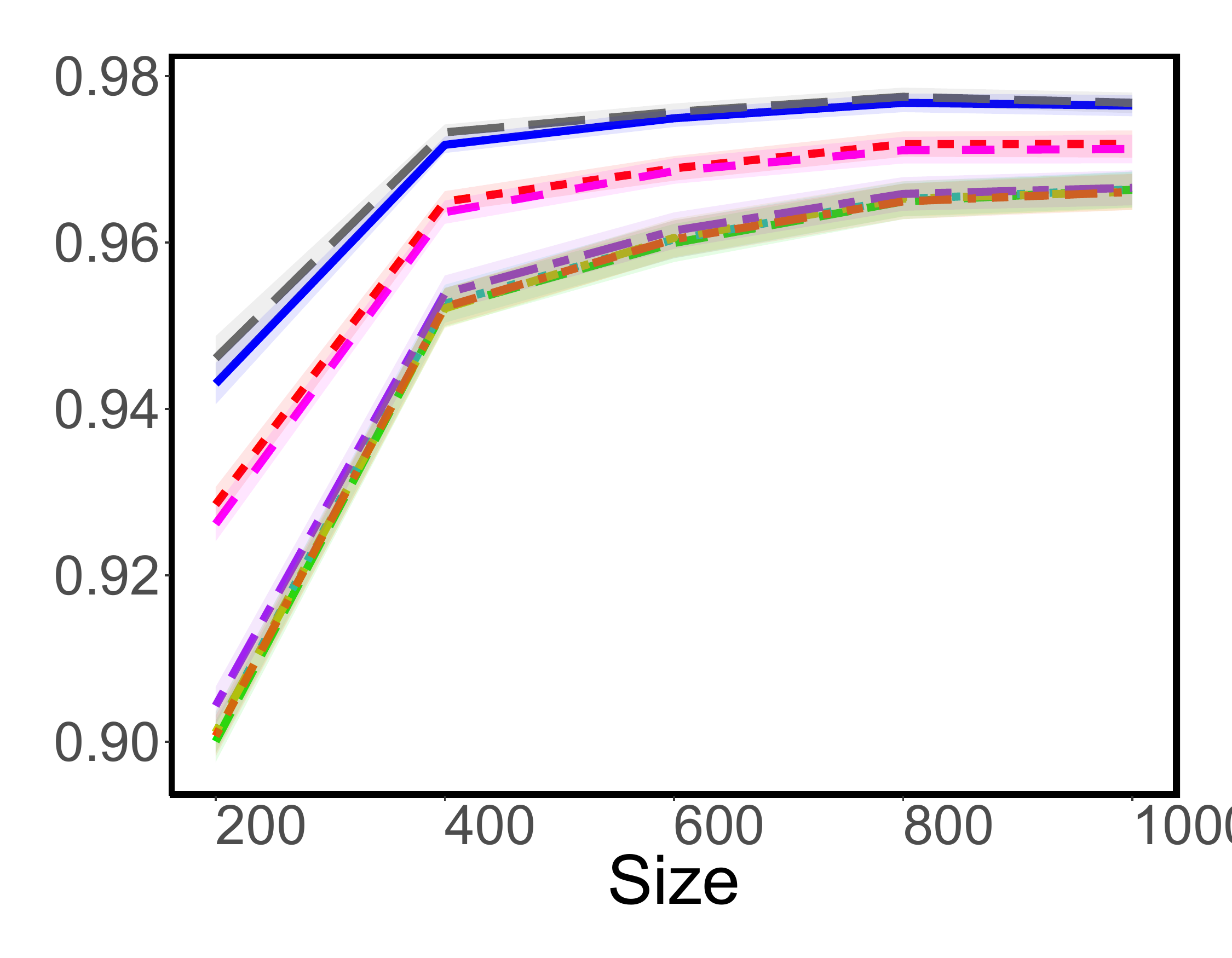} &
\includegraphics[width=\linewidth]{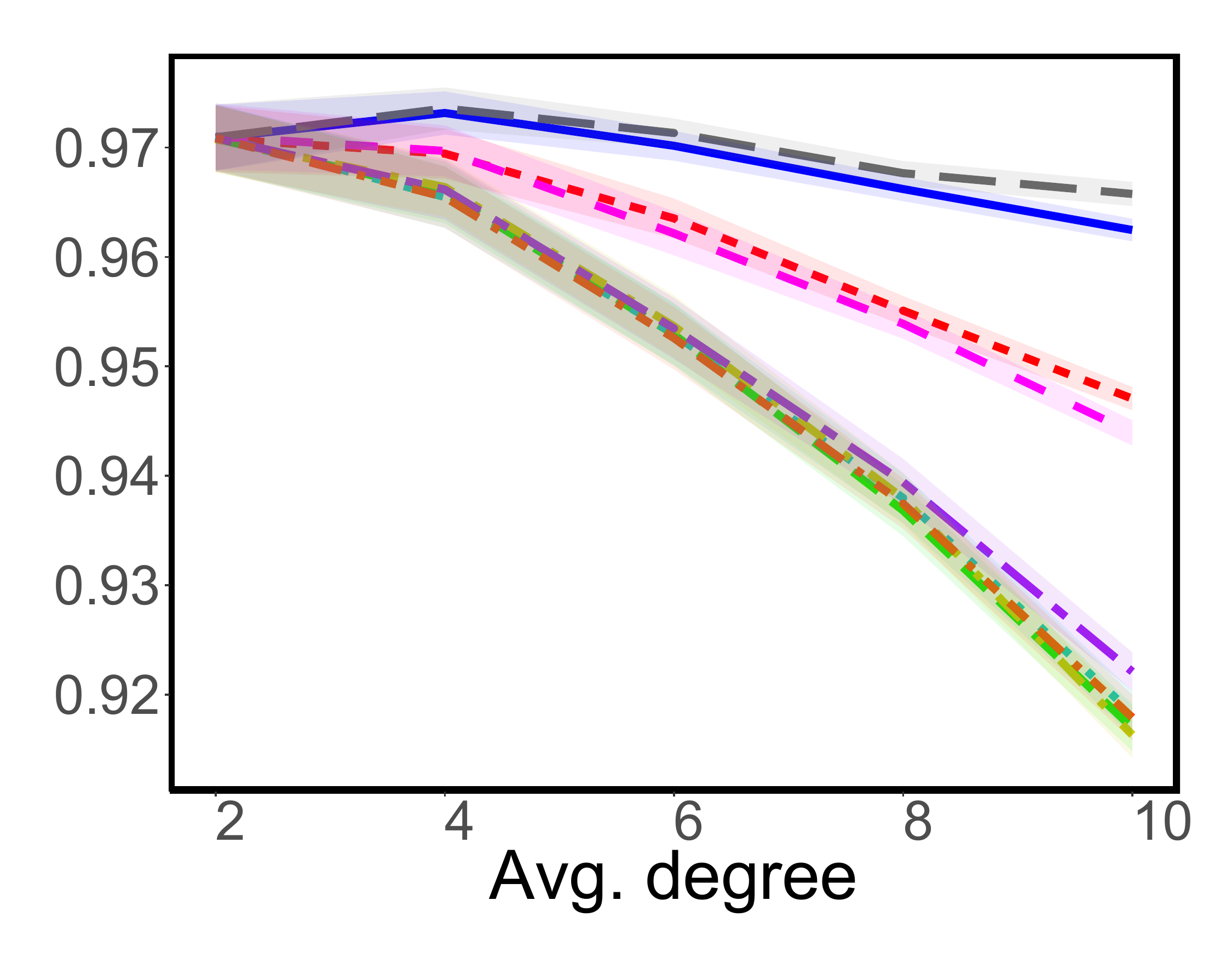} \\
\rotatebox{90}{\scriptsize CTR-$\ROC$} &
\includegraphics[width=\linewidth]{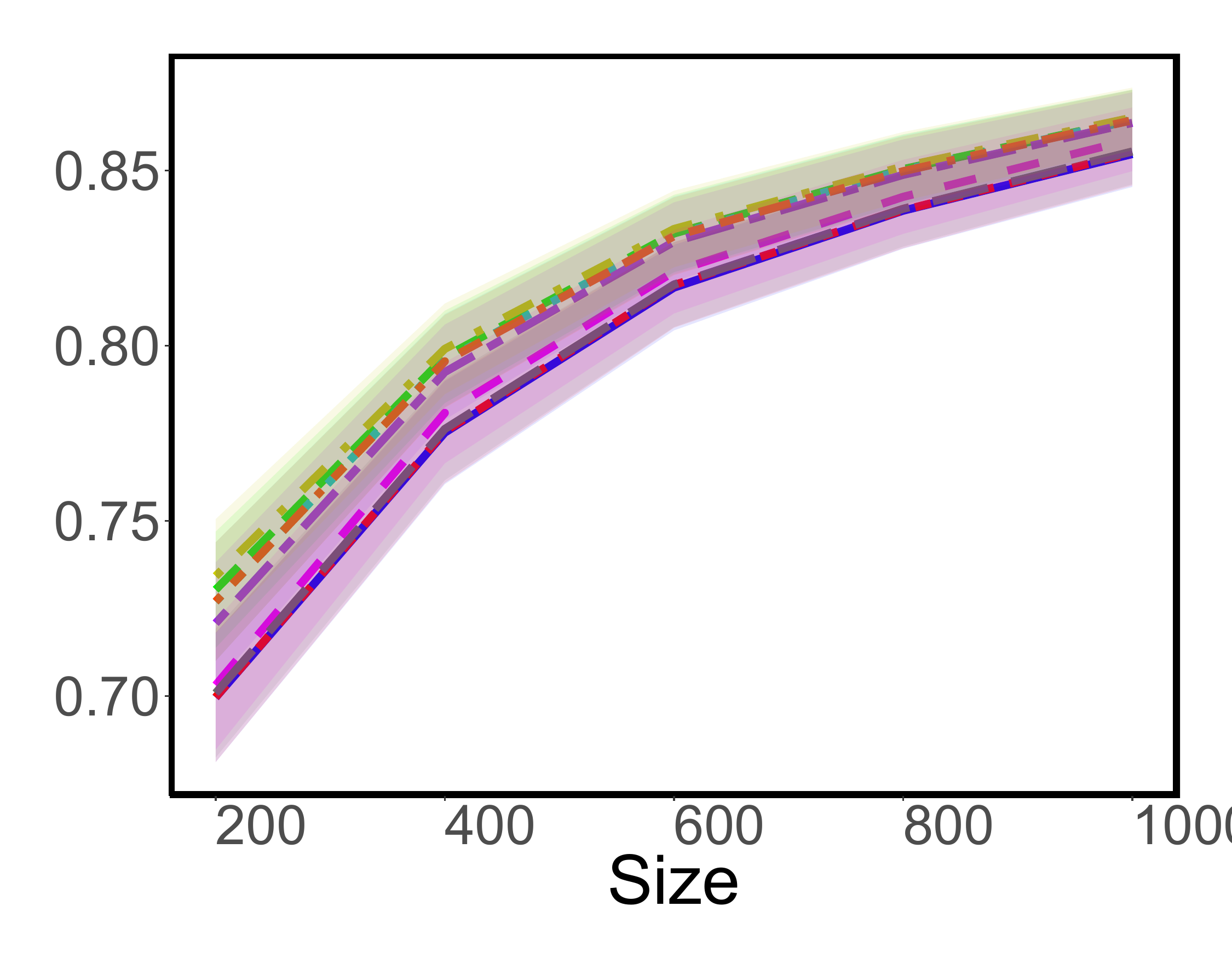} &
\includegraphics[width=\linewidth]{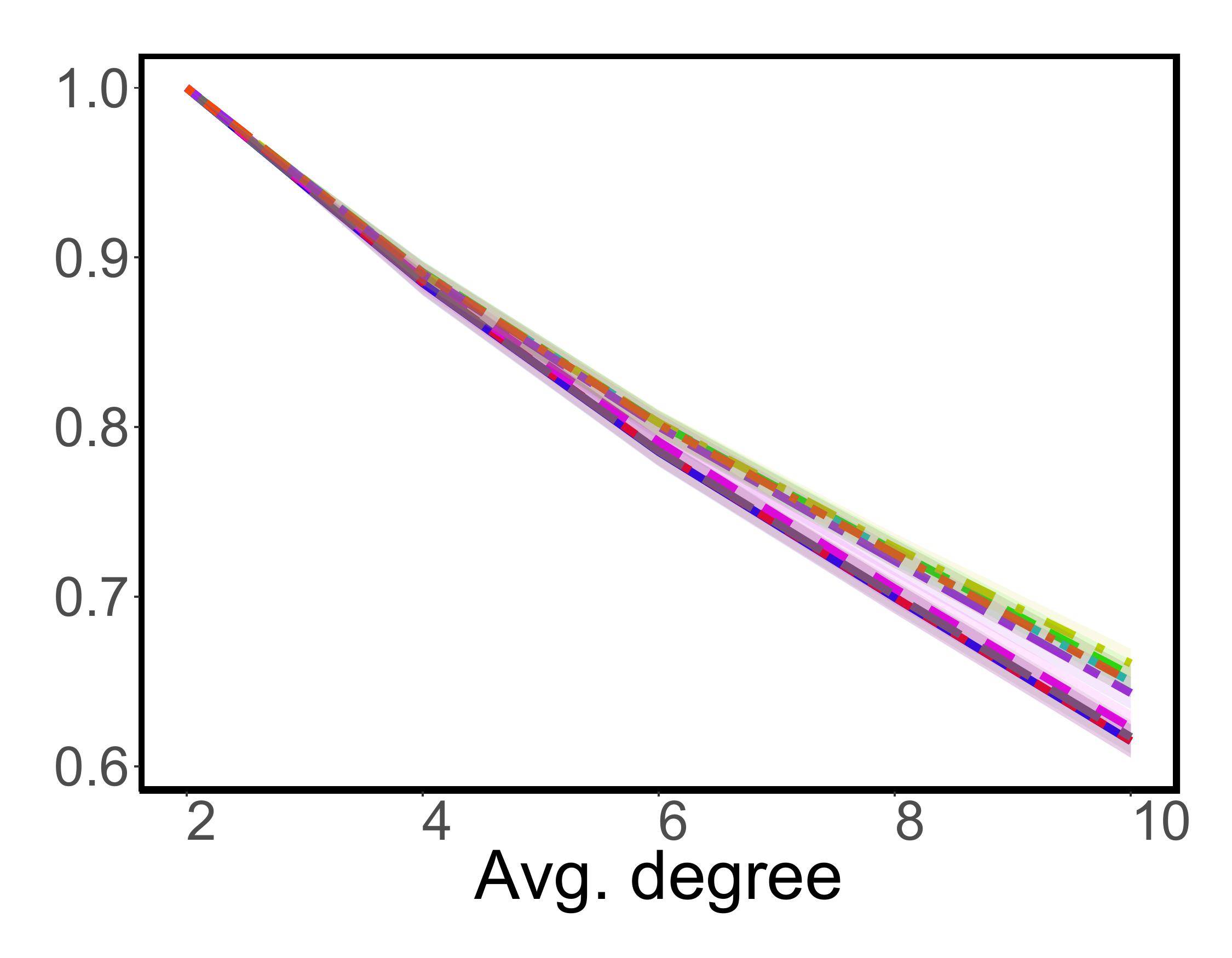} \\
\rotatebox{90}{\scriptsize OTC-$\AP$} &
\includegraphics[width=\linewidth]{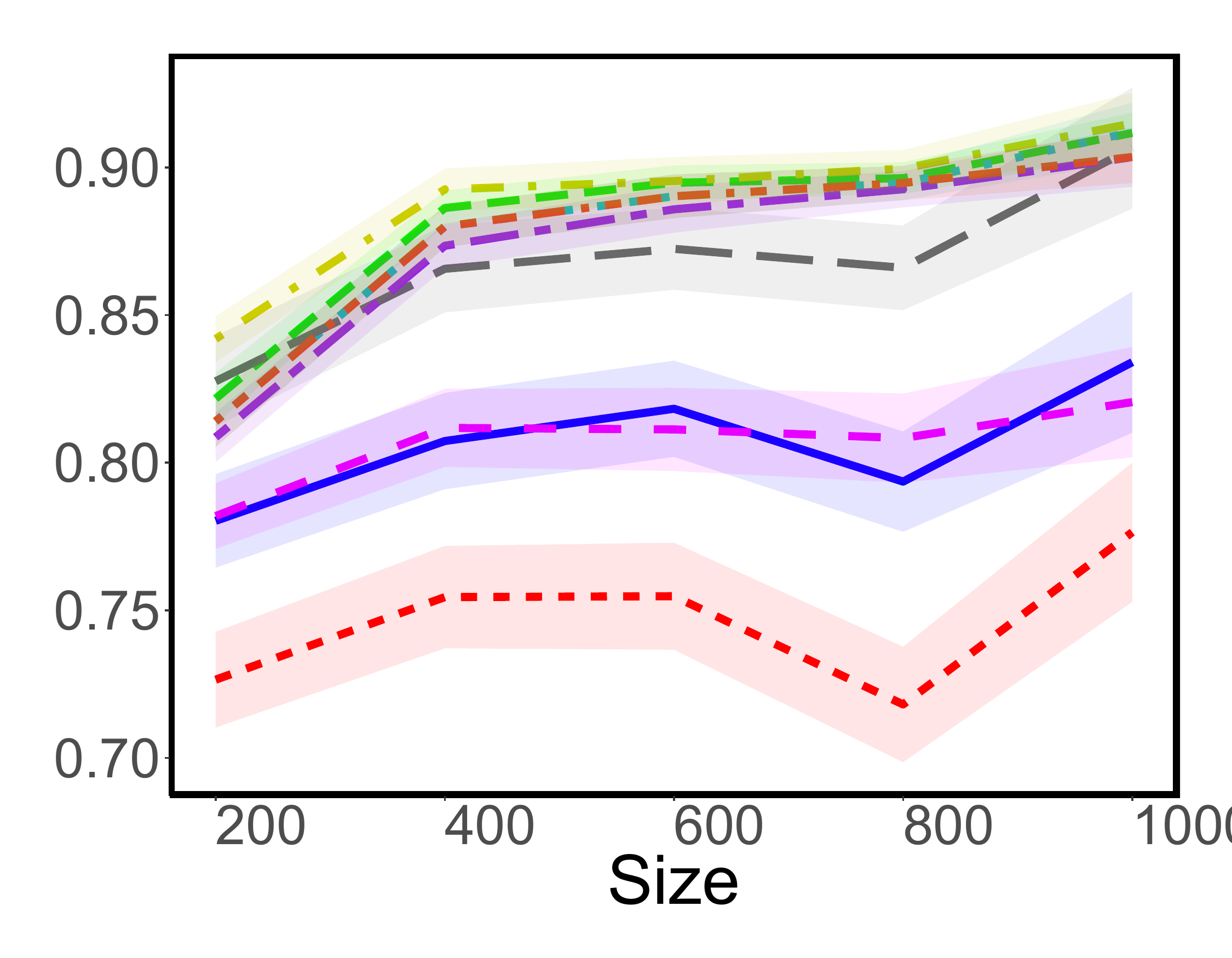} &
\includegraphics[width=\linewidth]{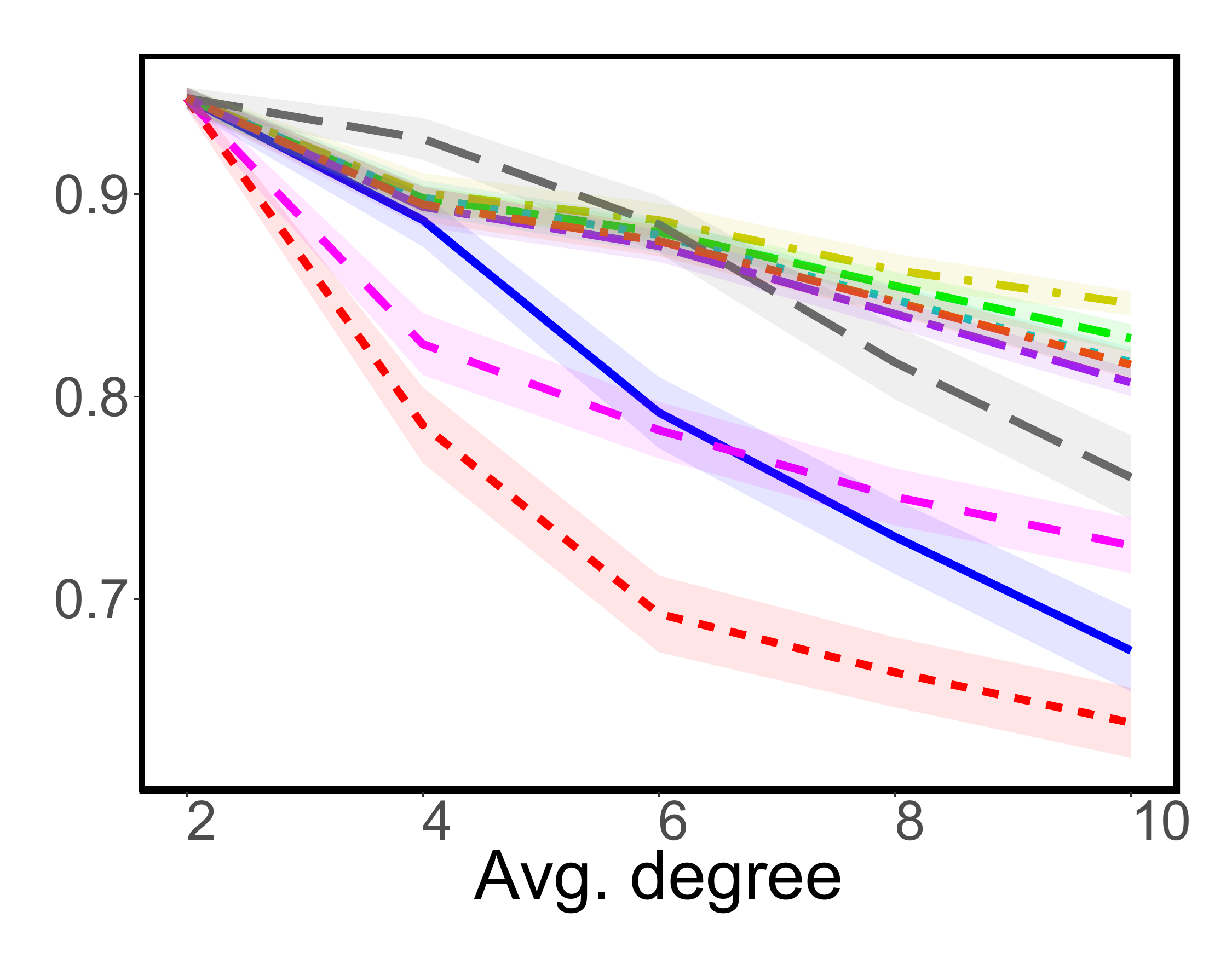} \\
\rotatebox{90}{\scriptsize CTR-$\AP$} &
\includegraphics[width=\linewidth]{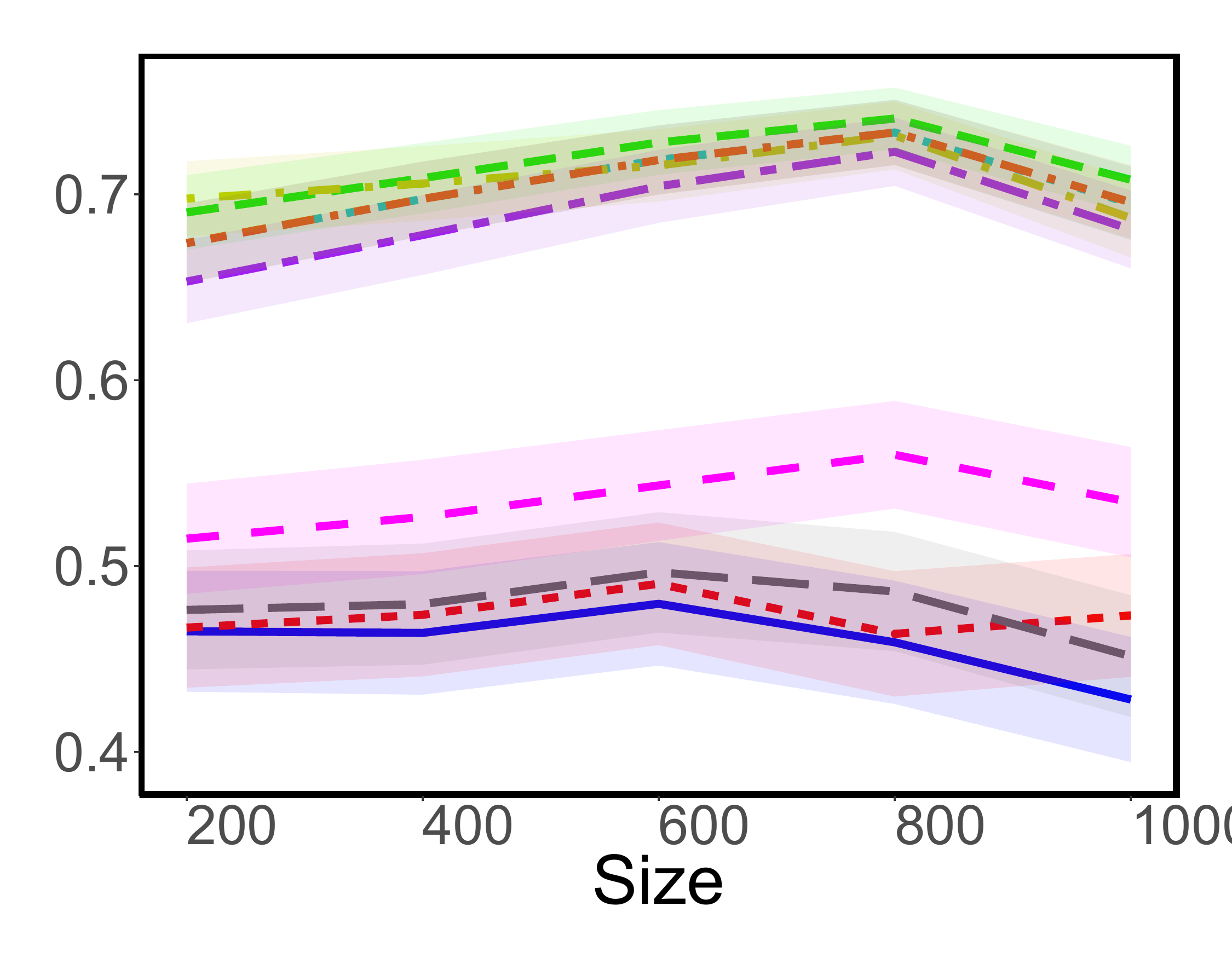} &
\includegraphics[width=\linewidth]{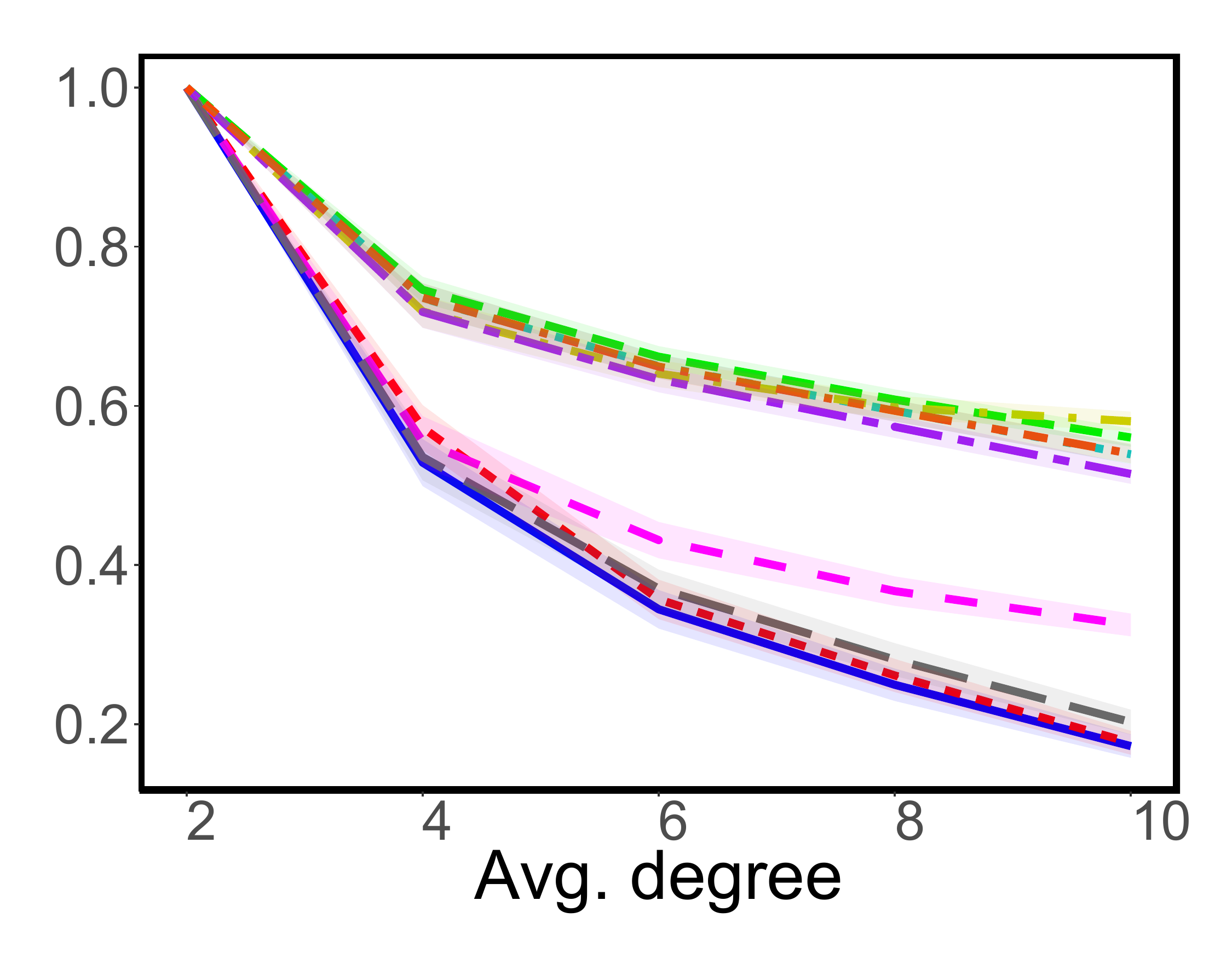} \\
\multicolumn{3}{c}{\includegraphics[width=\linewidth]{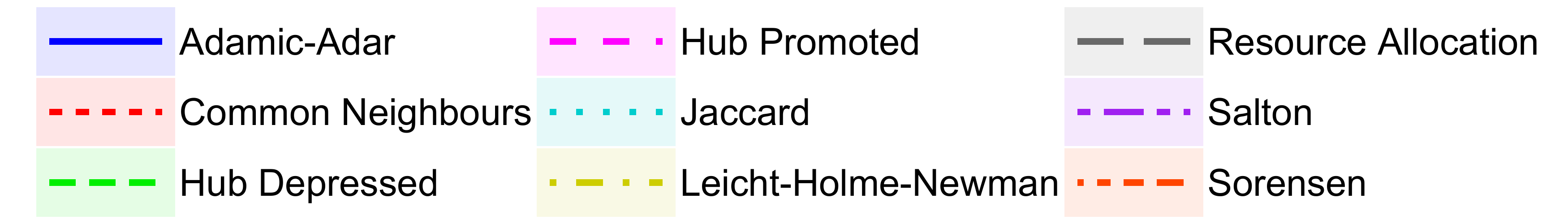}}
\end{tabular}
\caption{Evaluating the attack tolerance of the similarity indices outlined in Theorem~\ref{theorem:NPcompleteness} against OTC (which adds edges) and CTR (which removes edges) in scale-free networks while varying the number of nodes, $n=200,400,\ldots, 1000$, and the average degree, $d=2,4,\ldots, 10$. Here, attack tolerance is measured in terms of the relative change in $\ROC$ and $\AP$. For each $n$, we report the average over different values of $d$. Likewise, for each $d$, we report the average over different values of $n$. The links in $\Hide$ are chosen at random, where $|\Hide|=100$ and $b=4|\Hide|$. Each experiment is repeated $50$ times, and the figure depicts the average result with coloured areas representing the $95\%$ confidence intervals.
}
\label{fig:attack:tolerance}
\end{figure}

Next, we evaluate the attack tolerance of the similarity indices outlined in Theorem~\ref{theorem:NPcompleteness} based on two performance metrics---$\ROC$ and $\AP$---while varying the number of nodes, $n$, and the average degree, $d$, in scale-free networks; see Figure~\ref{fig:attack:tolerance}. Overall, the attack tolerance of those similarity indices tends to increase with $n$ (especially in terms of $\ROC$) and decreases with $d$ (especially when facing CTR). Similar trends were observed when experimenting with Small-World networks and Erdos-Renyi random graphs; see Section~S8.3.

\begin{figure}[tbhp]
\centering
\setlength\tabcolsep{1pt}
\renewcommand{\arraystretch}{0.01}
\begin{tabular}{m{.03\linewidth}m{.315\linewidth}m{.315\linewidth}m{.315\linewidth}}
&
\multicolumn{1}{c}{\footnotesize OTC} &
\multicolumn{1}{c}{\hspace*{0.3cm}\footnotesize OTC \& CTR} &
\multicolumn{1}{c}{\footnotesize CTR} \\
\rotatebox{90}{\footnotesize $\ROC$} &
\includegraphics[width=1.03\linewidth]{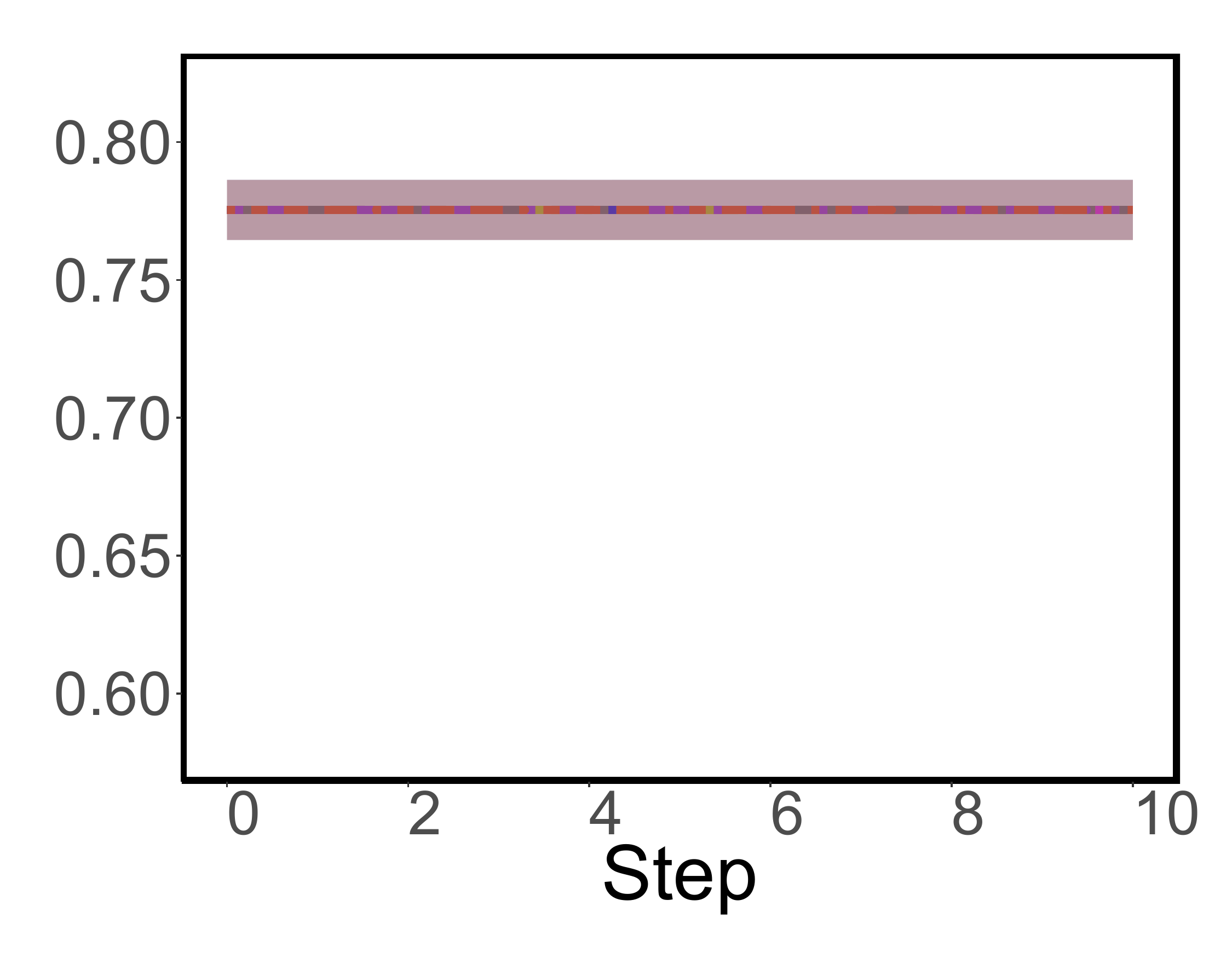} &
\includegraphics[width=1.03\linewidth]{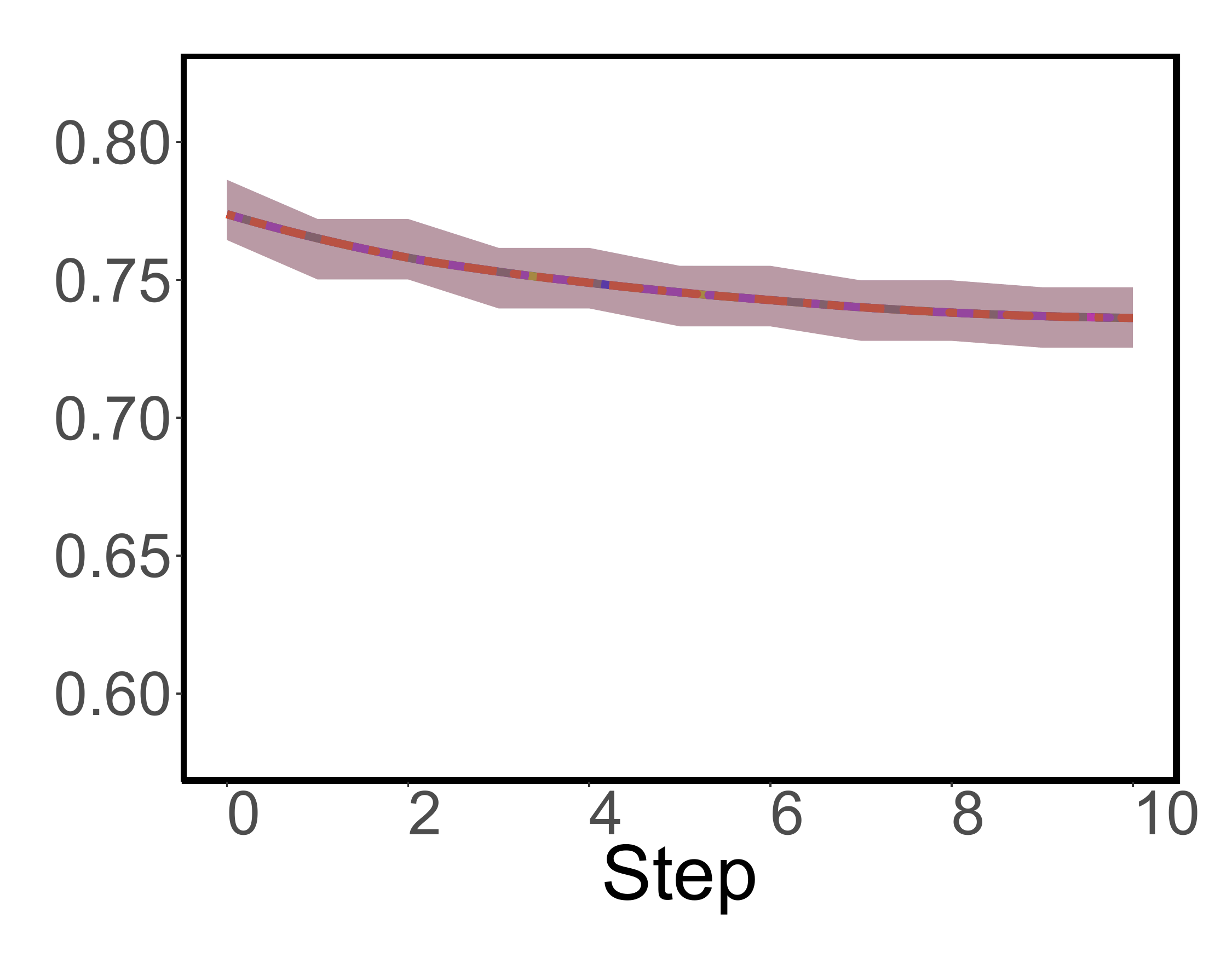} &
\includegraphics[width=1.03\linewidth]{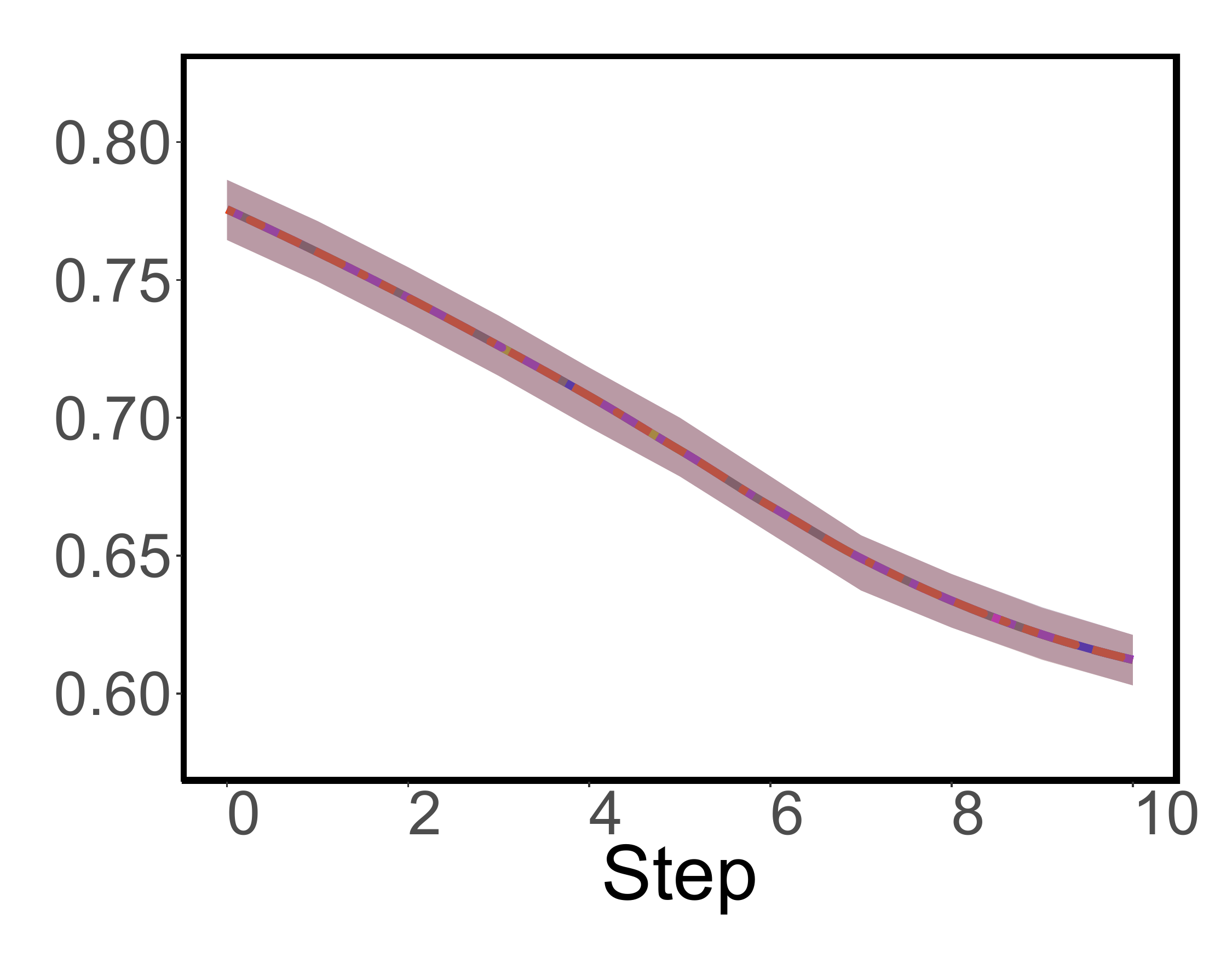} \\
\rotatebox{90}{\footnotesize $\AP$} &
\includegraphics[width=1.03\linewidth]{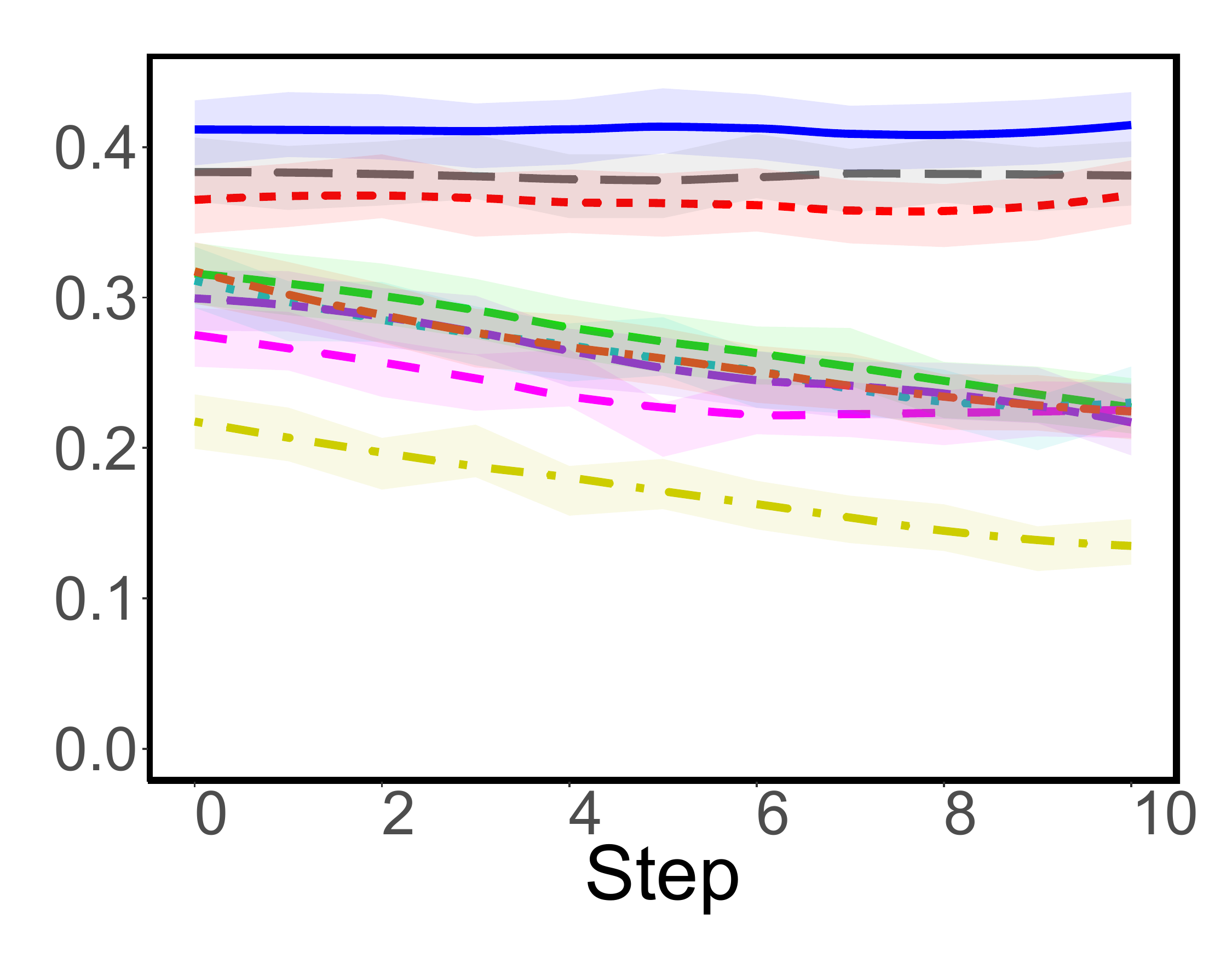} &
\includegraphics[width=1.03\linewidth]{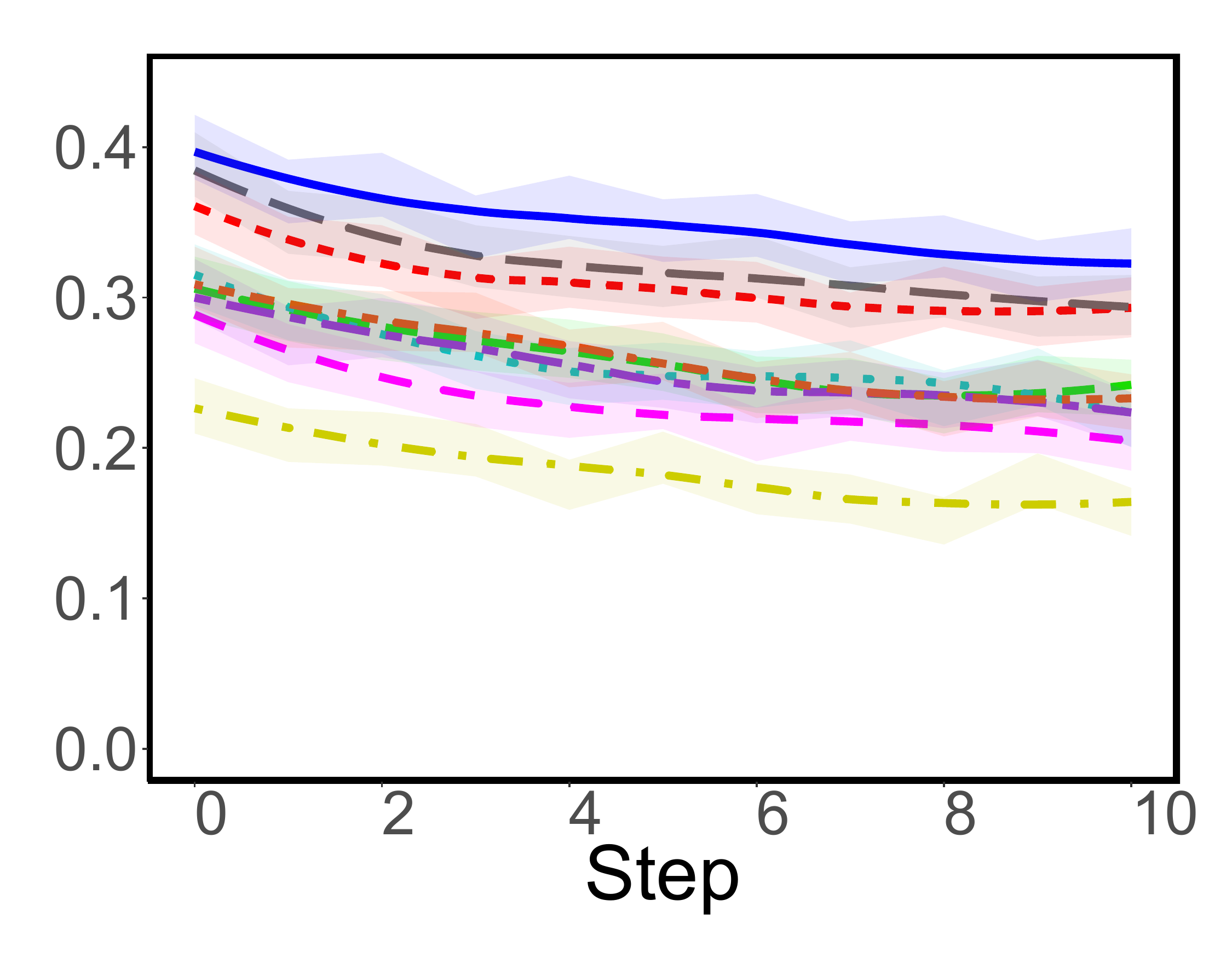} &
\includegraphics[width=1.03\linewidth]{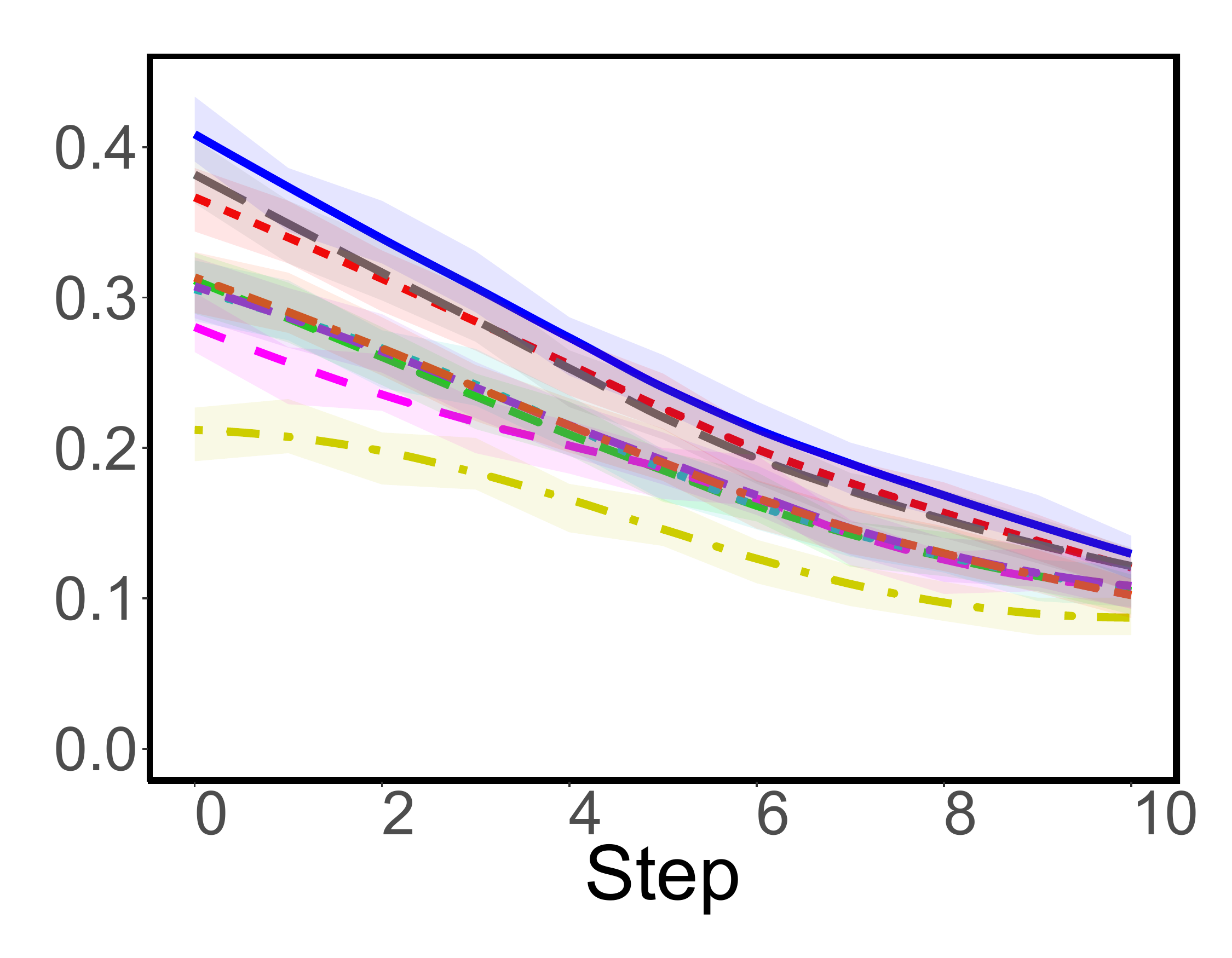} \\
\multicolumn{4}{c}{\includegraphics[width=\linewidth]{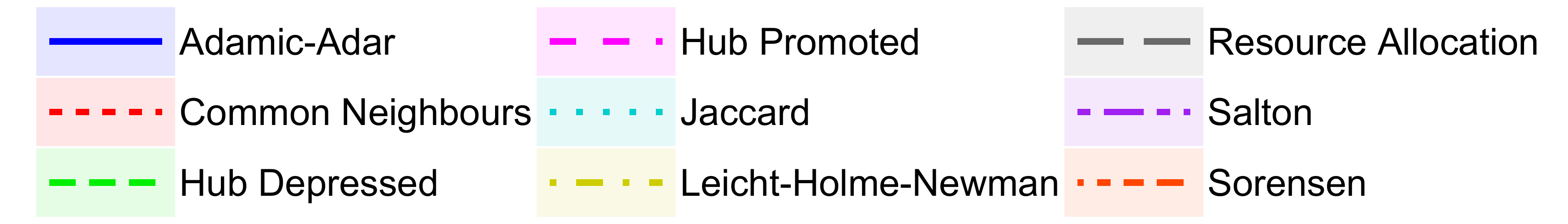}}
\end{tabular}
\caption{Given the similarity indices outlined in Theorem~\ref{theorem:NPcompleteness} and a telecommunication network consisting of $248,763$ nodes and $829,725$ edges, the figure depicts the average $\ROC$ and $\AP$ during the execution of OTC and CTR given a budget $b=10$, where $H$ contains just a single link. More specifically, for each similarity index, we consider the $1,000$ highest-ranked links, and for each such link, $(v,u)$, we run the heuristic once where the evader is $v$ and another where the evader is $u$. This entire process is repeated 10 times, and the average results are reported with the coloured areas representing the $95\%$ confidence intervals.
}
\label{fig:single-telecommunication-large}
\end{figure}

Finally, we consider a practical scenario that could be faced by any individual whose goal is to hide just a single relationship using only 10 modifications in a massive telecommunication network. To this end, we consider a network consisting of all $829,725$ phone calls between the $248,763$ users of a particular European telecom operator who live in 4 geographically continuous districts \cite{miritello2013limited}. Figure~\ref{fig:single-telecommunication-large} depicts the results for OTC (which adds edges) and CTR (which removes edges), and also shows what happens when the budget is split between the two heuristics (by alternating between adding and removing edges). As can be seen, CTR is effective in terms of both $\AP$ and $\ROC$. In contrast, OTC is less effective in terms of $\AP$, and not effective at all in terms of $\ROC$. Mixing the two heuristics does not seem to produce any synergistic effects. Similar trends were observed when considering just 1 instead of 4 districts; see Section~S8.4.

\section*{Materials and Methods}

\subsection*{Analyzing CTR and OTC}
\label{sec:analyzingCTRandOTC}

Let $N_G(v)$ denote the set of neighbours of node $v$, i.e., $N_G(v) = \{w \in V : (v,w) \in E\}$, and let $N_G(v,w)$ denote the set of common neighbours of $v$ and $w$, i.e., $N_G(v,w)= N_G(v) \cap N_G(w)$. The \textit{degree} of $v$ will be denoted by $d_G(v)$, i.e., $d_G(v)=|N_G(v)|$. Whenever it is clear from the context, we will omit the graph subscript, e.g., by writing $N(v)$ instead of $N_G(v)$.
Now, let $\mathcal{S}$ denote the set of all the similarity indices outlined in Theorem~\ref{theorem:NPcompleteness}; the formula for each of these indices is specified in Section~S1. Looking at these formulae, one can see that the similarity score of every non-edge, $(x,w)\in\ER$, depends solely on (some of) the following factors:

\begin{itemize}
\item \textit{Factor 1:} \textit{the number of common neighbours of the non-edge}. More specifically, for every $s\in\mathcal{S}$, the score $s(x,w)$ \emph{increases} with $|N(x,w)|$.

\item \textit{Factor 2:} \textit{the degree of each end of the non-edge, but only if both ends have some common neighbours}. Specifically, for every similarity index, $s\in\mathcal{S}\setminus\{\sCN, \sAA, \sRA\}$, the score $s(x,w)$ \emph{decreases} with $d(x)$ and with $d(w)$ if $N(x,w)\neq \emptyset$.\footnote{\footnotesize The \emph{Jaccard} index \cite{jaccard1901etude}, $\sJac$, is affected by $d(x)$ and $d(w)$, since: $|N(x) \cup N(w)| = d(x)+d(w)-|N(x,w)|$.} Otherwise, if $N(x,w)=\emptyset$, then $s(x,w)$ is not affected by $d(x)$ nor by $d(w)$. As for the remaining similarity indices, i.e., those in $\{\sCN, \sAA, \sRA\}$, their scores are not affected by $d(x)$ nor by $d(w)$, regardless of whether $N(x,w)=\emptyset$.

\item \textit{Factor 3:} \textit{the degree of every common neighbour of the non-edge}. More specifically, for every similarity index $s\in\{\sAA,\sRA\}$ and every common neighbour $v\in N(x,w)$, the score $s(x,w)$ \emph{decreases} with $d(v)$. As for the remaining similarity indices, i.e., those in $\mathcal{S}\setminus\{\sAA,\sRA\}$, their scores are not affected by any $d(v):v\in N(x,w)$.
\end{itemize}

Therefore, the \emph{addition} of an edge, $(v,w)$, can only affect the scores of the following types of non-edges:

\begin{itemize}
\item \textit{Type 1:} $(x,w):x\in N(v)\setminus N(w)$.
Such a non-edge is affected by the addition of $(v,w)$, which adds $v$ to $N(x,w)$, thereby increasing $|N(x,w)|$. This, in turn, \emph{increases} $s(x,w)$ for every similarity index $s\in\mathcal{S}$; see \textit{Factor~1}.

\item \textit{Type 2:} $(x,w):N(x,w)\neq \emptyset$.
Such a non-edge is affected by the addition of $(v,w)$, which increases $d(w)$. This, in turn, \emph{decreases} $s(x,w)$ for every $s\in\mathcal{S}\setminus\{\sCN, \sAA, \sRA\}$; see \textit{Factor~2}.

\item \textit{Type 3:} $(x,y):x,y\in N(w)$.
Such a non-edge is affected by the addition of $(v,w)$, which increases the degree of a common neighbour of $x$ and $y$, namely $w$. This, in turn, \emph{decreases} $s(x,y)$ for every $s\in\{\sAA,\sRA\}$; see \textit{Factor~3}.
\end{itemize}

Note that a non-edge $(x,v)$ can be of both \textit{Type~1} and \textit{Type~2} simultaneously; this happens when $x\in N(w)\setminus N(v)$ and $N(x,v)\neq \emptyset$. In this case, $(x,v)$ is affected by \textit{Factor~1}---which increases $s(x,v)$---as well as \textit{Factor~2}---which decreases $s(x,v)$. Since these two factor have opposite effects, whether $s(x,v)$ increases depends on whether the effect of \textit{Factor~1} outweighs that of \textit{Factor~2}.

Finally, note that the impact of removing $(v,w)$ is exactly the opposite to that of adding $(v,w)$. For instance, suppose that $(v,x)$ is a non-edge of Type~1 and not of Type~2. Then, by adding $(v,w)$ to a network $(V,E):(v,w)\not\in E$, we \textit{increase} $s(v,x)$ for every $s\in\mathcal{S}$. In contrast, by removing $(v,w)$ from a network $(V,E):(v,w)\in E$, we \textit{decrease} $s(v,x)$.

With these observations in mind, let us analyse our heuristics, starting with CTR. Recall that this heuristic removes an edge, $(v,w)\in E$, where:
$$
\exists x\in V: \big((v,x)\in E\big) \wedge \big((x,w)\in\Hide\big).
$$
Importantly, by removing $(v,w)$:

\begin{itemize}
\item the node $v$ is removed from the common neighbours of $w$ and $x$, thereby reducing $|N(x,w)|$. As a result, the similarity score of $(x,w)$ \textit{decreases} according to \emph{Factor~1}.

\item the degree of node $w$ decreases. As a result, the similarity score of $(x,w)$ can only \textit{increase} according to \emph{Factor~2}.
\end{itemize}

To put it differently, by removing $(v,w)$, the similarity score of $(x,w)$ is subjected to two opposing forces; one that decreases it, and another that increases it, Nevertheless, the following theorem implies that the latter force never outweighs the former one. In other words, by removing $(v,w)$, the similarity score of $(x,w)$ can only \textit{decrease} given the similarity indices in $\mathcal{S}$; see the proof in Section~S4.

\begin{theorem}\label{theorem:comparingPositiveAndNegativeEffects}
Let $G'=(V,E')$ be a network, and let $(x,w)$ be a non-edge in $G'$. Furthermore, let $v$ be a node in $G'$ such that $v\in N_{G'}(x)$ and $v\not\in N_{G'}(w)$. Finally, let $G$ be the network that results from adding $(v,w)$ to $G'$, i.e., $G=(V,E)$ where $E=E'\cup\{(v,w)\}$. Then, for every similarity index, $s\in\mathcal{S}$, we have:
$$
s_{G'}(x,w) \leq s_{G}(x,w) \smallskip\smallskip
$$
\end{theorem}

Moving on to OTC, recall that this heuristic adds to the network a non-edge $(v,w)$ such that, \textit{after the addition of} $(v,w)$:
\begin{itemize}\itemsep-0.25em
\item $\exists u\in V\!:\!(w,u)\in \Hide$;
\item $\exists x\in V\!:\!\big(\{(x,v),(v,w)\}\subseteq E\big) \wedge \big((x,w)\in\ER\setminus \Hide\big)$.
\end{itemize}

Based on this, by adding $(v,w)$:
\begin{itemize}\itemsep-0.25em
\item the degree of $w$ increases, which can only decrease the similarity score of $(w,u)$ according to \emph{Factor~2}.
\item the similarity scores of $(x,w)$ and $(y,v)$ can only increase according to Theorem~\ref{theorem:comparingPositiveAndNegativeEffects}.
\end{itemize}

Thus, given the similarity indices in $\mathcal{S}$, the addition of $(v,w)$ can only decrease the position of $(w,u)$ in the similarity-based ranking of all non-edges.

\section*{Conclusion}
We studied the attack tolerance of link prediction algorithms when an individual is strategically rewiring the network to hide some of his/her relations. We analyzed the corresponding optimization problem, and showed that an optimal solution is hard to compute. Based on this finding, we focused our attention on developing two heuristics, called OTC (which adds edges) and CTR (which removes edges). Both heuristics can readily be executed by lay people on existing social media platforms, without requiring extensive computational power nor full knowledge of the entire network topology. Our empirical evaluation showed that both heuristics are effective in practice, although CTR seems more effective than OTC, suggesting that in order to hide a relationship, ``unfriending'' carefully-chosen individuals can provide a better disguise than befriending new ones. Next, we evaluated the attack tolerance of various similarity indices while varying the number of nodes, $n$, and the average degree, $d$. We found that the attack tolerance of these indices tends to increase with $n$ and decreases with $d$. Finally, we consider a practical scenario where the goal is to hide a single relation in a massive telecommunication network. In this scenario, we found that OTC has no impact according to a certain performance measure, unlike CTR.

Our study demonstrates the fragility of existing link prediction algorithms in the face of a strategic evader, and highlights the need to develop new algorithms that are harder to fool.

\section*{Acknowledgments}
Marcin Waniek was supported by the Polish National Science Centre grant 2015/17/N/ST6/03686.
Tomasz Michalak was supported by the European Research Council under Advanced Grant 291528 (``RACE'') and by the Polish National Science Centre grant 2014/13/B/ST6/01807.

{\fontsize{8}{8}\selectfont{
\bibliographystyle{abbrv}
\bibliography{bibliography}
}}

\appendix

\onecolumn

\section*{Organization of the Appendix}

This document is structured as follows:
\smallskip\smallskip
\begin{itemize}\itemsep0.5em
\item \textbf{Section~\ref{sec:linkPredictionAlgorithms}} (\emph{page~\pageref{sec:linkPredictionAlgorithms}}) presents link prediction algorithms;
\item \textbf{Section~\ref{sec:evaluationmetrics}} (\emph{page~\pageref{sec:evaluationmetrics}}) presents performance evaluation metrics; 
\item \textbf{Section~\ref{appendix:proof:NPcompleteness}} (\emph{page~\pageref{appendix:proof:NPcompleteness}}) presents the proof of Theorem~\ref{theorem:NPcompleteness};
\item \textbf{Section~\ref{appendix:proof:theorem:comparingPositiveAndNegativeEffects}} (\emph{page~\pageref{appendix:proof:theorem:comparingPositiveAndNegativeEffects}}) presents the proof of Theorem~\ref{theorem:comparingPositiveAndNegativeEffects};
\item \textbf{Section~\ref{sec:CTR:pseudocode}} (\emph{page~\pageref{sec:CTR:pseudocode}}) presents the pseudocode of CTR;
\item \textbf{Section~\ref{sec:OTC:pseudocode}} (\emph{page~\pageref{sec:OTC:pseudocode}}) presents the pseudocode of OTC;
\item \textbf{Section~\ref{sec:WTC}} (\emph{page~\pageref{sec:WTC}}) illustrates the workings of CTR on the 9/11 terrorist network;
\item \textbf{Section~\ref{sec:allFigures}} (\emph{page~\pageref{sec:allFigures}}) evaluates the \textit{effectiveness} of CTR and OTC. More specifically:
\begin{itemize}
\item \textbf{Section~\ref{sec:networks}} (\emph{page~\pageref{sec:networks}}) describes the networks considered in our experiments;
\item \textbf{Section~\ref{sec:supplementary:local:evaluation}} (\emph{page~\pageref{sec:supplementary:local:evaluation}}) evaluates our heuristics against \emph{local} similarity indices;
\item \textbf{Section~\ref{sec:supplementary:global:evaluation}} (\emph{page~\pageref{sec:supplementary:global:evaluation}}) evaluates our heuristics against \emph{global} similarity indices;
\item \textbf{Section~\ref{sec:supplementary:Telecommunication}} (\emph{page~\pageref{sec:supplementary:Telecommunication}}) presents results for a practical telecommunications scenario;
\end{itemize}
\item \textbf{Section~\ref{sec:runtime}} (\emph{page~\pageref{sec:runtime}}) empirically evaluates the \textit{runtime} of both heuristics;
\item \textbf{Section~\ref{sec:evaluatingAttackTolerance}} (\emph{page~\pageref{sec:evaluatingAttackTolerance}}) evaluates the attack tolerance of different similarity indices.
\end{itemize}

\newpage


\section{Link Prediction Algorithms}\label{sec:linkPredictionAlgorithms}

\noindent For any network, and any pair of nodes that are not connected in that network, a link prediction algorithm estimates the likelihood that there exists a not-yet-discovered edge between those two nodes, or that an edge will form between the two nodes in the future~\cite{getoor2005link}.
Many link prediction algorithms are based on \textit{similarity indices}, also known as \textit{kernels}~\cite{shawe2004kernel}. 
Formally, given a network, $G=(V,E)$, a \textit{similarity index} is a function, $s_G: \ER \rightarrow \R$, that assigns to each non-edge $(v,w)\in\ER$ a score indicating the probability of $(v,w)$ forming in the future, or the probability of $(v,w)$ being a not-yet-discovered edge in the network~\cite{getoor2005link}. For any similarity index, $s_G$, and any non-edge, $(v,w)\in\ER$, we will often write $s_G(v,w)$ instead of $s_G((v,w))$ to improve readability, and we will omit the graph subscript when it is clear from the context. Furthermore, following common practice in the literature, we will not consider self-loops, i.e., edges or non-edges of the form $(v,v):v\in V$.

\subsection{Local Similarity Indices}\label{sec:linkPredictionAlgorithms:local}
\noindent An important class of link prediction algorithms are those based on \textit{local} similarity indices, i.e., indices that account for only \emph{local information} pertaining to the non-edge in question.
As such, the algorithms based on local similarity indices are typically computationally tractable and can be used even with massive networks. In our study, we consider the following local similarity indices, taken from the survey by L{\"u} and Zhou~\cite{lu2011link}:\footnote{\footnotesize The only local similarity index in \cite{lu2011link} that is excluded from our analysis is the Preferential Attachment Index. Unlike the other indices in \cite{lu2011link}, the Preferential Attachment index is based on the assumption that the degree distribution follows a power law---an assumption that does not hold for many of the networks on which we conduct our experiments.}

\begin{itemize}
\item \emph{Common Neighbours} \cite{newman2001clustering}: {\fontsize{10}{10}\selectfont{$\sCN(v,w)=|N(v,w)|$}}
\item \emph{Salton} \cite{salton1986introduction}: {\fontsize{10}{10}\selectfont{$\sSal(v,w)=\frac{|N(v,w)|}{\sqrt{d(v) d(w)}}$}}
\item \emph{Jaccard} \cite{jaccard1901etude}: {\fontsize{10}{10}\selectfont{$\sJac(v,w)=\frac{|N(v,w)|}{|N(v) \cup N(w)|}$}}
\item \emph{S{\o}rensen} \cite{sorensen1948method}: {\fontsize{10}{10}\selectfont{$\sSor(v,w)=\frac{2|N(v,w)|}{d(v) + d(w)}$}}
\item \emph{Hub Promoted} \cite{ravasz2002hierarchical}: {\fontsize{10}{10}\selectfont{$\sHPI(v,w)=\frac{|N(v,w)|}{\min(d(v),d(w))}$}}
\item \emph{Hub Depressed} \cite{ravasz2002hierarchical}: {\fontsize{10}{10}\selectfont{$\sHDI(v,w)=\frac{|N(v,w)|}{\max(d(v),d(w))}$}}
\item \emph{Leicht-Holme-Newman} \cite{leicht2006vertex}: {\fontsize{10}{10}\selectfont{$\sLHN(v,w)=\frac{|N(v,w)|}{d(v) d(w)}$}}
\item \emph{Adamic-Adar} \cite{adamic2003friends}: {\fontsize{10}{10}\selectfont{$\sAA(v,w)=\sum\limits_{u \in N(v,w)} \frac{1}{\log(d(u))}$}}
\item \emph{Resource Allocation} \cite{zhou2009predicting}: {\fontsize{10}{10}\selectfont{$\sRA(v,w)=\sum\limits_{u \in N(v,w)} \frac{1}{d(u)}$}}
\end{itemize}
The set consisting of all those similarity indices will be denoted by $\mathcal{S}$. More formally:
$$
\mathcal{S}=\{\sCN, \sSal, \sJac, \sSor, \sHPI, \sHDI, \sLHN, \sAA, \sRA\}.
$$

\subsection{Global Similarity Indices}

\noindent Another important class of link prediction algorithms are those categorized by L{\"u} and Zhou~\cite{lu2011link} as \textit{global} similarity indices. Before presenting those indices, we need to introduce some additional notation. Let $A$ denote the adjacency matrix of a network, let $\lambda^*$ denote the largest eigenvalue of the adjacency matrix, let $L^+$ denote the pseudoinverse of the Laplacian matrix, and let $I$ denote a unit matrix. Now, for any global similarity index, $s$, let $S$ denote the corresponding \textit{similarity matrix}, whereby the similarity of any pair of nodes, $v_i,v_j\in V$, is specified at the $i$-th row and $j$-th column of $S$. More formally, $\forall v_i,v_j\in V,\ s(v_i,v_j)=S_{i,j}$. With this notation in place, we can now present the global similarity indices outlined in \cite{lu2011link}:

\begin{itemize}

\item \emph{Katz}~\cite{katz1953new} is based on the number of paths between the two nodes, where longer paths are taken with lesser weight according to the dampening factor. Formally, the similarity matrix of this index is:
$$
\SKatz = (I - \beta A)^{-1} - I,
$$
\noindent where $\beta$ is the dampening factor.
In our experiments we set $\beta=\frac{1}{2 \lambda^*}$, as the value has to be smaller than the reciprocal of the largest eigenvalue of the adjacency matrix.

\item \emph{Leicht-Holme-Newman Global}~\cite{leicht2006vertex} is based on the idea that two nodes are similar if their neighbourhoods are similar. More formally, the similarity matrix of this index is:
$$
\SLHNG = 2 |E| \lambda^* D^{-1} (I - \frac{\phi A}{\lambda^*})^{-1} D^{-1},
$$
\noindent where $D$ is the degree matrix, i.e., a diagonal matrix where $D_{i,i}=d(v_i)$, and $\phi$ is a free parameter.
In our experiments we set $\phi = \frac{97}{100}$, as in the original article.

\item \emph{Average Commute Time}~\cite{gobel1974random} is based on the assumption that two nodes are more similar if a random walker can travel between them in a shorter average time. Formally, it is defined as follows:
$$
\sACT(v_i,v_j) = \frac{1}{L^+_{i,i}+L^+_{j,j}-2L^+_{i,j}}.
$$

\item \emph{Cosine}~\cite{fouss2007random} is based on the cosine of the angle between the vectors representing the two nodes. More formally, it is defined as follows:
$$
\sCos(v_i,v_j) = \frac{L^+_{i,j}}{\sqrt{L^+_{i,i} L^+_{j,j}}}.
$$

\item \emph{Random Walk with Restart}~\cite{brin1998anatomy} is based on the idea that node $v_i$ is more similar to node $v_j$ if node $v_i$ is visited with higher frequency by a random walker who starts at node $v_j$ and iteratively moves to a random neighbor with probability $c$ and returns to node $v_j$ with probability $1-c$. Formally, this index is defined as follows:
$$
\sRWR(v_i,v_j) = Q_{i,j}+Q_{j,i},
$$
\noindent with matrix $Q$ being:
$$
Q=(1-c)(I-cP^T)^{-1},
$$
\noindent where $P$ is the transition matrix: $P_{i,j}=\frac{1}{d(v_i)}$ if $v_j \in N(v_i)$ and $P_{i,j}=0$ otherwise.
In our experiments we set $c=\frac{3}{4}$.

\item \emph{SimRank}~\cite{jeh2002simrank} is based on the idea that two nodes are more similar if two random walkers starting at those nodes are expected to meet faster. This index can be computed iteratively as follows:
$$
\sSR(v_i,v_j) = \frac{c \sum_{v \in N(v_i)} \sum_{w \in N(v_j)} \sSR(v,w)}{d(v_i) d(v_j)},
$$
\noindent where $\forall{v \in V} \sSR(v,v) = 1$ and $c$ is the decay factor. In our experiments we set $c=\frac{8}{10}$.

\item \emph{Matrix Forest Index}~\cite{chebotarev2006matrix} assumes that two nodes are more similar if there is a higher probability that they belong to the same tree in a spanning rooted forest. Formally, the similarity matrix of this index is defined as follows:
$$
\SMFI = (I + L)^{-1},
$$
\noindent where $L$ is the Laplacian matrix.

\end{itemize}


\section{Performance Evaluation Metrics}\label{sec:evaluationmetrics}

\noindent Arguably, the most common metrics for evaluating the performance of a similarity index are: \textit{Area under the ROC curve ($\ROC$)}~\cite{fawcett2006introduction} and \textit{Area under the Precision-Recall curve ($\PR$)}~\cite{manning1999foundations}.
To compute any of these metrics for a given similarity index, $s$, we are given a training set, $E$, and a probe set, $Q$, such that $E \cap Q = \emptyset$, i.e., $Q \subset \ER$.
The probe set $Q$ is considered the correct solution of link prediction, i.e., similarity indices are expected to assign high scores to non-edges from $Q$.
The network $(V,E)$ serves as input to the similarity index, $s$, which produces a ranking of the elements of $\ER$.
One can express the quality of this ranking using either $\ROC$ or $\PR$.
To explain how these metrics are computed, we need some additional notation.
Let $\sigma_k$ denote the top $k$ elements of $\ER$ when ranked according to $s$, and let $X= \ER \setminus Q$.
Next, we explain how $\ROC$ or $\PR$ are computed, and then explain an alternative metric called \emph{average precision} ($\AP$).

\smallskip\smallskip
\noindent\textbf{Area under the ROC curve ($\ROC$):} For any given $E$ and $Q$,  $\ROC(E,Q, s)$ is the area under the plot consisting of the following points:
$$
\left\{\left(\frac{|\sigma_k \cap X|}{|X|}, \frac{|\sigma_k \cap Q|}{|Q|}\right)\right\}_{k=1}^{|\ER|}
$$

\noindent $\ROC(E,Q)$ can be interpreted as the probability that the similarity index, $s$, assigns a greater score to a randomly chosen non-edge from $Q$ than to a randomly chosen non-edge from $X$ (ties broken at random), i.e.:
$$
\ROC(E,Q,s)=\frac{|\{(e_1,e_2) \in Q \times X : s(e_1) > s(e_2)\}| + \frac{1}{2}|\{(e_1,e_2) \in Q \times X : s(e_1) = s(e_2)\}|}{|Q||X|}.
$$


\smallskip\smallskip
\noindent\textbf{Area under the Precision-Recall curve ($\PR$):} For any given $E$ and $Q$, $\PR(E,Q, s)$ is the area under the plot consisting of the following points:
$$
\left\{\left(\frac{|\sigma_k \cap Q|}{|Q|}, \frac{|\sigma_k \cap Q|}{k}\right)\right\}_{k=1}^{|\ER|}
$$

\smallskip\smallskip
\noindent\textbf{Average precision ($\AP$):} Since the $\PR$ value is not well-defined for plots that are not continuous, we use instead the \emph{average precision}, $\AP$, described by Boyd et al.~\cite{boyd2013area} as one of the most robust estimators of the area under the Precision-Recall curve. Taking into account the possibility of equal scores, the average precision value is computed as follows:
$$
\AP(E,Q,s) = \frac{1}{|Q|} \sum_{\widehat{e} \in Q} \frac{|\{e \in Q : s(e) > s(\widehat{e})\}| + 1 + \frac{1}{2}|\{e \in Q \setminus \{\widehat{e}\} : s(e) = s(\widehat{e})\}|}{|\{e \in \ER : s(e) > s(\widehat{e})\}| +  1 + \frac{1}{2}|\{e \in \ER \setminus \{\widehat{e}\} : s(e) = s(\widehat{e})\}|}.
$$



\clearpage
\section{Proof of Theorem~\ref{theorem:NPcompleteness}}\label{appendix:proof:NPcompleteness}

\noindent We will prove that the problem of Evading Link Prediction is NP-complete for all the similarity indices described in Section~\ref{sec:linkPredictionAlgorithms:local}, and that is for both the $\ROC$ and $\AP$ metrics which were described in Section~\ref{sec:evaluationmetrics}. To this end, we need to first define a certain network, which we denote by $\Gamma(c,P)$; this network will be used later on in our proofs.

\begin{definition}[The $\Gamma(c,P)$ Network]\label{def:GammNetwork} Let $U=\{u_1,\ldots, u_m\}$ be a set of $m$ elements, and let $P = \{P_1, \ldots, P_q\}$ be a cover of $U$ containing $q$ subsets that are each smaller than $U$. That is, $\forall_i P_i \subset U$ and $\bigcup_{P_i \in P}P_i=U$. Then, given a constant, $c \in \N$, the network $\Gamma(c,P)$ is created as follows:
\begin{itemize}[leftmargin=*]
\item \textbf{The set of nodes:}: For every $P_i \in P$, we create a single node, denoted by $P_i$. Moreover, for every $u_i \in \{u_0,\ldots, u_m\}$, we create a node denoted by $u_i$, as well as $c$ nodes denoted by $a_{i,1},\ldots,a_{i,c}$, and $q-|P(u_i)|$ nodes denoted by $d_{i,1},\ldots,d_{i,q-|P(u_i)|}$, where $P(u_i)=\{P_j \in P : u_i \in P_j\}$.
Additionally, we create three nodes, $v_0$, $v_1$, and $u_0$, as well as $c$ nodes, $a_{0,1},\ldots,a_{0,c}$, and $q$ nodes, $d_{0,1},\ldots,d_{0,q}$.
\item \textbf{The set of edges:} For every $P_j\in P$ we create the edge $(P_j,v_1)$, as well as the edges $(P_j,u_i)$ for every $u_i \in P_j$. Moreover, for every $u_i\in U$ we create the edge $(u_i,v_1)$, as well as the edges $(u_i,u_j)$ for every $u_j\in \{u_{i+1},\ldots, u_m\}$ (this way, the nodes in $\{u_0, \ldots, u_m\}$ form an $(m+1)$-clique). Furthermore, for every $d_{i,j}$ we create the edges $(d_{i,j},u_i)$ and $(d_{i,j},v_1)$. Finally, for every $a_{i,j}$ we create the edges $(a_{i,j},u_i)$, $(a_{i,j},v_0)$ and $(a_{i,j},v_1)$.
\end{itemize}
\end{definition}

\begin{figure}[thb]
\centering
\includegraphics[width=0.6\linewidth]{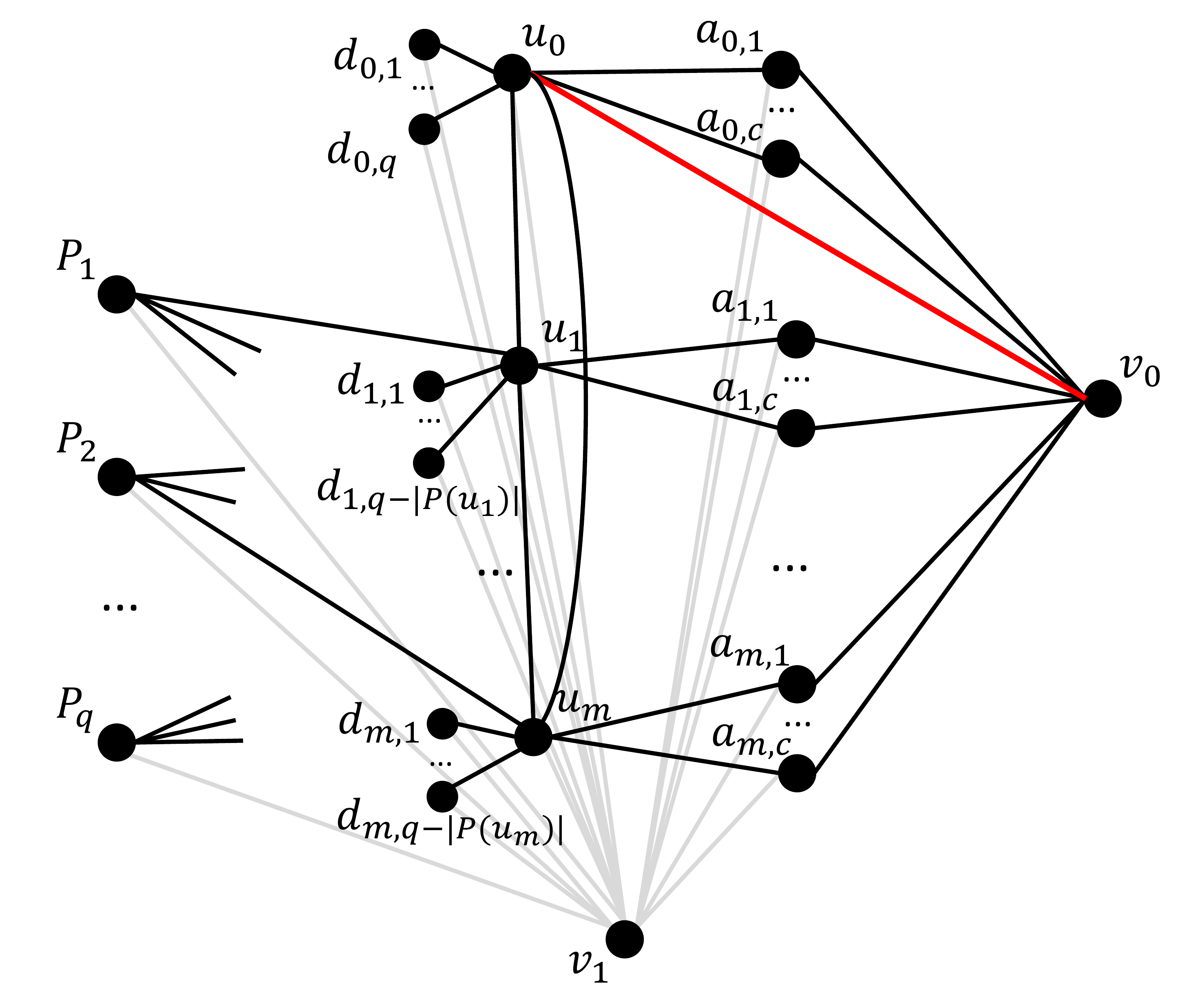}
\caption{An illustration of the $\Gamma(c,P)$ network. Edges connecting $v_1$ with other nodes are grayed out to improve readability. The red non-edge, $(u_0,v_0)$, is the one to be hidden.}
\label{fig:auc-nphard}
\end{figure}

\noindent An illustration of the $\Gamma(c,P)$ network is provided in Figure~\ref{fig:auc-nphard}. Now, suppose that we want to hide a particular non-edge in this network, which is $(u_0,v_0)$. Suppose further that, in order to hide $(u_0,v_0)$, we were only allowed to add edges of the form $(P_i,v_0)$. Then, for any given similarity index, $s\in\mathcal{S}$, we need to understand how the addition of those $(P_i,v_0)$ edges affects the position of $s(u_0,v_0)$ in the similarity-based ranking of all non-edges; if the position of $s(u_0,v_0)$ decreases in this ranking, then $(u_0,v_0)$ becomes more hidden. The following lemma implies that for every similarity index in $\mathcal{S}$ there exists some constant, $c \in \N$, such that the position of $s(u_0,v_0)$ decreases if we add edges of the form $(P_i,v_0)$ to the network $\Gamma(c,P)$. 

\begin{lemma}
\label{lem:linkpred-fixed-relation}
Consider a network $(V,E)=\Gamma(c,P)$ for which $m \geq 5$ and $|P_i|=3$ for all $P_i\in P$. Furthermore, let $\FA=\{(P_i,v_0) : P_i \in P\}$, and for every $A \subseteq \FA$ let $\PA = \{P_i\in P:(P_i,v_0) \in A\}$, and let $\PA(u_j)=\{P_i \in \PA : u_j \in P_i\}$. Then, for every $A \subseteq \FA$, and every similarity index, $s \in \mathcal{S}$, there exists some constant, $c \in \N$, such that:

\begin{enumerate}[label=(\alph*)]
\item \label{pt:linkpred-req1} for every non-edge of the form $(u_i,v_0):i\in\{0,\ldots, m\}$, we have:
\begin{itemize}
\item $s(u_i,v_0)=s(u_0,v_0)$ in the network $(V,E)$.
\item $s(u_i,v_0)=s(u_0,v_0)$ in the network $(V,E\cup A)$ if $\PA(u_i) =  \emptyset$.
\item $s(u_i,v_0)>s(u_0,v_0)$ in the network $(V,E\cup A)$ if $\PA(u_i)\neq\emptyset$.
\end{itemize}

\item \label{pt:linkpred-req2} for every non-edge of the form $(P_i,v_0) : i\in\{0,\ldots,q\}$, we have:
\begin{itemize}
\item $s(P_i,v_0) < s(u_0,v_0)$ in the network $(V,E)$.
\item $s(P_i,v_0) < s(u_0,v_0)$ in the network $(V,E\cup A)$ if $(P_i,v_0) \notin A$.\footnote{\footnotesize{Otherwise, if $(P_i,v_0) \in A$, then $(P_i,v_0)$ will not be a non-edge in $(V,E\cup A)$, and therefore we cannot compute $s(P_i,v_0)$.}}
\end{itemize}

\item \label{pt:linkpred-req3} for every other non-edge, $e \in \ER \setminus \{(u_0,v_0),\ldots,(u_m,v_0),(P_1,v_0),\ldots,(P_q,v_0)\}$:
\begin{itemize}
        \item if $s(e)>s(u_0,v_0)$ in network $(V,E)$, then we also have $s(e)>s(u_0,v_0)$ in network $(V,E\cup A)$.
        \item if $s(e)=s(u_0,v_0)$ in network $(V,E)$, then we also have $s(e)=s(u_0,v_0)$ in network $(V,E\cup A)$.
        \item if $s(e)<s(u_0,v_0)$ in network $(V,E)$, then we also have $s(e)<s(u_0,v_0)$ in network $(V,E\cup A)$.
\end{itemize}
\end{enumerate}
\end{lemma}

\noindent Before we prove the correctness of Lemma~\ref{lem:linkpred-fixed-relation}, let us first provide an example. Suppose that $U=\{u_1, \ldots, u_7\}$, and $P=\{P_1, P_2, P_3\}$ where $P_1=\{u_1,u_2,u_3\}$, $P_2=\{u_3,u_4,u_5\}$ and $P_3=\{u_5,u_6,u_7\}$. Then:
\begin{itemize}
%
\item The set $\FA$ consist of every edge of the form $(P_i,v_0)$. That is, $\FA=\{(P_1,v_0), (P_2,v_0), (P_3,v_0)\}$. Note that none of the edges in $\FA$ appear in the network $(V,E)=\Gamma(c,P)$.
\item The set $A$ is a subset of $\FA$. Suppose that $A=\{(P_2,v_0),(P_3,v_0)\}$. Then:
    \begin{itemize}
    \item $\PA$ consists of every $P_i$ that appears in $A$, i.e., $\PA=\{P_2,P_3\}$;
    \item $\PA(u_i)$ consists of every $P_i$ that contains $u_i$ and appears in $\PA$. For instance, we have: $\PA(u_1) = \emptyset$ and $\PA(u_5) = \{P_2,P_3\}$;
    \end{itemize}
\item Using Lemma~\ref{lem:linkpred-fixed-relation}, we can analyse how the similarity of the non-edges in $\Gamma(c,P)=(V,E)$ would change if we add the edges $(P_2,v_0)$ and $(P_3,v_0)$. To this end, we simply set $A=\{(P_2,v_0),(P_3,v_0)\}$ and analyse the network $(V,E \cup A)$. Now, based on Lemma~\ref{lem:linkpred-fixed-relation}, we know that for every similarity index $s\in\mathcal{S}$ there exists some constant, $c \in \N$, such that:
    \begin{itemize}
    \item Based on point~\ref{pt:linkpred-req1} of Lemma~\ref{lem:linkpred-fixed-relation}:
        \begin{itemize}
        \item for $i\in\{1,2\}$ we have $s(u_i,v_0)=s(u_0,v_0)$ in $(V,E)$ and $s(u_i,v_0)=s(u_0,v_0)$ in $(V,E\cup A)$, because $\PA(u_i) = \emptyset$. 
        \item for $i\in\{3,\ldots,7\}$ we have $s(u_i,v_0)=s(u_0,v_0)$ in $(V,E)$ and $s(u_i,v_0)>s(u_0,v_0)$ in $(V,E\cup A)$, because $\PA(u_i) \neq\emptyset$.
        \end{itemize}
    \item Based on point~\ref{pt:linkpred-req2} of Lemma~\ref{lem:linkpred-fixed-relation}, we have: $s(P_1,v_0) < s(u_0,v_0)$ both in $(V,E)$ and in $(V, E\cup A)$, since $(P_1,v_0)\not\in A$.
    \item Based on point~\ref{pt:linkpred-req3} of Lemma~\ref{lem:linkpred-fixed-relation}, for every non-edge $e$ whose form is neither $(u_i,v_0)$ nor $(P_i,v_0)$:
        \begin{itemize}
        \item if we had $s(e)>s(u_0,v_0)$ in  $(V,E)$, then we will also have $s(e)>s(u_0,v_0)$ in  $(V,E\cup A)$.
        \item if we had $s(e)=s(u_0,v_0)$ in  $(V,E)$, then we will also have $s(e)=s(u_0,v_0)$ in  $(V,E\cup A)$.
        \item if we had $s(e)<s(u_0,v_0)$ in  $(V,E)$, then we will also have $s(e)<s(u_0,v_0)$ in  $(V,E\cup A)$.
        \end{itemize}        
    \end{itemize}
\end{itemize}

\noindent Thus, before the addition of $\{(P_2,v_0),(P_3,v_0)\}$, the position of $(u_0,v_0)$ in the similarity-based ranking was the same as that of any non-edge of the form $(u_i,v_0)$. However, after the addition of $\{(P_2,v_0),(P_3,v_0)\}$, the edge $(u_0,v_0)$ has a ranking lower than that of any $(u_i,v_0):i\in\{3,\ldots,7\}$; as for the remaining non-edges, their relative rankings compared to that of $(u_0,v_0)$ remain unchanged after the addition of $\{(P_2,v_0),(P_3,v_0)\}$. Based on this, by adding $\{(P_2,v_0),(P_3,v_0)\}$ to the network $\Gamma(c,P)=(V,E)$, we decrease the position of $s(u_0,v_0)$ in the similarity-based ranking of all non-edges, i.e., we make $(u_0,v_0)$ more hidden.

Having explained Lemma~\ref{lem:linkpred-fixed-relation} through an example, we will now prove the correctness of this lemma, before presenting our main theorem.

\begin{proof}
First, note that the following holds:
\begin{itemize}
\item for every $P_i$ and every network $(V,E\cup A):A\subseteq \FA$ where $(P_i,v_0)\notin A$, we have $ N(P_i,v_0) = \emptyset$, i.e., $P_i$ and $v_0$ have no common neighbours;
\item for every $d_{i,j}$ and every network $(V,E\cup A):A\subseteq \FA$, we have $N(v_0,d_{i,j}) = \emptyset$.
\end{itemize}
This implies that for every similarity index, $s \in \mathcal{S}$, we have:
\begin{itemize}
\item for every $P_i$ and every network $(V,E\cup A):A\subseteq \FA$ where $(P_i,v_0)\notin A$, we have $s(P_i,v_0) = 0$;
\item for every $d_{i,j}$ and every network $(V,E\cup A):A\subseteq \FA$, we have $s(v_0,d_{i,j}) = 0$.
\end{itemize}
\noindent One can also verify that for every $s \in \mathcal{S}$ and every network $(V,E\cup A):A\subseteq \FA$ it holds that $s(u_0,v_0)>0$. This implies that point~\ref{pt:linkpred-req2} of Lemma~\ref{lem:linkpred-fixed-relation} holds, and that point~\ref{pt:linkpred-req3} holds for every non-edge of the form $(v_0,d_{i,j})$.
We still need to prove the correctness of point~\ref{pt:linkpred-req1}, as well as the correctness of point~\ref{pt:linkpred-req3} for every non-edge of the form:

\begin{enumerate}[label=(\roman*)]
\item $(v_0,v_1)$
\item $(u_i,P_j)$ for $u_i \notin P_j$
\item $(u_i,a_{j,l})$ for $i \neq j$
\item $(u_i,d_{j,l})$ for $i \neq j$
\item $(P_i,P_j)$ for $i \neq j$
\item $(P_i,a_{j,l})$
\item $(P_i,d_{j,l})$
\item $(a_{i_1,j_1},a_{i_2,j_2})$
\item $(a_{i_1,j_1},d_{i_2,j_2})$
\item $(d_{i_1,j_1},d_{i_2,j_2})$
\end{enumerate}

\noindent Next, for every similarity index in $\mathcal{S}$, we will prove the correctness of point~\ref{pt:linkpred-req1}, as well as the correctness of point~\ref{pt:linkpred-req3} for each of the above types of non-edges. To this end, first note that the following holds for every network $(V,E\cup A):A\subseteq \FA$ and every $a_{i,j}$, $d_{i,j}$, $P_i$ in that network:

\begin{itemize}
\item $d(a_{i,j})=3$ (because $a_{i,j}$ is connected to $v_0$, $v_1$ and $u_i$);
\item $d(d_{i,j})=2$  (because $d_{i,j}$ is connected to $v_1$ and $u_i$);
\item $4 \leq d(P_i) \leq 5$ (because $P_i$ is connected to $v_1$ and to every $u_j\in P_i$, where we assumed that $|P_i|=3$; also, if $P_i\in A$, then $P_i$ is connected to $v_0$).
\end{itemize}

\noindent Also note that $u_0 \notin P_i$ for every $P_i\in P$.
Therefore, for any given $A \subseteq \FA$, we have: $\PA(u_0)=0$.
In what follows, we will use the aforementioned facts without referring back to them. We will also use $r$ to denote the number of $a_{i,j}$ nodes, i.e., $r=c(m+1)$, and use $h$ to denote the number of  $d_{i,j}$ nodes, i.e., $h=(m+1)q - 3q=mq - 2q$.
\ \\\\
\noindent \textbf{Common Neighbours ($\sCN$):} We choose $c=6$. Then, to prove the correctness of point~\ref{pt:linkpred-req1}, it suffices to note that for every network $(V,E\cup A):A\subseteq \FA$ we have:
$$
\forall_{u_j\in\{u_0,\ldots, u_m\}} \sCN(u_j,v_0)=6+|\PA(u_j)|.
$$
Moving on to point~\ref{pt:linkpred-req3}, note that for every network $(V,E\cup A):A\subseteq \FA$ we have $\sCN(u_0,v_0)=6$ and that the following holds:

\begin{enumerate}[label=(\roman*)]
\item $\sCN(v_0,v_1)= 6(m+1) + |A| > \sCN(u_0,v_0)$, because the common neighbours of $u_0$ and $v_1$ are all the nodes $a_{i,j}$ and all the nodes $P_i$ where $(P_i,v_0)\in A$.

\item $\sCN(u_i,P_j) \leq 4 < \sCN(u_0,v_0)$, because the common neighbours of $u_i$ and $P_j$ consist of $v_1$ and every $u_l \in P_j : l \neq i$ (note that we assumed that $|P_j|=3$, and $u_i$ may or may not be an element of $P_j$). 

\item $\sCN(u_i,a_{j,l}) = 2 < \sCN(u_0,v_0)$, because the common neighbours of $u_i$ and $a_{j,l}$ are $v_1$ and $u_j$.

\item $\sCN(u_i,d_{j,l}) = 2 < \sCN(u_0,v_0)$, because the common neighbours of $u_i$ and $d_{j,l}$ are $v_1$ and $u_j$.

\item $\sCN(P_i,P_j) \leq 5 < \sCN(u_0,v_0)$, because the common neighbours of $P_i$ and $P_j$ consist of $v_1$, and possibly $v_0$ (if $\{(P_i,v_0),(P_j,v_0)\}\subseteq A$), as well as the every element in $P_i\cap P_j$ (there can be at most 3 such elements, since we assumed that $|P_i|=|P_j|=3$, and we place no restrictions on having $P_i=P_j$).

\item $\sCN(P_i,a_{j,l}) \leq 3 < \sCN(u_0,v_0)$, because the common neighbours of $P_i$ and $a_{j,l}$ consist of $v_1$, and possibly $v_0$ (if $(P_i,v_0)\in A$) and possibly $u_j$ (if $i=j$).

\item $\sCN(P_i,d_{j,l}) \leq 2 < \sCN(u_0,v_0)$, because the common neighbours of $P_i$ and $d_{j,l}$ consist of $v_1$ and possibly $u_j$ (if $i=j$).

\item $\sCN(a_{i_1,j_1},a_{i_2,j_2}) \leq 3 < \sCN(u_0,v_0)$, because the common neighbours of $a_{i_1,j_1}$ and $a_{i_2,j_2}$ consist of $v_1$ and $v_0$ and possibly $u_{i_1}$ (if $i_2=i_1$).

\item $\sCN(a_{i_1,j_1},d_{i_2,j_2}) \leq 2 < \sCN(u_0,v_0)$, because the common neighbours of $a_{i_1,j_1}$ and $d_{i_2,j_2}$ consist of $v_1$ and possibly $u_{i_1}$ (if $i_2=i_1$).

\item $\sCN(d_{i_1,j_1},d_{i_2,j_2}) \leq 2 < \sCN(u_0,v_0)$, because the common neighbours of $d_{i_1,j_1}$ and $d_{i_2,j_2}$ consist of $v_1$ and possibly $u_{i_1}$ (if $i_2=i_1$).
\end{enumerate}

\noindent \textbf{Salton similarity index ($\sSal$):} We choose $c=1$. Then, to prove the correctness of point~\ref{pt:linkpred-req1}, it suffices to note that for every network $(V,E\cup A):A \subseteq \FA$ we have:
$$
\sSal(u_j,v_0)=\frac{1+|\PA(u_j)|}{\sqrt{(r+|A|)(m+q+2)}},\ \ \forall u_j\in\{u_0,\ldots, u_m\}.
$$
Moving on to point~\ref{pt:linkpred-req3}, note that $\sSal(u_0,v_0) \leq \frac{1}{\sqrt{42}}$ (since $m \geq 5$), and that the following holds for every $(V,E\cup A):A \subseteq \FA$:

\begin{enumerate}[label=(\roman*),leftmargin=*]\itemsep0.5em
\item $\sSal(v_0,v_1)=\frac{\sqrt{r + |A|}}{\sqrt{r+h+q+m+1}} > \sSal(u_0,v_0)$
\item $\sSal(u_i,P_j) \geq \frac{3}{\sqrt{(m+q+2)5}} > \sSal(u_0,v_0)$
\item $\sSal(u_i,a_{j,l}) \geq \frac{1}{\sqrt{(m+q+2)3}} > \sSal(u_0,v_0)$
\item $\sSal(u_i,d_{j,l}) \geq \frac{1}{\sqrt{(m+q+2)2}} > \sSal(u_0,v_0)$
\item $\sSal(P_i,P_j) \geq \frac{1}{\sqrt{20}} > \sSal(u_0,v_0)$
\item $\sSal(P_i,a_{j,l}) \geq \frac{1}{2\sqrt{3}} > \sSal(u_0,v_0)$
\item $\sSal(P_i,d_{j,l}) \geq \frac{1}{\sqrt{10}} > \sSal(u_0,v_0)$
\item $\sSal(a_{i_1,j_1},a_{i_2,j_2}) \geq \frac{2}{3} > \sSal(u_0,v_0)$
\item $\sSal(a_{i_1,j_1},d_{i_2,j_2}) \geq \frac{1}{\sqrt{6}} > \sSal(u_0,v_0)$
\item $\sSal(d_{i_1,j_1},d_{i_2,j_2}) \geq \frac{1}{2} > \sSal(u_0,v_0)$
\end{enumerate}

\noindent \textbf{Jaccard similarity index ($\sJac$):} We choose $c=1$. Then, to prove the correctness of point~\ref{pt:linkpred-req1}, it suffices to note that for every $(V,E\cup A):A \subseteq \FA$ we have: $$
\sJac(u_j,v_0)=\frac{1+|\PA(u_j)|}{r+q+m+|A|+2-|\PA(u_j)|},\ \ \forall u_j\in\{u_0,\ldots, u_m\}.
$$
Moving on to point~\ref{pt:linkpred-req3}, note that $\sJac(u_0,v_0) \leq \frac{1}{13}$ (since $m \geq 5$), and that the following holds for every $(V,E\cup A):A \subseteq \FA$:

\begin{enumerate}[label=(\roman*),leftmargin=*]\itemsep0.5em
\item $\sJac(v_0,v_1)=\frac{r + |A|}{r+h+q+m+1} > \sJac(u_0,v_0)$
\item $\sJac(u_i,P_j) \geq \frac{3}{m+q+3} > \sJac(u_0,v_0)$
\item $\sJac(u_i,a_{j,l}) \geq \frac{1}{m+q+3} > \sJac(u_0,v_0)$
\item $\sJac(u_i,d_{j,l}) \geq \frac{1}{m+q+2} > \sJac(u_0,v_0)$
\item $\sJac(P_i,P_j) \geq \frac{1}{8} > \sJac(u_0,v_0)$
\item $\sJac(P_i,a_{j,l}) \geq \frac{1}{6} > \sJac(u_0,v_0)$
\item $\sJac(P_i,d_{j,l}) \geq \frac{1}{6} > \sJac(u_0,v_0)$
\item $\sJac(a_{i_1,j_1},a_{i_2,j_2}) \geq \frac{2}{4} > \sJac(u_0,v_0)$
\item $\sJac(a_{i_1,j_1},d_{i_2,j_2}) \geq \frac{1}{4} > \sJac(u_0,v_0)$
\item $\sJac(d_{i_1,j_1},d_{i_2,j_2}) \geq \frac{1}{3} > \sJac(u_0,v_0)$
\end{enumerate}

\noindent \textbf{S{\o}rensen similarity index ($\sSor$):} We choose $c=1$. Then, to prove the correctness of point~\ref{pt:linkpred-req1}, it suffices to note that for every $(V,E\cup A):A \subseteq \FA$ we have: $$
\sSor(u_j,v_0)=\frac{2+2|\PA(u_j)|}{r+q+m+|A|+2},\ \ \forall u_j\in\{u_0,\ldots, u_m\}.
$$
Moving on to point~\ref{pt:linkpred-req3}, note that $\sSor(u_0,v_0) \leq \frac{2}{13}$ (since $m \geq 5$), and that the following holds for every $(V,E\cup A):A \subseteq \FA$:

\begin{enumerate}[label=(\roman*),leftmargin=*]\itemsep0.5em
\item $\sSor(v_0,v_1)=\frac{2r + 2|A|}{2r+h+q+m+1+|A|} > \sSor(u_0,v_0)$
\item $\sSor(u_i,P_j) \geq \frac{6}{m+q+7} > \sSor(u_0,v_0)$
\item $\sSor(u_i,a_{j,l}) \geq \frac{2}{m+q+5} > \sSor(u_0,v_0)$
\item $\sSor(u_i,d_{j,l}) \geq \frac{2}{m+q+4} > \sSor(u_0,v_0)$
\item $\sSor(P_i,P_j) \geq \frac{2}{9} > \sSor(u_0,v_0)$
\item $\sSor(P_i,a_{j,l}) \geq \frac{2}{8} > \sSor(u_0,v_0)$
\item $\sSor(P_i,d_{j,l}) \geq \frac{2}{7} > \sSor(u_0,v_0)$
\item $\sSor(a_{i_1,j_1},a_{i_2,j_2}) \geq \frac{4}{6} > \sSor(u_0,v_0)$
\item $\sSor(a_{i_1,j_1},d_{i_2,j_2}) \geq \frac{2}{5} > \sSor(u_0,v_0)$
\item $\sSor(d_{i_1,j_1},d_{i_2,j_2}) \geq \frac{2}{4} > \sSor(u_0,v_0)$
\end{enumerate}

\noindent \textbf{Hub Promoted similarity index ($\sHPI$):} We choose $c=1$. Then, to prove the correctness of point~\ref{pt:linkpred-req1}, it suffices to note that for every $(V,E\cup A):A \subseteq \FA$ we have: $$
\sHPI(u_j,v_0)=\frac{1+|\PA(u_j)|}{r+|A|},\ \ \forall u_j\in\{u_0,\ldots, u_m\}.
$$
Moving on to point~\ref{pt:linkpred-req3}, note that $\sHPI(u_0,v_0) \leq \frac{1}{6}$ (since $m \geq 5$), and that the following holds for every $(V,E\cup A):A \subseteq \FA$:

\begin{enumerate}[label=(\roman*),leftmargin=*]\itemsep0.4em
\item $\sHPI(v_0,v_1)=\frac{r+|A|}{r+|A|}=1 > \sHPI(u_0,v_0)$
\item $\sHPI(u_i,P_j) \geq \frac{3}{5} > \sHPI(u_0,v_0)$
\item $\sHPI(u_i,a_{j,l}) \geq \frac{1}{3} > \sHPI(u_0,v_0)$
\item $\sHPI(u_i,d_{j,l}) \geq \frac{1}{2} > \sHPI(u_0,v_0)$
\item $\sHPI(P_i,P_j) \geq \frac{1}{4} > \sHPI(u_0,v_0)$
\item $\sHPI(P_i,a_{j,l}) \geq \frac{1}{3} > \sHPI(u_0,v_0)$
\item $\sHPI(P_i,d_{j,l}) \geq \frac{1}{2} > \sHPI(u_0,v_0)$
\item $\sHPI(a_{i_1,j_1},a_{i_2,j_2}) \geq \frac{2}{3} > \sHPI(u_0,v_0)$
\item $\sHPI(a_{i_1,j_1},d_{i_2,j_2}) \geq \frac{1}{2} > \sHPI(u_0,v_0)$
\item $\sHPI(d_{i_1,j_1},d_{i_2,j_2}) \geq \frac{1}{2} > \sHPI(u_0,v_0)$
\end{enumerate}

\noindent \textbf{Hub Depressed similarity index ($\sHDI$):} We choose $c=1$. Then, to prove the correctness of point~\ref{pt:linkpred-req1}, it suffices to note that for every $(V,E\cup A):A \subseteq \FA$ we have: $$
\sHDI(u_j,v_0)=\frac{1+|\PA(u_j)|}{m+q+2},\ \ \forall u_j\in\{u_0,\ldots, u_m\}.
$$
Moving on to point~\ref{pt:linkpred-req3}, note that $\sHDI(u_0,v_0) \leq \frac{1}{7}$ (since $m \geq 5$), and that the following holds for every $(V,E\cup A):A \subseteq \FA$:

\begin{enumerate}[label=(\roman*),leftmargin=*]\itemsep0.4em
\item $\sHDI(v_0,v_1)=\frac{r+|A|}{r+h+q+m+1} > \sHDI(u_0,v_0)$
\item $\sHDI(u_i,P_j) \geq \frac{3}{m+q+2} > \sHDI(u_0,v_0)$
\item Either $\sHDI(u_i,a_{j,l}) = \frac{1}{m+q+2} = \sHDI(u_0,v_0)$ (if $i=j$) or $\sHDI(u_i,a_{j,l}) = \frac{2}{m+q+2} > \sHDI(u_0,v_0)$ (otherwise)
\item Either $\sHDI(u_i,d_{j,l}) = \frac{1}{m+q+2} = \sHDI(u_0,v_0)$ (if $i=j$) or $\sHDI(u_i,d_{j,l}) = \frac{2}{m+q+2} > \sHDI(u_0,v_0)$ (otherwise)
\item $\sHDI(P_i,P_j) \geq \frac{1}{5} > \sHDI(u_0,v_0)$
\item $\sHDI(P_i,a_{j,l}) \geq \frac{1}{5} > \sHDI(u_0,v_0)$
\item $\sHDI(P_i,d_{j,l}) \geq \frac{1}{5} > \sHDI(u_0,v_0)$
\item $\sHDI(a_{i_1,j_1},a_{i_2,j_2}) \geq \frac{2}{3} > \sHDI(u_0,v_0)$
\item $\sHDI(a_{i_1,j_1},d_{i_2,j_2}) \geq \frac{1}{3} > \sHDI(u_0,v_0)$
\item $\sHDI(d_{i_1,j_1},d_{i_2,j_2}) \geq \frac{1}{2} > \sHDI(u_0,v_0)$
\end{enumerate}

\noindent \textbf{Leicht-Holme-Newman similarity index ($\sLHN$):} We choose $c=1$.
Then, to prove the correctness of point~\ref{pt:linkpred-req1}, it suffices to note that for every $(V,E\cup A):A \subseteq \FA$:
$$
\sLHN(u_j,v_0)=\frac{1+|\PA(u_j)|}{(r+|A|)(m+q+2)},\ \ \forall u_j\in\{u_0,\ldots, u_m\}.
$$
Moving on to point~\ref{pt:linkpred-req3}, note that $\sLHN(u_0,v_0) \leq \frac{1}{42}$ (since $m \geq 5$), and that the following holds for every $(V,E\cup A):A \subseteq \FA$:

\begin{enumerate}[label=(\roman*),leftmargin=*]\itemsep0.4em
\item $\sLHN(v_0,v_1)=\frac{1}{r+h+q+m+1} > \sLHN(u_0,v_0)$
\item $\sLHN(u_i,P_j) \geq \frac{3}{(m+q+2)5} > \sLHN(u_0,v_0)$
\item $\sLHN(u_i,a_{j,l}) \geq \frac{1}{(m+q+2)3} > \sLHN(u_0,v_0)$
\item $\sLHN(u_i,d_{j,l}) \geq \frac{1}{(m+q+2)2} > \sLHN(u_0,v_0)$
\item $\sLHN(P_i,P_j) \geq \frac{1}{20} > \sLHN(u_0,v_0)$
\item $\sLHN(P_i,a_{j,l}) \geq \frac{1}{12} > \sLHN(u_0,v_0)$
\item $\sLHN(P_i,d_{j,l}) \geq \frac{1}{10} > \sLHN(u_0,v_0)$
\item $\sLHN(a_{i_1,j_1},a_{i_2,j_2}) \geq \frac{2}{9} > \sLHN(u_0,v_0)$
\item $\sLHN(a_{i_1,j_1},d_{i_2,j_2}) \geq \frac{1}{6} > \sLHN(u_0,v_0)$
\item $\sLHN(d_{i_1,j_1},d_{i_2,j_2}) \geq \frac{1}{4} > \sLHN(u_0,v_0)$
\end{enumerate}

\noindent \textbf{Adamic-Adar similarity index ($\sAA$):} We choose $c=3$. Then, to prove the correctness of point~\ref{pt:linkpred-req1}, it suffices to note that for every $(V,E\cup A):A \subseteq \FA$ we have: $$
\sAA(u_j,v_0)=\frac{3}{\log(3)}+\frac{|\PA(u_j)|}{\log(5)},\ \ \forall u_j\in\{u_0,\ldots, u_m\}.
$$
Moving on to point~\ref{pt:linkpred-req3}, note that $\sAA(u_0,v_0) = \frac{3}{\log(3)} > 6$ and that the following holds for every $(V,E\cup A):A \subseteq \FA$:

\begin{enumerate}[label=(\roman*),leftmargin=*]\itemsep0.4em
\item $\sAA(v_0,v_1)=\frac{r}{\log(3)}+\frac{|A|}{\log(5)} > \sAA(u_0,v_0)$
\item $\sAA(u_i,P_j) = \frac{1}{\log(r+h+q+m+1)} + \frac{3}{\log(q+m+4)} < \sAA(u_0,v_0)$
\item $\sAA(u_i,a_{j,l}) \leq \frac{1}{\log(r+h+q+m+1)} + \frac{1}{\log(q+m+4)} < \sAA(u_0,v_0)$
\item $\sAA(u_i,d_{j,l}) \leq \frac{1}{\log(r+h+q+m+1)} + \frac{1}{\log(q+m+4)} < \sAA(u_0,v_0)$
\item $\sAA(P_i,P_j) \leq \frac{1}{\log(r+h+q+m+1)} + \frac{1}{\log(r+|A|)} + \frac{3}{\log(q+m+4)} < \sAA(u_0,v_0)$
\item $\sAA(P_i,a_{j,l}) \leq \frac{1}{\log(r+h+q+m+1)} + \frac{1}{\log(r+|A|)} + \frac{1}{\log(q+m+4)} < \sAA(u_0,v_0)$
\item $\sAA(P_i,d_{j,l}) \leq \frac{1}{\log(r+h+q+m+1)} + \frac{1}{\log(q+m+4)} < \sAA(u_0,v_0)$
\item $\sAA(a_{i_1,j_1},a_{i_2,j_2}) \leq \frac{1}{\log(r+h+q+m+1)} + \frac{1}{\log(r+|A|)} + \frac{1}{\log(q+m+4)} < \sAA(u_0,v_0)$
\item $\sAA(a_{i_1,j_1},d_{i_2,j_2}) \leq \frac{1}{\log(r+h+q+m+1)} + \frac{1}{\log(q+m+4)} < \sAA(u_0,v_0)$
\item $\sAA(d_{i_1,j_1},d_{i_2,j_2}) \leq \frac{1}{\log(r+h+q+m+1)} + \frac{1}{\log(q+m+4)} < \sAA(u_0,v_0)$
\end{enumerate}

\noindent \textbf{Resource Allocation similarity index ($\sRA$):} We choose $c=3$.
Then, to prove the correctness of point~\ref{pt:linkpred-req1}, it suffices to note that for every $(V,E\cup A):A \subseteq \FA$ we have: $$
\sRA(u_j,v_0)=\frac{3}{3}+\frac{|\PA(u_j)|}{5}
$$
Moving on to point~\ref{pt:linkpred-req3}, note that $\sRA(u_0,v_0) = 1$ and that the following holds for every $(V,E\cup A):A \subseteq \FA$:

\begin{enumerate}[label=(\roman*),leftmargin=*]\itemsep0.4em
\item $\sRA(v_0,v_1)=\frac{r}{3}+\frac{|A|}{5} > \sRA(u_0,v_0)$
\item $\sRA(u_i,P_j) = \frac{1}{r+h+q+m+1} + \frac{3}{q+m+4} < \sRA(u_0,v_0)$
\item $\sRA(u_i,a_{j,l}) \leq \frac{1}{r+h+q+m+1} + \frac{1}{q+m+4} < \sRA(u_0,v_0)$
\item $\sRA(u_i,d_{j,l}) \leq \frac{1}{r+h+q+m+1} + \frac{1}{q+m+4} < \sRA(u_0,v_0)$
\item $\sRA(P_i,P_j) \leq \frac{1}{r+h+q+m+1} + \frac{1}{r+|A|} + \frac{3}{q+m+4} < \sRA(u_0,v_0)$
\item $\sRA(P_i,a_{j,l}) \leq \frac{1}{r+h+q+m+1} + \frac{1}{r+|A|} + \frac{1}{q+m+4} < \sRA(u_0,v_0)$
\item $\sRA(P_i,d_{j,l}) \leq \frac{1}{r+h+q+m+1} + \frac{1}{q+m+4} < \sRA(u_0,v_0)$
\item $\sRA(a_{i_1,j_1},a_{i_2,j_2}) \leq \frac{1}{r+h+q+m+1} + \frac{1}{r+|A|} + \frac{1}{q+m+4} < \sRA(u_0,v_0)$
\item $\sRA(a_{i_1,j_1},d_{i_2,j_2}) \leq \frac{1}{r+h+q+m+1} + \frac{1}{q+m+4} < \sRA(u_0,v_0)$
\item $\sRA(d_{i_1,j_1},d_{i_2,j_2}) \leq \frac{1}{r+h+q+m+1} + \frac{1}{q+m+4} < \sRA(u_0,v_0)$\hspace*{\fill}
\end{enumerate}

\noindent This concludes the proof of Lemma~\ref{lem:linkpred-fixed-relation}.
\end{proof}

Having defined the $\Gamma(c,P)$ network, and having proven the correctness of Lemma~\ref{lem:linkpred-fixed-relation}, we are now ready to move to the proof of Theorem~\ref{theorem:NPcompleteness}.

Before we present our proof, let us first explain the intuition behind it. Specifically, the proof is based on a reduction from the NP-complete \emph{3-Set Cover problem} to a particular instance of our problem of \emph{Evading Link Prediction}. Recall that the 3-Set Cover problem is defined by (i) a universe $U=\{u_1, \ldots, u_l\}$; (ii) a collection of subsets $P = \{P_1, \ldots, P_m\}$ such that $\forall_j P_j\subset U$ and $\forall_j \left|P_j\right|=3$; and (iii) an integer $k\leq m$. The goal is then to determine whether there exist $k$ elements of $P$ the union of which equals $U$. In our proof, the 3-Set Cover problem will be reduced to the problem of \emph{Evading Link Prediction} (see Definition~1) where:
\begin{itemize}\itemsep-0.3em
\item the network under consideration is: $G=\Gamma(c,P)$, where $c \in \N$ satisfies the conditions in Lemma~\ref{lem:linkpred-fixed-relation};
\item the set of non-edges to be hidden is: $H=\{(u_0,v_0)\}$;
\item the set of edges that can be added is: $\FA=\{(P_i,v_0):P_i\in P\}$;
\item the set of edges that can be removed is: $\FR=\emptyset$;
\item the budget that specifies the number of edges that can be modified (i.e., added or removed) is: $b=k$.
\end{itemize}

\noindent Note that the above instance of the problem of Evading Link Prediction is exactly the same as the instance considered in Lemma~\ref{lem:linkpred-fixed-relation}. We already know from this lemma that, in the similarity-based ranking of all non-edges in $\Gamma(c,P)$, the position of $(u_0,v_0)$ is the same as that of any non-edge of the form $(u_i,v_0)$. However, after adding some edges, $A\subseteq \FA$, the position of $(u_0,v_0)$ becomes lower than that of any $(u_i,v_0)$ such that $\exists P_j\in A : u_i\in P_j$. As for the remaining non-edges, their relative ranking compared to that of $(u_0,v_0)$ remains unchanged after the addition of $A$. Based on this, in order to decrease the position of $(u_0,v_0)$ in the similarity-based ranking as much as possible, we need to add some edges, $A\subseteq \FA$, such that: $\exists P_j\in A : u_i\in P_j$ for every $u_i\in U$. That is, we need to find a subset of $P$ that covers all the elements in $U$, which leads us to the 3-Set Cover problem.


\begin{proof}
The problem of Evading Link Prediction is trivially in NP, since computing $\ROC$ and $\AP$ before and after the addition of a given set of edges $A \subseteq \FA$ and the removal of a given set of edges $R \subseteq \FR$ can be done in polynomial time for every similarity index in $\mathcal{S}$.

Next, we will prove that the problem is NP-hard. To this end, we will give a reduction from the NP-complete 3-Set Cover problem. This problem is defined by (i) a universe $U=\{u_1, \ldots, u_l\}$; (ii) a collection of subsets $P = \{P_1, \ldots, P_m\}$ such that $\forall_j P_j\subset U$ and $\forall_j \left|P_j\right|=3$; and (iii) an integer $k\leq m$. The goal is then to determine whether there exist $k$ elements of $P$ the union of which equals $U$.

Let us assume that $m \geq 5$, as all other cases can be easily solved in polynomial time. Now, for any given similarity index, $s \in \mathcal{S}$, consider the following instance of the problem of Evading Link Prediction $(G,s,f,\Hide,b,\FA,\FR)$, where:

\begin{itemize}\itemsep-0.3em
\item $G = (V,E) = \Gamma(c,P)$, where $c \in \N$ is chosen to be a constant that satisfies the conditions in Lemma~\ref{lem:linkpred-fixed-relation} (the lemma states that such a constant exists);
\item $s$ is the similarity index under consideration;
\item $f$ is either the $\ROC$ or the $\AP$ metric;
\item $\Hide = \{(u_0,v_0)\}$;
\item $b=k$, where $k$ is the parameter of the 3-Set Cover problem, and the goal is to determine whether there exist $k$ elements of $P$ the union of which equals $U$;
\item $\FA=\{(P_i,v_0) : P_i \in P\}$;
\item $\FR=\emptyset$.
\end{itemize}

\noindent Let us also introduce the following notation:

\begin{itemize}\itemsep-0.3em
\item $\Upsilon^<(G) = \{e \in \ER: s(e) < s(u_0,v_0)\}$ in network $G=(V,E)$;
\item $\Upsilon^=(G) = \{e \in \ER \setminus \{(u_0,v_0)\}: s(e) = s(u_0,v_0)\}$ in network $G=(V,E)$;
\item $\Upsilon^>(G) = \{e \in \ER: s(e) > s(u_0,v_0)\}$ in network $G=(V,E)$.
\end{itemize}

\noindent Note that $\ER$ is the set of non-edges in $G=(V,E)$, whereas $\ER \setminus A$ is the set of non-edges in $(V,E\cup A)$. For every network $G'=(V,E \cup A):A \subseteq \FA$, we know from the definition of $\ROC$ in Section~\ref{sec:evaluationmetrics} that:

\begin{equation}
\label{eqn:linkpred-theorem-1}
\ROC(E \cup A, \Hide)=\frac{|\Upsilon^<(G')| + \frac{1}{2} |\Upsilon^=(G')|}{|\ER \setminus A|-1}.
\end{equation}

\noindent We also know from the definition of $\AP$ in Section~\ref{sec:evaluationmetrics} that:

\begin{equation}
\label{eqn:linkpred-theorem-2}
\AP(E \cup A,\Hide) = \frac{1}{|\Upsilon^>(G')| + 1 + \frac{1}{2} |\Upsilon^=(G')|} .
\end{equation}

\noindent Now, let $U_A = \{u_i:\exists_{P_j\in \PA} u_i \in P_j\}$.
Point~\ref{pt:linkpred-req2} of Lemma~\ref{lem:linkpred-fixed-relation} implies that:

\begin{equation}
\label{eqn:linkpred-theorem-3}
|\Upsilon^<(G')| = |\Upsilon^<(G)| - |A|.
\end{equation}

\noindent On the other hand, point~\ref{pt:linkpred-req1} of Lemma~\ref{lem:linkpred-fixed-relation} implies that:

\begin{equation}
\label{eqn:linkpred-theorem-4}
|\Upsilon^=(G')| = |\Upsilon^=(G)| - |U_A|,
\end{equation}


\begin{equation}
\label{eqn:linkpred-theorem-5}
|\Upsilon^>(G')| = |\Upsilon^>(G)| + |U_A|,
\end{equation}

\noindent Equations \eqref{eqn:linkpred-theorem-1}, \eqref{eqn:linkpred-theorem-3} and \eqref{eqn:linkpred-theorem-4} imply that:

\begin{equation}
\label{eqn:linkpred-theorem-6}
\ROC(E \cup A, \Hide)=\frac{|\Upsilon^<(G)| - |A| + \frac{1}{2}(|\Upsilon^=(G)|-|U_A|)}{|\ER|-|A|-1}
\end{equation}

\noindent On the other hand, equations \eqref{eqn:linkpred-theorem-2} and \eqref{eqn:linkpred-theorem-5} imply that:

\begin{equation}
\AP(E \cup A,\Hide) = \frac{1}{|\Upsilon^>(G)| + |U_A| + 1 + \frac{1}{2}(|\Upsilon^=(G)|-|U_A|)}
\end{equation}

\noindent This, in turn, implies that:

\begin{equation}
\label{eqn:linkpred-theorem-7}
\AP(E \cup A,\Hide) = \frac{1}{|\Upsilon^>(G)| + 1 + \frac{1}{2}(|\Upsilon^=(G)|+|U_A|)} .
\end{equation}

\noindent Equations \eqref{eqn:linkpred-theorem-6} and \eqref{eqn:linkpred-theorem-7} imply that both $\ROC$ and $\AP$ decrease with $|U_A|$. Thus, for each of these two metrics an optimal choice of $A$ is one that maximizes $|U_A|$. This happens when $U_A=U$. For any choice of $A$ such that $U_A=U$, the following holds: $\forall_{u_j \in U} \exists_{(P_i,v_0) \in A} u_j \in P_i$. Such an optimal choice of $A$ constitutes a solution to our instance of the problem of Evading Link Prediction. It also corresponds directly to a solution to the 3-Set Cover problem.
\end{proof}

\clearpage
\section{Proof of Theorem~\ref{theorem:comparingPositiveAndNegativeEffects}}\label{appendix:proof:theorem:comparingPositiveAndNegativeEffects}

\begin{proof}
From the definitions of $G$ and $G'$, we know that:
\begin{itemize}
\item $d_{G'}(x) = d_{G}(x)$;
\item $d_{G'}(w) = d_{G}(w)-1$;
\item $\left|N_{G'}(x,w)\right| = \left|N_{G}(x,w)\right|-1$;
\item $\forall u\in N_{G'}(x,w): d_{G'}(u) = d_G(u)$.
\end{itemize}
With these facts in mind, we will now handle each similarity index in $\mathcal{S}$ separately. In particular:

\begin{itemize}
\item For $\sCN(x,w)=|N(x,w)|$, we know that $\sCN_{G'}(x,w)< \sCN_{G}(x,w)$ because $\left|N_{G'}(x,w)\right| = \left|N_{G}(x,w)\right| - 1$.
\item For $\sSal(x,w)=\frac{|N(x,w)|}{\sqrt{d(x) d(w)}}$, to prove that $\sSal_{G'}(x,w)\leq \sSal_G(x,w)$, it suffices to prove that:
$$
\frac{|N_G(x,w)|-1}{\sqrt{d_G(x) (d_G(w)-1)}}\leq \frac{|N_G(x,w)|}{\sqrt{d_G(x) d_G(w)}}
$$
This holds if and only if: $|N_G(x,w)|^2 \leq d_G(w)(2|N_G(x,w)|-1)$. This, in turn, always holds since $|N_G(x,w)| \leq d_G(w)$ and $|N_G(x,w)| \leq 2|N_G(x,w)|-1$.
\item For $\sJac(x,w)=\frac{|N(x,w)|}{|N(x) \cup N(w)|}$, we know that the following holds in any network: $|N(x) \cup N(w)| = d(x)+d(w)-|N(x,w)|$. Based on this, to prove that $\sJac_{G'}(x,w)< \sJac_G(x,w)$, it suffices to note that:
$$
\frac{|N_G(x,w)|-1}{d_G(x)+(d_G(w)-1)-(|N_G(x,w)|-1)}\ <\ \frac{|N_G(x,w)|}{d_G(x)+d_G(w)-|N_G(x,w)|}
$$
\item For $\sSor(x,w)=\frac{2|N(x,w)|}{d(x) + d(w)}$, to prove that $\sSor_{G'}(x,w)\leq \sSor_G(x,w)$, it suffices to prove that:
$$
\frac{2|N_G(x,w)|-2}{d_G(x) + d_G(w)-1}\leq \frac{2|N_G(x,w)|}{d_G(x) + d_G(w)}
$$
This holds if and only if: $|N_G(x,w)|\leq d_G(x) + d_G(w)$. This, in turn, always holds since $N_G(x,w)\subseteq N_G(x)$ and $N_G(x,w)\subseteq N_G(w)$.
\item For $\sHPI(x,w)=\frac{|N(x,w)|}{\min(d(x),d(w))}$, let us first consider the case where $d_G(x)<d_G(w)$. In this case, we have: $\min(d_G(x),d_G(w))=d_G(x)$ and $\min(d_{G'}(x),d_{G'}(w))=\min(d_{G}(x),d_{G}(w)-1)=d_{G}(x)$. This implies that $\sHPI_{G'}(x,w)< \sHPI_G(x,w)$, since:
$$
\frac{|N_{G'}(x,w)|}{\min(d_{G'}(x),d_{G'}(w))} = \frac{|N_{G}(x,w)|-1}{d_G(x)} < \frac{|N_G(x,w)|}{d_{G}(x)} = \frac{|N_G(x,w)|}{\min(d_G(x),d_G(w))}
$$
On the other hand, if $d_G(x)\geq d_G(w)$, then $\min(d_G(x),d_G(w))=d_G(w)$ and $\min(d_{G'}(x),d_{G'}(w))=\min(d_{G}(x),d_{G}(w)-1)=d_{G}(w)-1$. Based on this, in order to prove that $\sHPI_{G'}(x,w)\leq \sHPI_G(x,w)$, we need to prove that:
$$
\frac{|N_G(x,w)|-1}{d_G(w)-1} \leq \frac{|N_G(x,w)|}{d_G(w)}
$$
This holds if and only if: $|N_G(x,w)|\leq d_G(w)$. This, in turn, always holds since $N_G(x,w)\subseteq N_G(w)$.
\item For $\sHDI(x,w)=\frac{|N(x,w)|}{\max(d(x),d(w))}$, let us first consider the case where $d_G(x)\geq d_G(w)$. In this case, we have: $\max(d_G(x),d_G(w))=d_G(x)$ and $\max(d_{G'}(x),d_{G'}(w))=\max(d_{G}(x),d_{G}(w)-1)=d_{G}(x)$. This implied that $\sHDI_{G'}(x,w)<\sHDI_{G}(x,w)$, since:
$$
\frac{|N_{G'}(x,w)|}{\max(d_{G'}(x),d_{G'}(w))} = \frac{|N_G(x,w)|-1}{d_G(x)} < \frac{|N_G(x,w)|}{d_G(x)} = \frac{|N_G(x,w)|}{\max(d_G(x),d_G(w))}
$$
On the other hand, if $d_G(x)<d_G(w)$, then $\max(d_G(x),d_G(w))=d_G(w)$ and $\max(d_{G'}(x),d_{G'}(w))=\max(d_{G}(x),d_{G}(w)-1)=d_{G}(w)-1$. Based on this, in order to prove that $\sHDI_{G'}(x,w)\leq \sHDI_G(x,w)$, we need to prove that:
$$
\frac{|N_G(x,w)|-1}{d_G(w)-1} \leq \frac{|N_G(x,w)|}{d_G(w)}
$$
This holds if and only if: $|N_G(x,w)|\leq d_G(w)$. This, in turn, always holds since $N_G(x,w)\subseteq N_G(w)$.
\item For $\sLHN(x,w)=\frac{|N(x,w)|}{d(x) d(w)}$, to prove that $\sLHN_{G'}(x,w)\leq \sLHN_G(x,w)$, it suffices to prove that:
$$
\frac{|N_G(x,w)|-1}{d_G(x) (d_G(w)-1)} \leq \frac{|N_G(x,w)|}{d_G(x) d_G(w)}
$$
This holds if and only if: $|N_G(x,w)|\leq d_G(w)$. This, in turn, always holds since $N_G(x,w)\subseteq N_G(w)$.
\item For $\sAA(x,w)=\sum_{u \in N(x,w)} \frac{1}{\log(d(u))}$, to prove that $\sAA_{G'}(x,w)\leq\sAA_{G}(x,w)$, it suffices to note that:
$$
\sum\limits_{u \in N_{G'}(x,w)}\frac{1}{\log(d_{G'}(u))} = \left(\sum\limits_{u \in N_{G}(x,w)}\frac{1}{\log(d_G(u))}\right) - \frac{1}{\log(d_G(v))} \leq  \sum\limits_{u \in N_{G}(x,w)}\frac{1}{\log(d_G(u))}
$$
\item For $\sRA(x,w)=\sum_{u \in N(x,w)} \frac{1}{d(u)}$, to prove that $\sRA_{G'}(x,w)\leq\sRA_{G}(x,w)$, it suffices to note that:
$$
\sum\limits_{u \in N_{G'}(x,w)}\frac{1}{d_{G'}(u)} = \left(\sum\limits_{u \in N_{G}(x,w)}\frac{1}{d_G(u)}\right) - \frac{1}{d_G(v)} \leq  \sum\limits_{u \in N_{G}(x,w)}\frac{1}{d_G(u)}
$$
\end{itemize}
\end{proof}

\clearpage

\section{The Pseudo-code of CTR}\label{sec:CTR:pseudocode}

\noindent The pseudo-code of CTR is presented in Algorithm~\ref{alg:heuristicCTR-basic}. Specifically, in Line~1, out of all the edges that can be removed (i.e., all the edges in $\FR$), the algorithm narrows the search to only the subset $R'\subseteq \FR$ in which every edge has at least one end that belongs to some non-edge in $\Hide$. After that, in Lines 3 to 13, the algorithm computes for every edge, $(v,w)\in R'$, a score, $\sigma_{(v,w)}$, which reflects the gain from removing $(v,w)$ from the network. More specifically, this score is computed by counting the number of closed triads that contain $(v,w)$ and two other edges, one of which is in $\Hide$. The edge with the greatest gain is chosen in Line~14, and removed from the network in Line~16. This entire process is repeated until the budget, $b$, runs out.

\IncMargin{1em}
\begin{algorithm}[hbtp]
\LinesNumbered
\KwIn{A network, $(V,E)$, a budget, $b\in\N$, a set of edges that can be removed, $\FR \subseteq E$, and a set of non-edges to be hidden, $\Hide \subset \ER$.}
\smallskip \smallskip
$R' \gets\ 
\left\{ (v,w)\in\FR :
  {\fontsize{12}{12}\selectfont{
  \begin{subarray}{l}
    \big(\exists x\in N(v):(x,v)\in H\big) 
    \vee \big(\exists x\in N(w):(x,w)\in H\big)
  \end{subarray}
  }}
\right\}
$\;
\For{$i=1,\ldots,b$}{
	\For{$(v,w) \in R'$}{
		$\sigma_{(v,w)} \gets\ 0$\;
    }
	\For{$(x,w) \in \Hide$}{
		\For{$v \in N(x,w)$}{
			\If{$\big((v,w) \in E\big) \land \big((v,x) \in E\big)$}{
            	\textbf{if} $(v,w) \in R'$ \ \textbf{then}\ $\sigma_{(v,w)}\ \gets\ \sigma_{(v,w)} + 1$\;
            	\textbf{if} $(v,x) \in R'$ \ \textbf{then}\ \ $\sigma_{(v,x)}\gets\ \sigma_{(v,x)} + 1$\;
			}
		}
	}
	$(v^*,w^*) \gets\ \argmax_{(v,w) \in R'}\sigma_{(v,w)}$\;

    \If{$\sigma_{(v^*,w^*)} > 0$}{
		$E \gets\ E \setminus (v^*,w^*)$\;
	}
}
\caption{\textit{Closed-Triad-Removal (CTR)}}
\label{alg:heuristicCTR-basic}
\end{algorithm}
\DecMargin{1em}

The complexity of such a naive implementation of CTR is $\mathcal{O}(b|\Hide||V|)$. This is because for every non-edge $(x,w)\in\Hide$ (there are $|H|$ such edges), the algorithm updates the score of every $(v,w):v\in N(w)$ (there are at most $|V|$ such edges) and updates the score of every $(v,x):v\in N(x)$ (again there are at most $|V|$ such edges); this process is repeated $b$ times.

Note that when $|\Hide|=\omega(\log(|V|))$, an implementation utilizing a priority queue would be faster, with a complexity of $\mathcal{O}(|\Hide||V|+b|V|\log(|V|))$; see Algorithm~\ref{alg:heuristicCTR-efficient}. More specifically, this implementation utilizes a priority queue such as, e.g., a heap~\cite{thomas2001introduction}. Such a priority queue can be built in time $\mathcal{O}(|\Hide||V|)$. The cost of all operations of extracting an element with maximal score is then $\mathcal{O}(b\log(|\Hide||V|))$, which equals $\mathcal{O}(b\log(|V|))$, since $|\Hide|$ is at most $\Theta(|V|^2)$. However, the cost of updating the scores becomes $\mathcal{O}(b|V|\log(|V|))$, since it can only involve decreasing the scores (decreasing scores is realized by removing an element and adding it with a lower score).

\begin{algorithm}[hbtp]
\SetAlgoNoLine
\KwIn{A network, $(V,E)$, a budget, $b\in\N$, a set of edges that can be removed, $\FR \subseteq E$, and a set of non-edges to be hidden, $\Hide \subset \ER$.}
\smallskip \smallskip $R' \gets\ \left\{(v,w)\in\FR: \big(\exists x\in N(v):(x,v)\in H\big) \vee \big(\exists x\in N(w):(x,w)\in H\big)\right\}$\;
\For{$(v,w) \in R'$}{
	$\sigma_{(v,w)} \gets\ 0$\;
}   

\For{$(x,w) \in \Hide$}{
	\For{$v \in N(x,w)$}{
		\If{$\big((v,w) \in E\big) \land \big((v,x) \in E\big)$}{
           	\textbf{if} $(v,w) \in R'$ \ \textbf{then}\ \ $\sigma_{(v,w)}\gets\ \sigma_{(v,w)} + 1$\;
           	\textbf{if} $(v,x) \in R'$ \ \textbf{then}\ \ $\sigma_{(v,x)}\ \gets\ \sigma_{(v,x)} + 1$\;
		}
	}
}

\For{$i=1,\ldots,b$}{
	$(v^*,w^*) \gets\ \argmax_{(v,w) \in R'}\sigma_{(v,w)}$\;
    \If{$\sigma_{(v^*,w^*)} > 0$}{
		$E = E \setminus (v^*,w^*)$\;
		\For{$z \in N(v^*) \cup N(w^*)$}{
        	\textbf{if} $(z,v^*)\ \in \Hide \land (z,w^*) \in R'$\ \textbf{then}\ \ $\sigma_{(z,w^*)} \gets\ \sigma_{(z,w^*)} - 1$\;
        	\textbf{if} $(z,w^*) \in \Hide \land (z,v^*)\ \in R'$\ \textbf{then}\ \ $\sigma_{(z,v^*)}\ \gets\ \sigma_{(z,v^*)} - 1$\;
		}
	}
}
\caption{\textit{Closed-Triad-Removal} (CTR) with priority queue}
\label{alg:heuristicCTR-efficient}
\end{algorithm}


\section{The Pseudo-code of OTC}\label{sec:OTC:pseudocode}
\noindent The pseudo-code of OTC is presented in Algorithm~\ref{alg:heuristic-basic}. In Line~1, out of all the non-edges that can be added (i.e., all the edges in $\FA$), the algorithm narrows the search to only the subset $A'\subseteq \FA$ in which every non-edge has at least one end that belongs to some non-edge in $\Hide$. In Lines 3 to 9, the algorithm computes for every non-edge $(v,w)\in A'$ a score, $\sigma_{(v,w)}$, which reflects the gain from adding $(v,w)$ to the network. Here, Lines 4 and 5 ensure that the algorithm does not increase the number of common neighbours of some non-edge in $\Hide$, whereas Line~7 counts the non-edges whose number of common neighbours will increase as a result of adding $(v,w)$. In Lines 10 to 12, the algorithm selects the non-edge with the highest score, and adds it to the network if it is beneficial to do so. This entire process is repeated until the budget, $b$, runs out.

\IncMargin{1em}
\begin{algorithm}[hbtp]
\LinesNumbered
\KwIn{A network, $(V,E)$, a budget, $b\in\N$, a set of edges that can be added, $\FA \subseteq \ER\setminus\Hide$, and a set of non-edges to be hidden, $\Hide \subset \ER$.}
\smallskip \smallskip
$A' \gets\ 
\left\{ (v,w)\in\FA : 
  {\fontsize{12}{12}\selectfont{
  \begin{subarray}{l}
    \big(\exists u\in N(v):(u,v)\in H\big) 
    \vee \big(\exists u\in N(w):(u,w)\in H\big)
  \end{subarray}
  }}
\right\}
$\;
\For{$i=1,\ldots,b$}{
 	\For{$(v,w) \in A'$}{
 		\eIf{$\exists\ u \in V: ${\fontsize{12}{12}\selectfont{ 
 		$\begin{subarray}{l} 
 		\big((v,u)\in E \ \wedge\ (w,u)\in\Hide\big) 
 		\vee \big((w,u)\in E \ \wedge\ (v,u)\in\Hide\big)
 		\end{subarray}$}}}{
 			$\sigma_{(v,w)}\gets\ -\infty$\;
 		
 		}{
 			$\sigma_{(v,w)} \gets \left|\big(N_{\widehat{G}}(v) \cup N_{\widehat{G}}(w)\big)  \setminus  N_{\widehat{G}}(v,w)\right|$, where $\widehat{G}= (V,E)$\;
 		}
 	}
 	$(v^*,w^*) \gets\ \argmax_{(v,w) \in A'}\sigma_{(v,w)}$\;
 	\If{$\sigma_{(v^*,w^*)} > -\infty$}{
 		$E \gets\ E \cup (v^*,w^*)$\;
 	}
 }
\caption{\textit{Open-Triad-Creation (OTC)}}
\label{alg:heuristic-basic}
\end{algorithm}
\DecMargin{1em}

The complexity of such a naive implementation of OTC is $\mathcal{O}(b|\Hide||V|^2)$. In more detail, computing a score, $\sigma_{(v,w)}$, for each non-edge, $(v,w)\in A'$, can be done in time linear in $|V|$ for each of the $|\Hide||V|$ non-edges. Searching for a non-edge in $A'$ with the maximal score takes $b|\Hide||V|$ operations. Finally, updating the scores after adding each of the $b$ edges can be done in time linear in $|V|$. Next, we will present a more efficient implementation, the complexity of which is $\mathcal{O}(|\Hide||V|^2+b|V|\log(|V|))$ when using a priority queue, and is $\mathcal{O}(|\Hide||V|^2+b|\Hide||V|)$ without a priority queue.


Algorithm~\ref{alg:heuristic-efficient} presents a more efficient implementation of OTC compared to Algorithm~\ref{alg:heuristic-basic}. The complexity of this implementation is $\mathcal{O}(|\Hide||V|^2+b|\Hide||V|)$. Here, the $|\Hide||V|^2$ term comes from computing an initial score, $\sigma_{(v,w)}$, for each non-edge, $(v,w)\in A'$ (this can be done in time linear in $|V|$ for each of the $|\Hide||V|$ non-edges). The $b|\Hide||V|$ term comes from searching for a non-edge in $\FA$ with the maximal score. Finally, updating the scores after adding each of the $b$ edges can be done in time linear in $|V|$.

Notice that when $|\Hide|=\omega(\log(|V|))$ and $b = \omega(|V|)$, an implementation utilizing a priority queue (such as, e.g., a heap~\cite{thomas2001introduction}) would be faster. The complexity of such an implementation is $\mathcal{O}(|\Hide||V|^2+b|V|\log(|V|))$. In more detail, a priority queue can be built in time $\mathcal{O}(|\Hide||V|^2)$. The cost of all operations of extracting an element with maximal score is then $\mathcal{O}(b\log(|\Hide||V|))$, which equals $\mathcal{O}(b\log(|V|))$, since $|\Hide|$ is at most $\Theta(|V|^2)$. However, the cost of updating the scores is now $\mathcal{O}(b|V|\log(|V|))$, since it could involve either increasing or decreasing the scores (decreasing scores is realized by removing an element and adding it with a lower score).

\begin{algorithm}[hbtp]
\SetAlgoNoLine
\KwIn{A network, $(V,E)$, a budget, $b\in\N$, a set of edges that can be added, $\FA \subseteq \ER\setminus\Hide$, and a set of non-edges to be hidden, $\Hide \subset \ER$.}
\smallskip \smallskip $A' \gets\ \left\{(v,w)\in\FA: \big(\exists u\in N(v):(u,v)\in H\big) \vee \big(\exists u\in N(w):(u,w)\in H\big)\right\}$\;
\For{$(v,w) \in A'$}{
	\eIf{$\exists_{u \in V}\big((v,u) \in E \land (w,u) \in \Hide\big) \lor \big((w,u) \in E \land (v,u) \in H\big)$}{
		$\sigma_{(v,w)}\gets\ -\infty$\;
	}{
		$\sigma_{(v,w)} \gets\  \left|\big(N_{(V,E)}(v) \ \cup\ N_{(V,E)}(w)\big)  \setminus  N_{(V,E)}(v,w)\right|$\;
	}
}
\For{$i=1,\ldots,b$}{
	$(v^*,w^*) \gets\ \argmax_{(v,w) \in A'}\sigma_{(v,w)}$\;
	\If{$\sigma_{(v^*,w^*)} > -\infty$}{
		$E \gets\ E \cup (v^*,w^*)$\;
		\For{$u \in V \setminus \{v^*,w^*\}$}{
			\If{$(u,v^*) \in A'$}{
				\textbf{if} $(u,w^*) \in \Hide$ \hspace*{2.51cm} \textbf{then} $\sigma_{(u,v^*)}\gets\ -\infty$\;
				\textbf{if} $(u,w^*) \in E \land \sigma_{(u,v^*)} > -\infty$ \hspace*{0.25cm} \textbf{then} $\sigma_{(u,v^*)}\gets\ \sigma_{(u,v^*)}-1$\;
				\textbf{if} $(u,w^*) \in \ER \land \sigma_{(u,v^*)} > -\infty$ \hspace*{0.25cm} \textbf{then} $\sigma_{(u,v^*)}\gets\ \sigma_{(u,v^*)}+1$\;
			}
			\If{$(u,w^*) \in A'$}{
				\textbf{if} $(u,v^*) \in \Hide$ \hspace*{2.57cm} \textbf{then} $\sigma_{(u,w^*)}\gets\ -\infty$\;
				\textbf{if} $(u,v^*) \in E \land \sigma_{(u,w^*)} > -\infty$ \hspace*{0.25cm} \textbf{then} $\sigma_{(u,w^*)}\gets\ \sigma_{(u,w^*)}-1$\;
				\textbf{if} $(u,v^*) \in \ER \land \sigma_{(u,w^*)} > -\infty$ \hspace*{0.25cm} \textbf{then} $\sigma_{(u,w^*)}\gets\ \sigma_{(u,w^*)}+1$\;
			}
		}
	}
}
\caption{A more efficient implementation of \textit{Open-Triad-Creation (OTC)}}
\label{alg:heuristic-efficient}
\end{algorithm}

\clearpage

\section{Illustrating the Workings of CTR on the 9/11 Terrorist Network}\label{sec:WTC}

\begin{figure}[ht!]
 \centering
 \includegraphics[width=0.65\columnwidth]{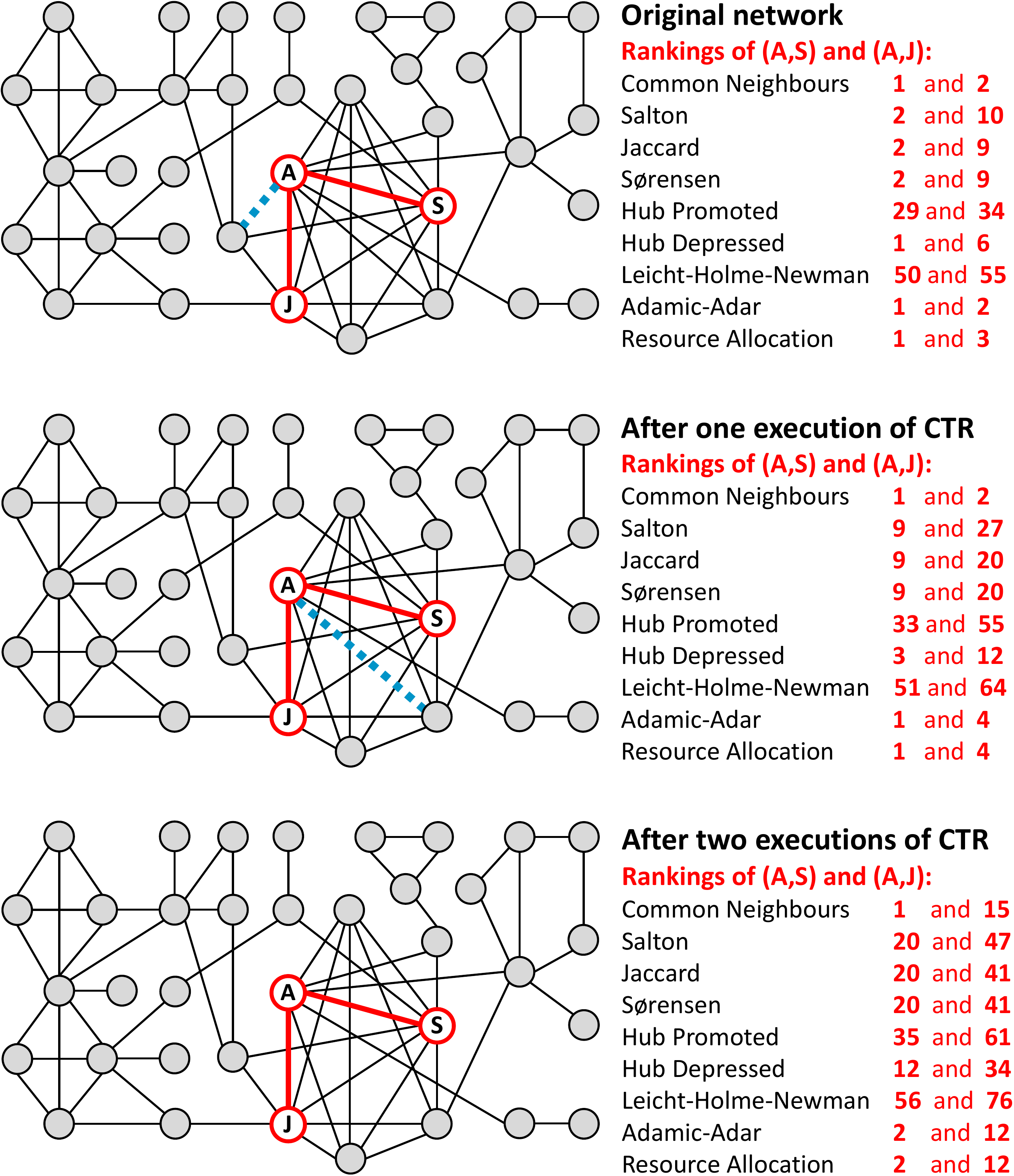}
\caption{Executing the CTR heuristic twice on the WTC 9/11 terrorist network in order to hide the links between Mohamed Atta---a hijacker-pilot and one of the ringleaders of the attack---and two other hijacker-pilots, namely Marwan al-Shehhi (node ``S'') and Ziad Jarrah (node ``J''). The red links are the ones to be to hidden, and the dashed links are the ones removed by the heuristic. The figure shows how the ranking of the red links (according to different link-prediction algorithms) decreases after each execution of the heuristic.}
 \label{fig:wtc}
\end{figure}

\noindent This section illustrates the workings of the CTR heuristic on the WTC 9/11 terrorist network, where the goal is to hide the links between Mohamed Atta---a hijacker-pilot and one of the ringleaders of the attack---and two other hijacker-pilots, namely Marwan al-Shehhi (node ``S'') and Ziad Jarrah (node ``J''). As shown in Figure~\ref{fig:wtc}, avoiding just a couple of contacts (the dashed links) can significantly alter the likelihood of a network analyzer exposing the links between those hijacker-pilots.

\clearpage
\section{Experimental Evaluation}\label{sec:allFigures}

\subsection{Networks Considered in Our Study}\label{sec:networks}

\noindent In our experiments, we considered both real-life networks as well as randomly-generated networks. As for the latter ones, they were generated using the following standard models:

\begin{itemize}[leftmargin=*]
\item \emph{Scale-free} networks, generated using the Barabasi-Albert model \cite{barabasi1999emergence}: We denote such a network by $\mathit{ScaleFree}(n,d)$, where $n$ is the number of nodes and $d$ is the number of links added with each node;
\item \emph{Small-world} networks, generated using the Watts-Strogatz model \cite{watts1998collective}: We denote every such network by $\mathit{SmallWorld}(n,d,p)$, where $n$ is the number of nodes, $d$ is the average degree, and $p$ is the rewiring probability;
\item \emph{Random graphs}, generated using the Erdos-Renyi model \cite{erdds1959random}: We denote every such network by $\mathit{RandomGraph}(n,d)$, with $n$ being the number of nodes, and $d$ being the expected average degree.
\end{itemize}

\noindent Next, we describe the real-life networks used in our experiments:

\begin{itemize}[leftmargin=*]
\item Facebook~\cite{leskovec2012learning}: we consider three fragments of Facebook's social network: (i) a ``small'' fragment consisting of 61 nodes and 272 edges; (ii) a ``medium'' fragment consisting of 333 nodes and 2,523 edges; and (iii) a ``large'' fragment consisting of 786 nodes and 14,027 edges;
\item Madrid terrorist network~\cite{hayes2006connecting}---the network of terrorists behind the 2004 Madrid bombing, consisting of 70 nodes and 98 edges;
\item Bali terrorist network~\cite{hayes2006connecting}---the network of terrorists behind the 2002 Bali attack, consisting of 17 nodes and 63 edges;
\item WTC terrorist network~\cite{krebs2002mapping}---the network of terrorists behind the 9/11 attacks, consisting of 36 nodes and 64 edges;
\item Zachary's Karate Club~\cite{zachary1977information}---the social network of participants of a university karate club, consisting of 34 nodes and 78 edges;
\item Les Mis\'erables~\cite{knuth1993stanford}---the network of co-occurances of characters in Victor Hugo's novel ``Les Mis\'erables'', consisting of 77 nodes and 254 edges;
\item Greek blogs~\cite{zafiropoulos2012connectivity}---a network of Greek political blogs, consisting of 142 nodes and 354 edges.
\end{itemize}

\begin{figure*}[ht!]
\centering
\setlength\tabcolsep{1pt}
\begin{tabular}{m{.03\textwidth}m{.40\textwidth}m{.40\textwidth}}
&
\multicolumn{1}{c}{OTC} &
\multicolumn{1}{c}{CTR} \\
\rotatebox{90}{\hspace*{1.7cm} Relative change in $\ROC$} &
\includegraphics[width=1\linewidth]{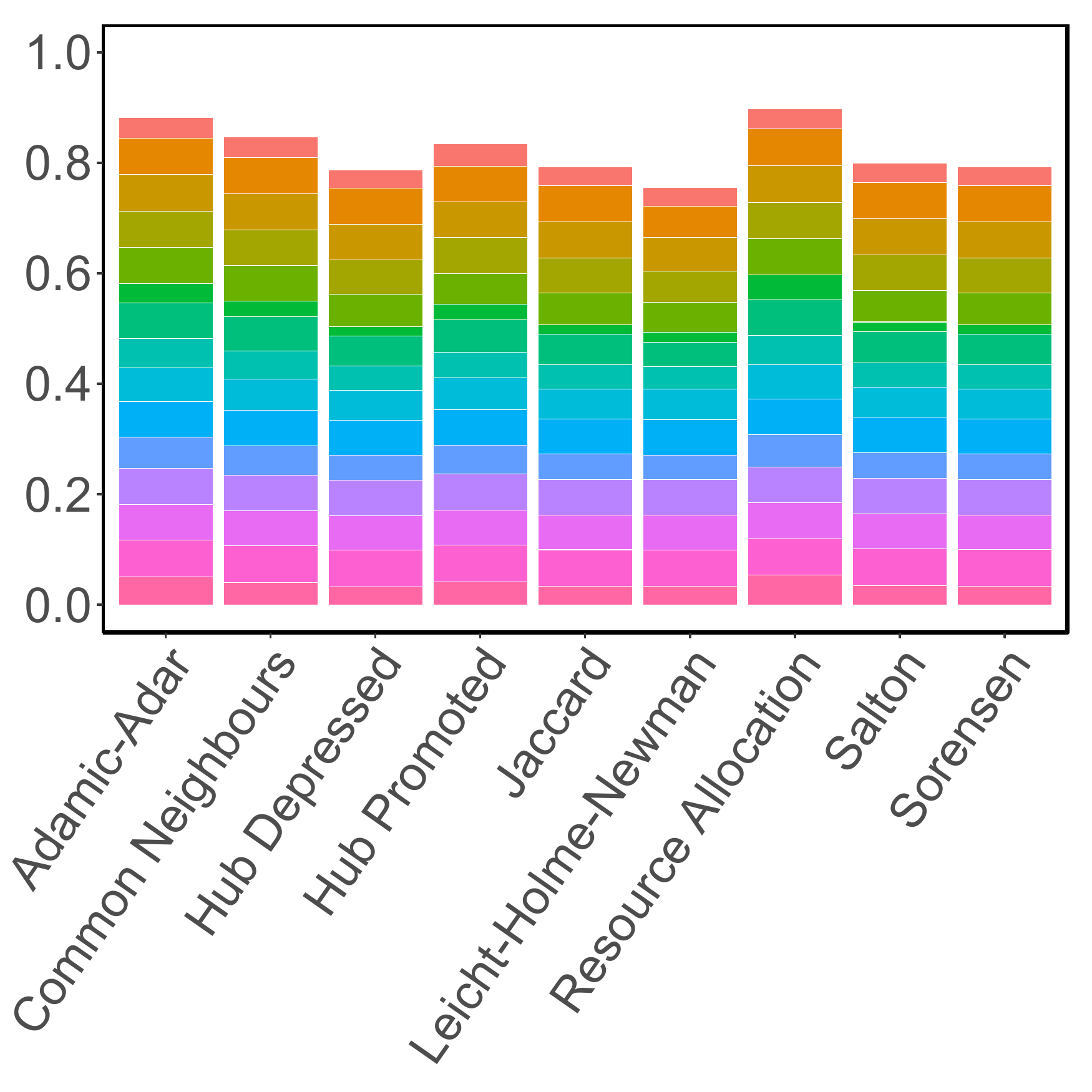} &
\includegraphics[width=1\linewidth]{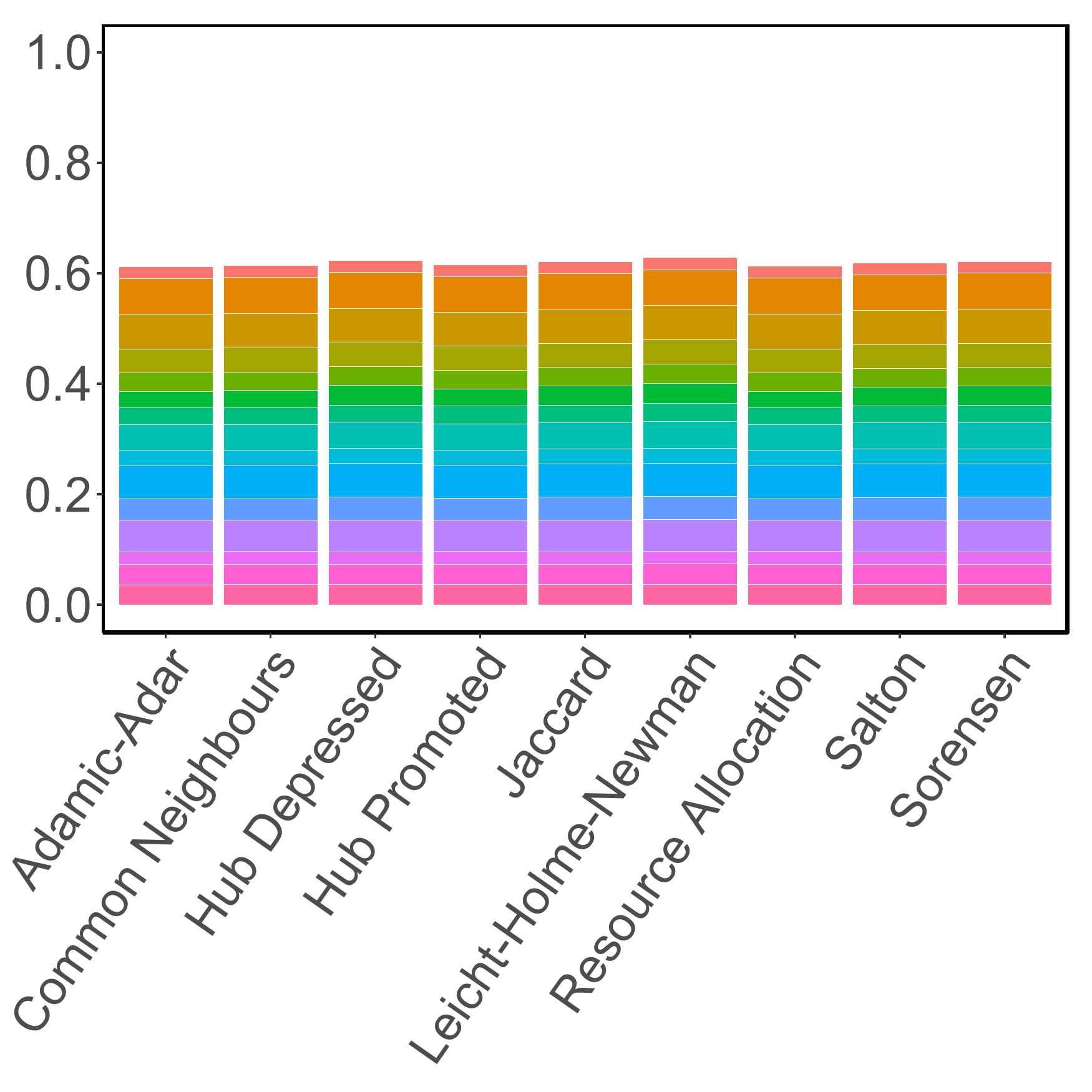} \\
\rotatebox{90}{\hspace*{1.7cm} Relative change in $\AP$} &
\includegraphics[width=1\linewidth]{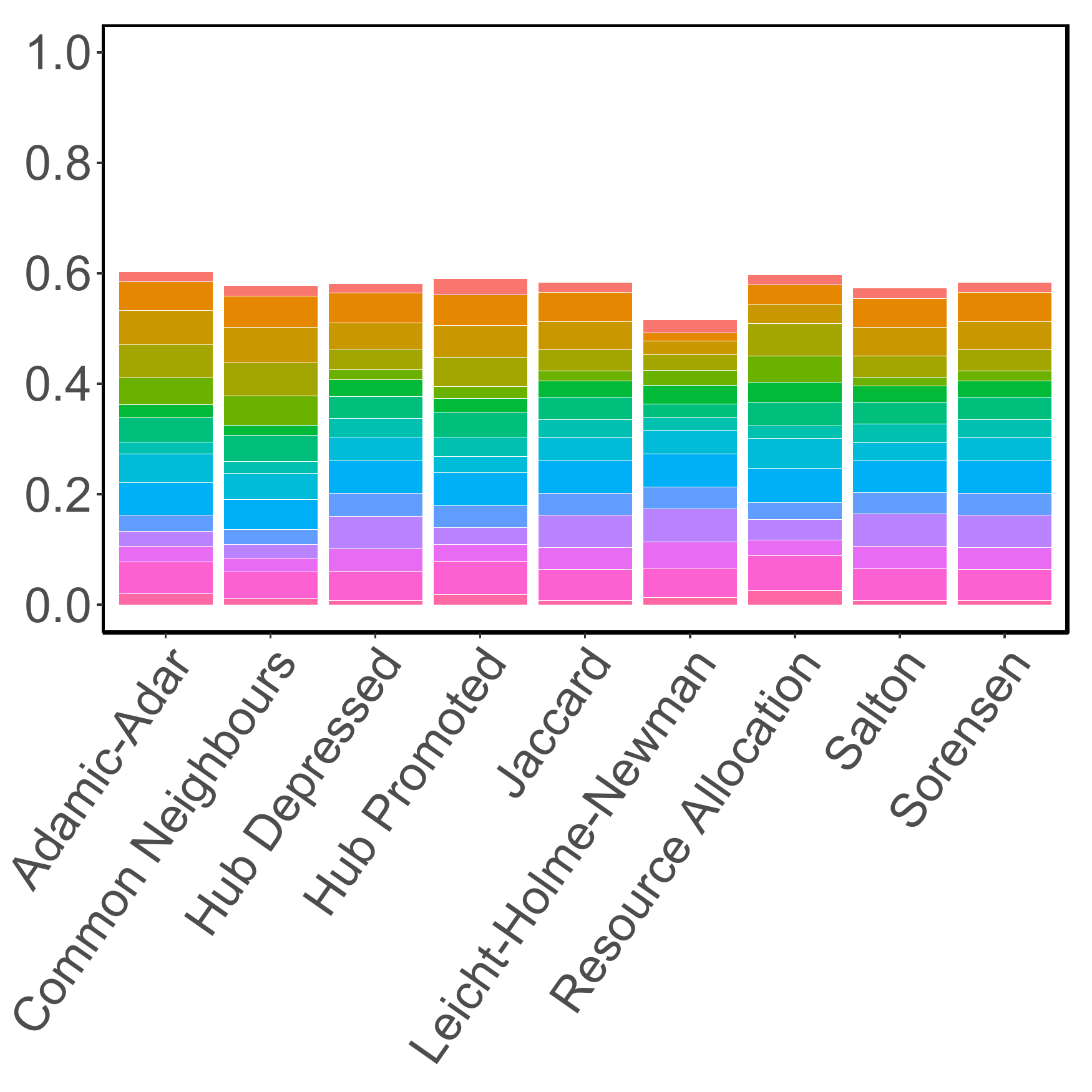} &
\includegraphics[width=1\linewidth]{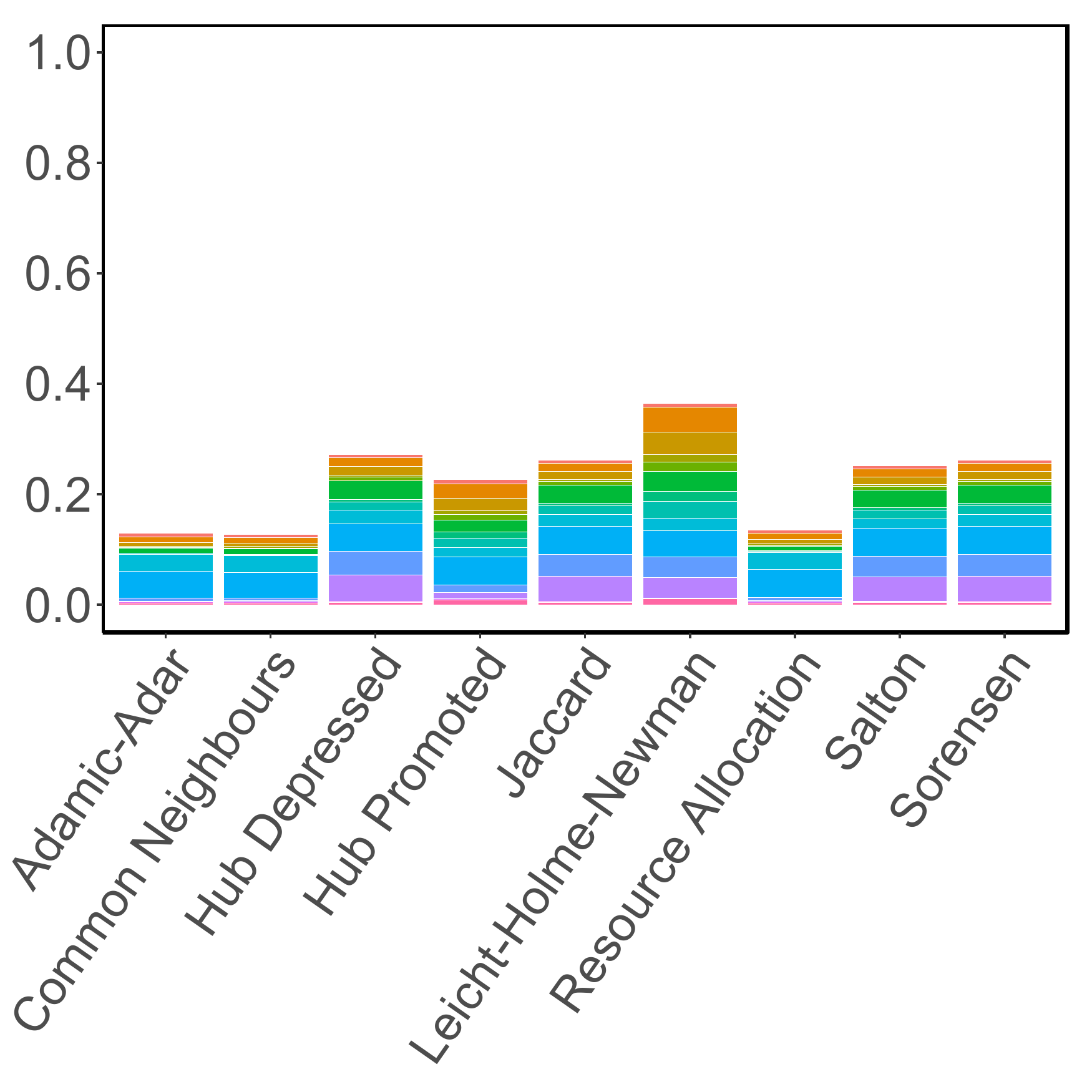} \\
\multicolumn{3}{c}{\includegraphics[width=0.8\linewidth]{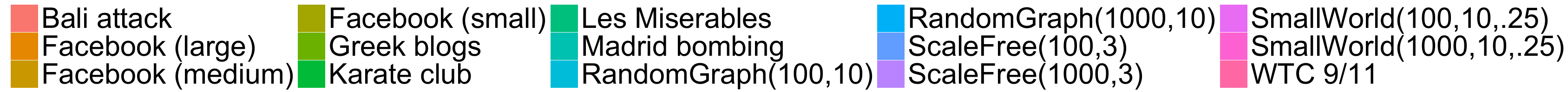}}
\end{tabular}
\caption{Given different \textbf{local similarity} indices, the figure depicts the relative change in $\ROC$ (the area under the ROC curve) and $\AP$ (the average precision) after running OTC and CTR in different networks, where $|\Hide|=\max(10,|E|/100)$ and $b=4|\Hide|$, and the links in $\Hide$ are chosen at random. For each similarity index, the height of the corresponding bar represents the average change taken over all networks, and the height of each segment in that bar is proportional to the change within the corresponding network.}
\label{fig:bars-relative-local-supplementary}
\end{figure*}

\begin{figure*}[ht!]
\centering
\setlength\tabcolsep{1pt}
\renewcommand{\arraystretch}{0.01}
\begin{tabular}{m{.03\textwidth}m{.27\textwidth}m{.27\textwidth}m{.27\textwidth}}
& \multicolumn{1}{c}{WTC 9/11 network}
& \multicolumn{1}{c}{ScaleFree$(100,3)$}
& \multicolumn{1}{c}{Facebook (medium)}\\
\rotatebox{90}{\footnotesize $\ROC$ values for OTC} &
\includegraphics[width=1.03\linewidth]{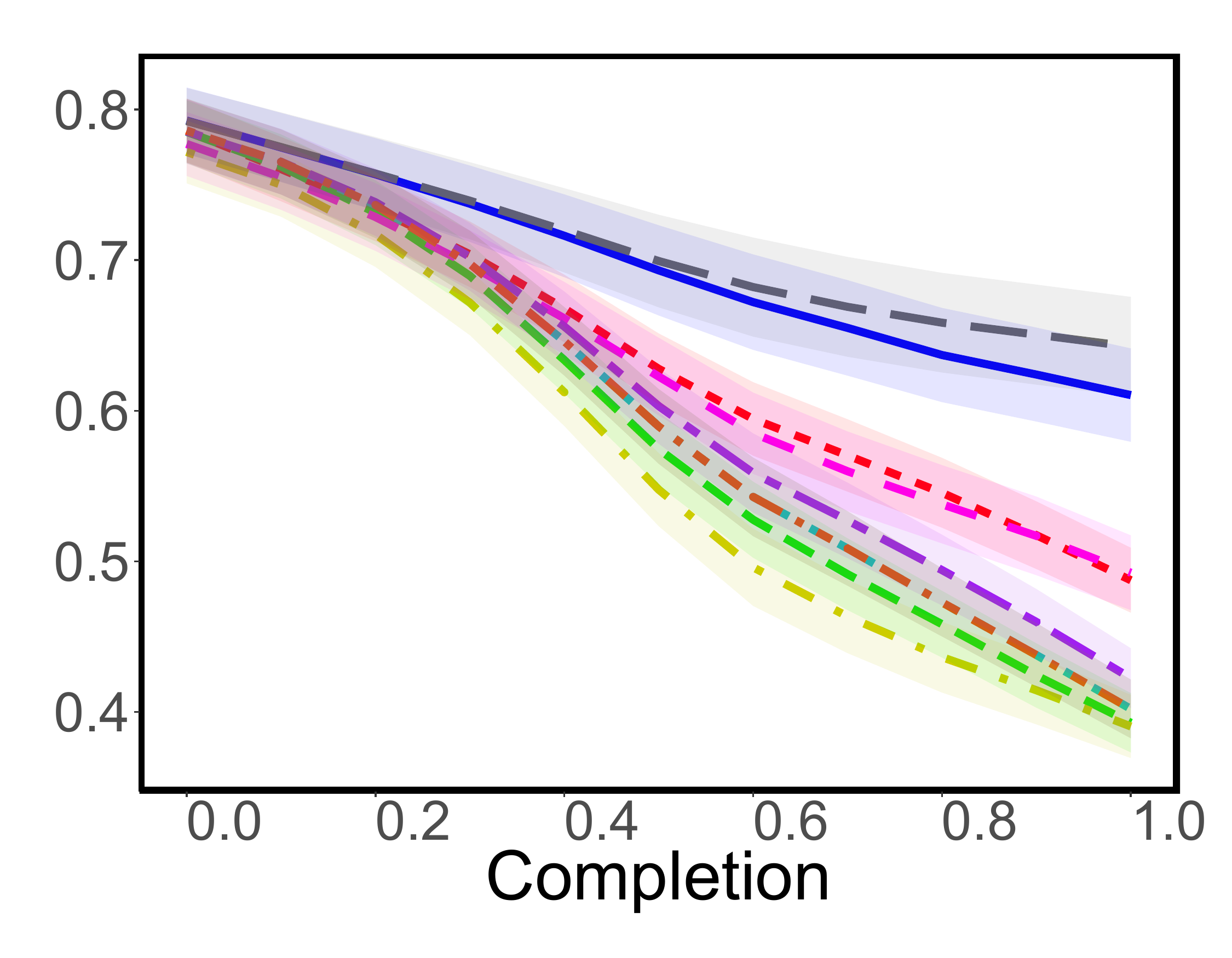} &
\includegraphics[width=1.03\linewidth]{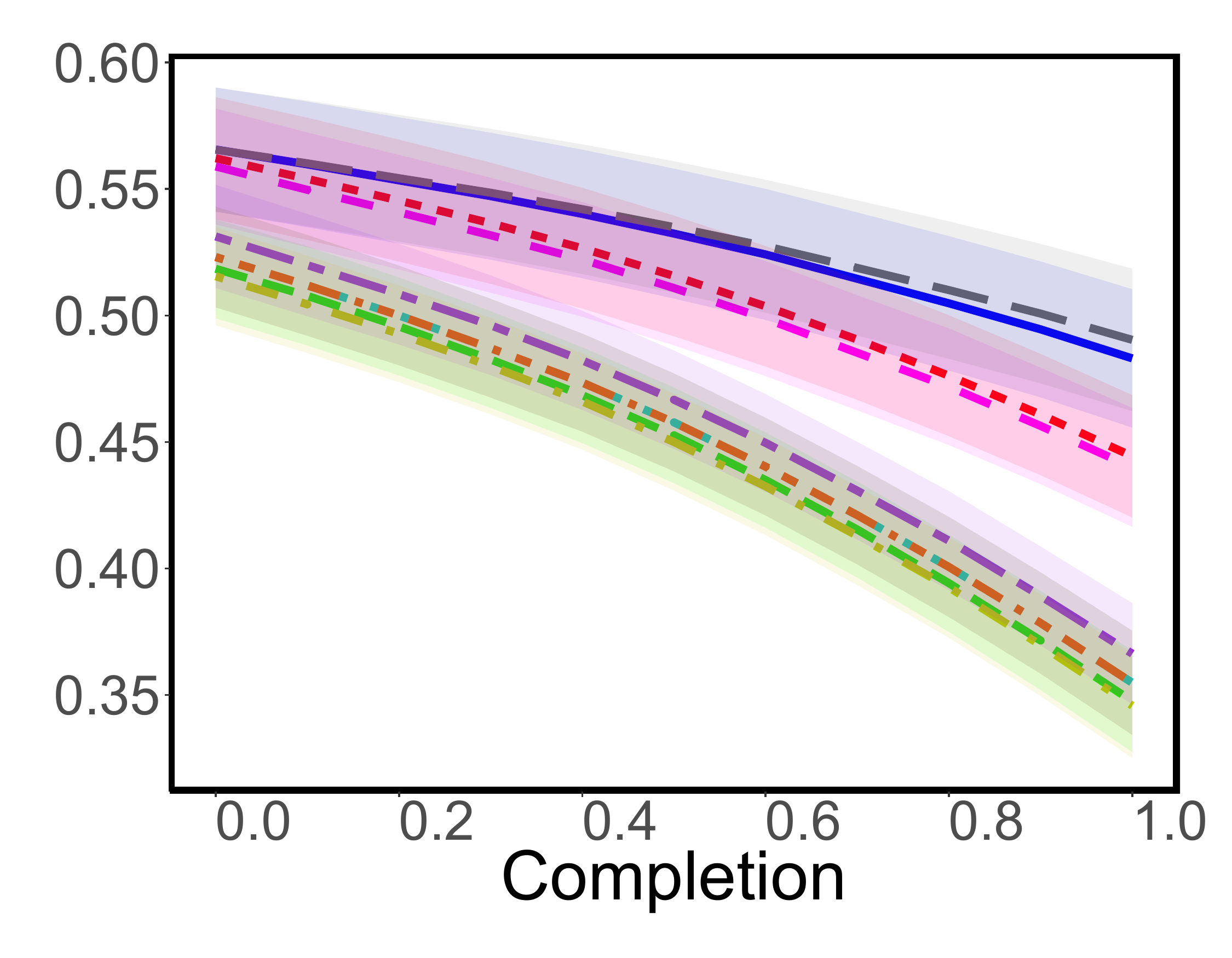} &
\includegraphics[width=1.03\linewidth]{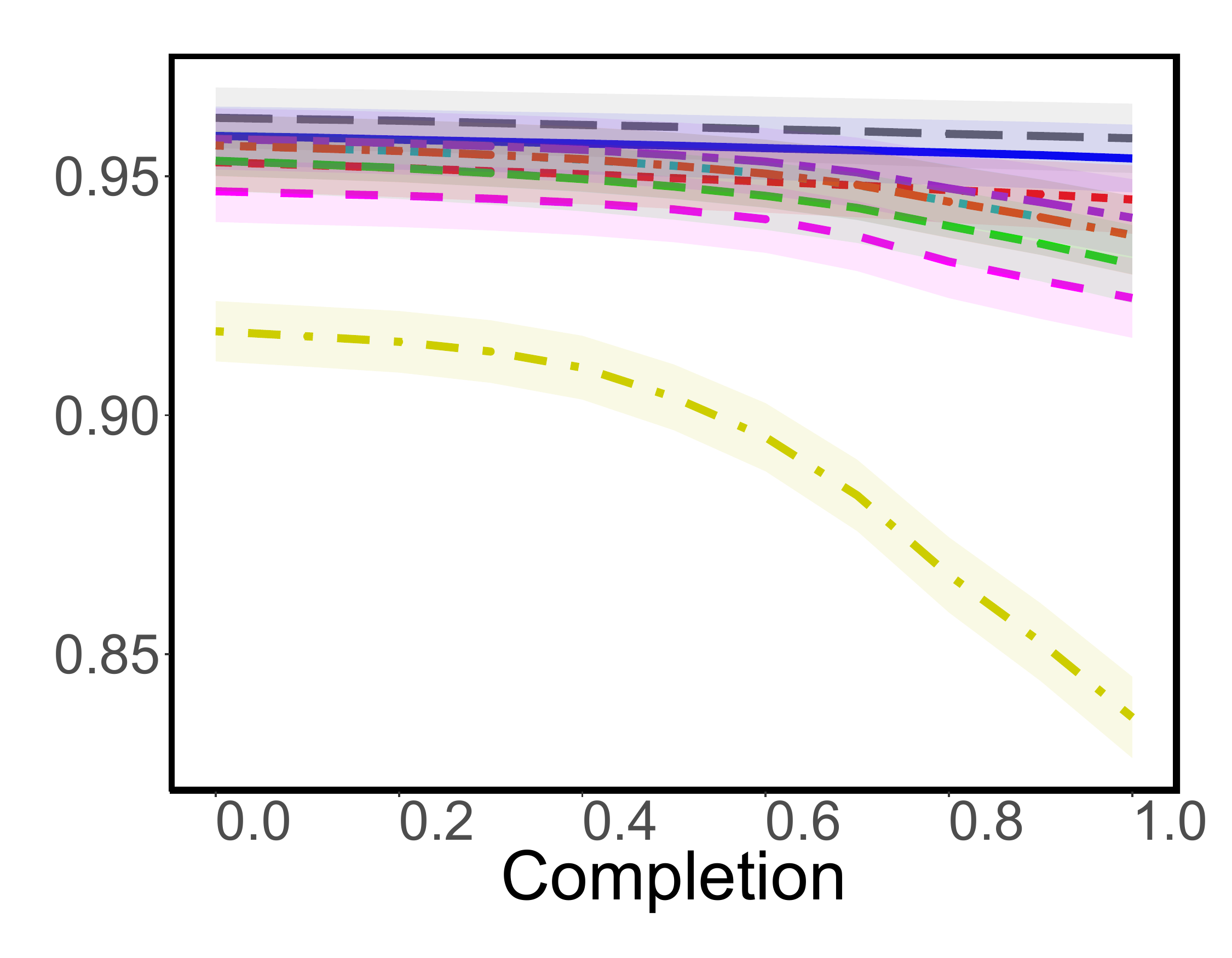}\\
\rotatebox{90}{\footnotesize $\ROC$ values for CTR} &
\includegraphics[width=1.03\linewidth]{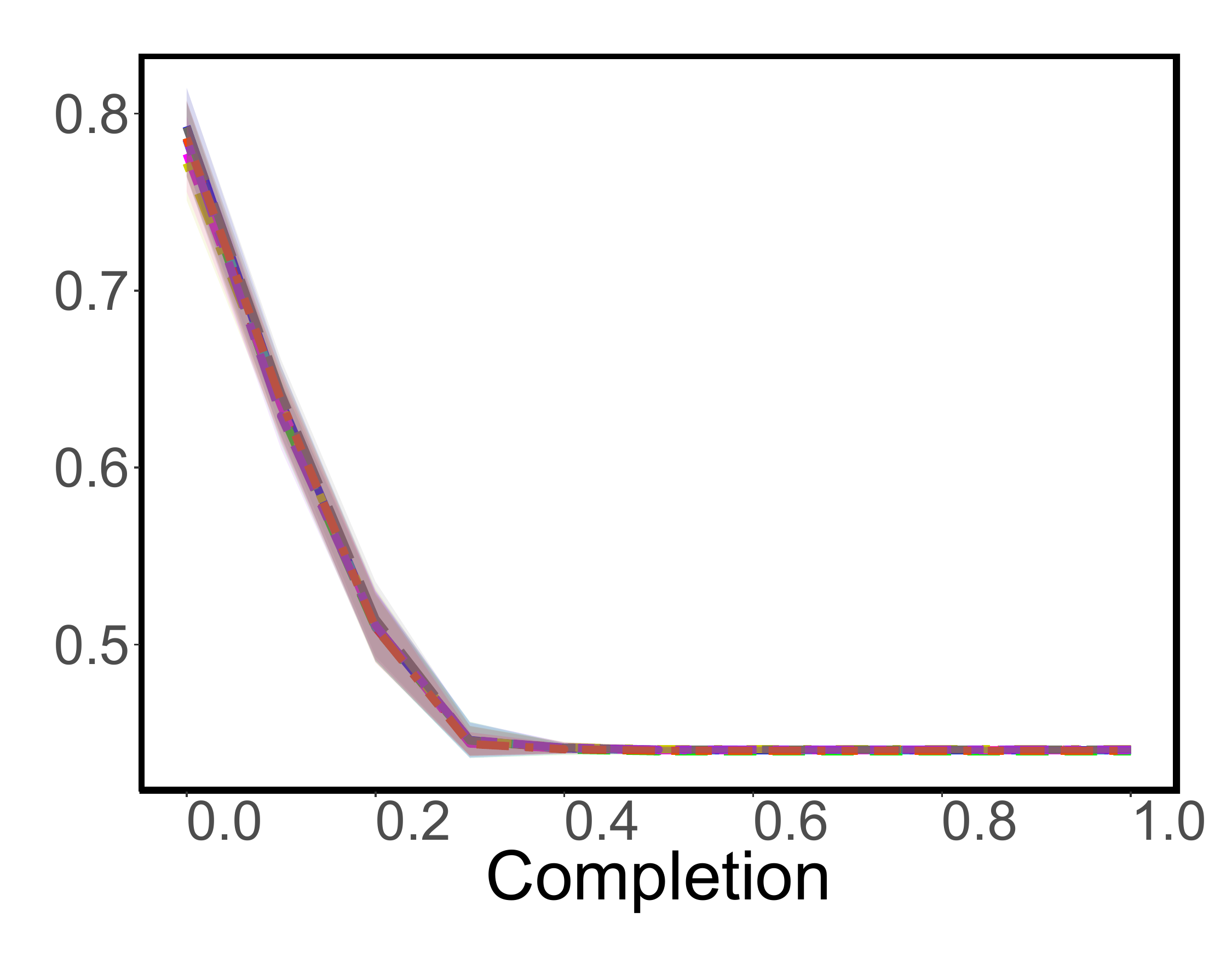} &
\includegraphics[width=1.03\linewidth]{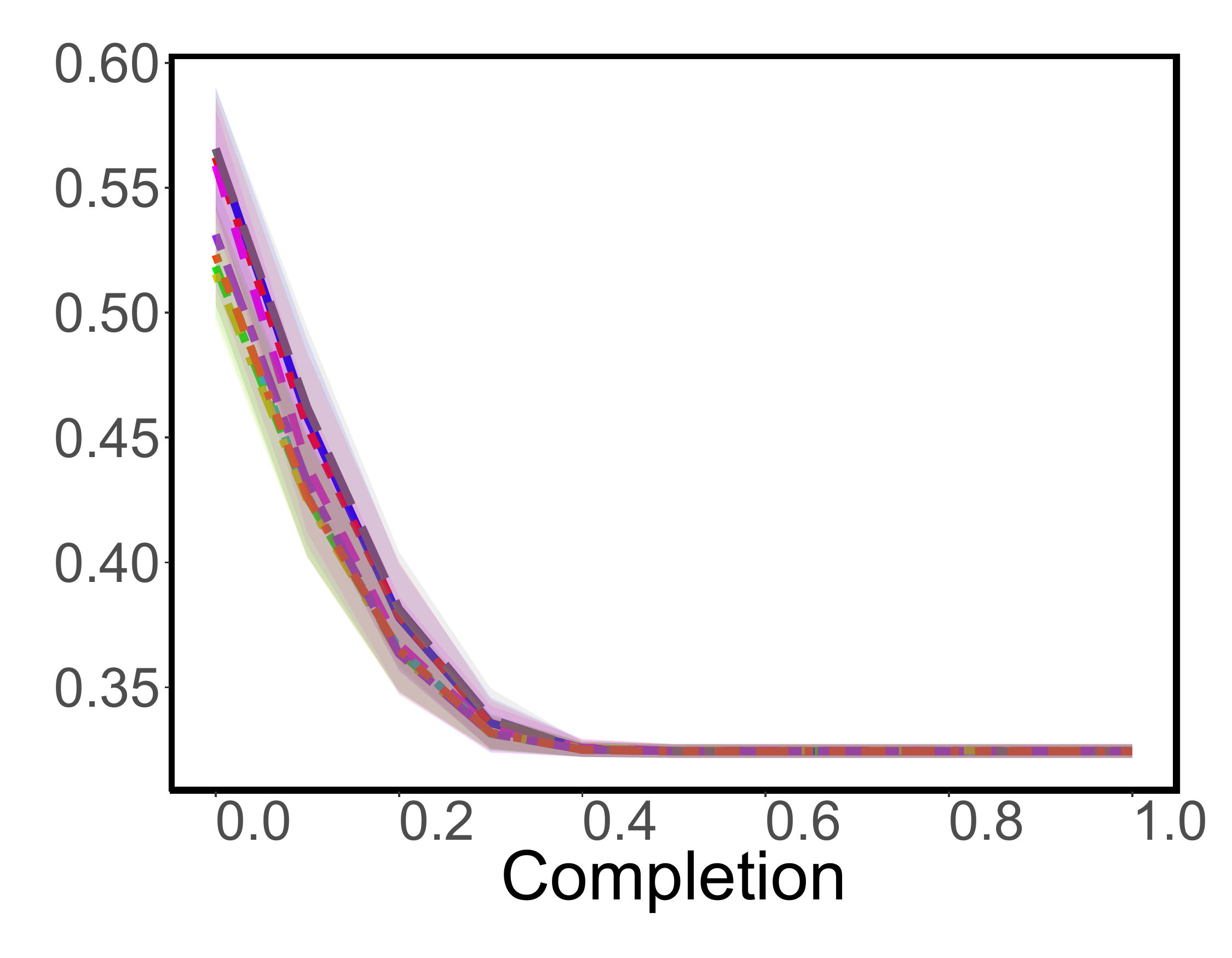} &
\includegraphics[width=1.03\linewidth]{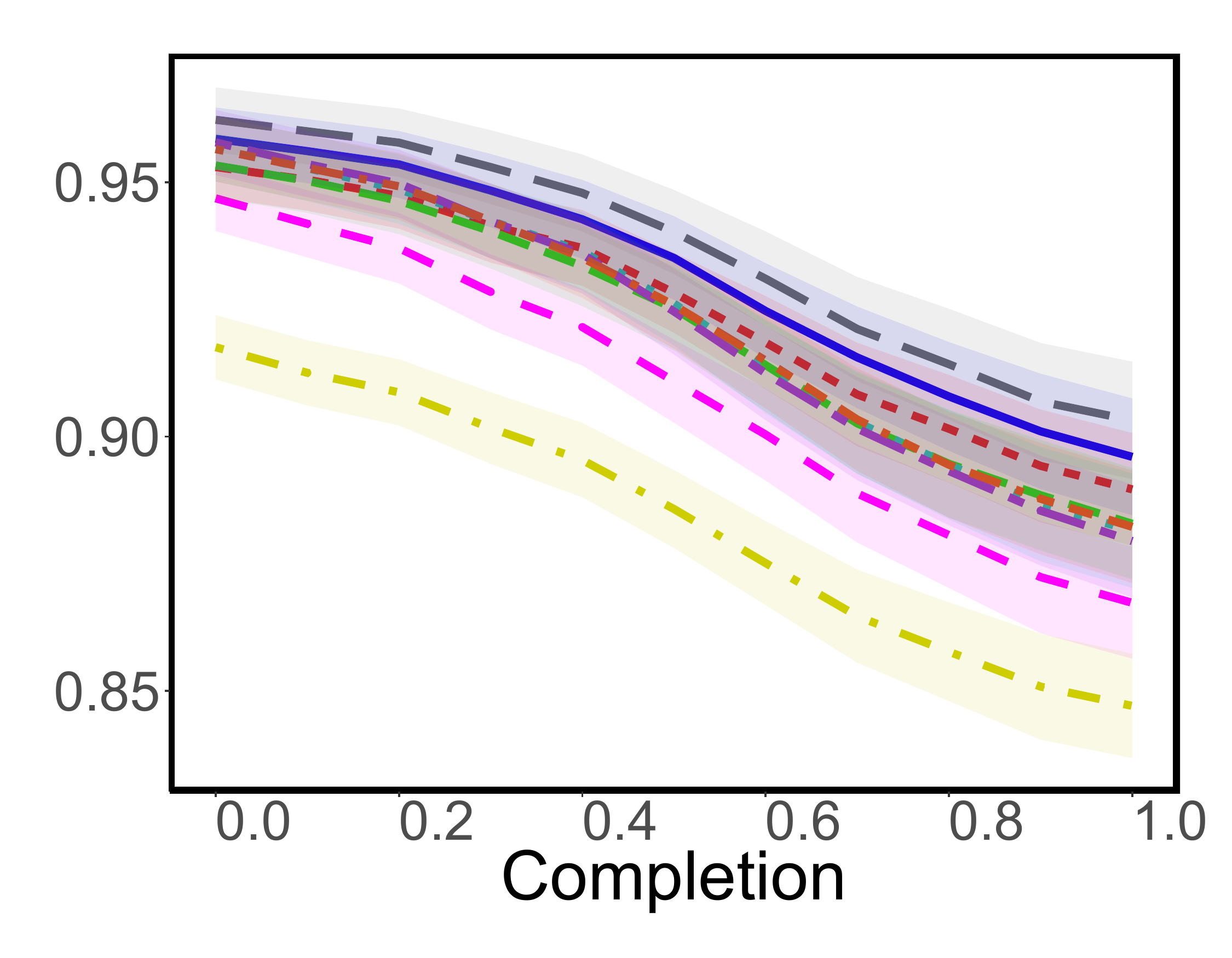} \\
\rotatebox{90}{\footnotesize $\AP$ values for OTC} &
\includegraphics[width=1.03\linewidth]{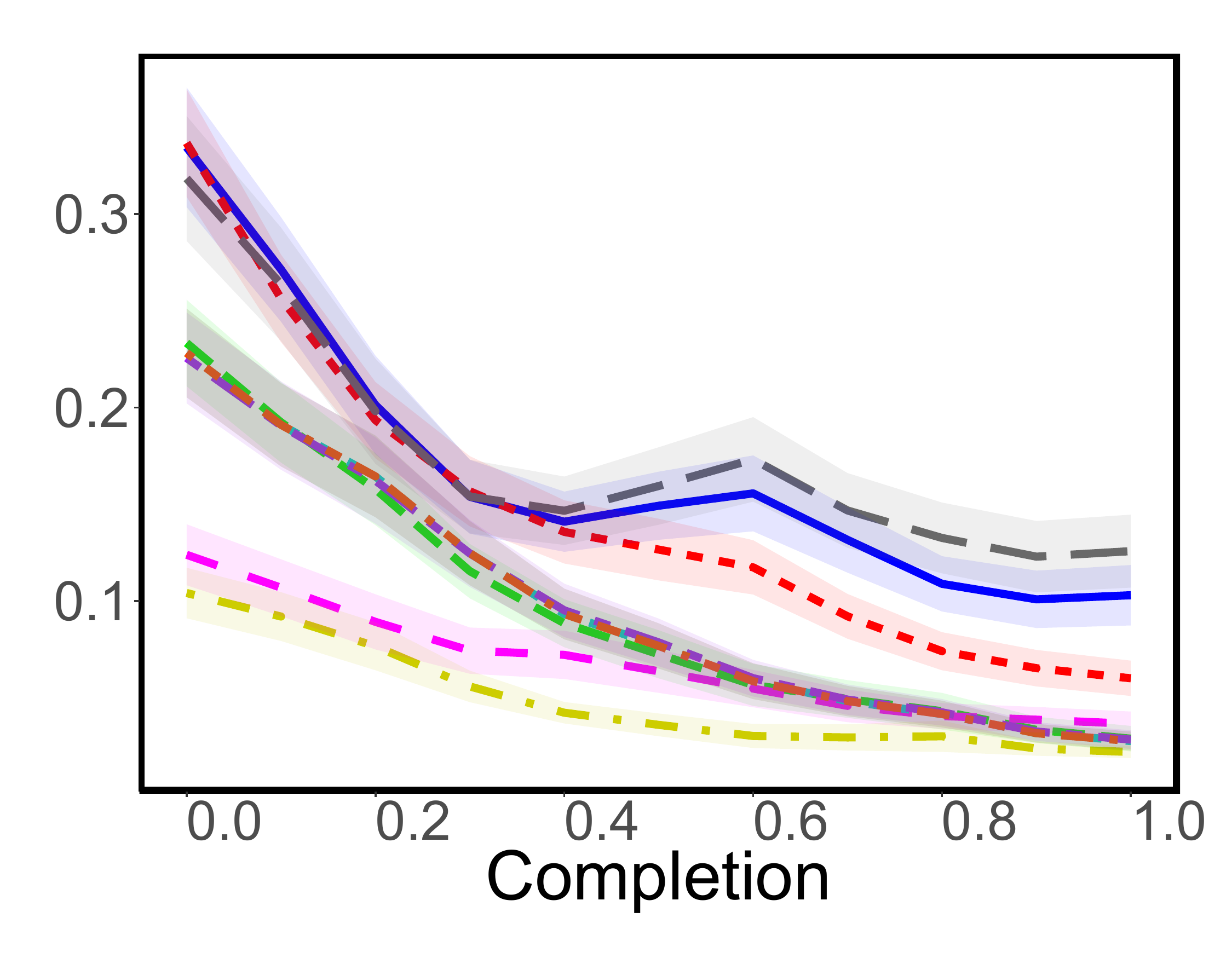} &
\includegraphics[width=1.03\linewidth]{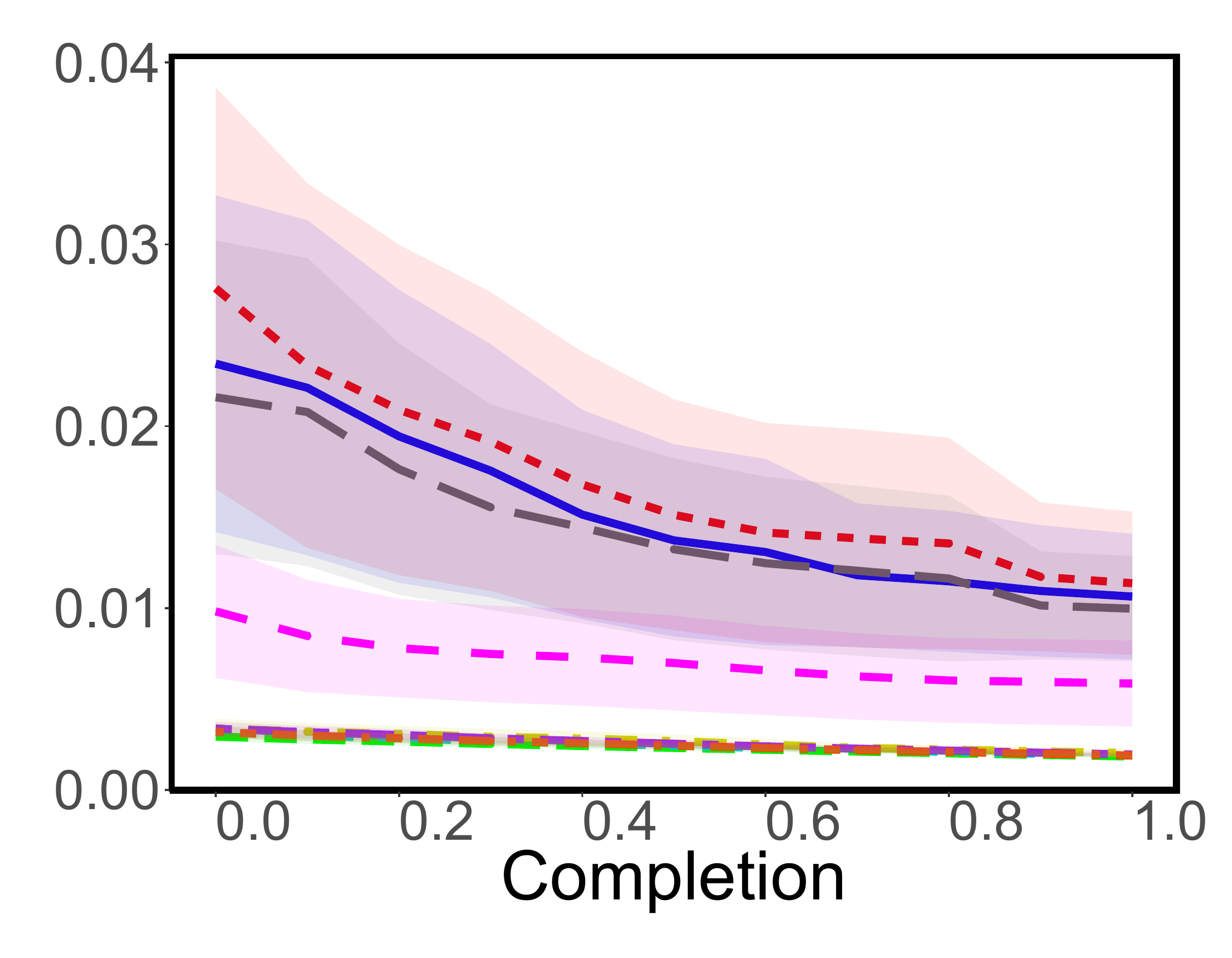} &
\includegraphics[width=1.03\linewidth]{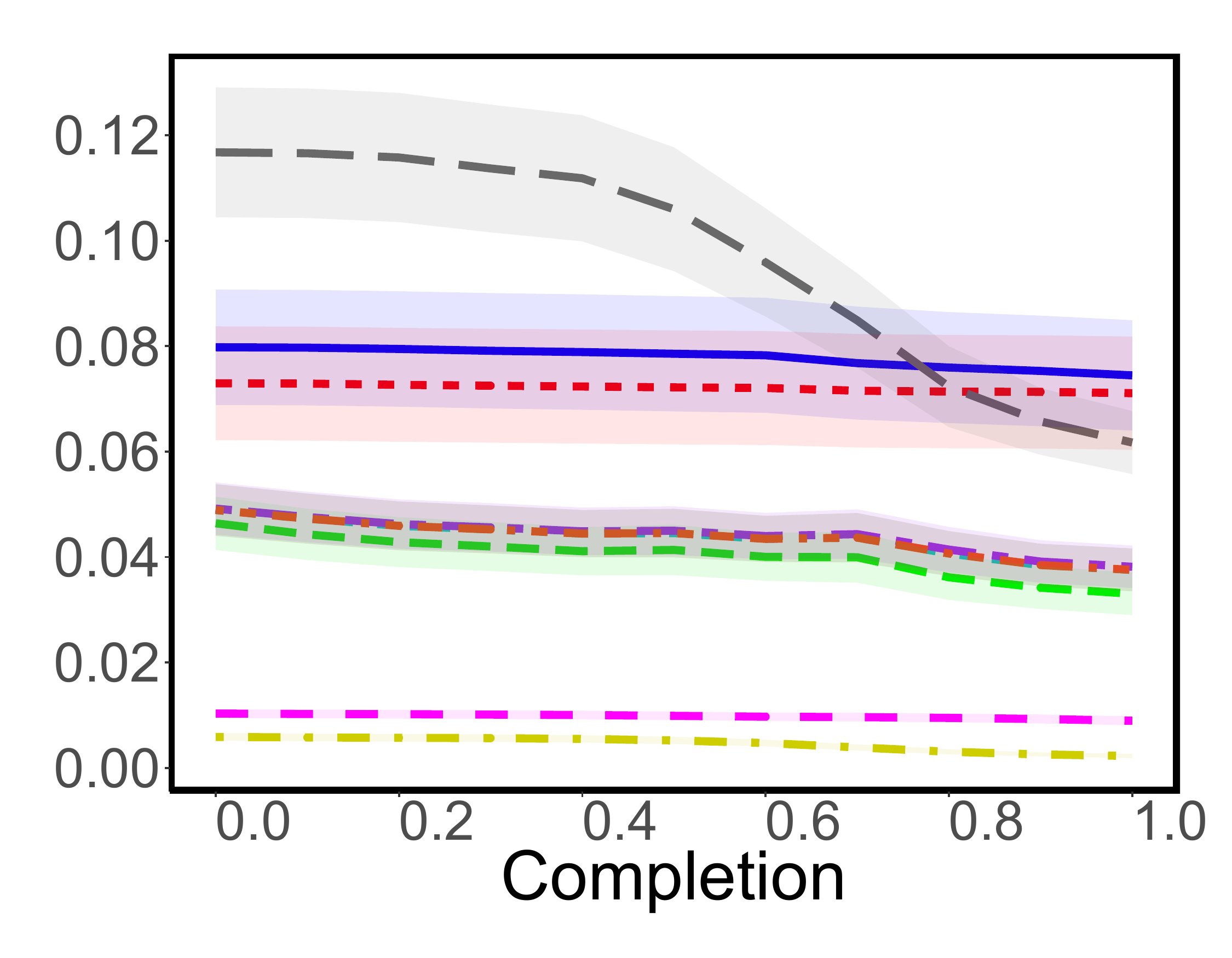} \\
\rotatebox{90}{\footnotesize $\AP$ values for CTR} &
\includegraphics[width=1.03\linewidth]{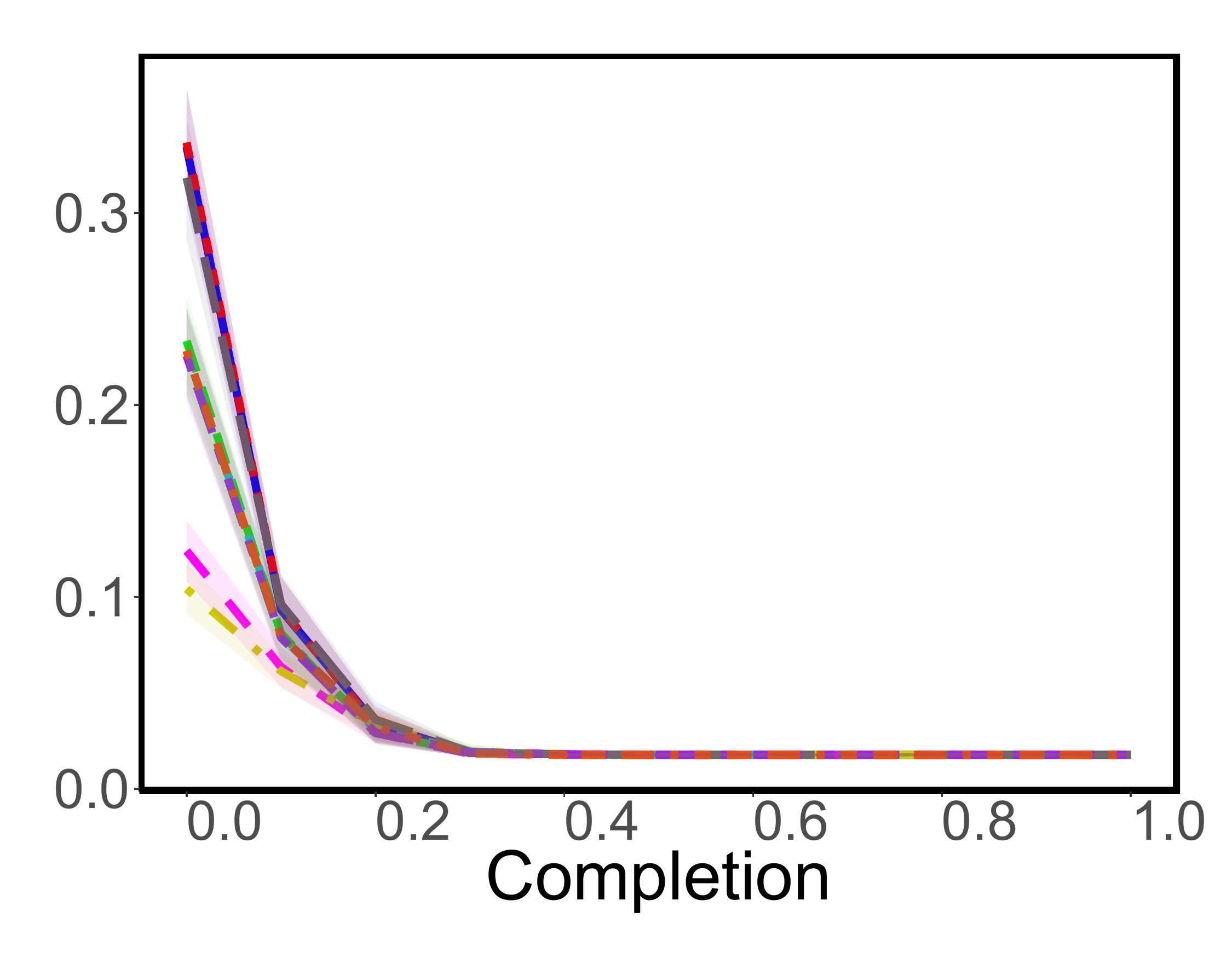} &
\includegraphics[width=1.03\linewidth]{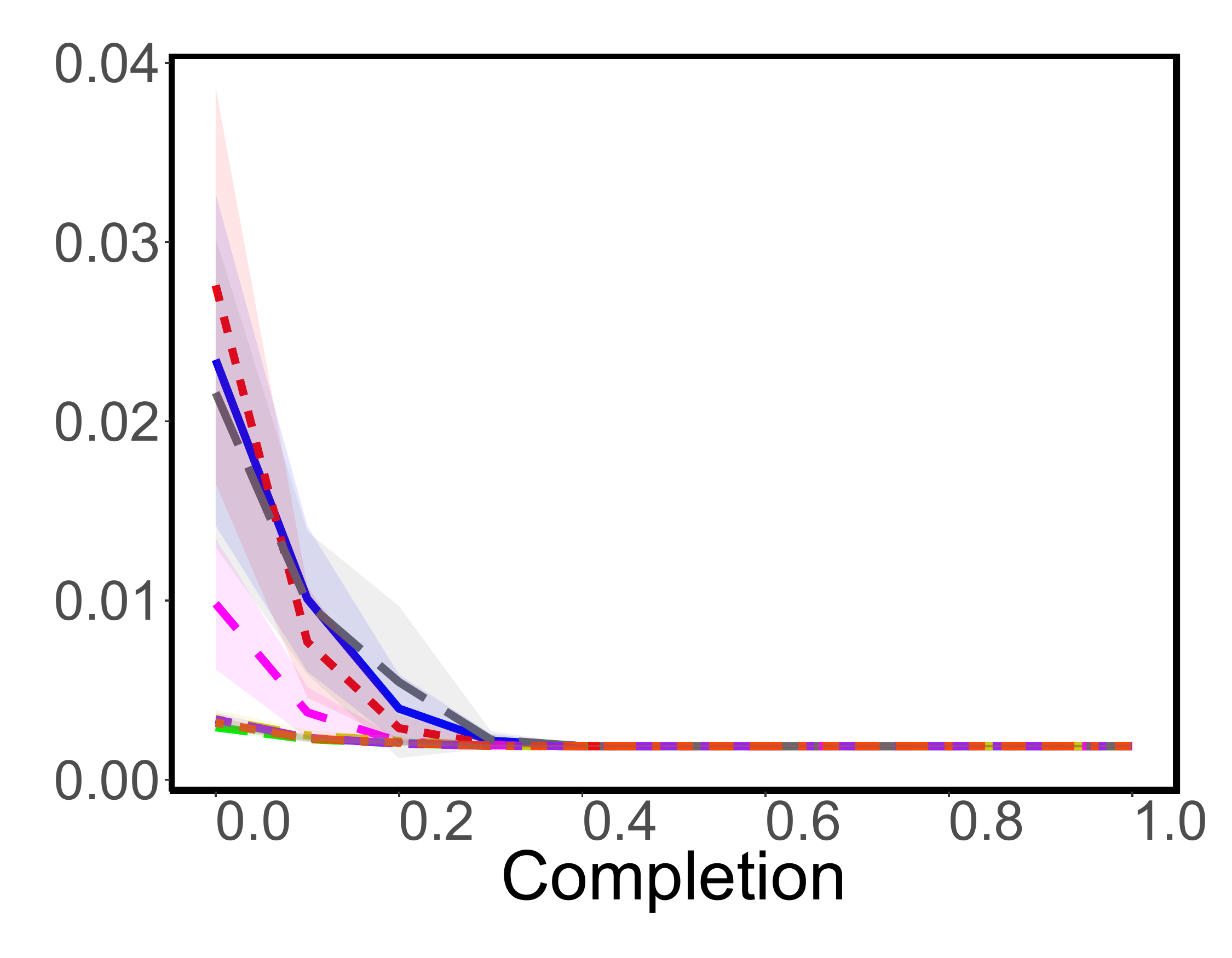} &
\includegraphics[width=1.03\linewidth]{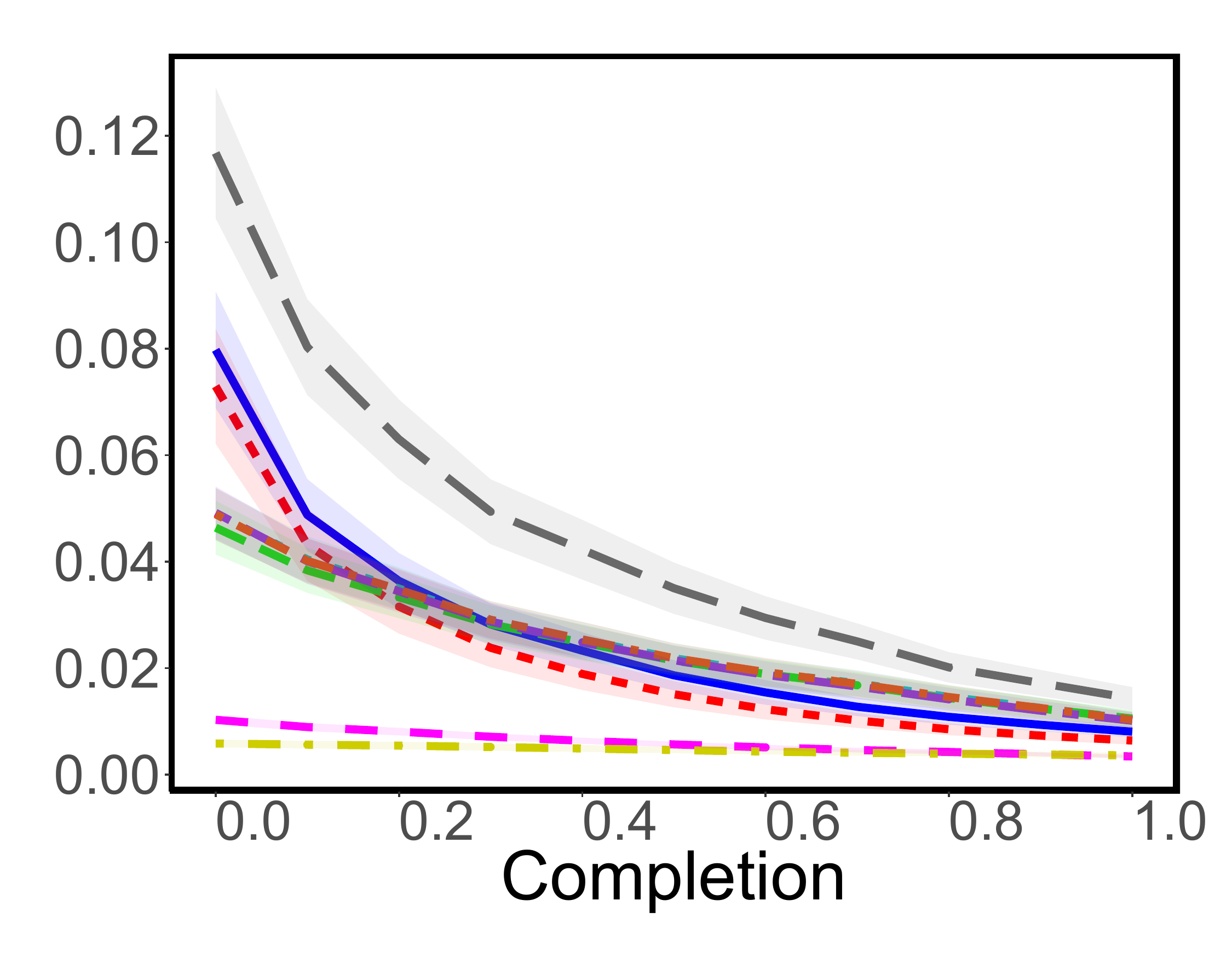} \\
\multicolumn{4}{c}{\includegraphics[width=0.65\linewidth]{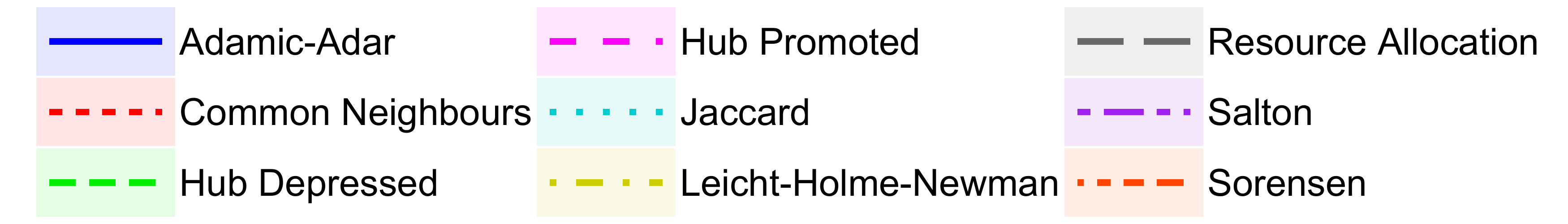}}
\end{tabular}
\caption{Given different \textbf{local similarity} indices, the figure depicts the values of $\ROC$ (the area under the ROC curve) and $\AP$ (the average precision) during the execution of OTC and CTR given $|\Hide|=\max(10,|E|/100)$ and $b=4|\Hide|$ in three networks: (i) the \textbf{WTC 9/11 terrorist network}; (ii) \textbf{ScaleFree(100,3)}; and (iii)  \textbf{a medium fragment of Facebook}.
In each execution, the links in $\Hide$ are chosen at random. Results are taken as the average over $50$ executions, with coloured areas representing the $95\%$ confidence intervals.}
\label{fig:local-0}
\end{figure*}

\begin{figure*}[tbhp]
\centering
\setlength\tabcolsep{1pt}
\renewcommand{\arraystretch}{0.01}
\begin{tabular}{m{.03\textwidth}m{.27\textwidth}m{.27\textwidth}m{.27\textwidth}}
& \multicolumn{1}{c}{ScaleFree$(1000,3)$}
& \multicolumn{1}{c}{RandomGraph$(100,10)$}
& \multicolumn{1}{c}{RandomGraph$(1000,10)$}\\
\rotatebox{90}{\footnotesize $\ROC$ values for OTC} &
\includegraphics[width=1.03\linewidth]{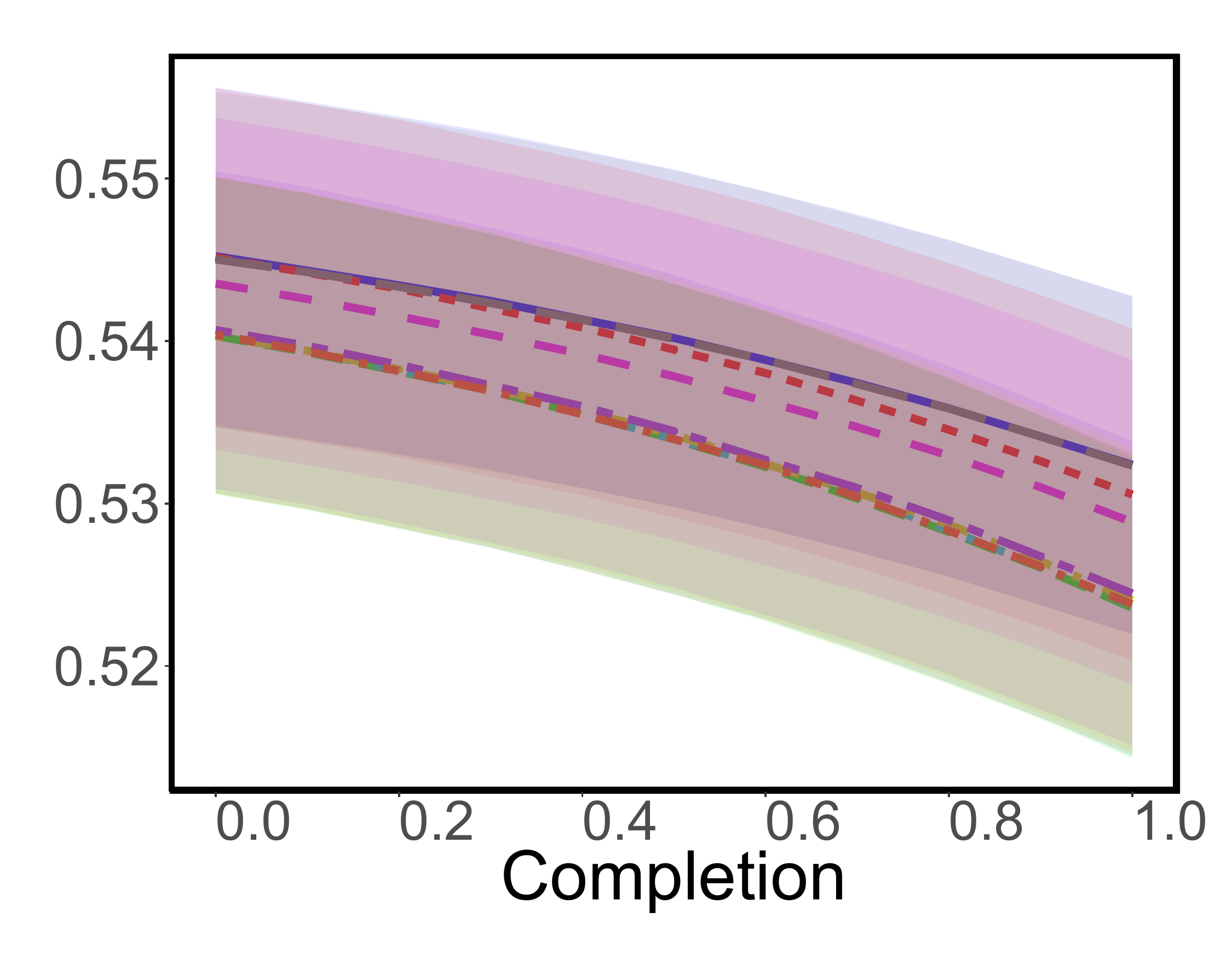} &
\includegraphics[width=1.03\linewidth]{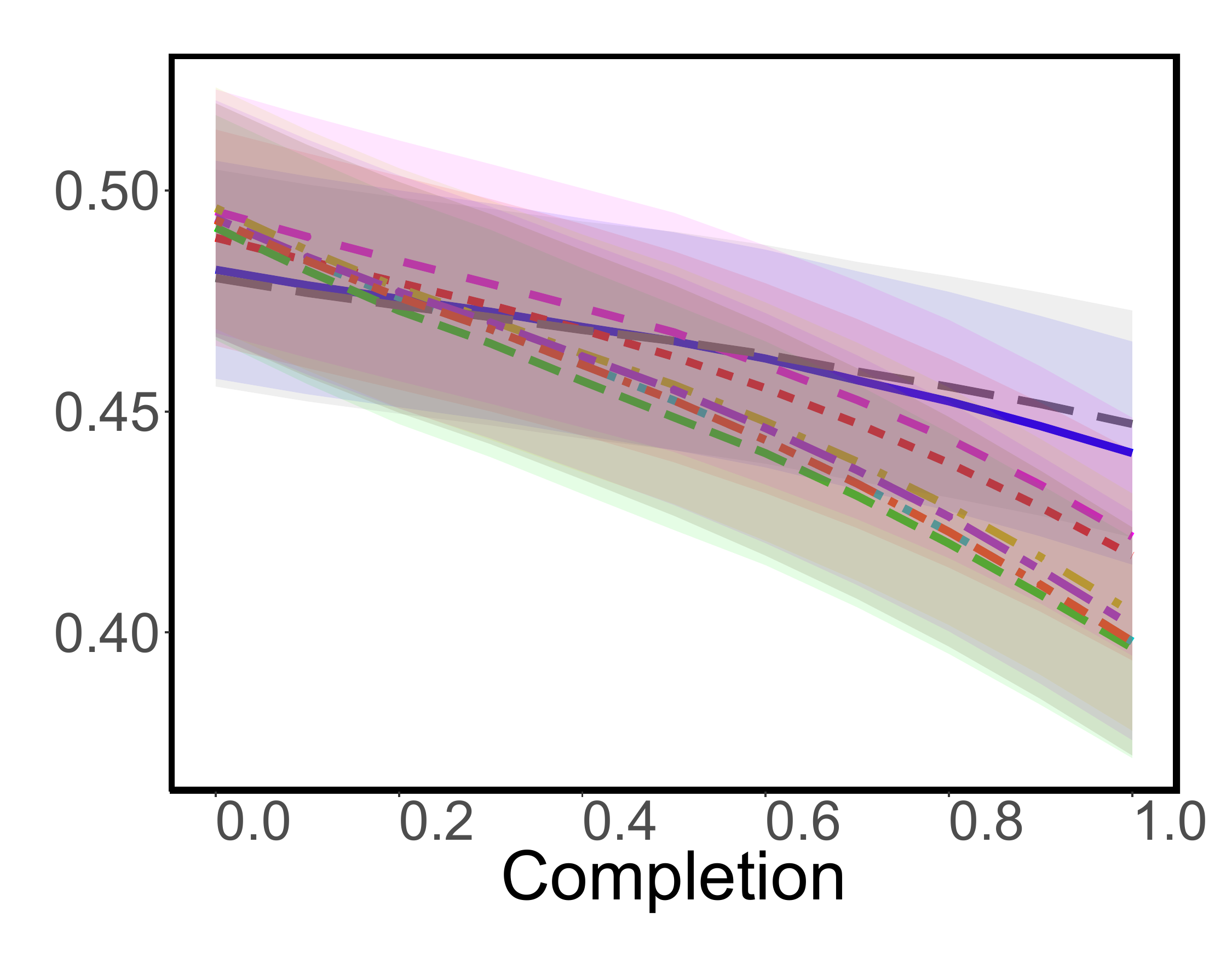} &
\includegraphics[width=1.03\linewidth]{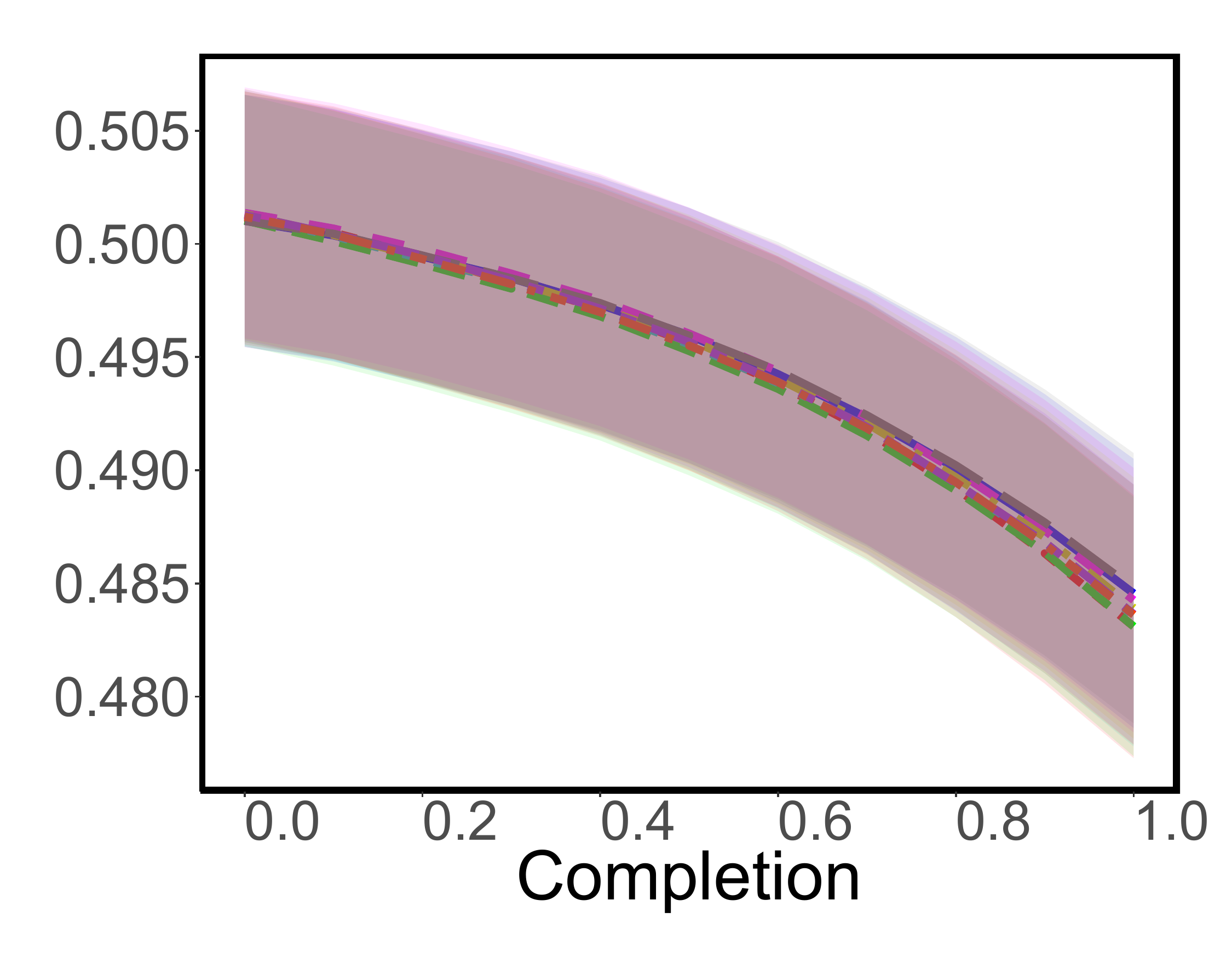}\\
\rotatebox{90}{\footnotesize $\ROC$ values for CTR} &
\includegraphics[width=1.03\linewidth]{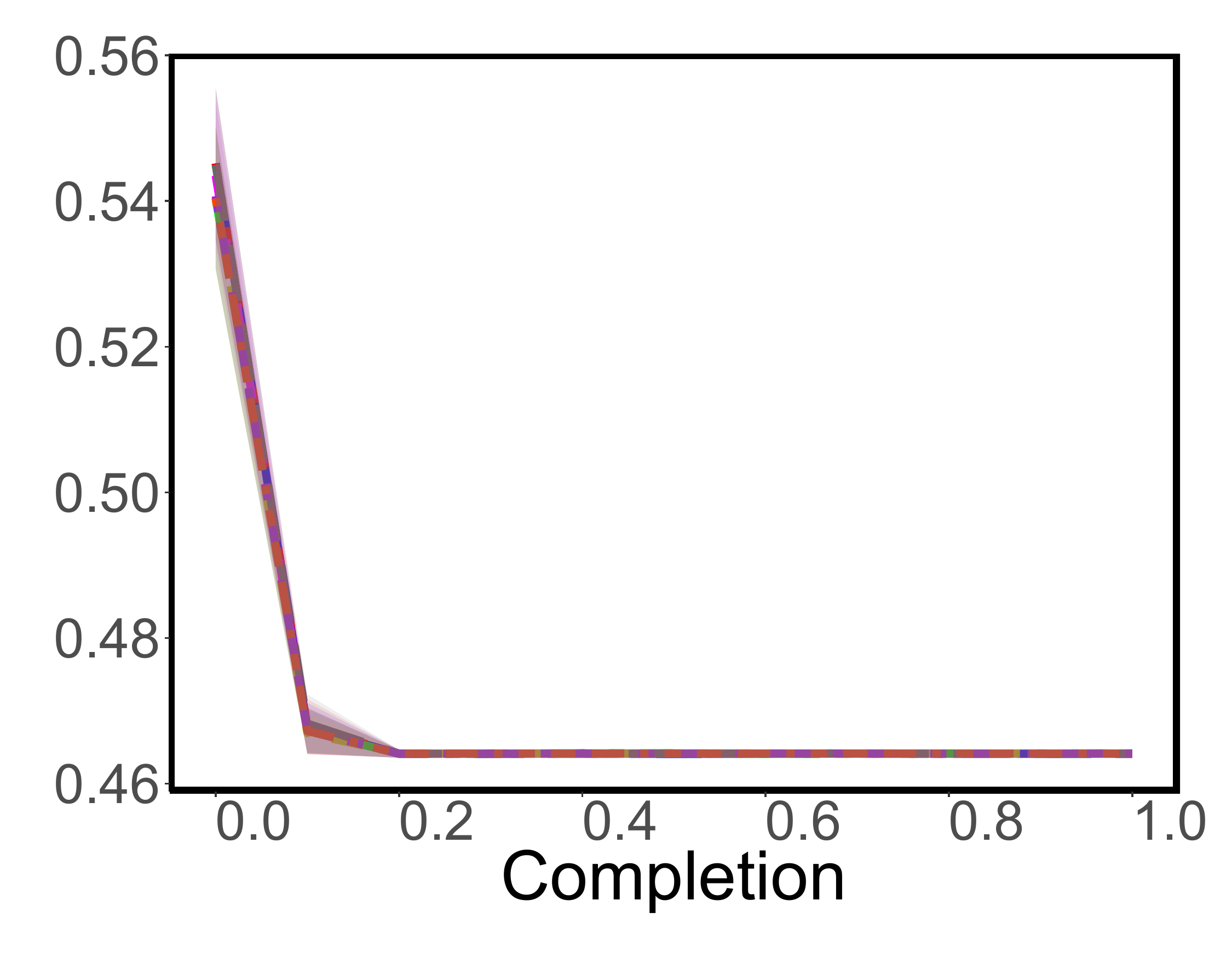} &
\includegraphics[width=1.03\linewidth]{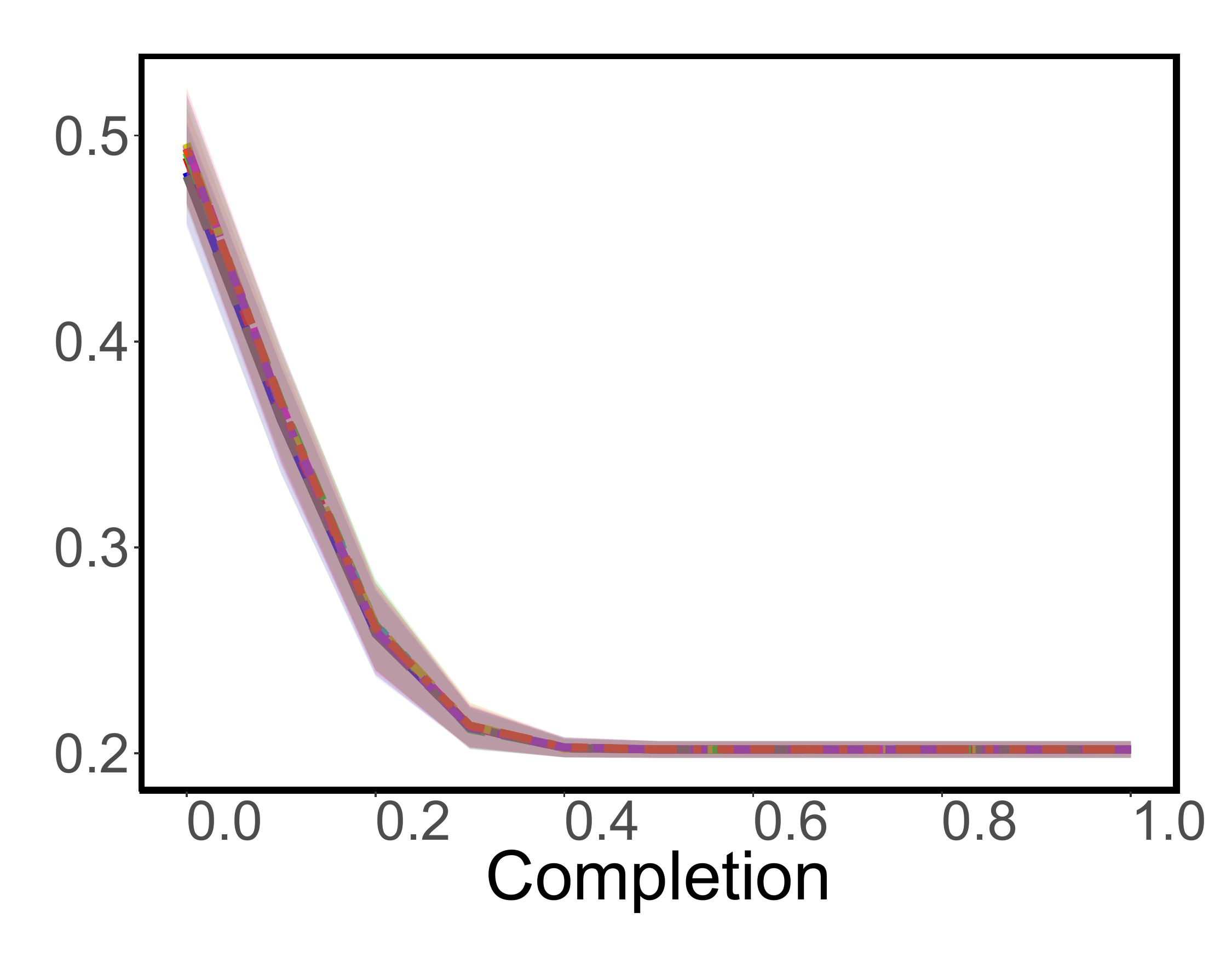} &
\includegraphics[width=1.03\linewidth]{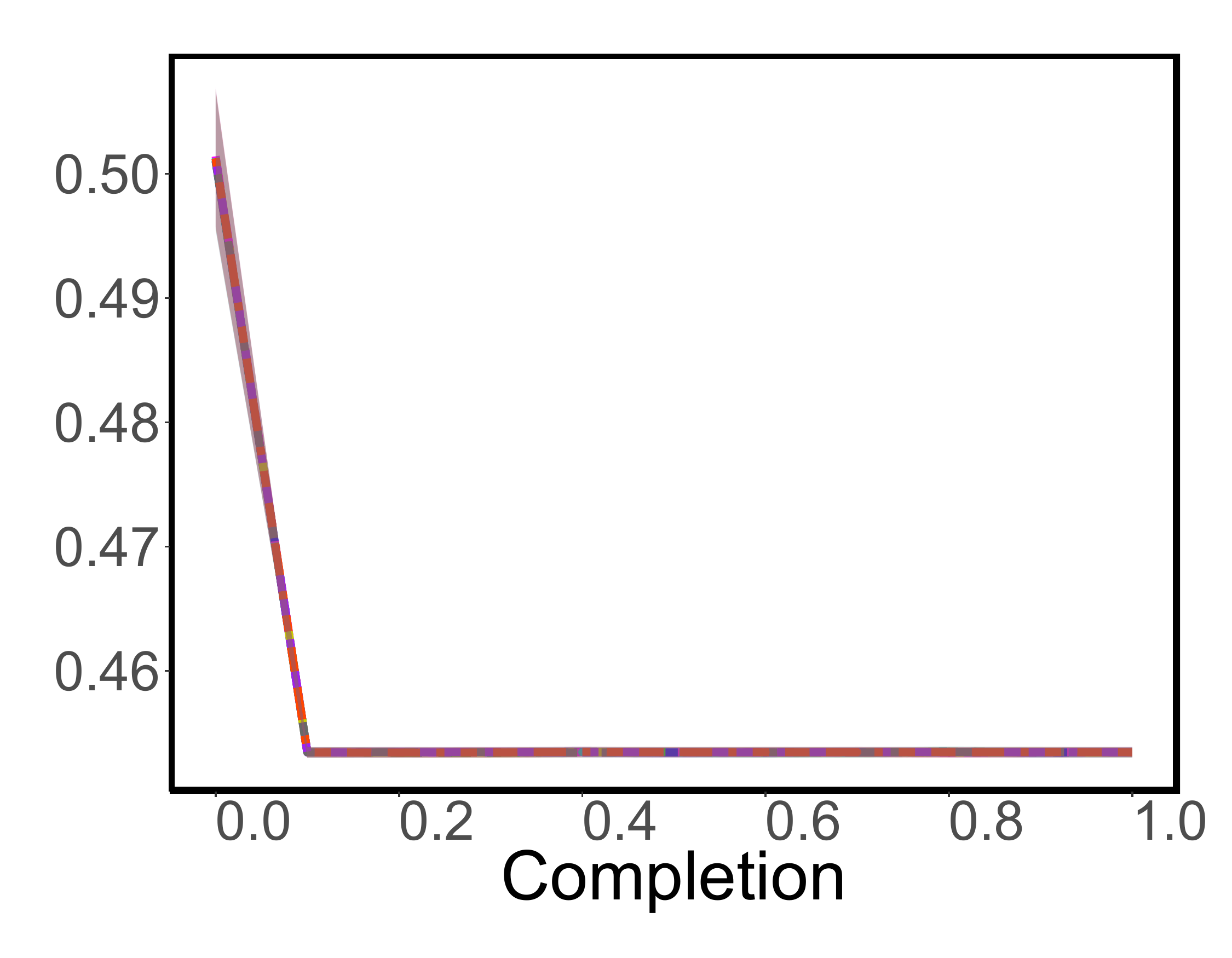} \\
\rotatebox{90}{\footnotesize $\AP$ values for OTC} &
\includegraphics[width=1.03\linewidth]{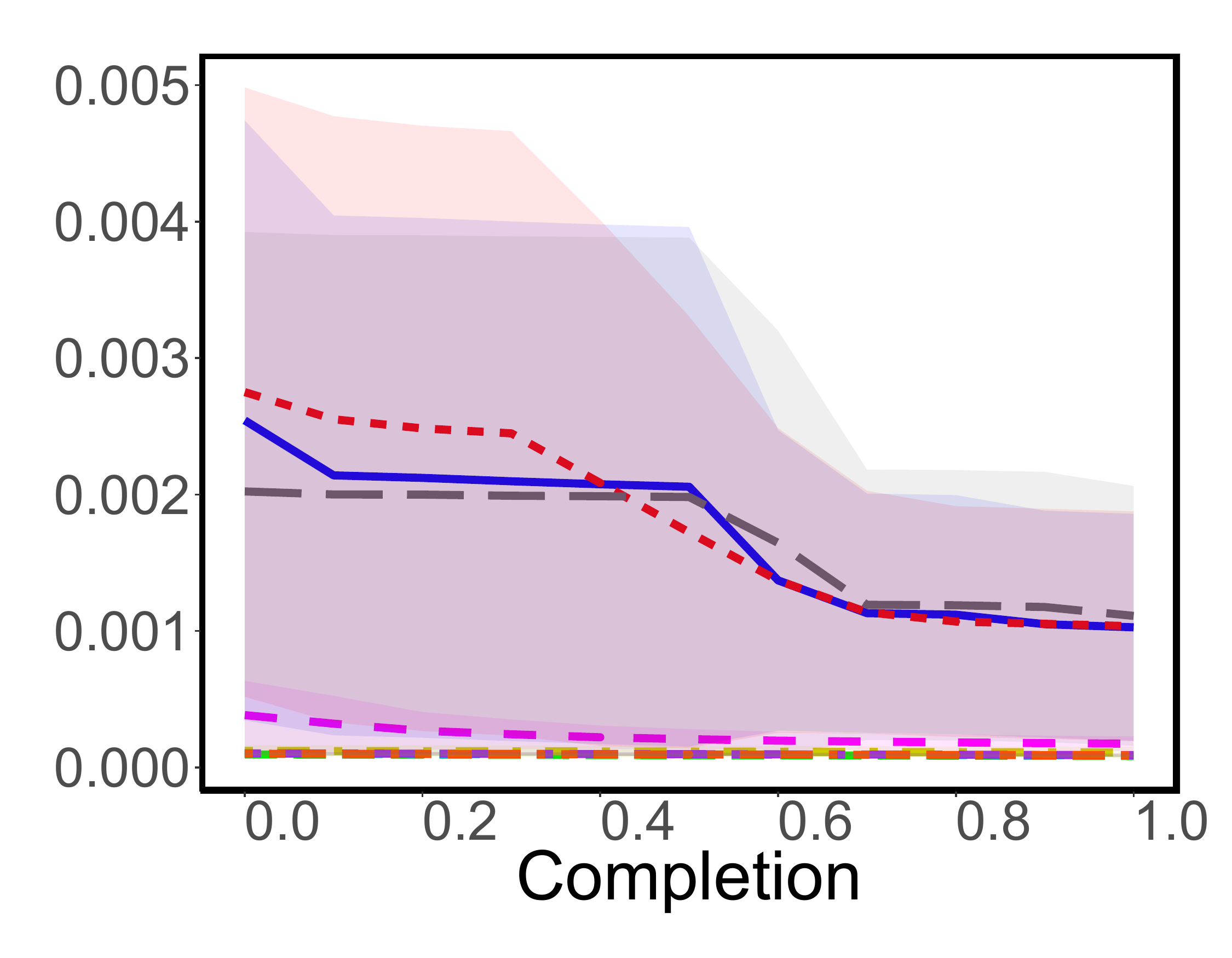} &
\includegraphics[width=1.03\linewidth]{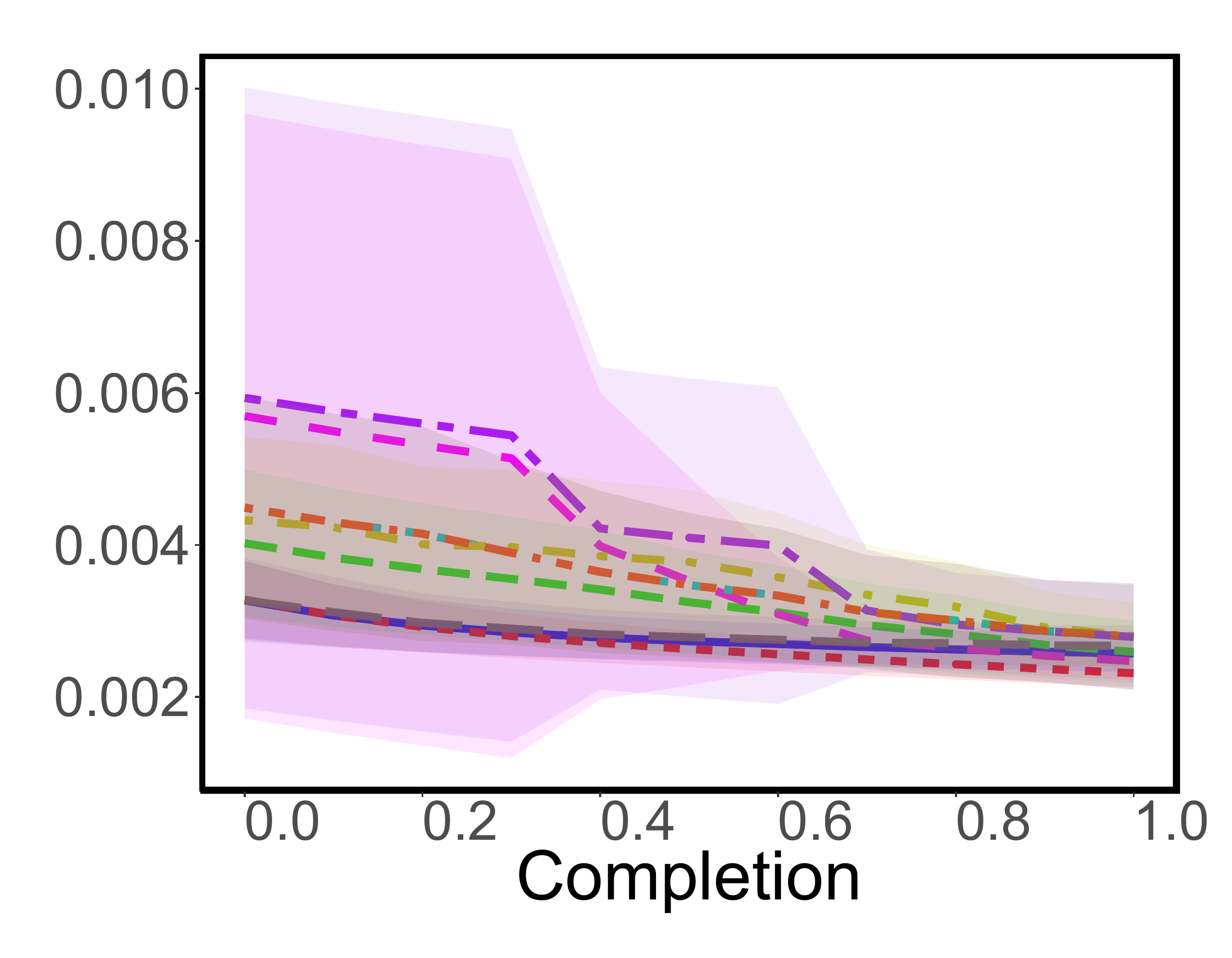} &
\includegraphics[width=1.03\linewidth]{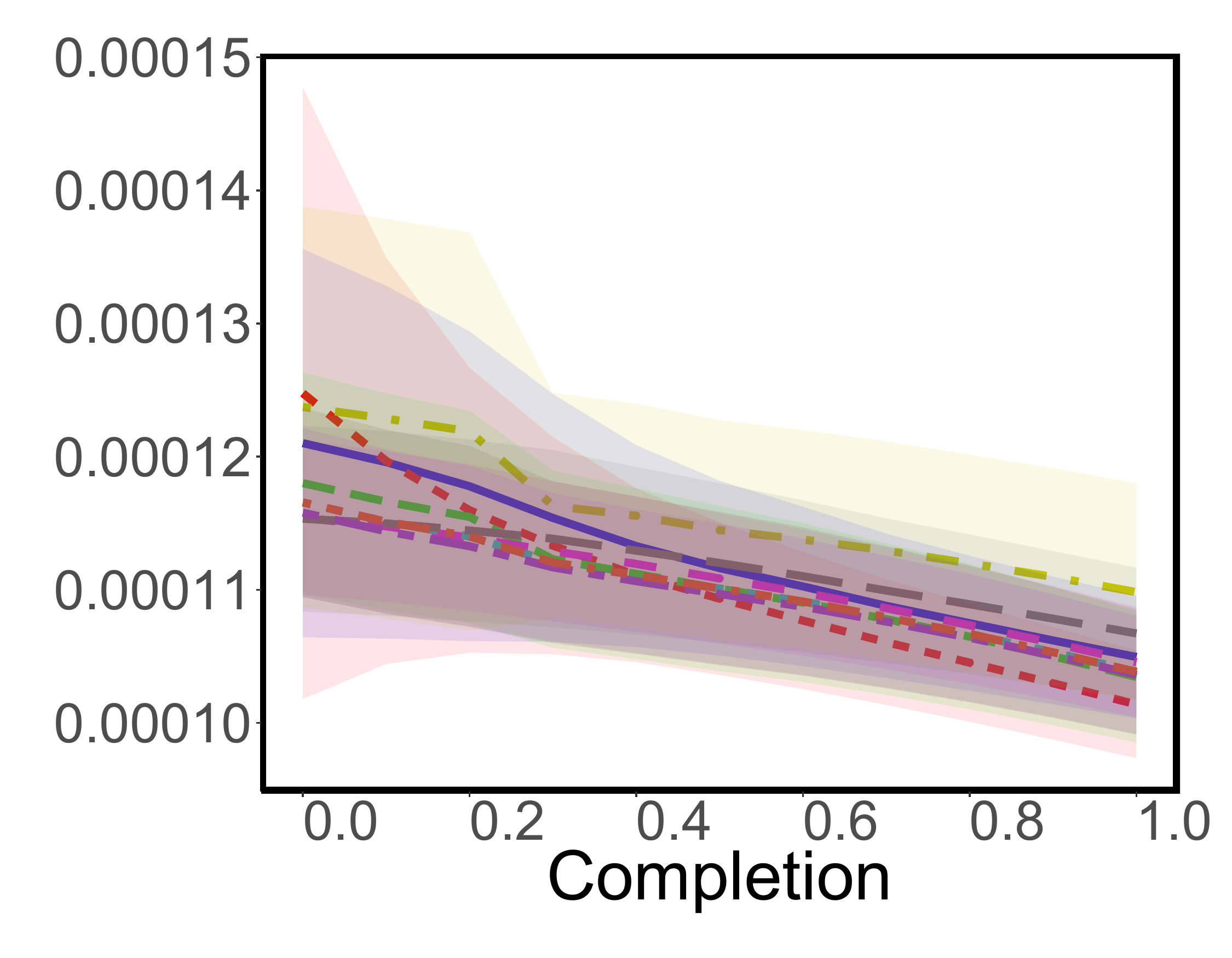} \\
\rotatebox{90}{\footnotesize $\AP$ values for CTR} &
\includegraphics[width=1.03\linewidth]{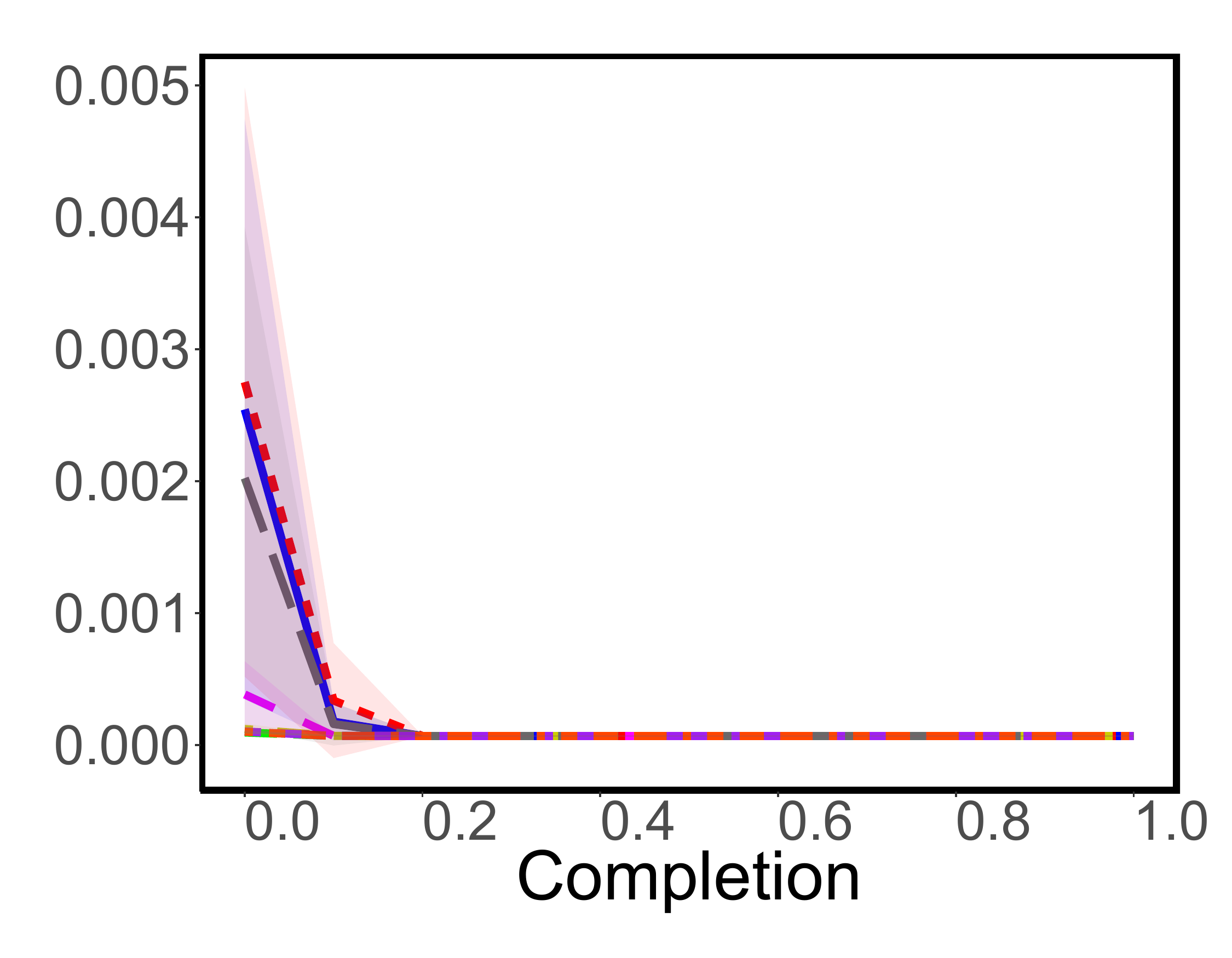} &
\includegraphics[width=1.03\linewidth]{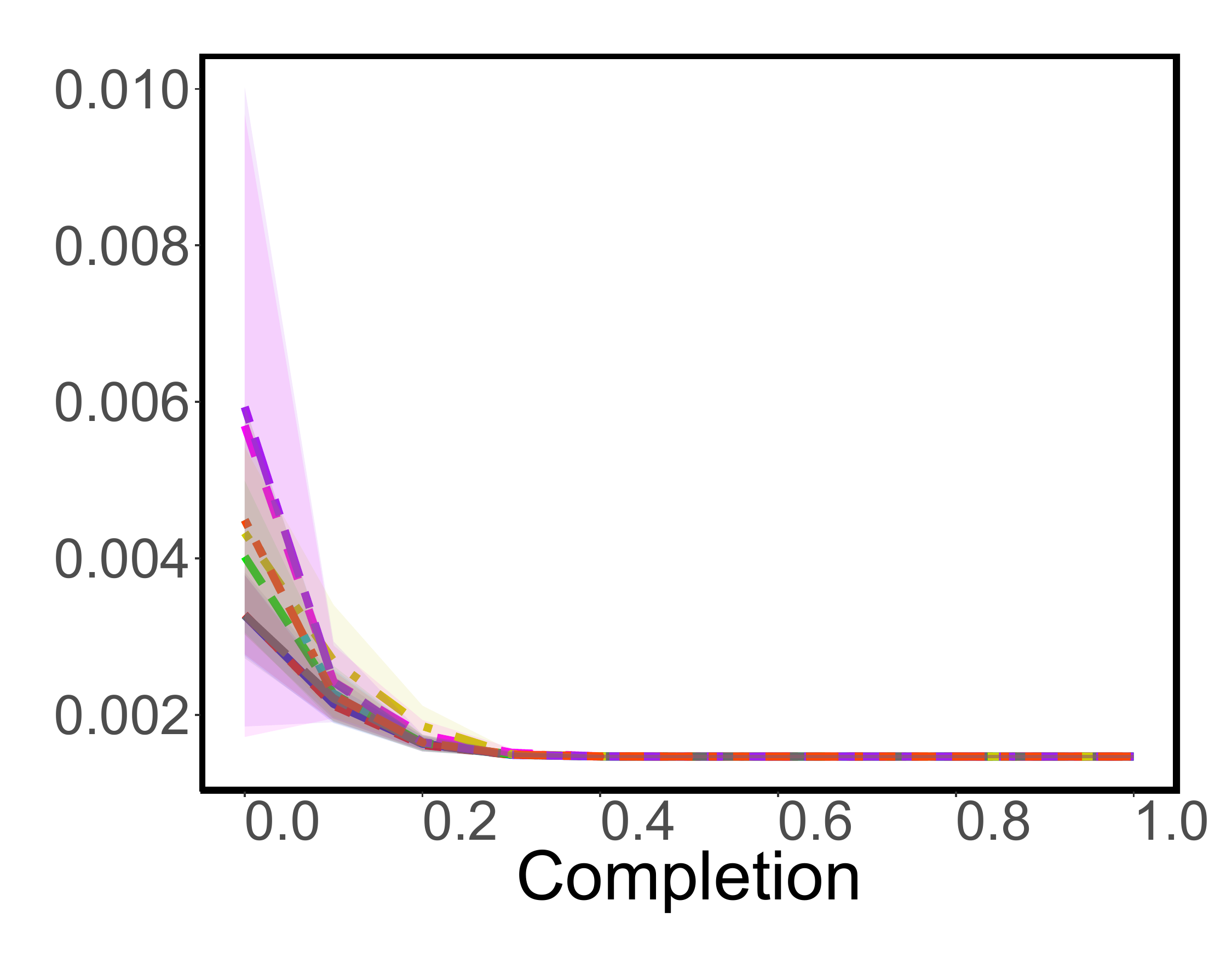} &
\includegraphics[width=1.03\linewidth]{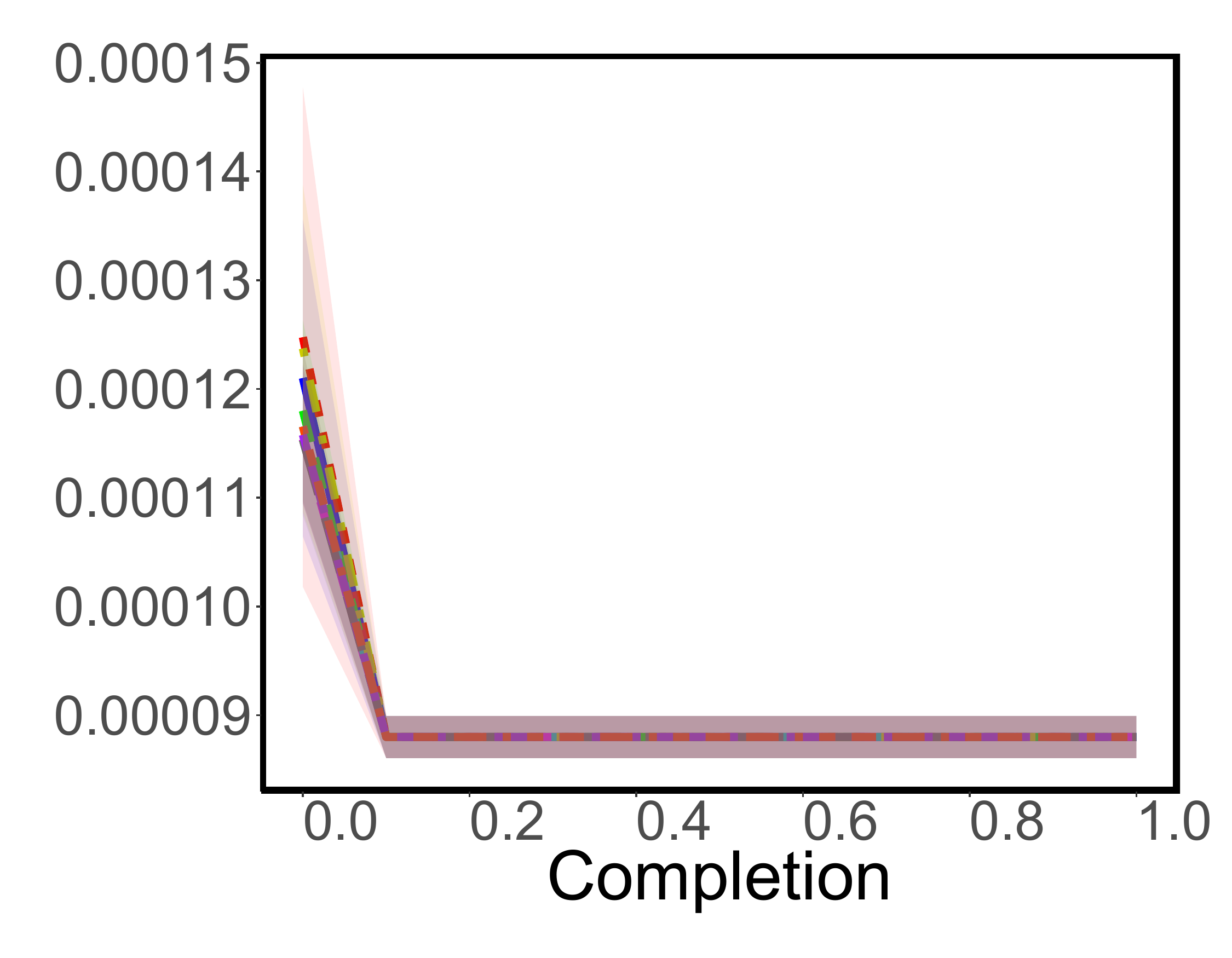} \\
\multicolumn{4}{c}{\includegraphics[width=0.65\linewidth]{figures/plots/local/legend}}
\end{tabular}
\caption{Given different \textbf{local similarity} indices, the figure depicts the values of $\ROC$ (the area under the ROC curve) and $\AP$ (the average precision) during the execution of OTC and CTR given $|\Hide|=\max(10,|E|/100)$ and $b=4|\Hide|$ in three networks: (i) \textbf{ScaleFree(1000,3)}; (ii) \textbf{RandomGraph(100,10)}; and (iii) \textbf{RandomGraph(1000,10)}.
In each execution, the links in $\Hide$ are chosen at random. Results are taken as the average over $50$ executions, with coloured areas representing the $95\%$ confidence intervals.}
\label{fig:local-1}
\end{figure*}

\begin{figure*}[tbhp]
\centering
\setlength\tabcolsep{1pt}
\renewcommand{\arraystretch}{0.01}
\begin{tabular}{m{.03\textwidth}m{.27\textwidth}m{.27\textwidth}m{.27\textwidth}}
& \multicolumn{1}{c}{SmallWorld$(100,10,0.25)$}
& \multicolumn{1}{c}{SmallWorld$(1000,10,0.25)$}
& \multicolumn{1}{c}{Les Mis\'erables network}\\
\rotatebox{90}{\footnotesize $\ROC$ values for OTC} &
\includegraphics[width=1.03\linewidth]{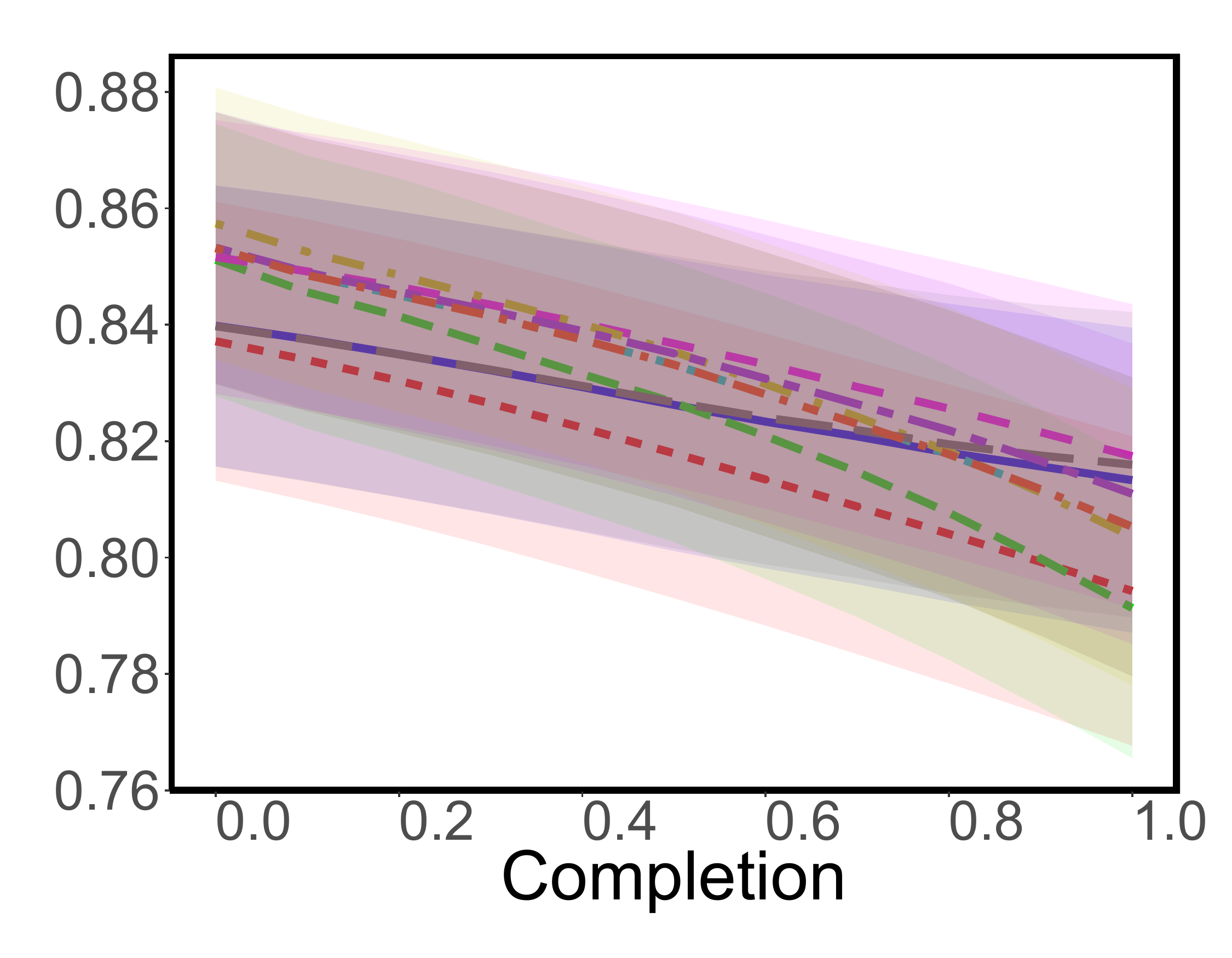} &
\includegraphics[width=1.03\linewidth]{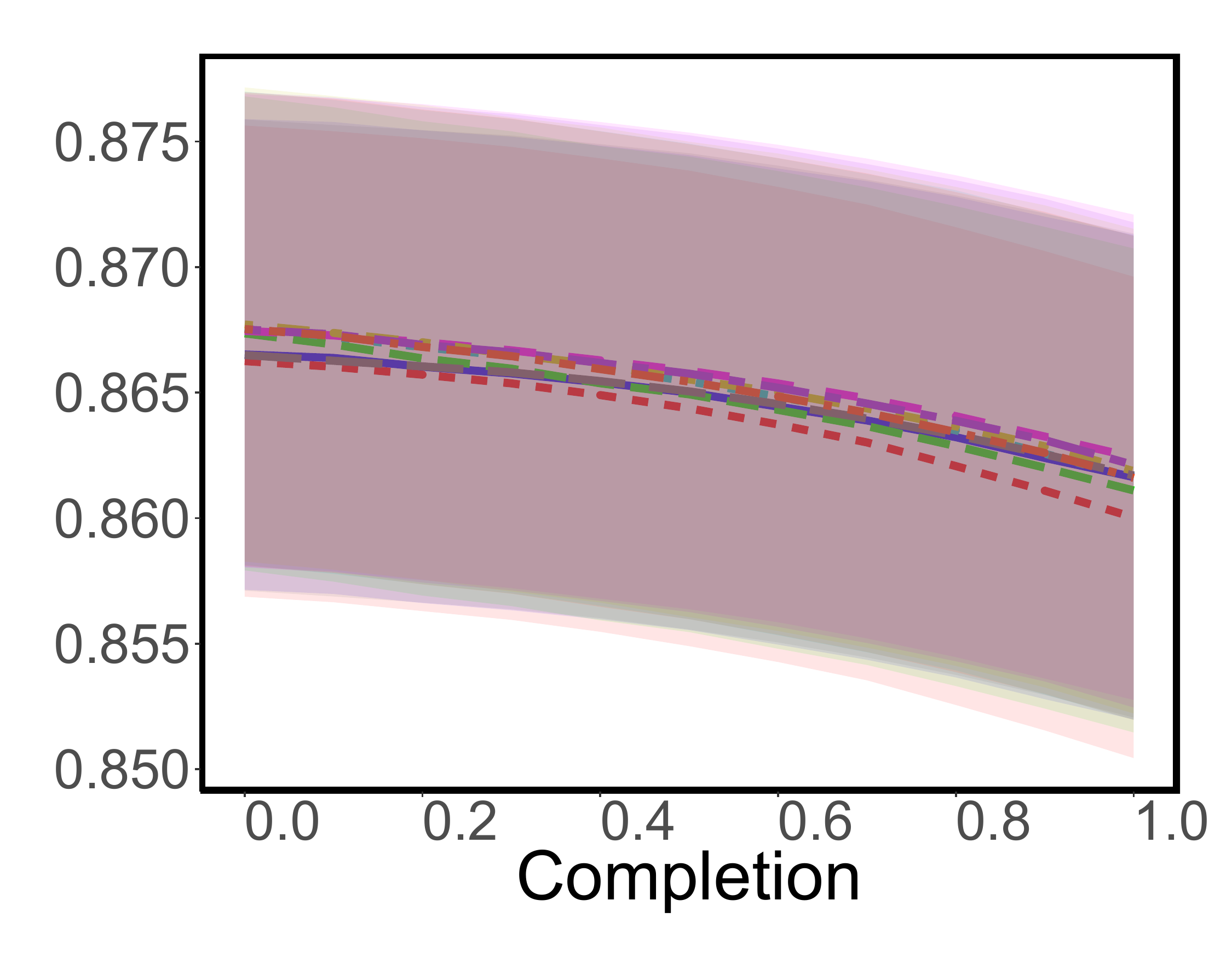} &
\includegraphics[width=1.03\linewidth]{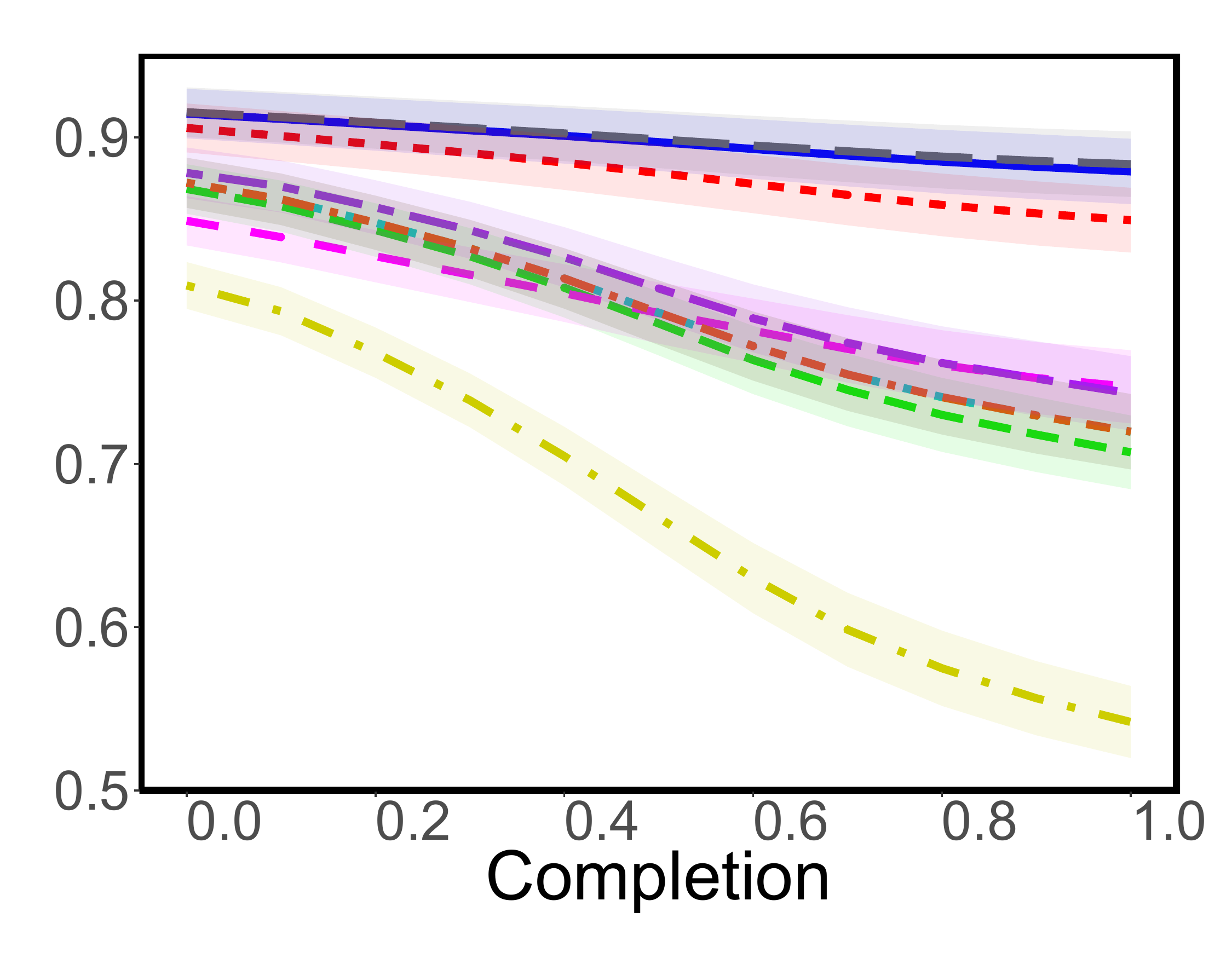}\\
\rotatebox{90}{\footnotesize $\ROC$ values for CTR} &
\includegraphics[width=1.03\linewidth]{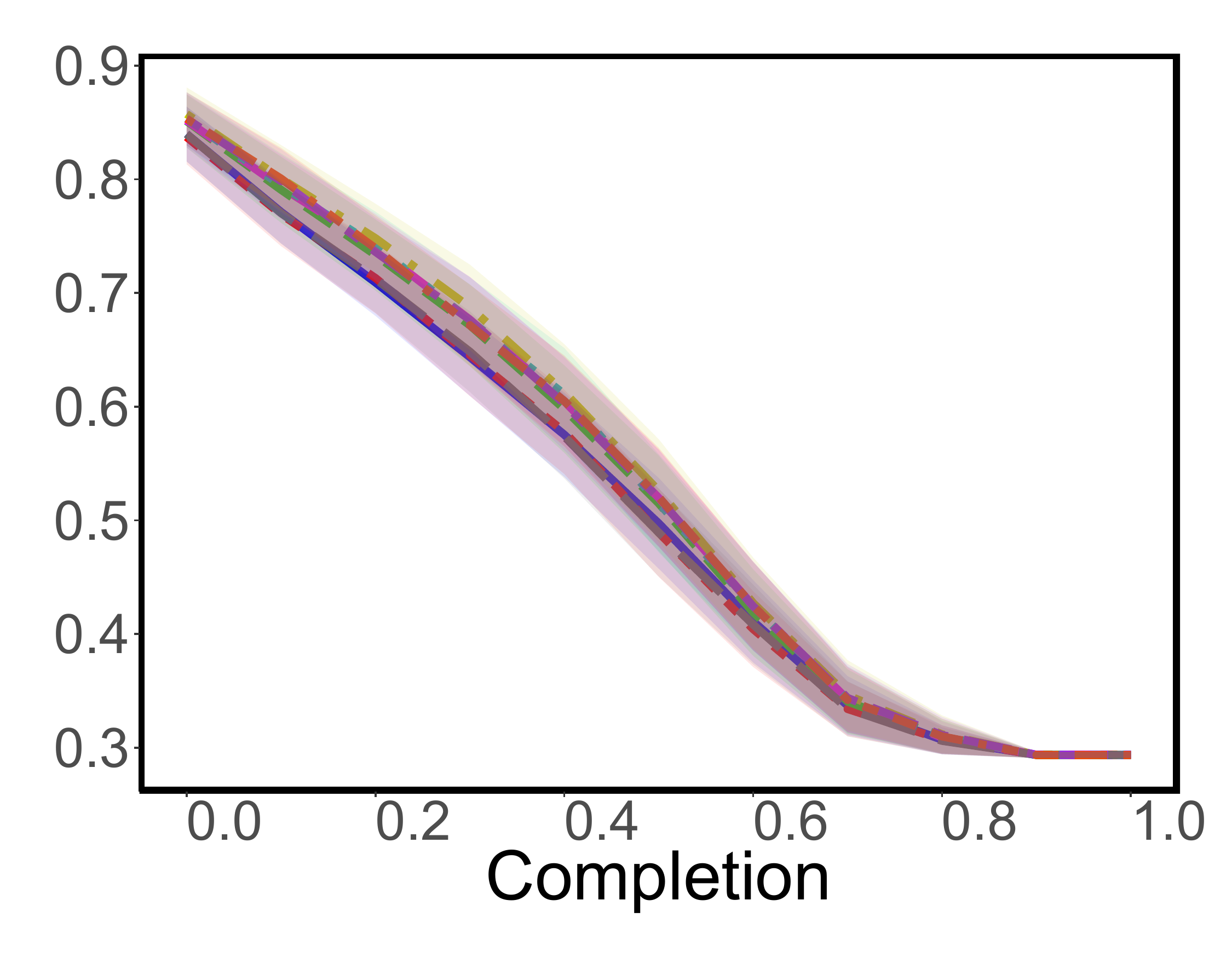} &
\includegraphics[width=1.03\linewidth]{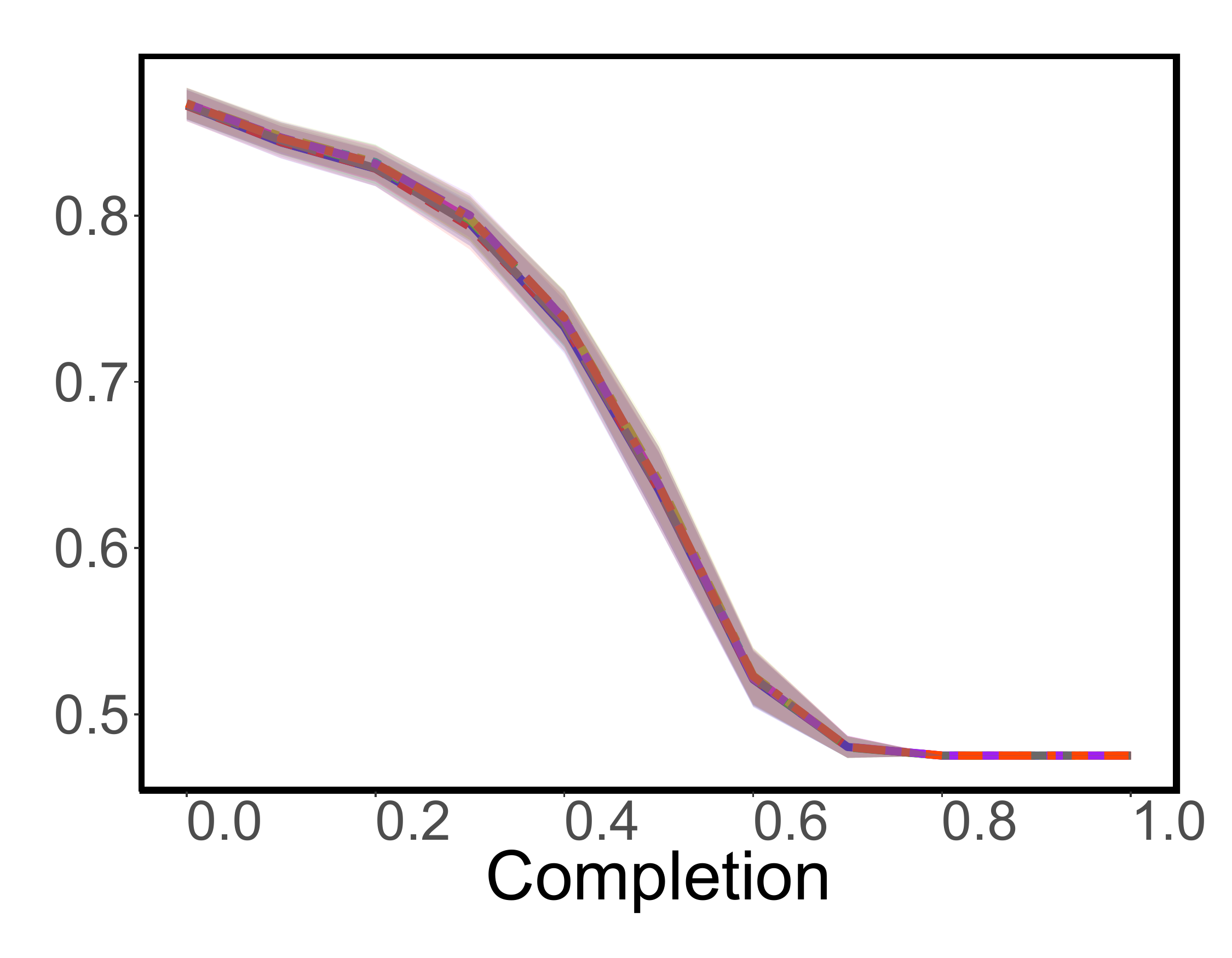} &
\includegraphics[width=1.03\linewidth]{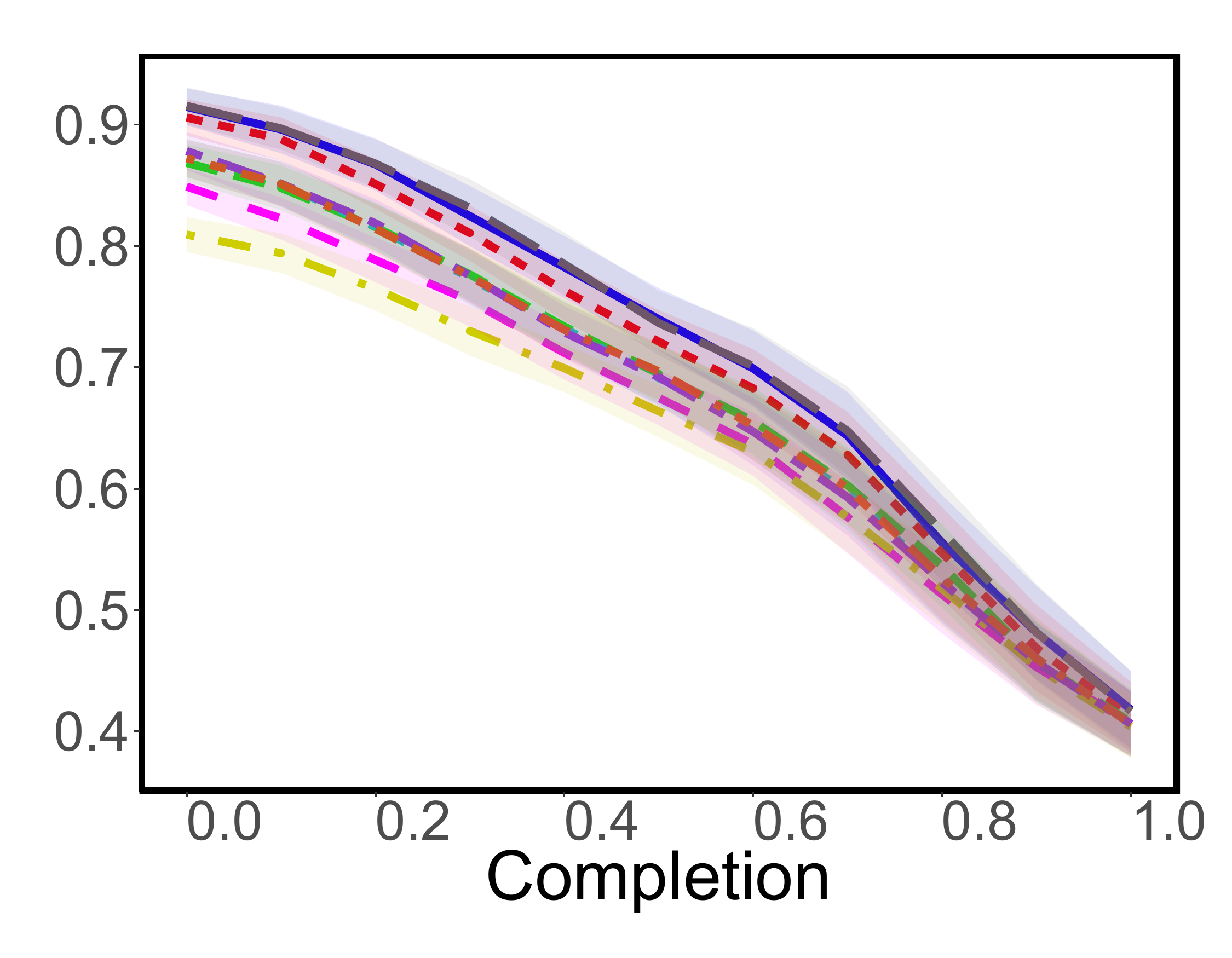} \\
\rotatebox{90}{\footnotesize $\AP$ values for OTC} &
\includegraphics[width=1.03\linewidth]{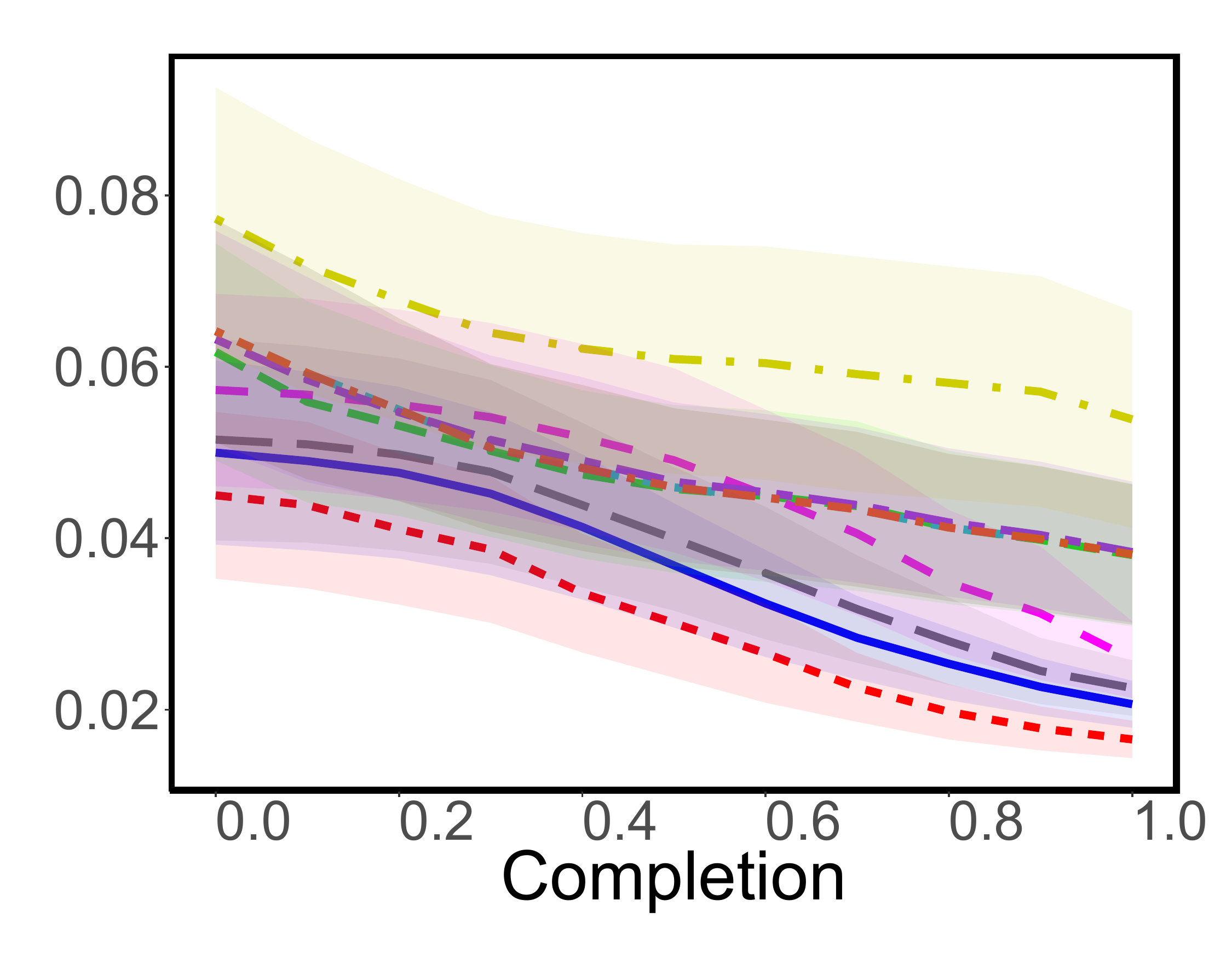} &
\includegraphics[width=1.03\linewidth]{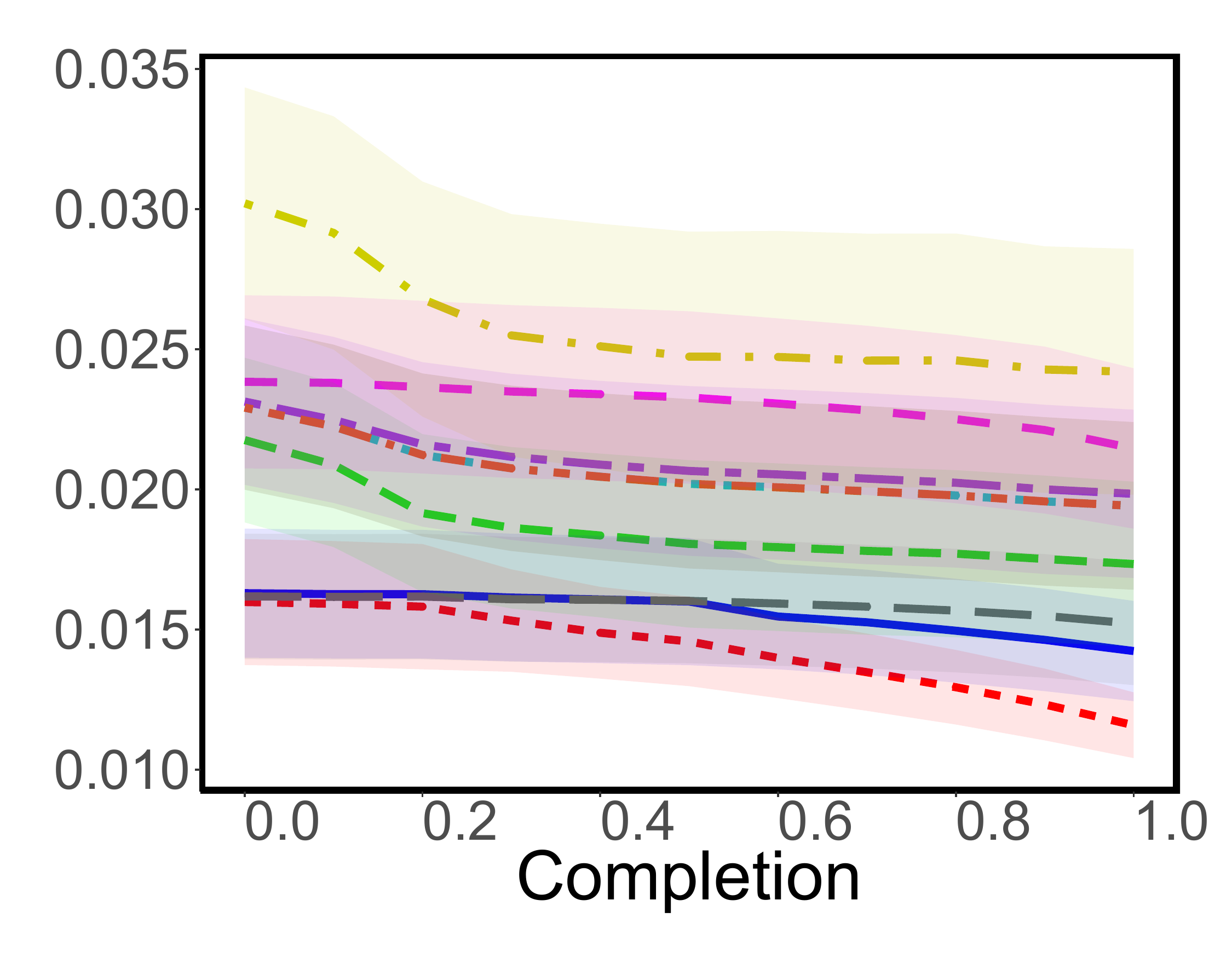} &
\includegraphics[width=1.03\linewidth]{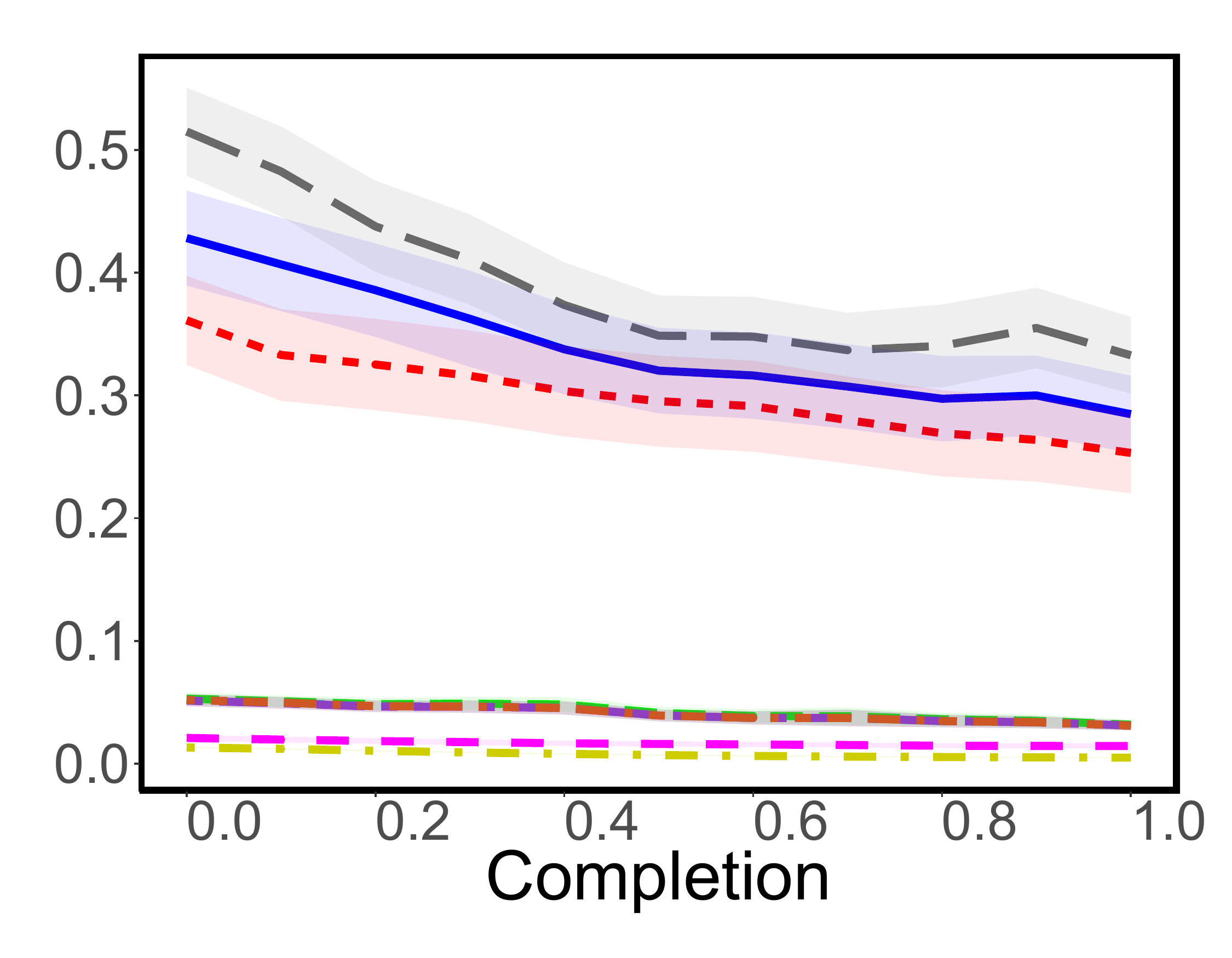} \\
\rotatebox{90}{\footnotesize $\AP$ values for CTR} &
\includegraphics[width=1.03\linewidth]{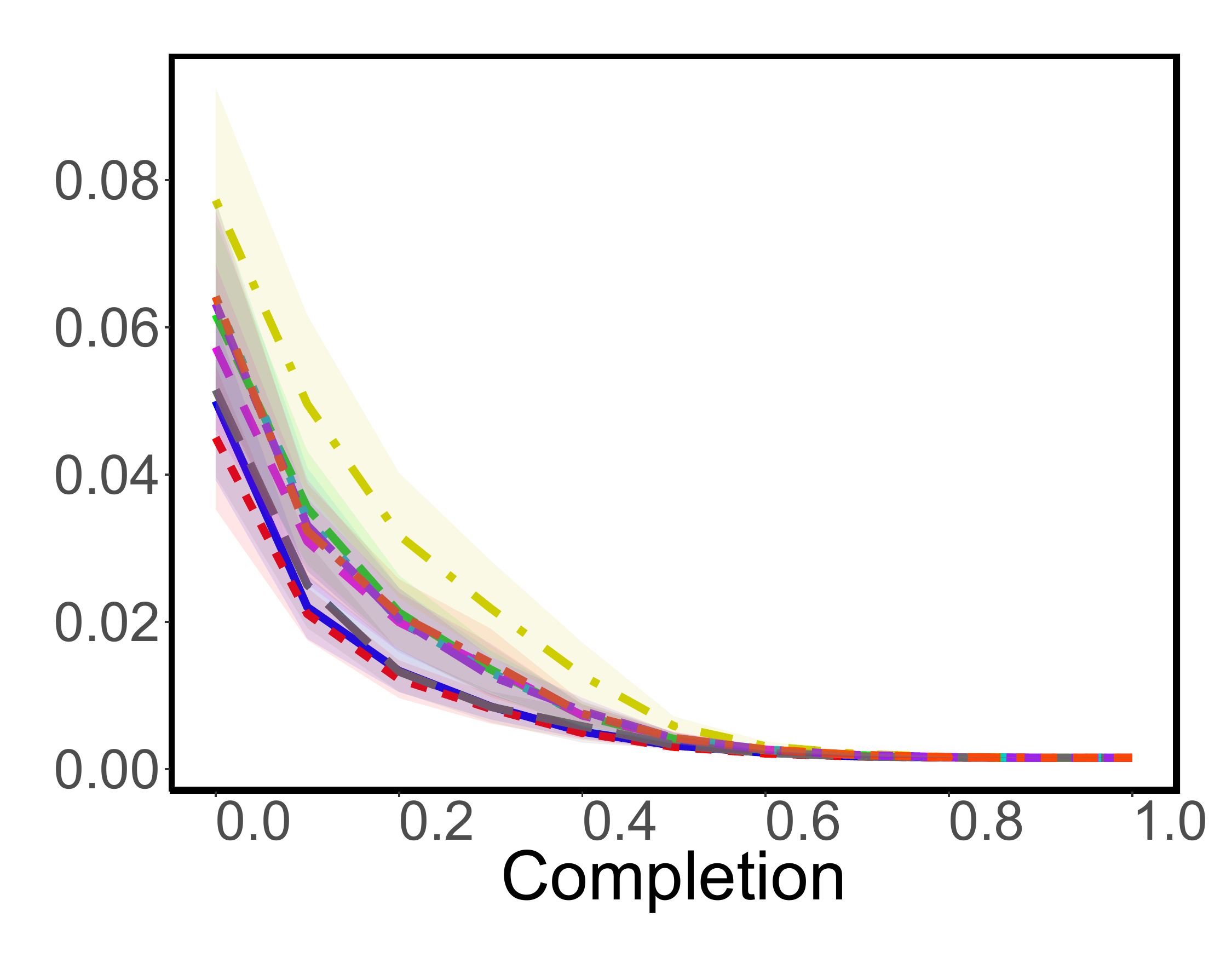} &
\includegraphics[width=1.03\linewidth]{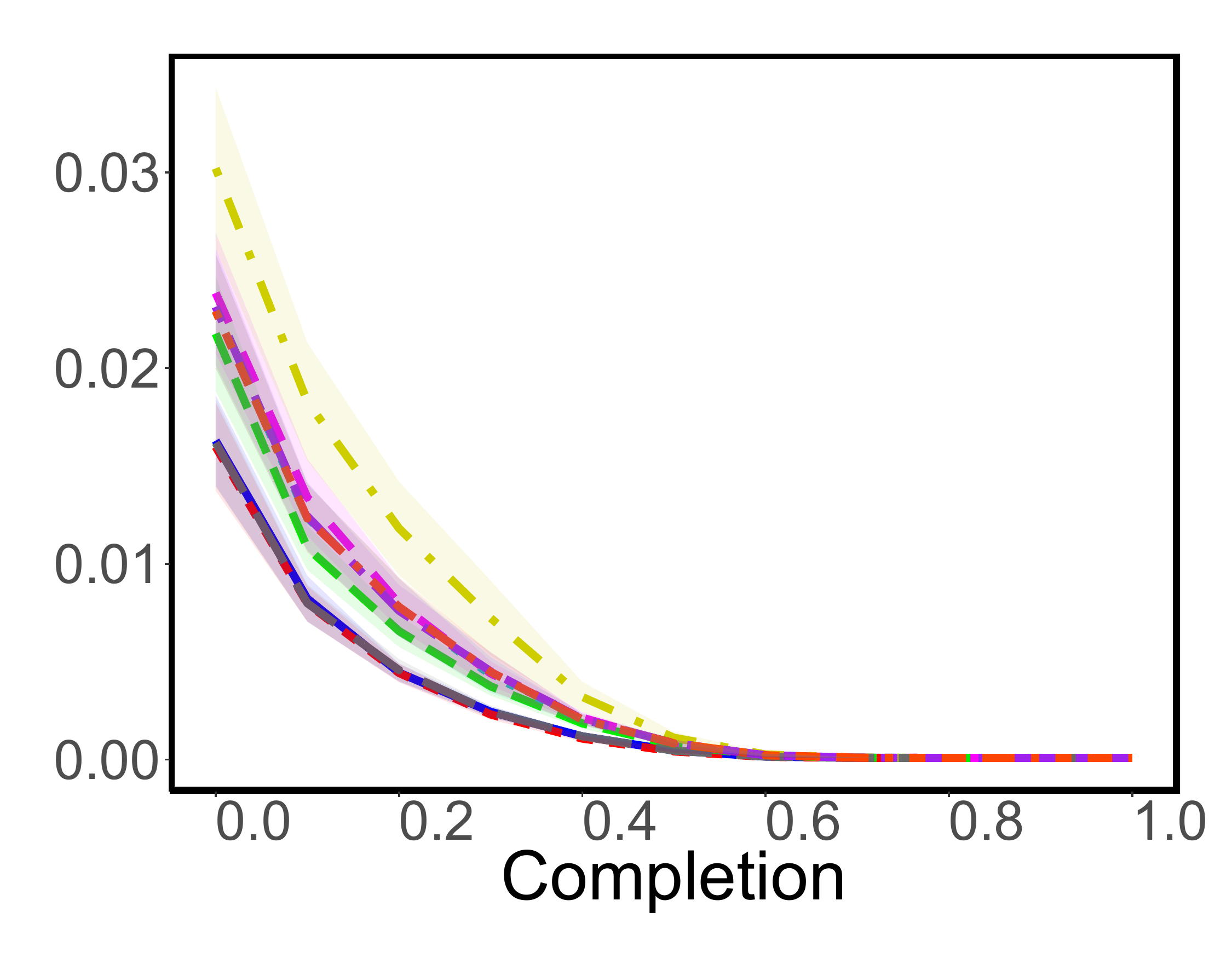} &
\includegraphics[width=1.03\linewidth]{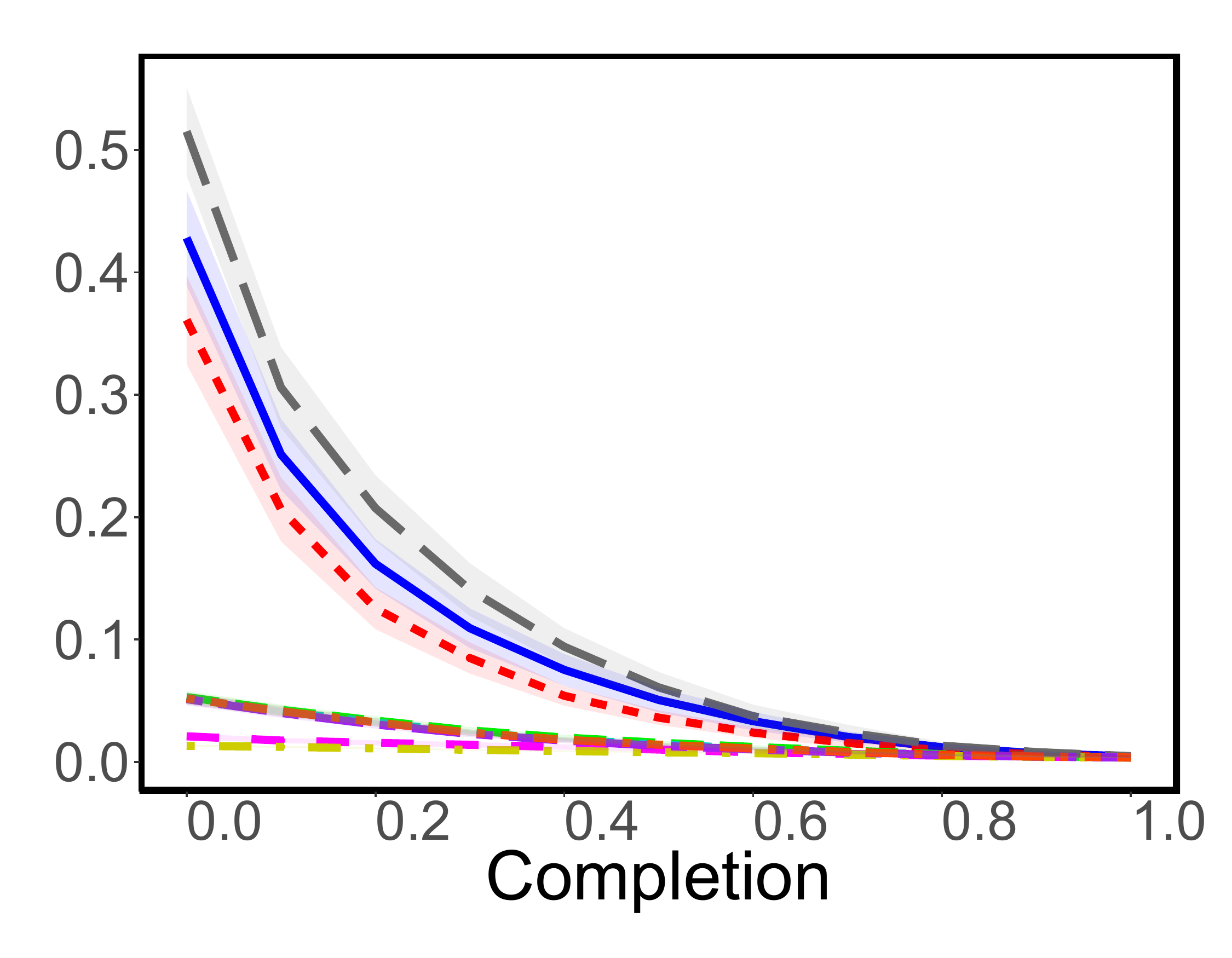} \\
\multicolumn{4}{c}{\includegraphics[width=0.65\linewidth]{figures/plots/local/legend}}
\end{tabular}
\caption{Given different local \textbf{similarity indices}, the figure depicts the values of $\ROC$ (the area under the ROC curve) and $\AP$ (the average precision) during the execution of OTC and CTR given $|\Hide|=\max(10,|E|/100)$ and $b=4|\Hide|$ in three networks: (i) \textbf{SmallWorld(100,10,0.25)}; (ii) \textbf{SmallWorld(1000,10,0.25)}; and (iii) \textbf{Les Mis\'erables network}.
In each execution, the links in $\Hide$ are chosen at random. Results are taken as the average over $50$ executions, with coloured areas representing the $95\%$ confidence intervals.}
\label{fig:local-2}
\end{figure*}

\begin{figure*}[tbhp]
\centering
\setlength\tabcolsep{1pt}
\renewcommand{\arraystretch}{0.01}
\begin{tabular}{m{.03\textwidth}m{.27\textwidth}m{.27\textwidth}m{.27\textwidth}}
& \multicolumn{1}{c}{Facebook fragment (small)}
& \multicolumn{1}{c}{Facebook fragment (large)}
& \multicolumn{1}{c}{Zachary Karate Club}\\
\rotatebox{90}{\footnotesize $\ROC$ values for OTC} &
\includegraphics[width=1.03\linewidth]{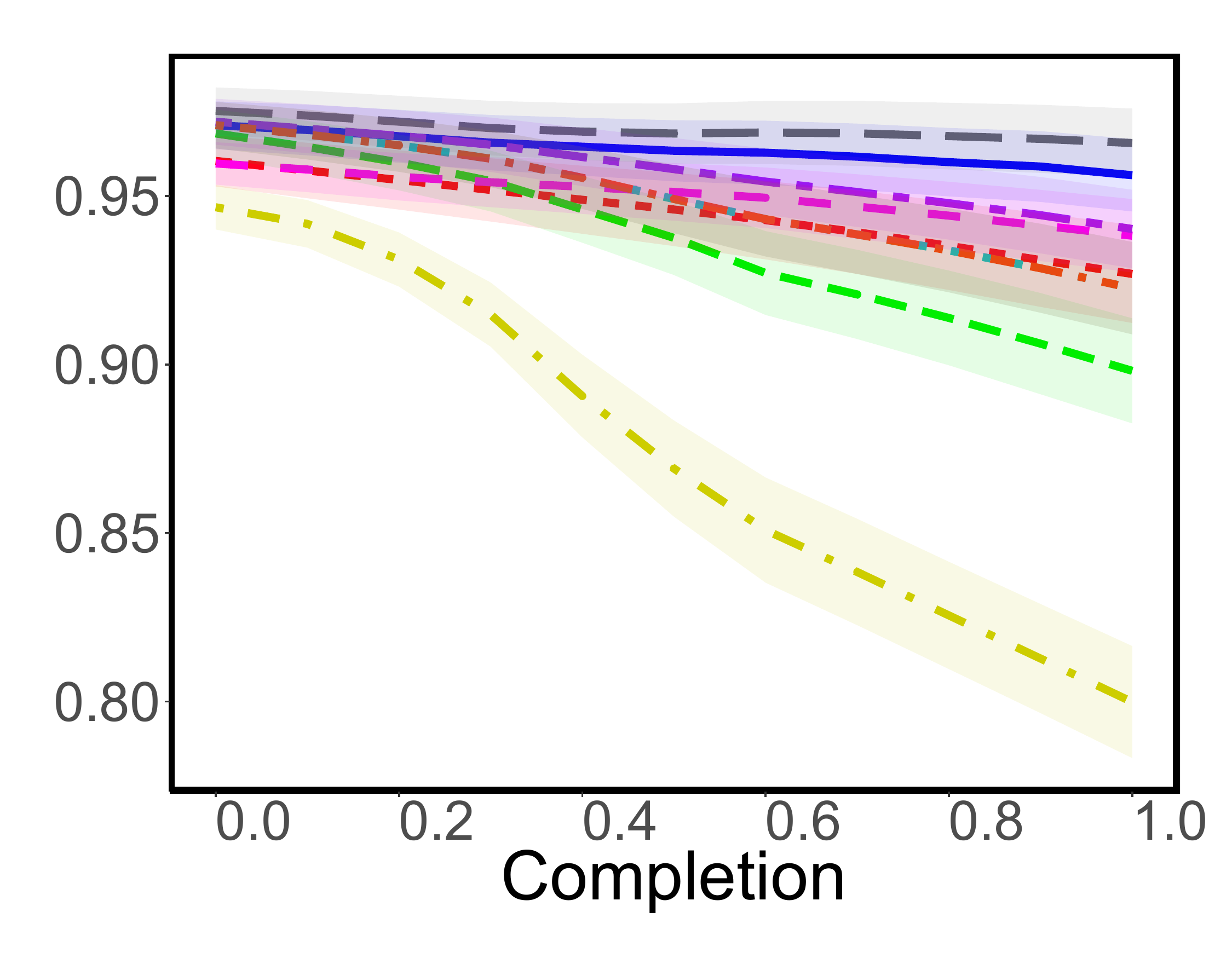} &
\includegraphics[width=1.03\linewidth]{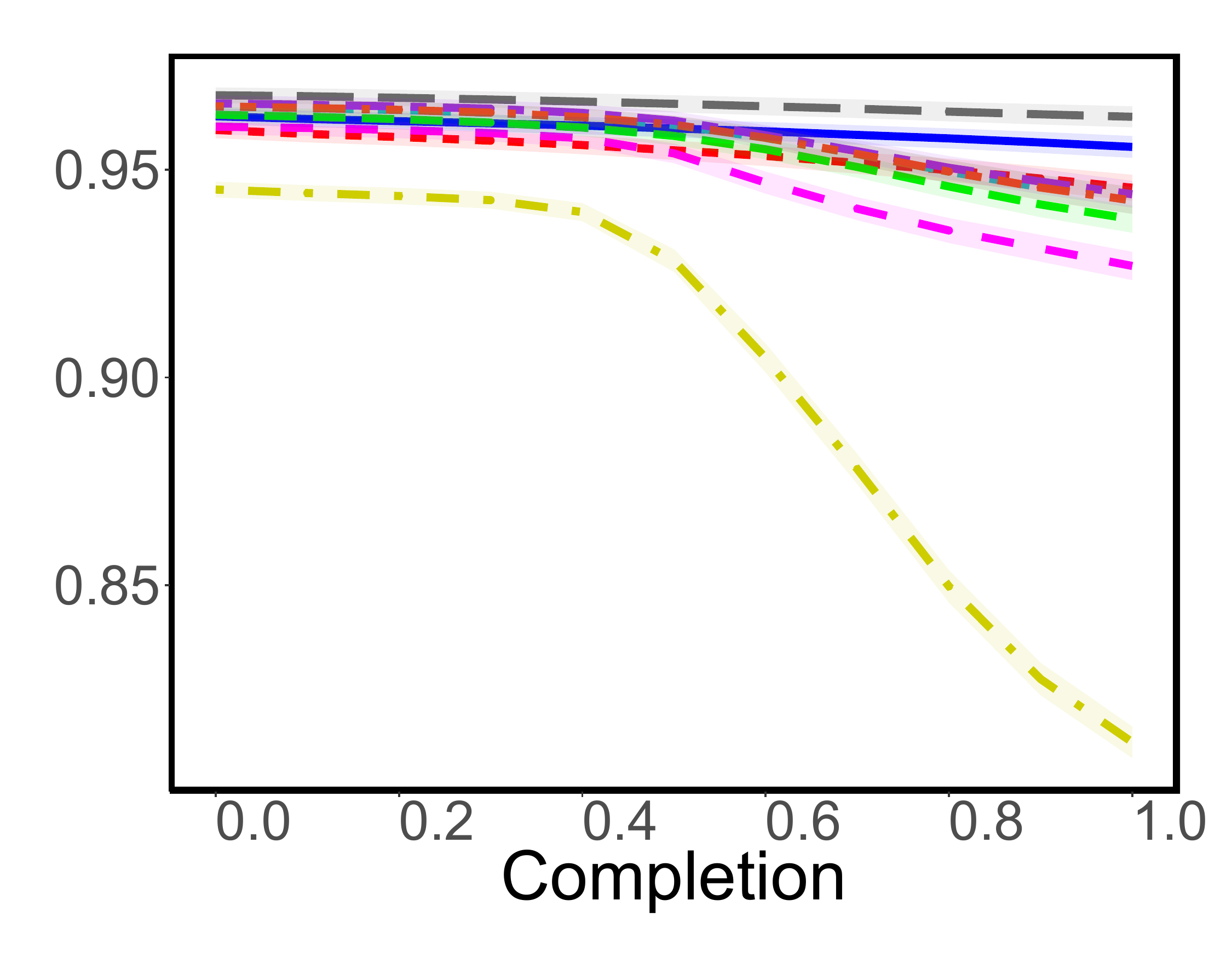} &
\includegraphics[width=1.03\linewidth]{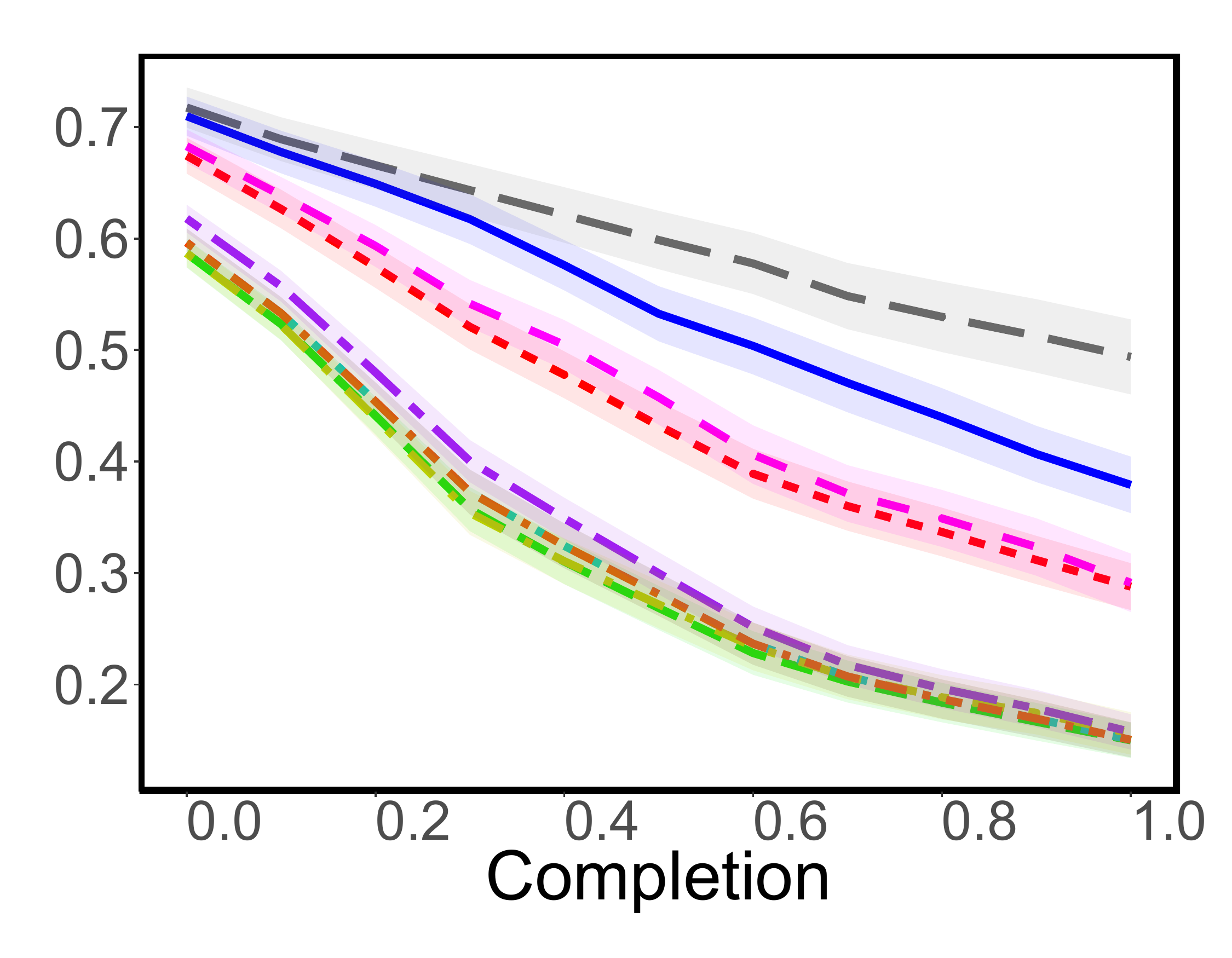}\\
\rotatebox{90}{\footnotesize $\ROC$ values for CTR} &
\includegraphics[width=1.03\linewidth]{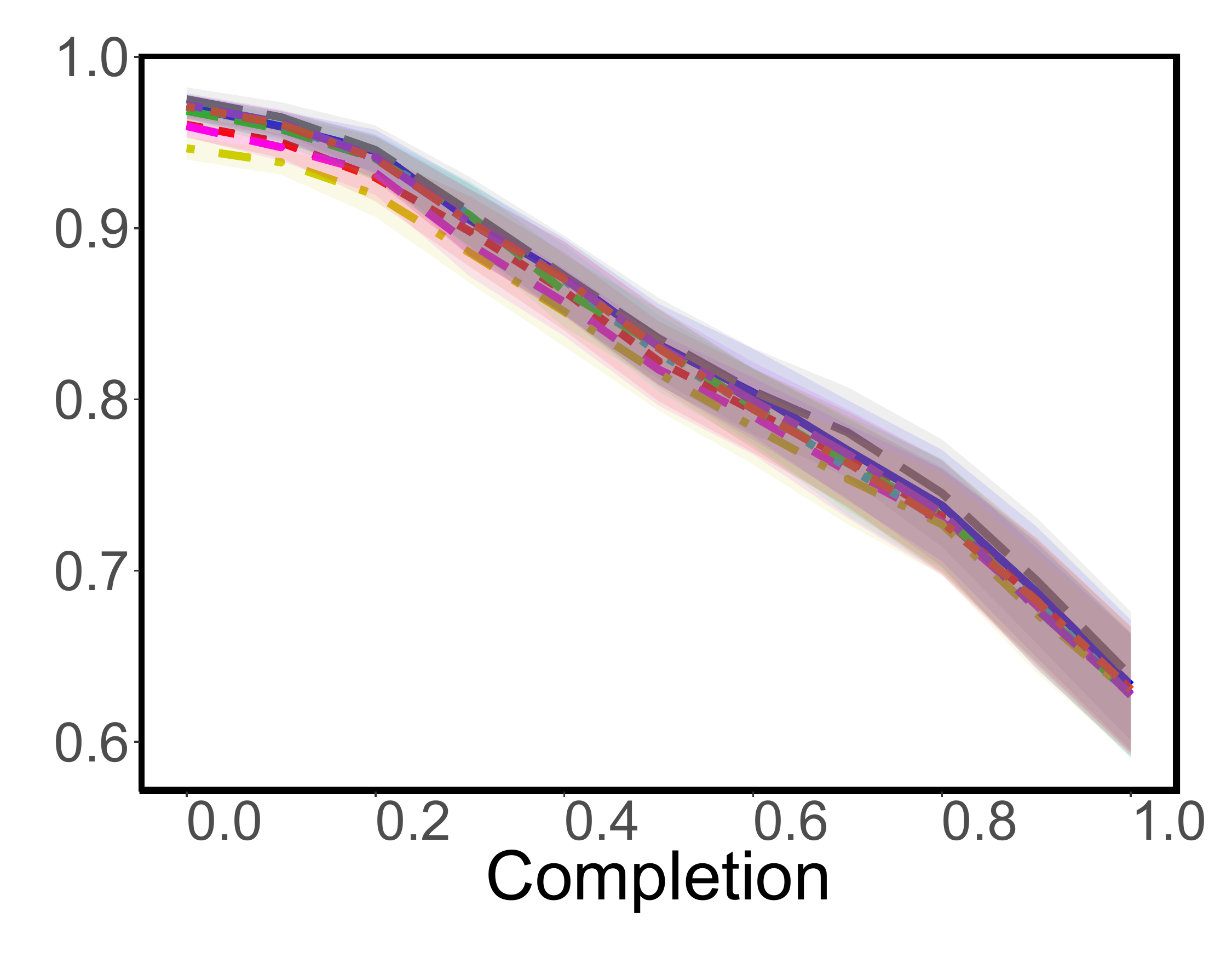} &
\includegraphics[width=1.03\linewidth]{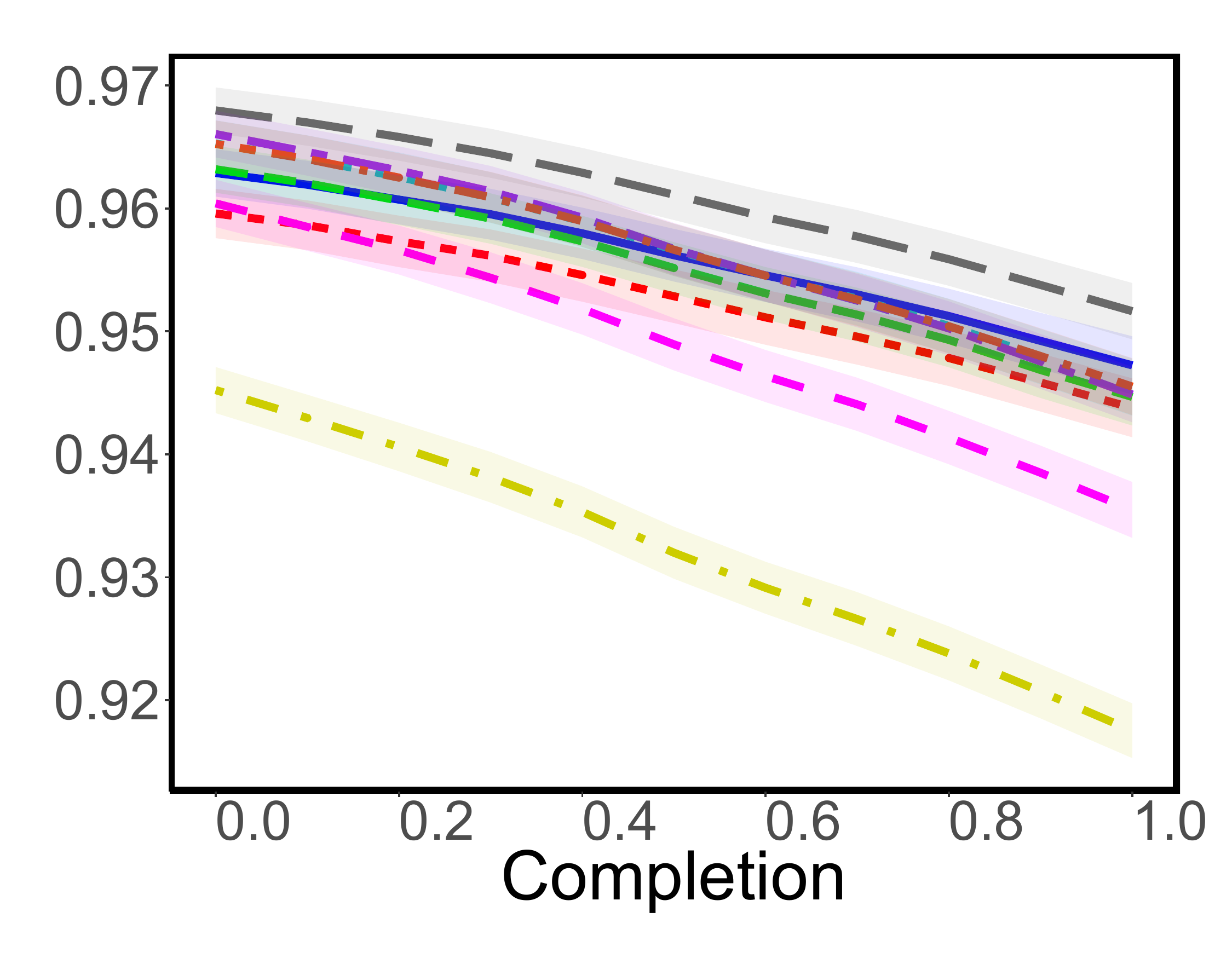} &
\includegraphics[width=1.03\linewidth]{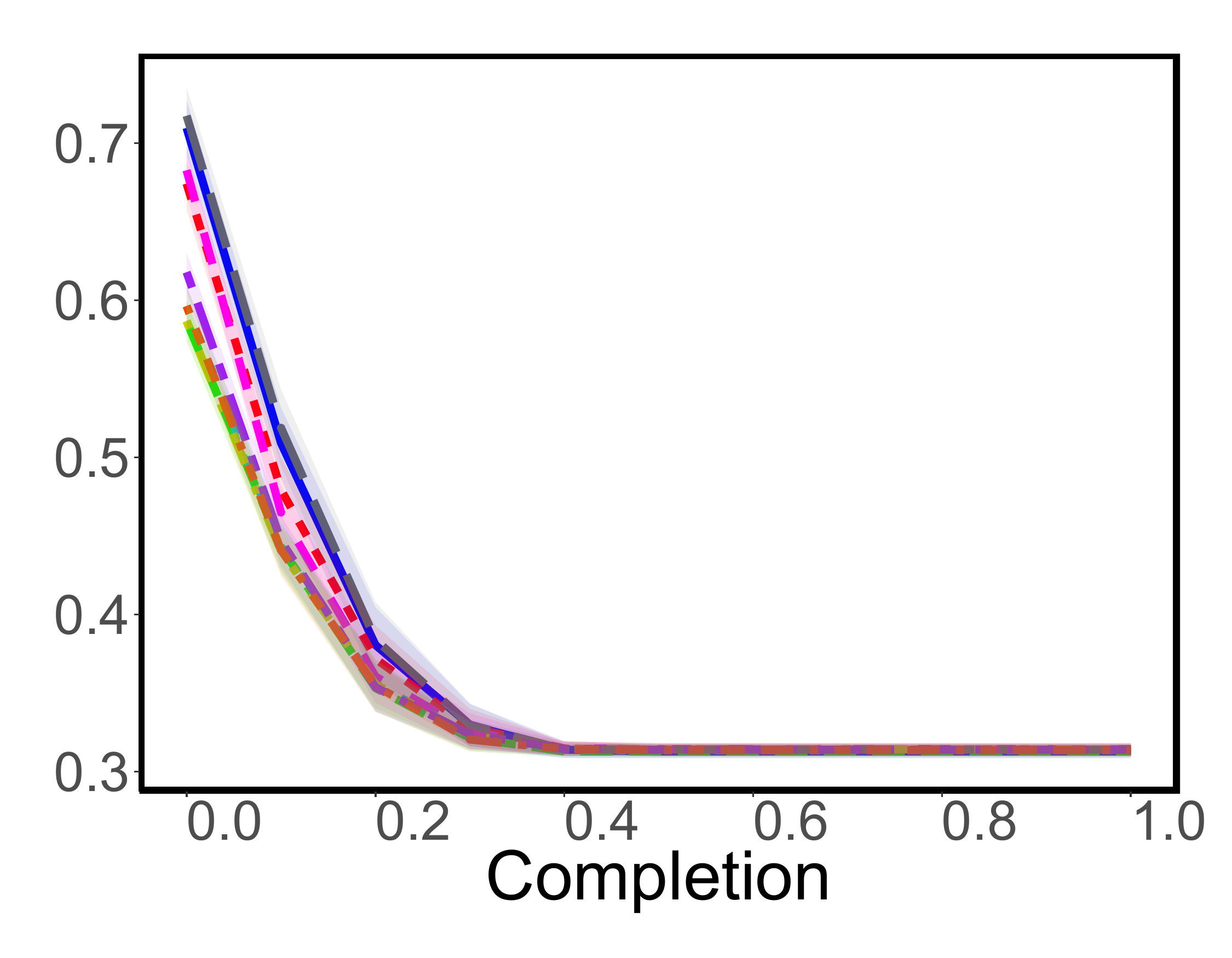} \\
\rotatebox{90}{\footnotesize $\AP$ values for OTC} &
\includegraphics[width=1.03\linewidth]{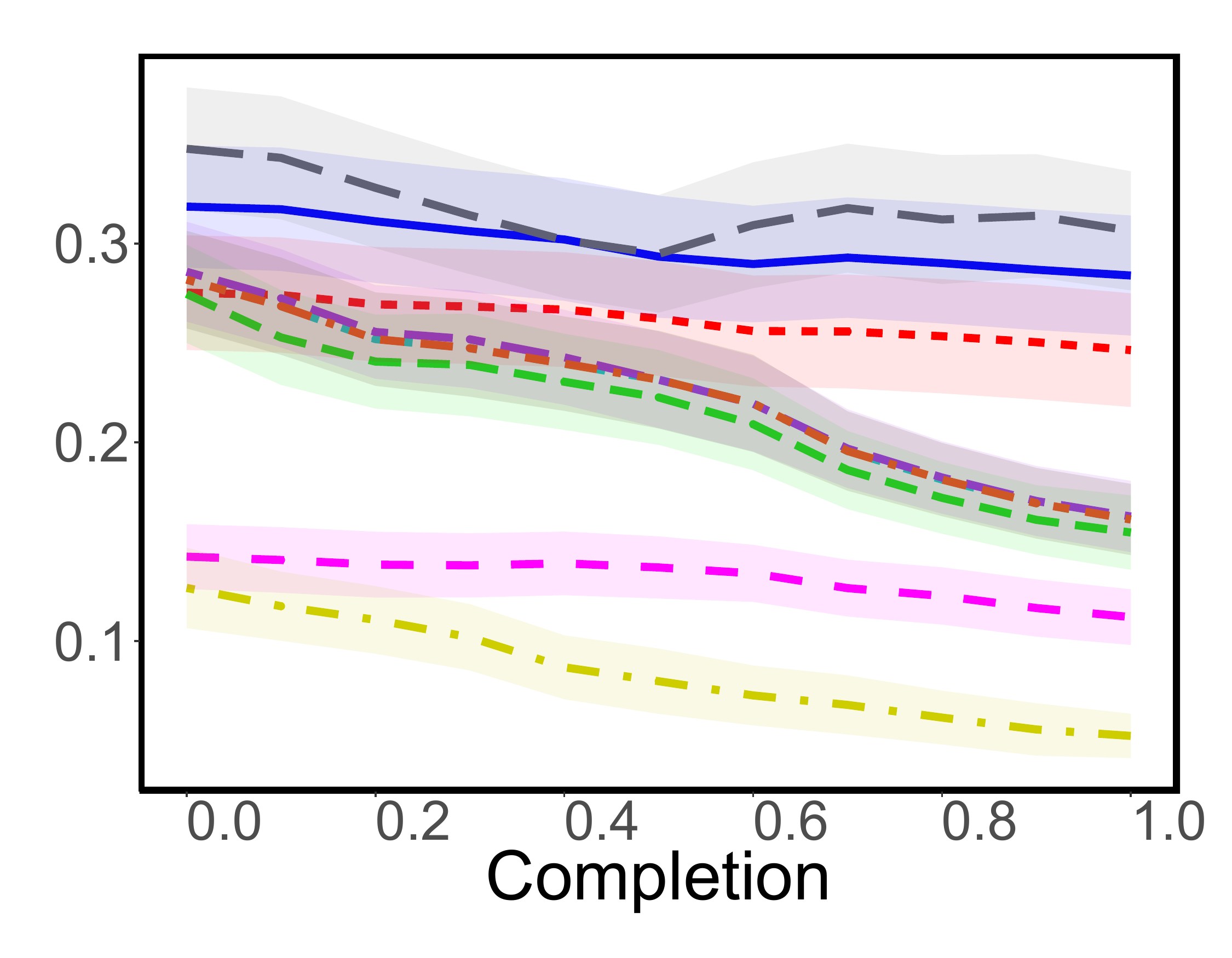} &
\includegraphics[width=1.03\linewidth]{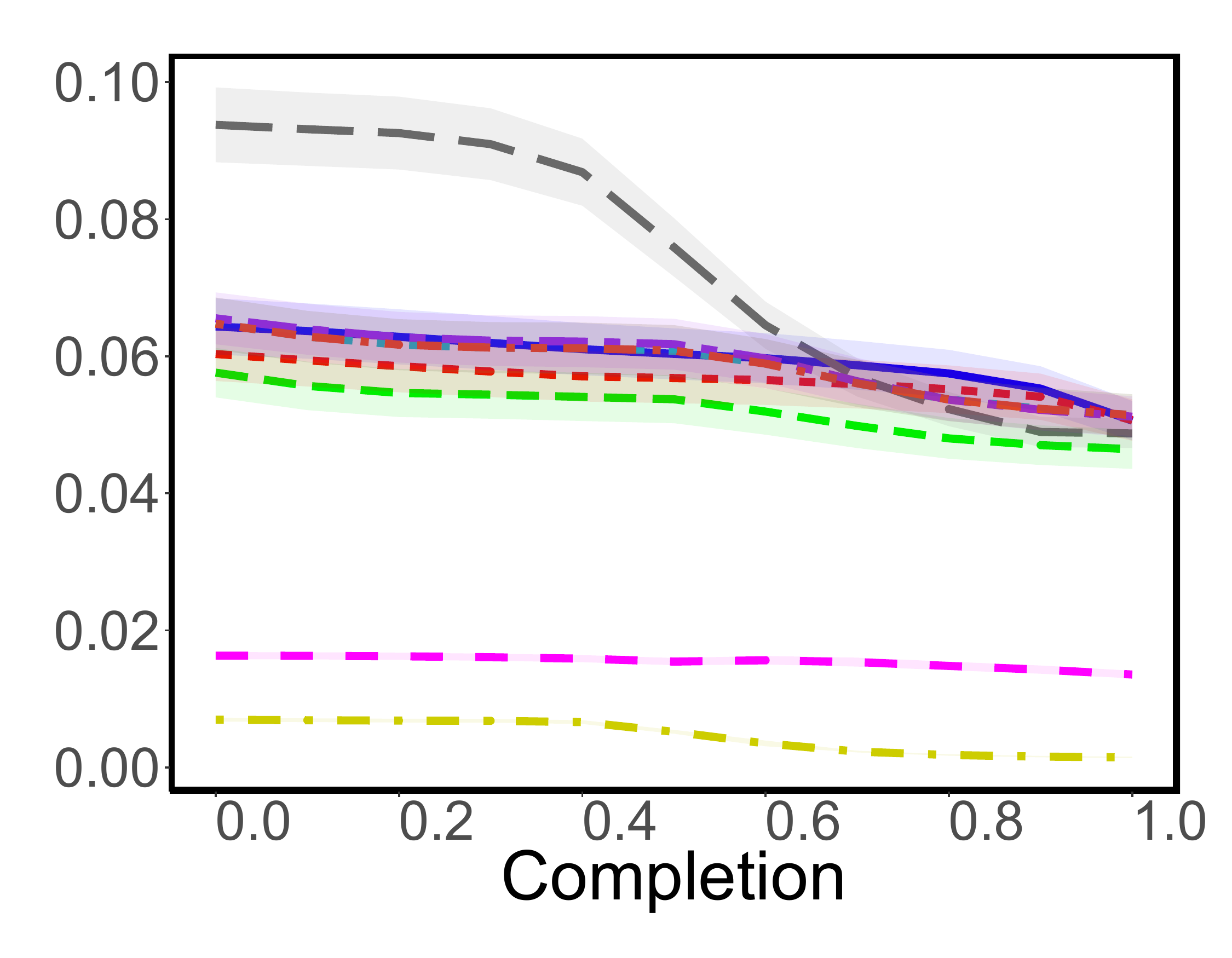} &
\includegraphics[width=1.03\linewidth]{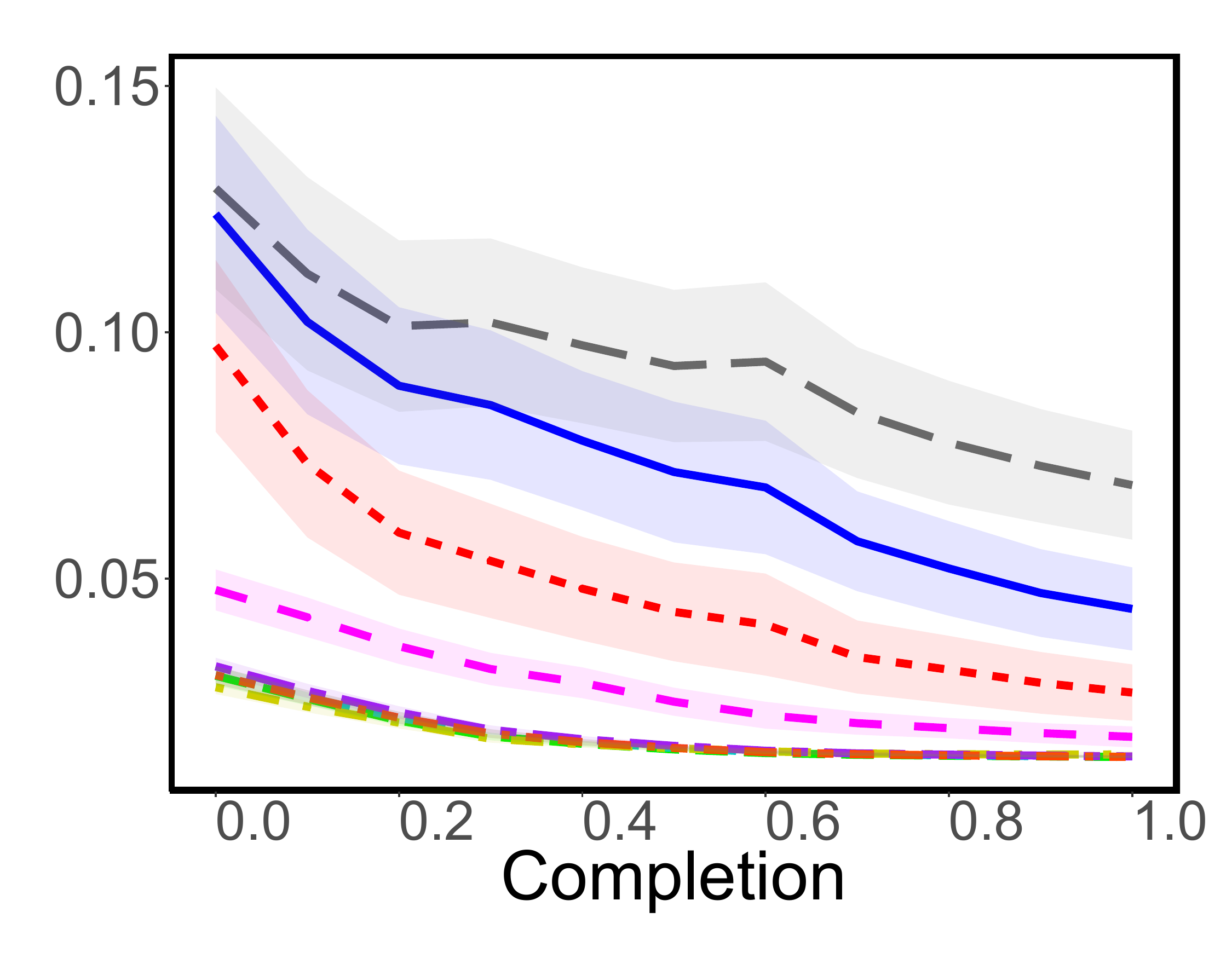} \\
\rotatebox{90}{\footnotesize $\AP$ values for CTR} &
\includegraphics[width=1.03\linewidth]{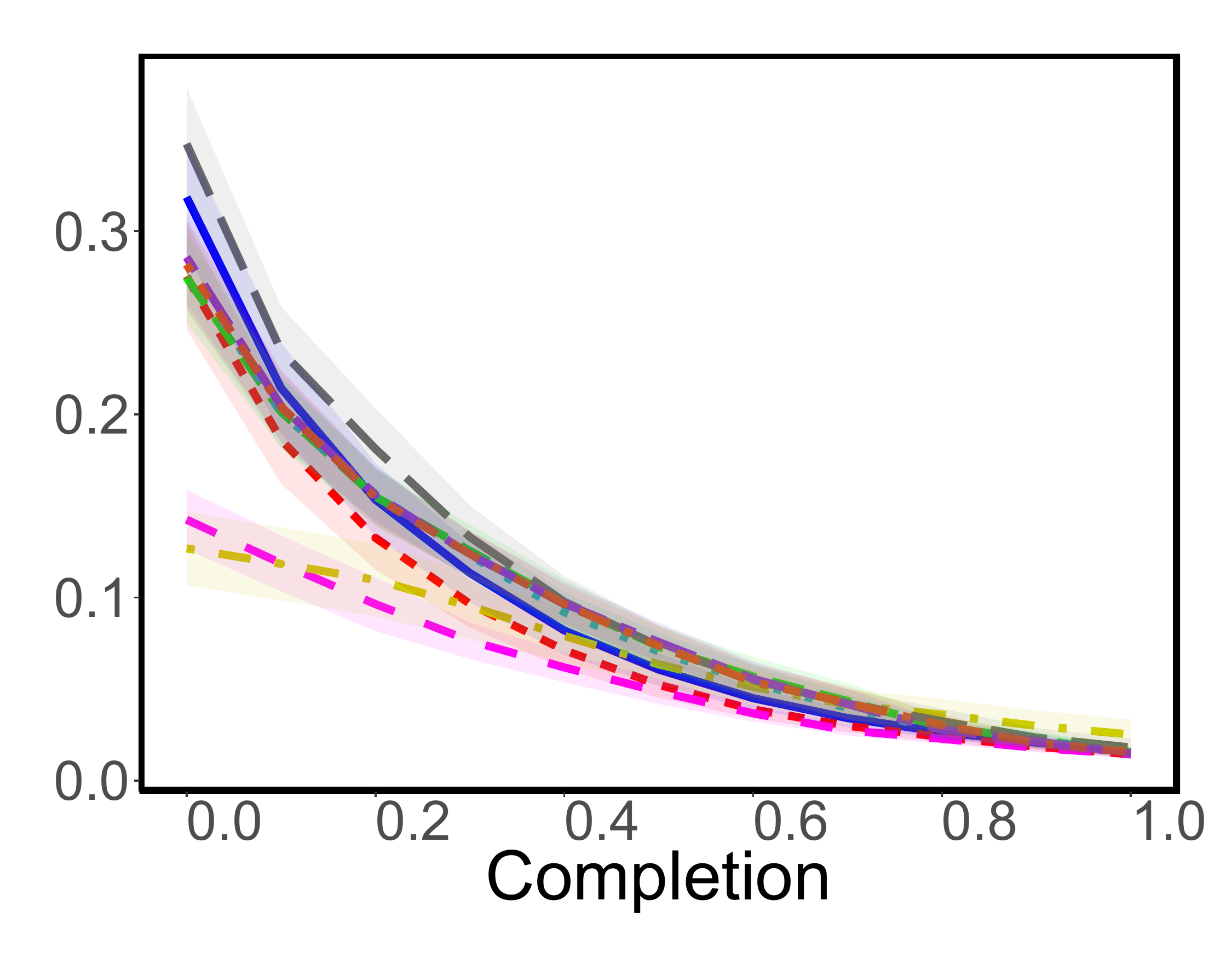} &
\includegraphics[width=1.03\linewidth]{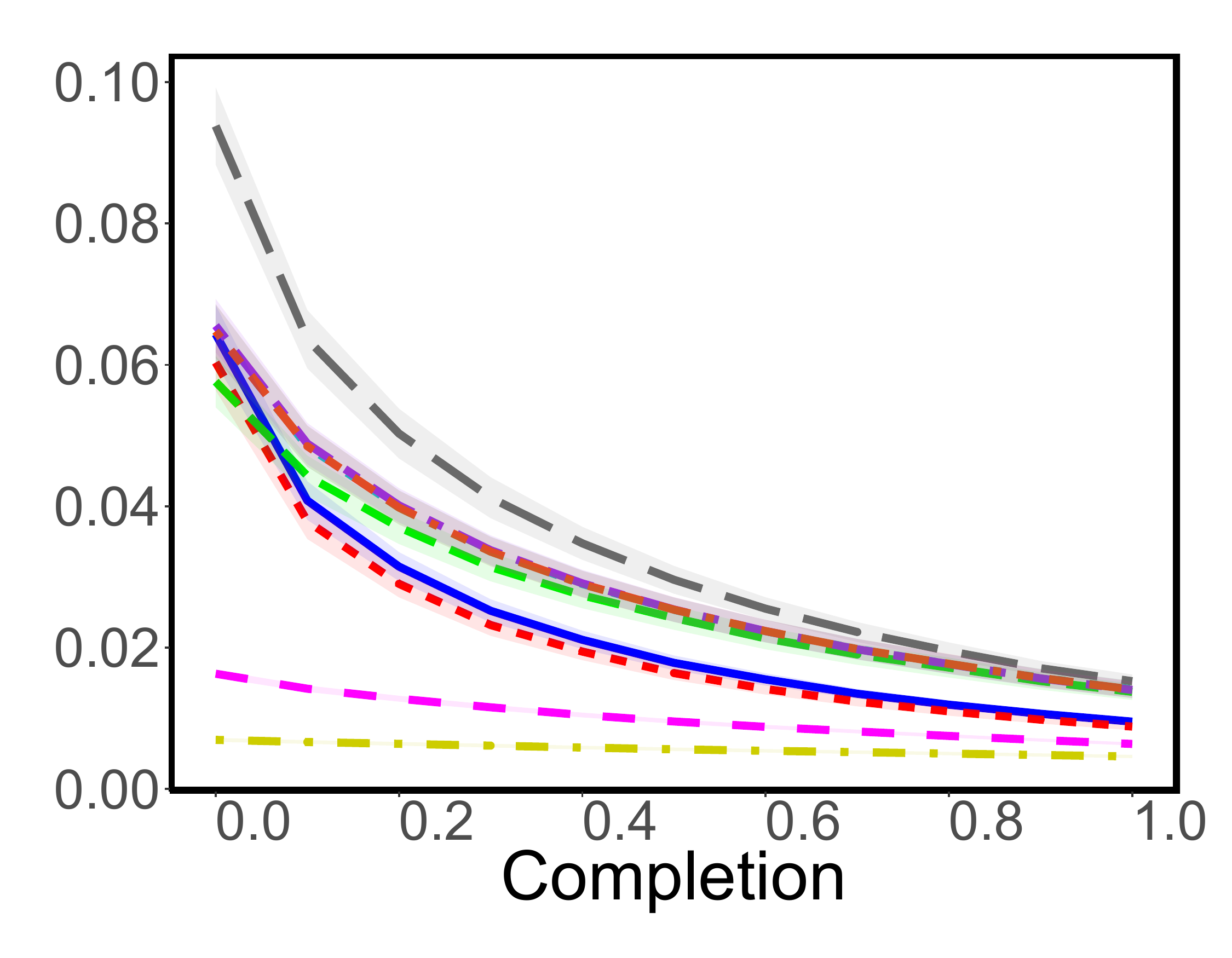} &
\includegraphics[width=1.03\linewidth]{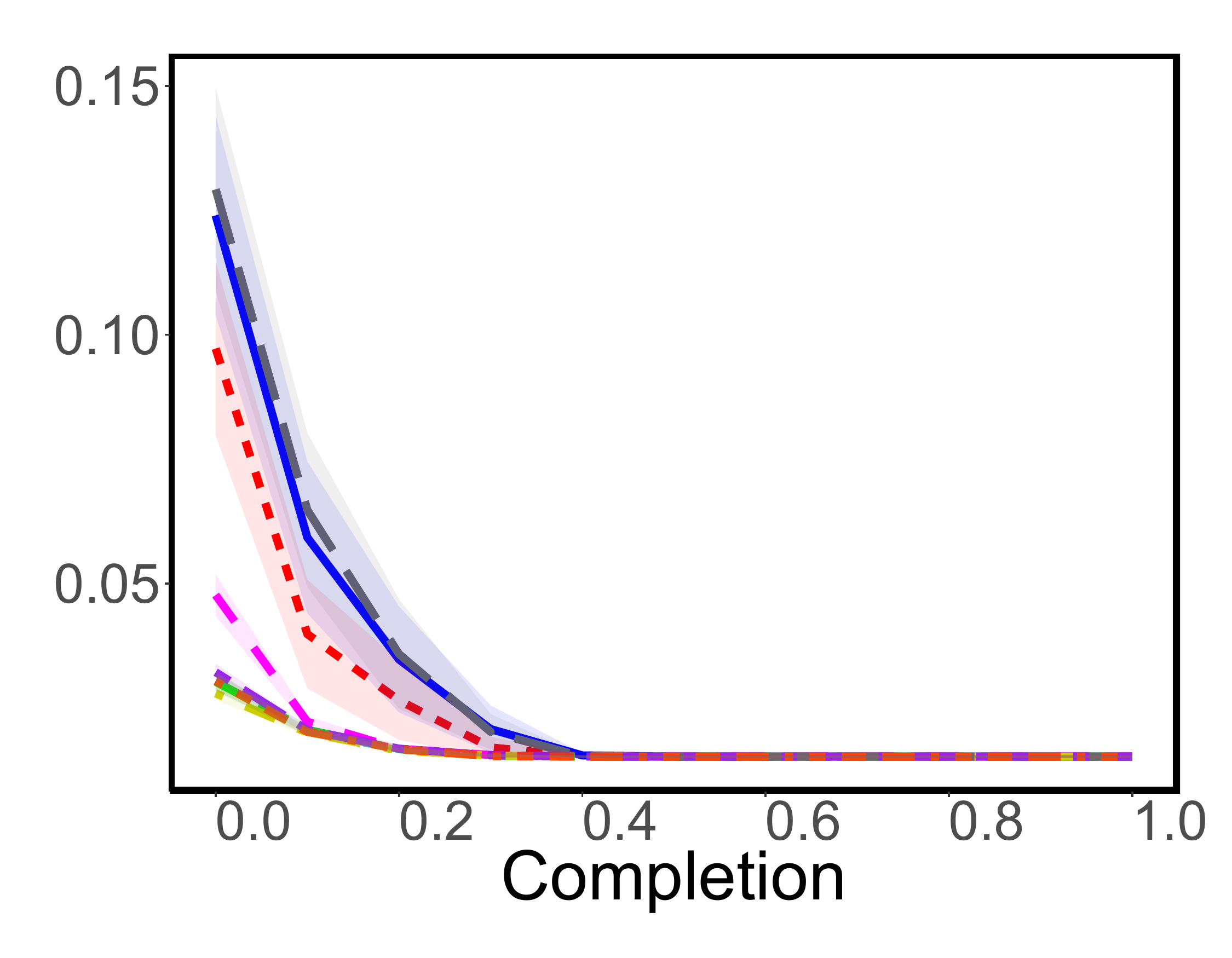} \\
\multicolumn{4}{c}{\includegraphics[width=0.65\linewidth]{figures/plots/local/legend}}
\end{tabular}
\caption{Given different \textbf{local similarity} indices, the figure depicts the values of $\ROC$ (the area under the ROC curve) and $\AP$ (the average precision) during the execution of OTC and CTR given $|\Hide|=\max(10,|E|/100)$ and $b=4|\Hide|$ in three networks: (i) \textbf{a small fragment of Facebook}; (ii) \textbf{a large fragment of Facebook}; and (iii) \textbf{the Zachary karate club network}.
In each execution, the links in $\Hide$ are chosen at random. Results are taken as the average over $50$ executions, with coloured areas representing the $95\%$ confidence intervals.}
\label{fig:local-3}
\end{figure*}

\begin{figure*}[tbhp]
\centering
\setlength\tabcolsep{1pt}
\renewcommand{\arraystretch}{0.01}
\begin{tabular}{m{.03\textwidth}m{.27\textwidth}m{.27\textwidth}m{.27\textwidth}}
& \multicolumn{1}{c}{Bali-attack network}
& \multicolumn{1}{c}{Madrid-bombing network}
& \multicolumn{1}{c}{Greek political blogs}\\
\rotatebox{90}{\footnotesize $\ROC$ values for OTC} &
\includegraphics[width=1.03\linewidth]{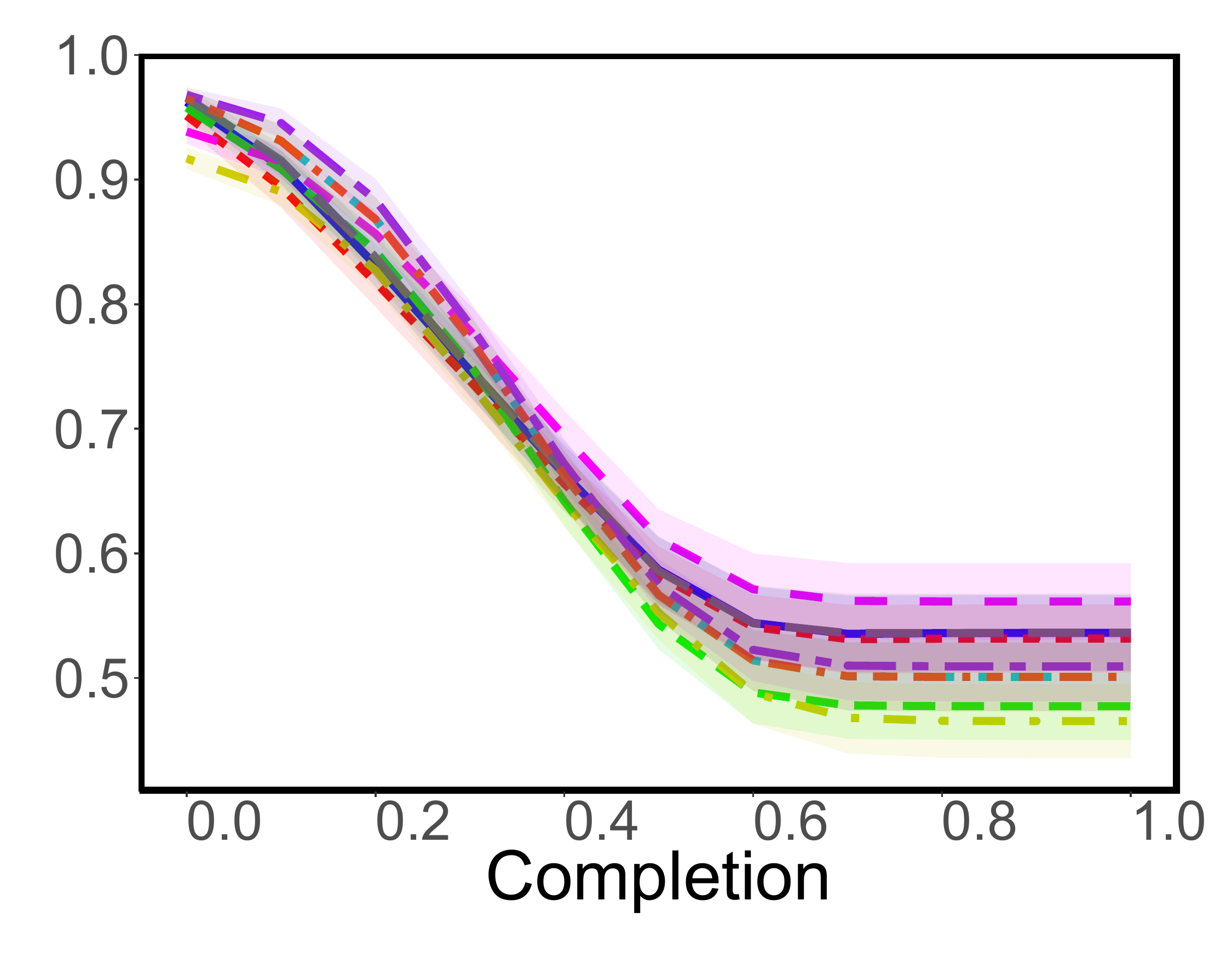} &
\includegraphics[width=1.03\linewidth]{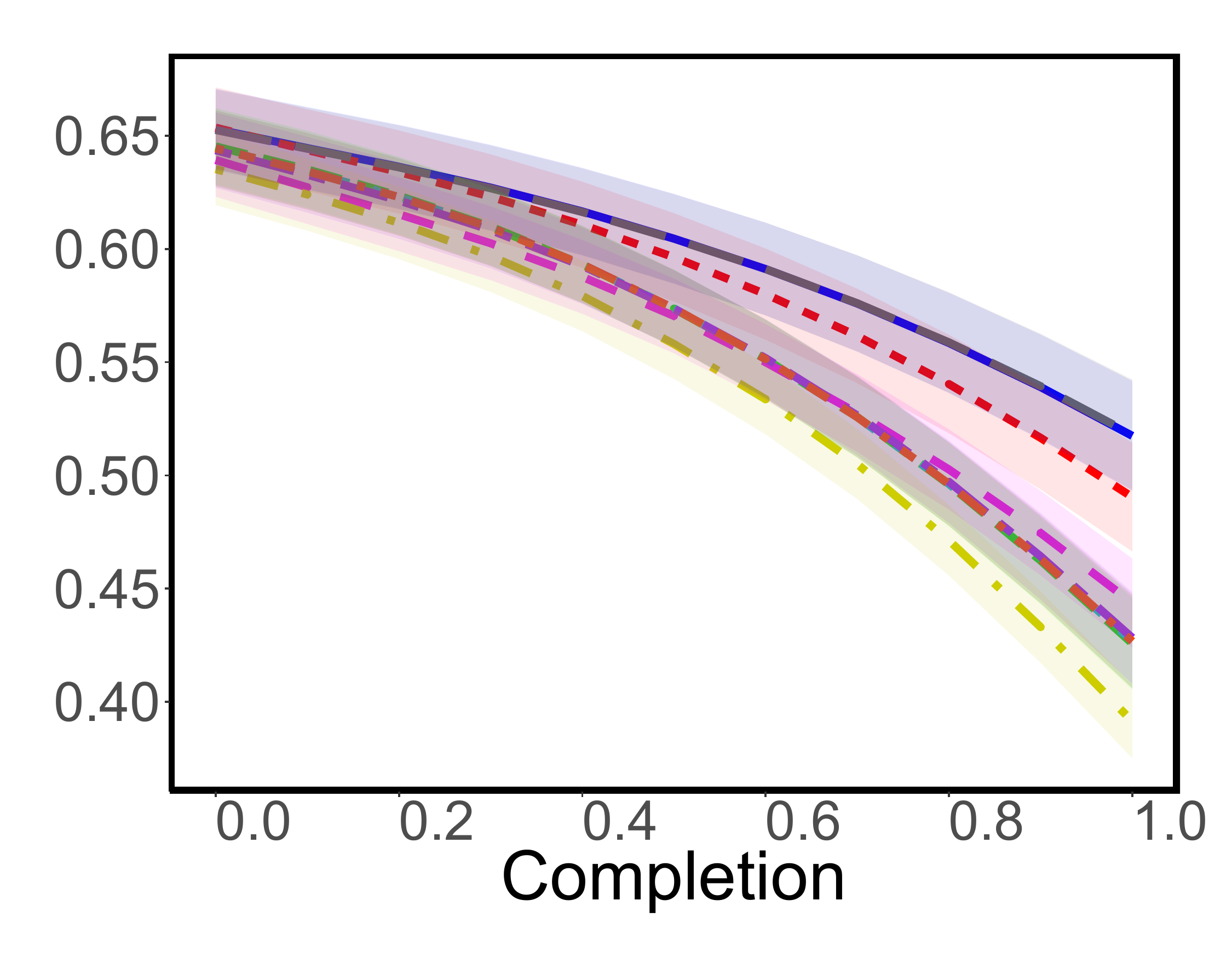} &
\includegraphics[width=1.03\linewidth]{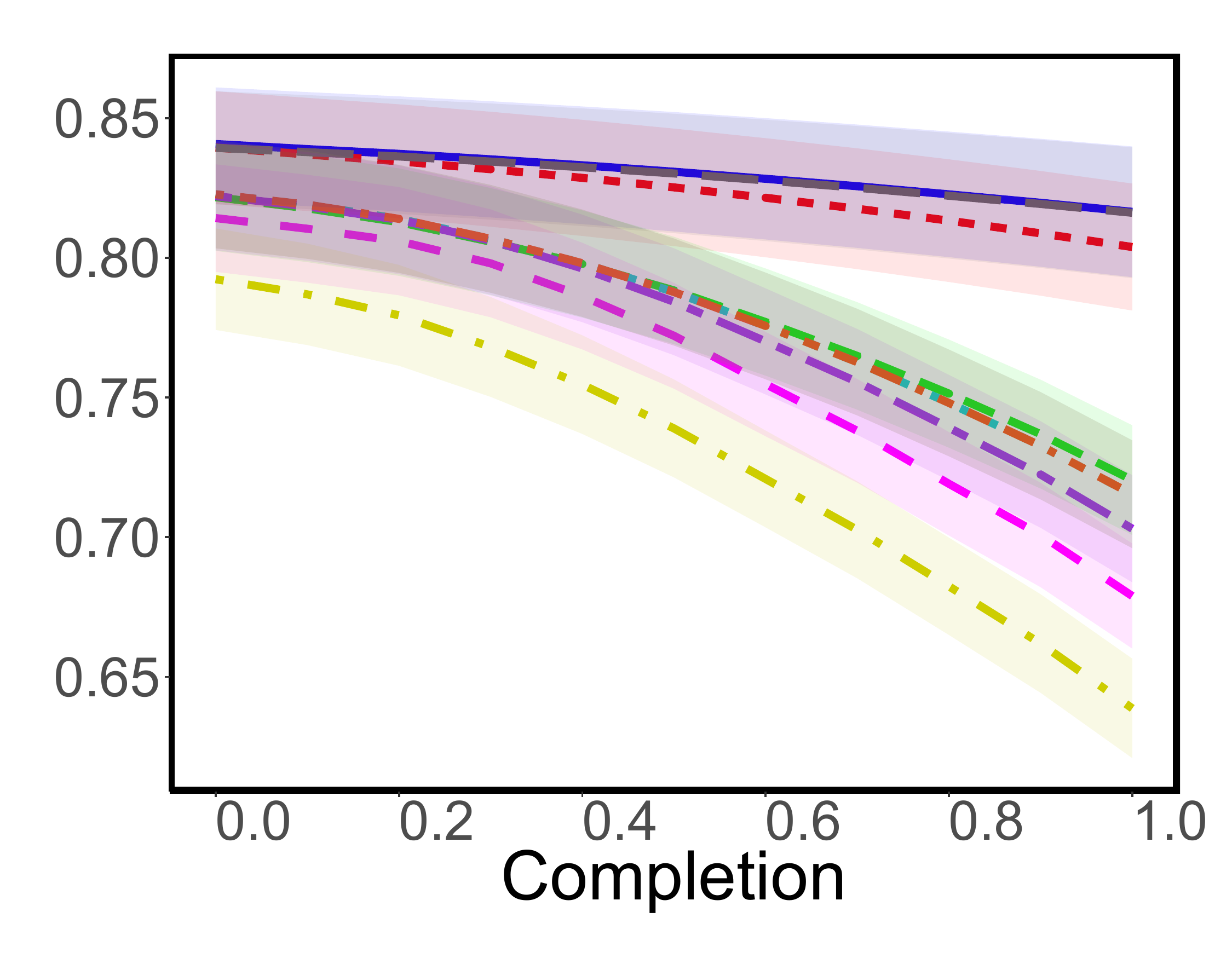}\\
\rotatebox{90}{\footnotesize $\ROC$ values for CTR} &
\includegraphics[width=1.03\linewidth]{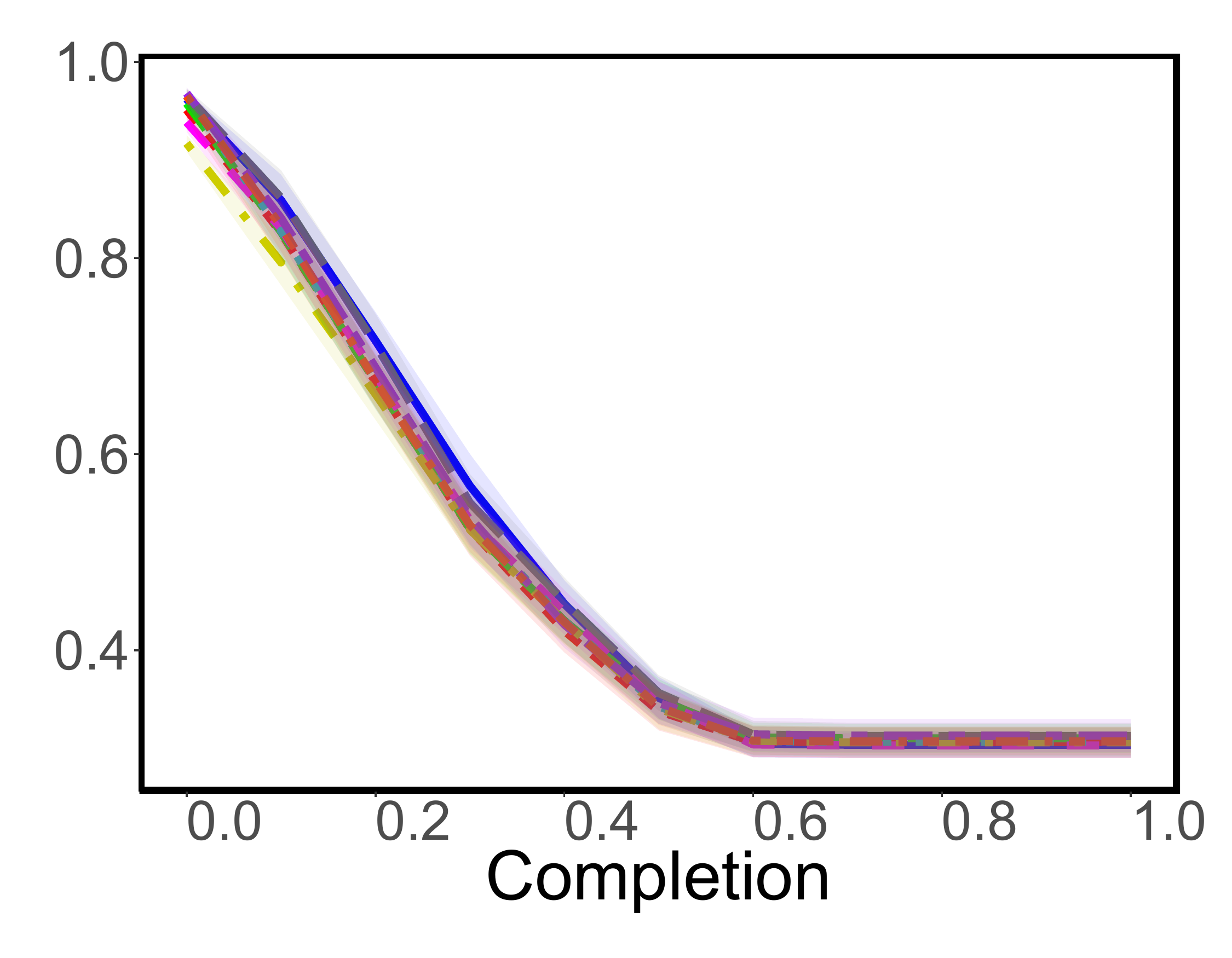} &
\includegraphics[width=1.03\linewidth]{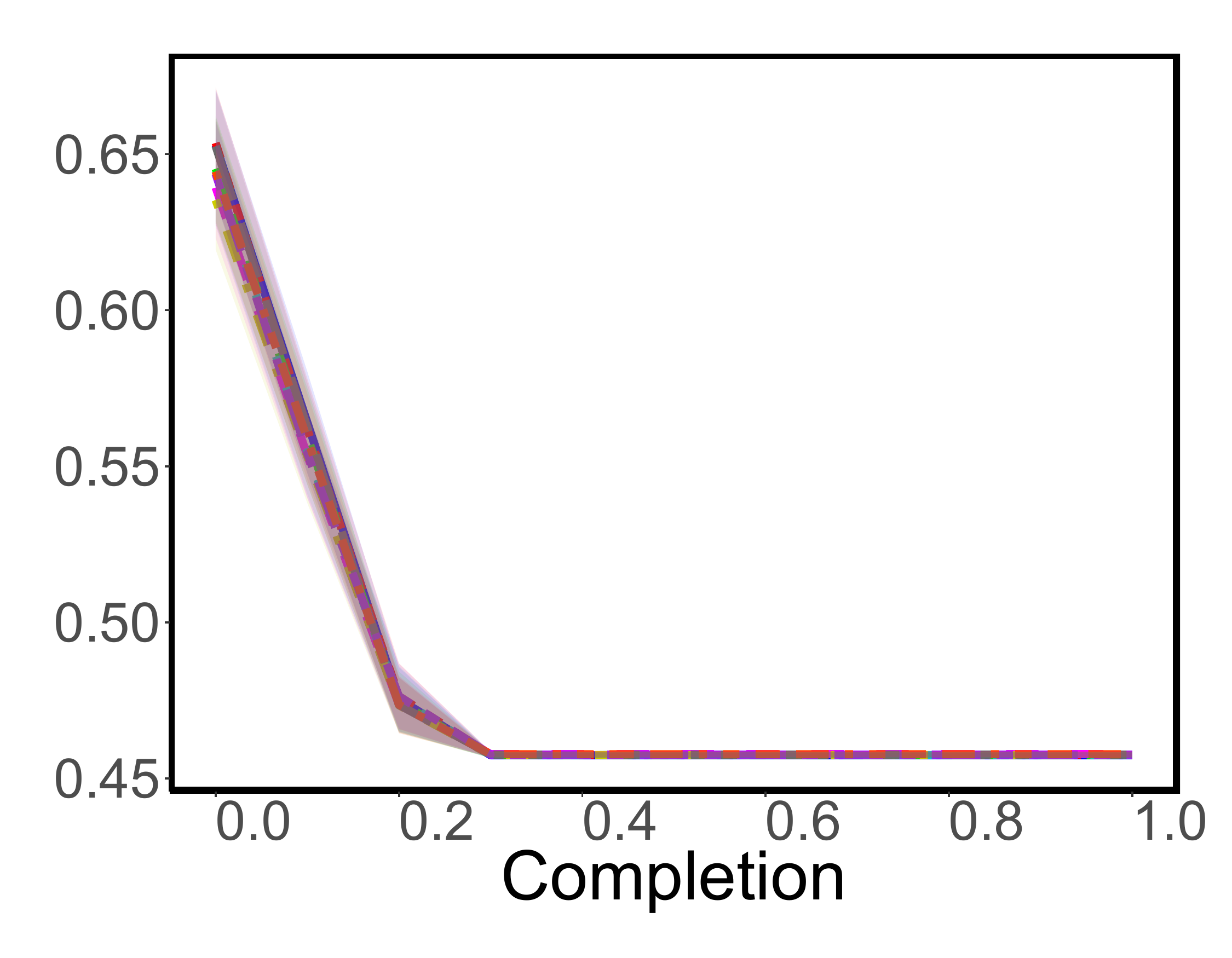} &
\includegraphics[width=1.03\linewidth]{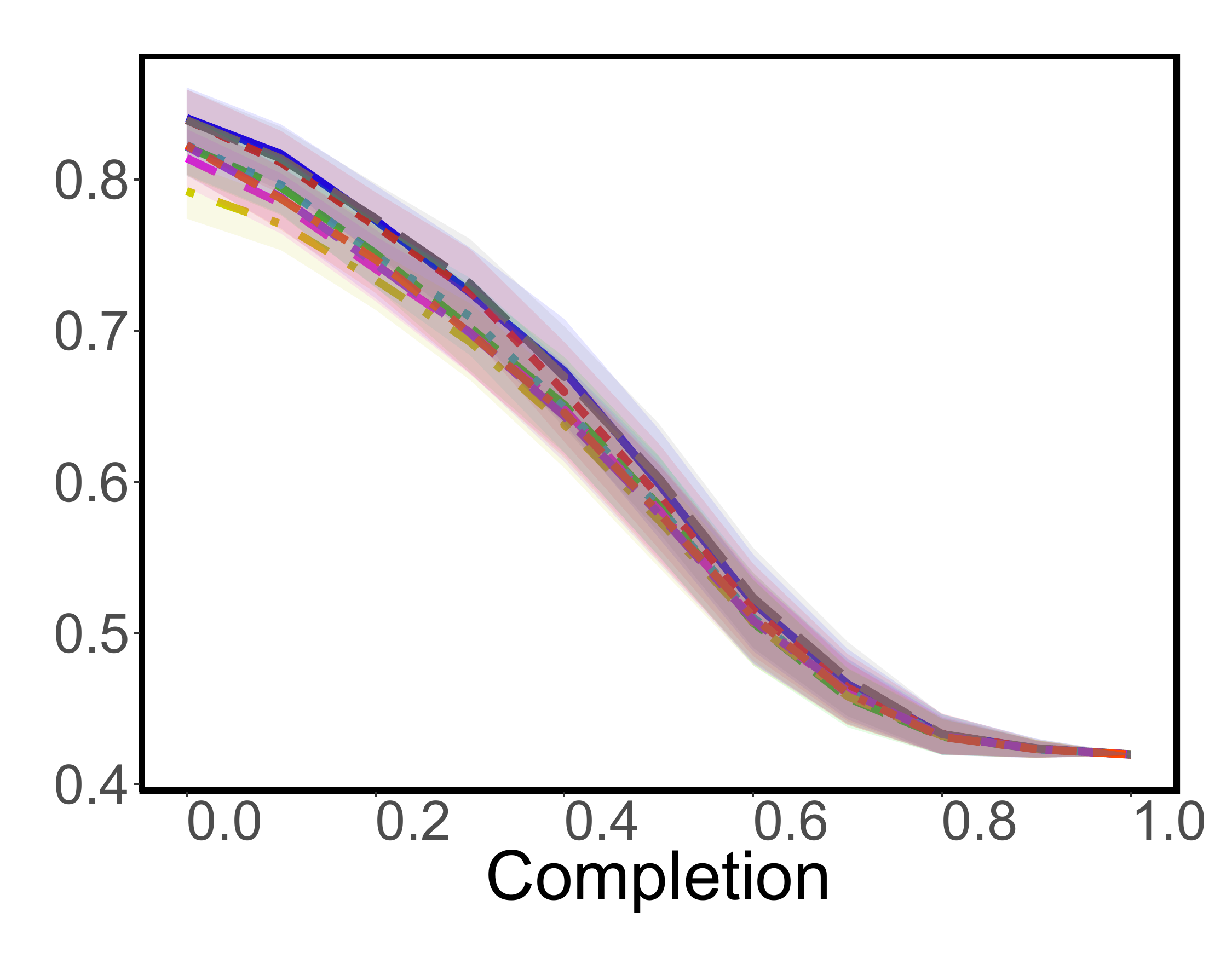} \\
\rotatebox{90}{\footnotesize $\AP$ values for OTC} &
\includegraphics[width=1.03\linewidth]{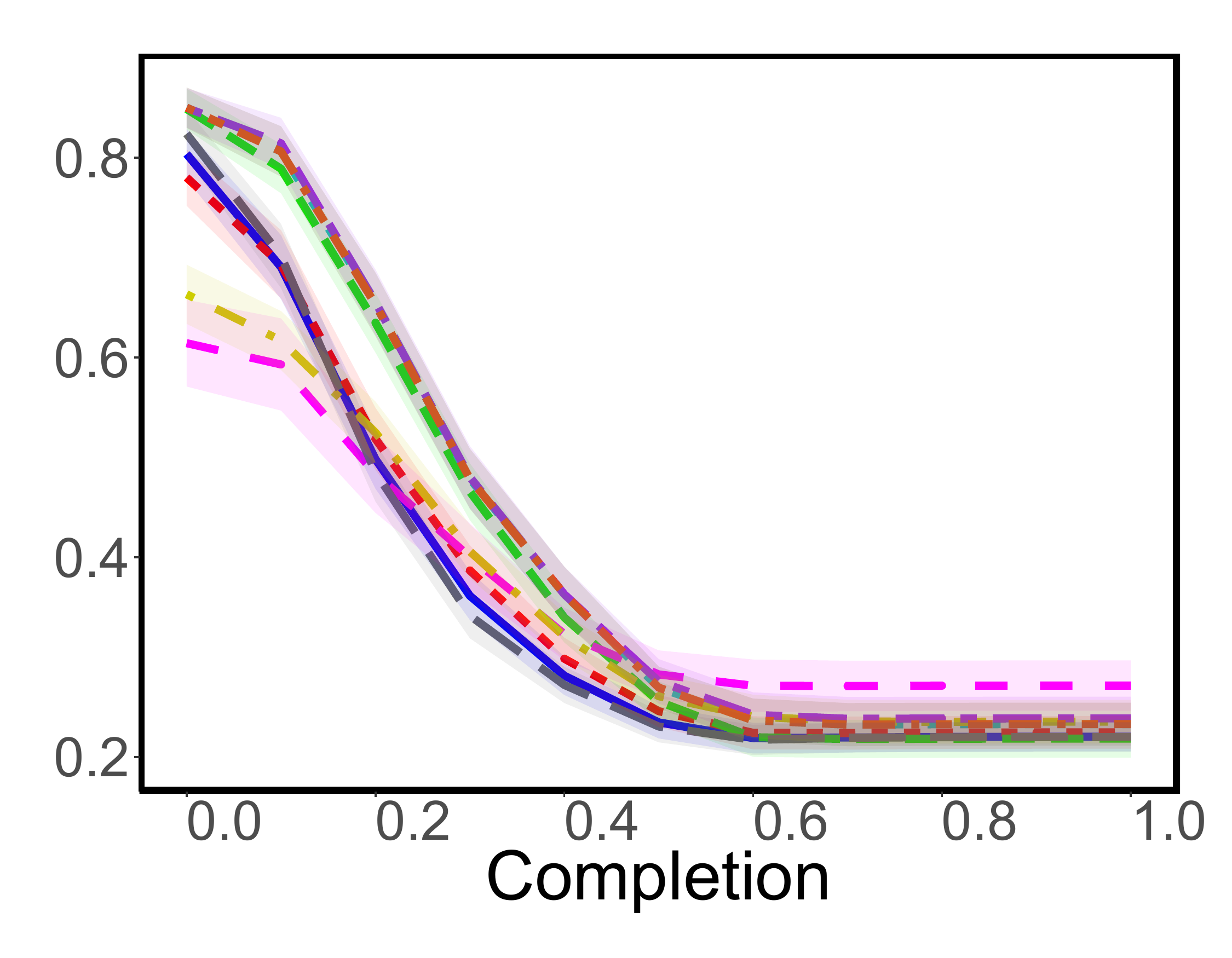} &
\includegraphics[width=1.03\linewidth]{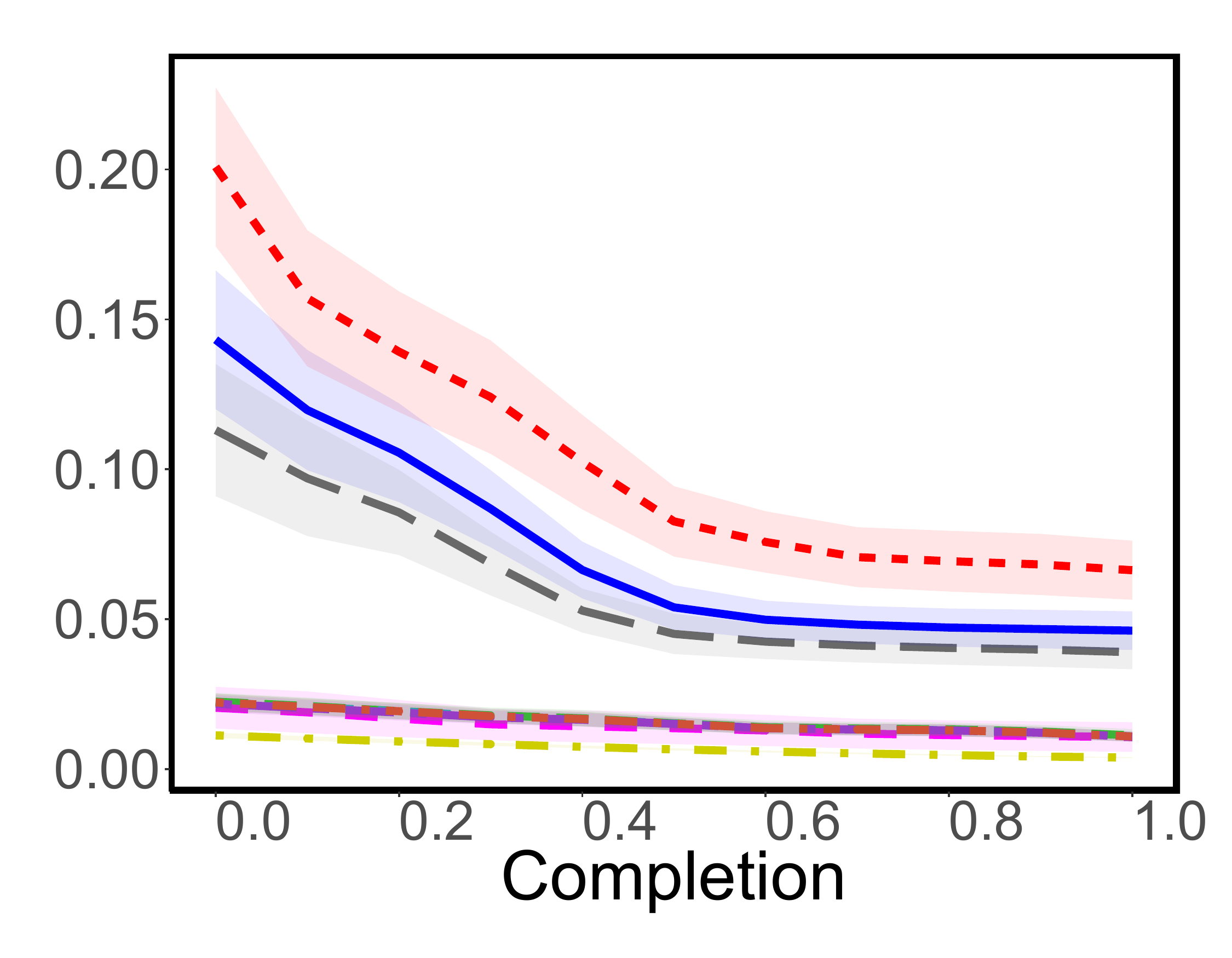} &
\includegraphics[width=1.03\linewidth]{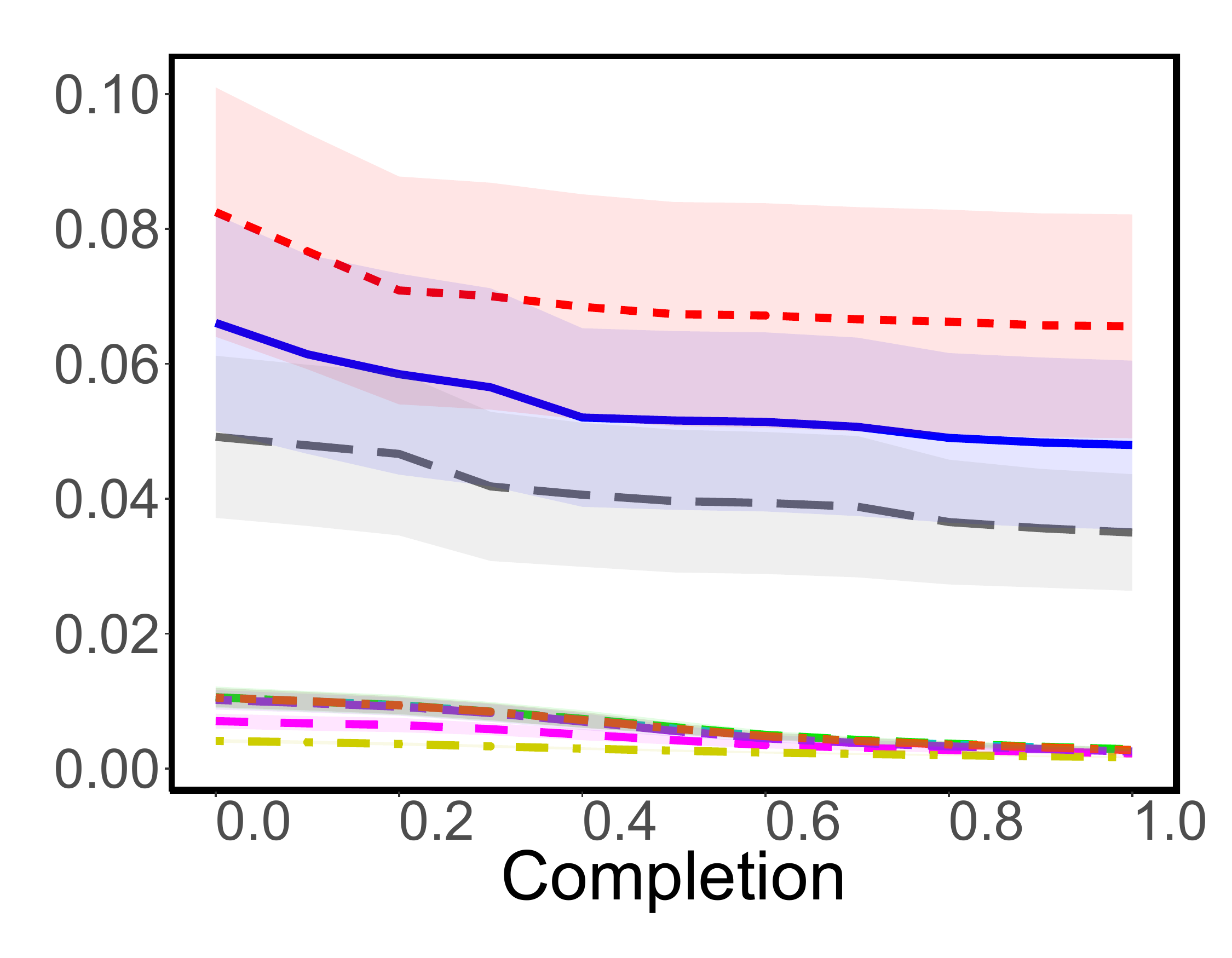} \\
\rotatebox{90}{\footnotesize $\AP$ values for CTR} &
\includegraphics[width=1.03\linewidth]{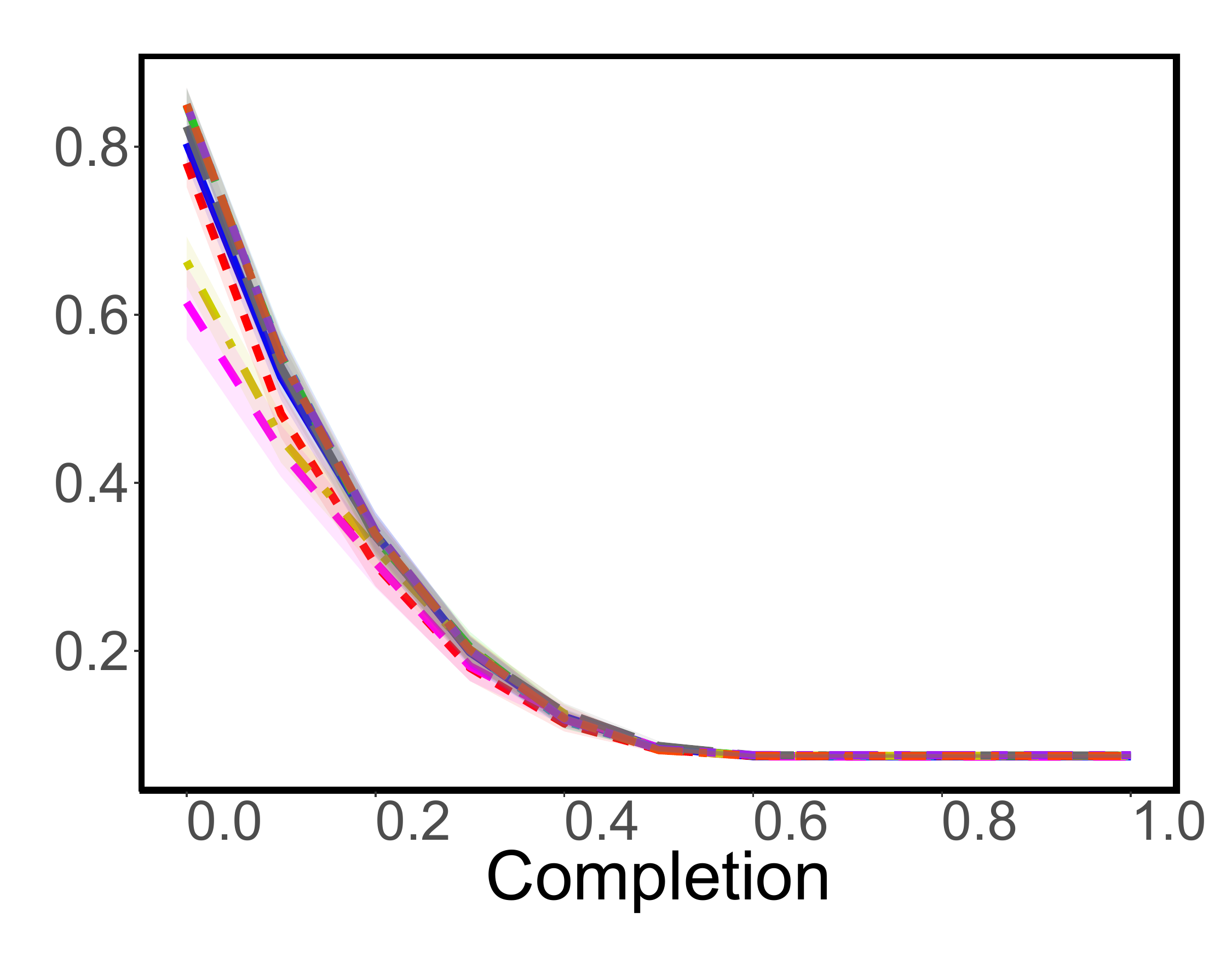} &
\includegraphics[width=1.03\linewidth]{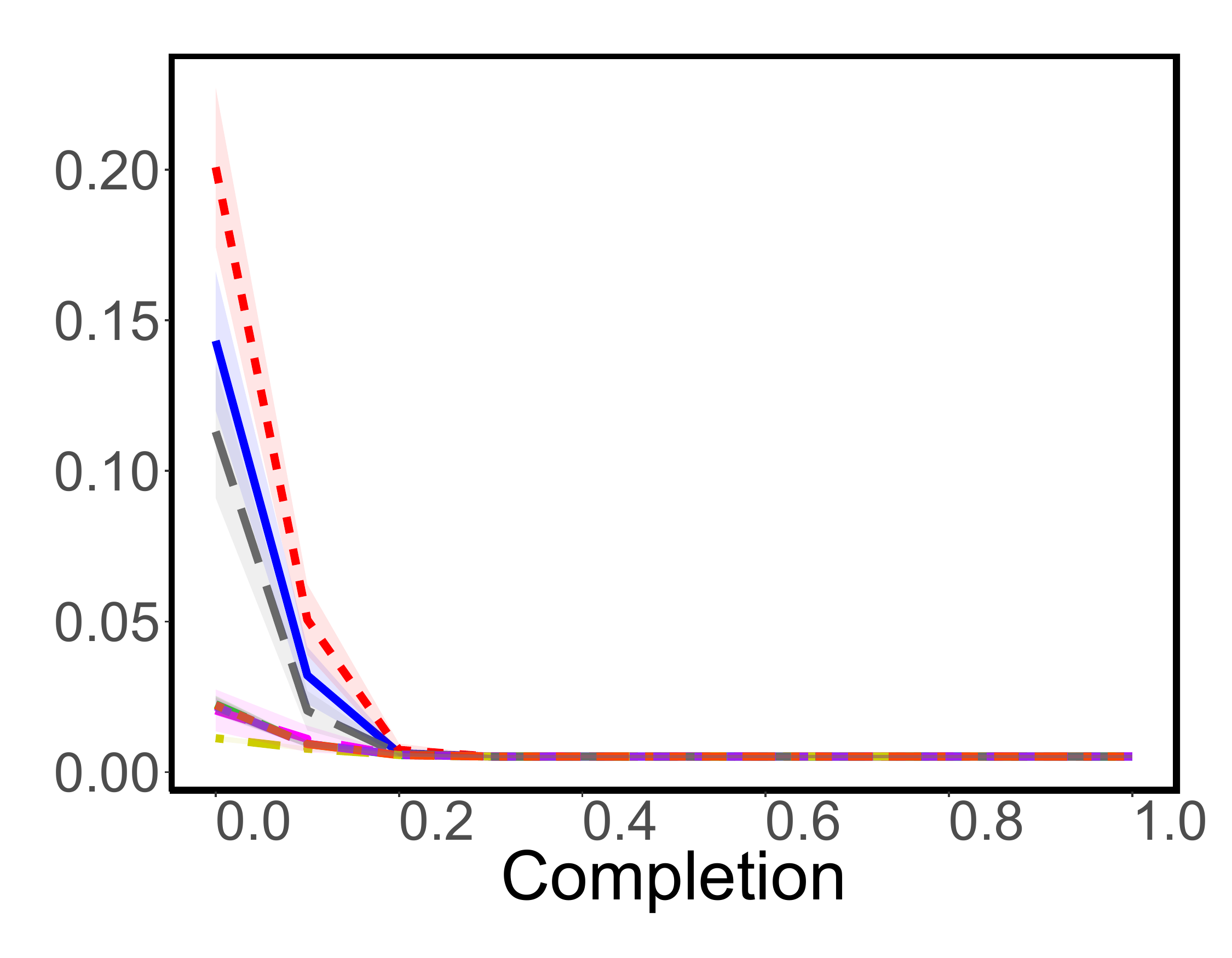} &
\includegraphics[width=1.03\linewidth]{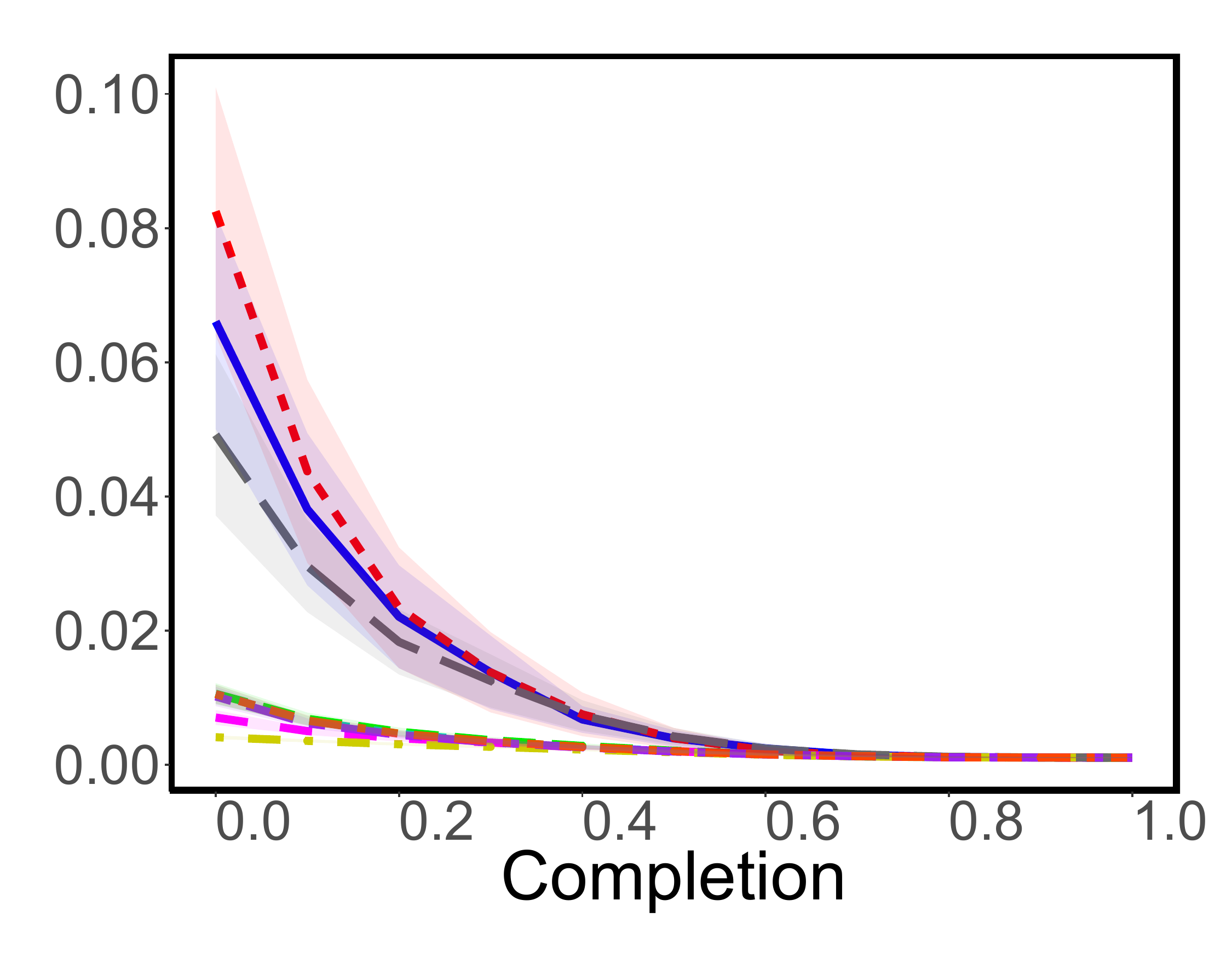} \\
\multicolumn{4}{c}{\includegraphics[width=0.65\linewidth]{figures/plots/local/legend}}
\end{tabular}
\caption{Given different \textbf{local similarity} indices, the figure depicts the values of $\ROC$ (the area under the ROC curve) and $\AP$ (the average precision) during the execution of OTC and CTR given $|\Hide|=\max(10,|E|/100)$ and $b=4|\Hide|$ in three networks: (i) \textbf{the Bali-attack network}; (ii) \textbf{the Madrid-bombing network}; and (iii) \textbf{the Greek political blog network}.
In each execution, the links in $\Hide$ are chosen at random. Results are taken as the average over $50$ executions, with coloured areas representing the $95\%$ confidence intervals.}
\label{fig:local-4}
\end{figure*}

\begin{figure*}[ht!]
\centering
\setlength\tabcolsep{1pt}
\begin{tabular}{m{.03\textwidth}m{.4\textwidth}m{.4\textwidth}}
&
\multicolumn{1}{c}{OTC} &
\multicolumn{1}{c}{CTR} \\
\rotatebox{90}{\hspace*{2.1cm} Relative change in $\ROC$} &
\includegraphics[width=1\linewidth]{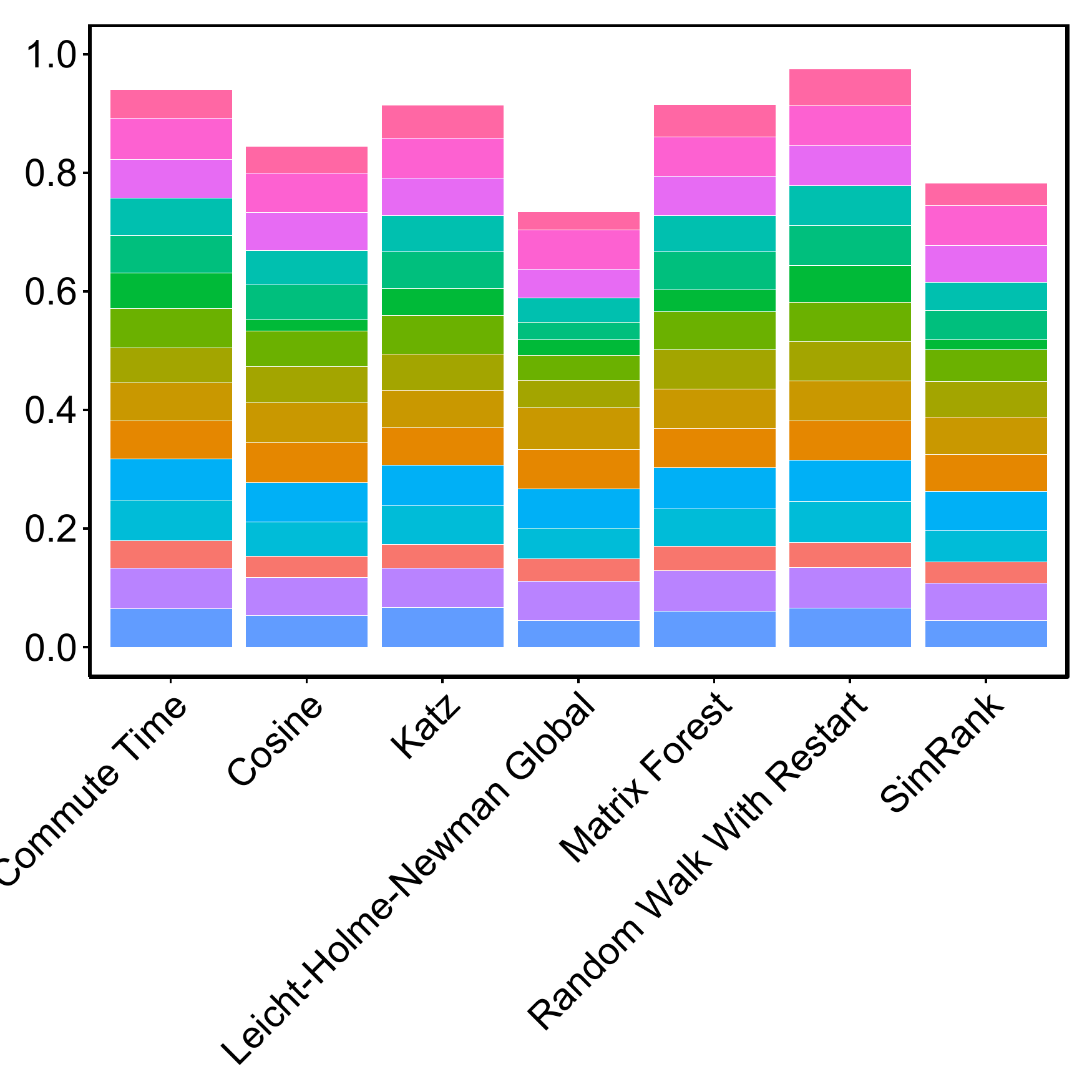} &
\includegraphics[width=1\linewidth]{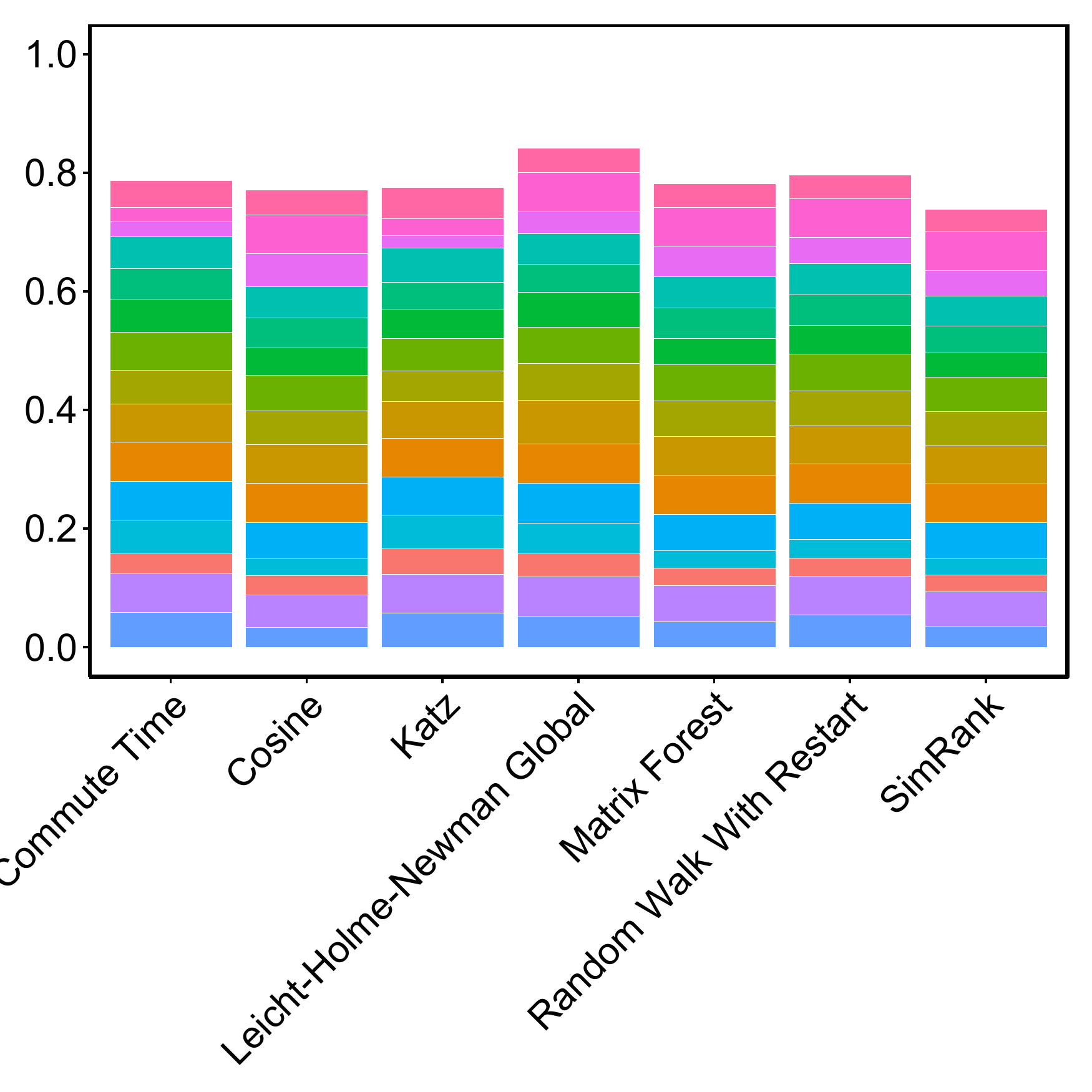} \\
\rotatebox{90}{\hspace*{2.1cm} Relative change in $\AP$} &
\includegraphics[width=1\linewidth]{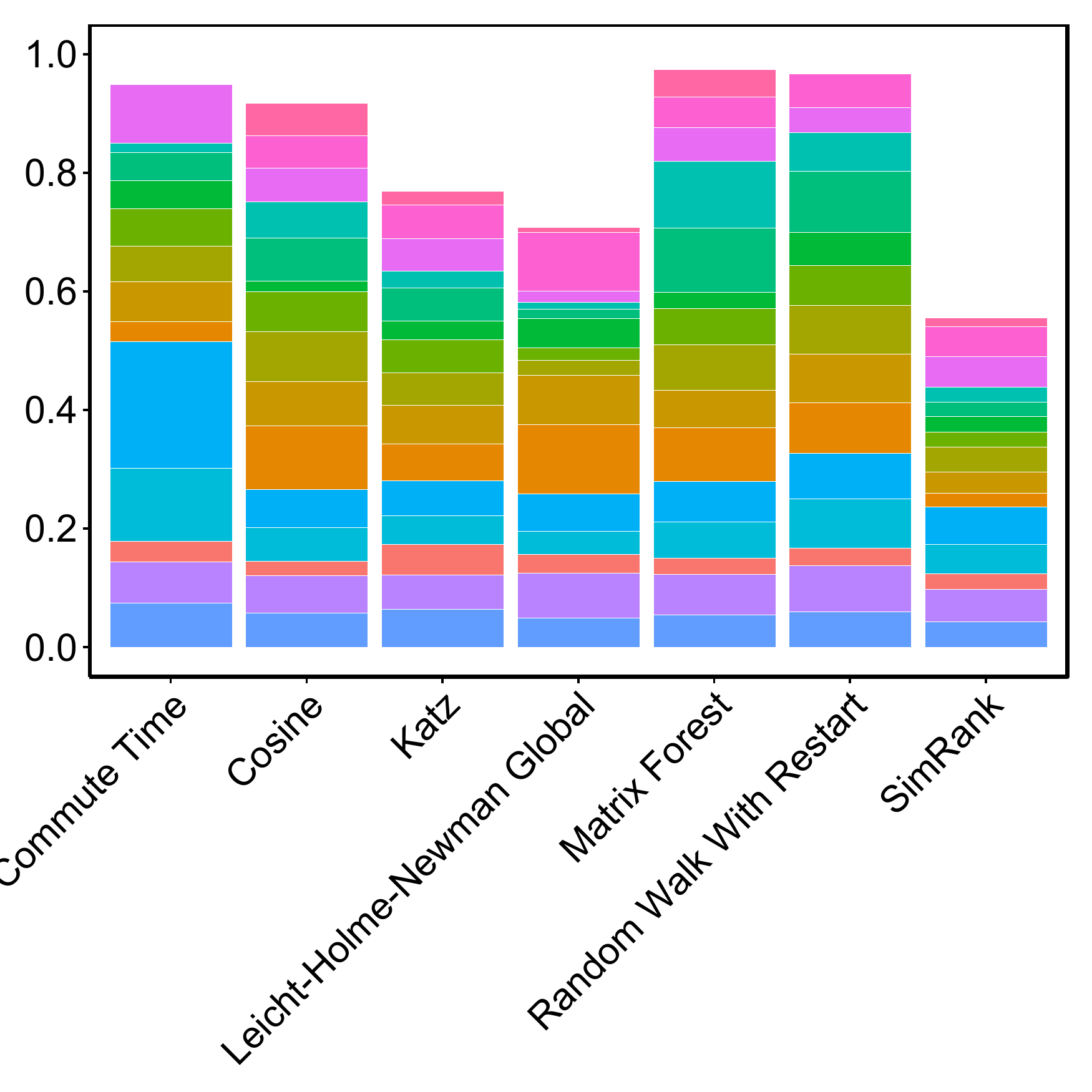} &
\includegraphics[width=1\linewidth]{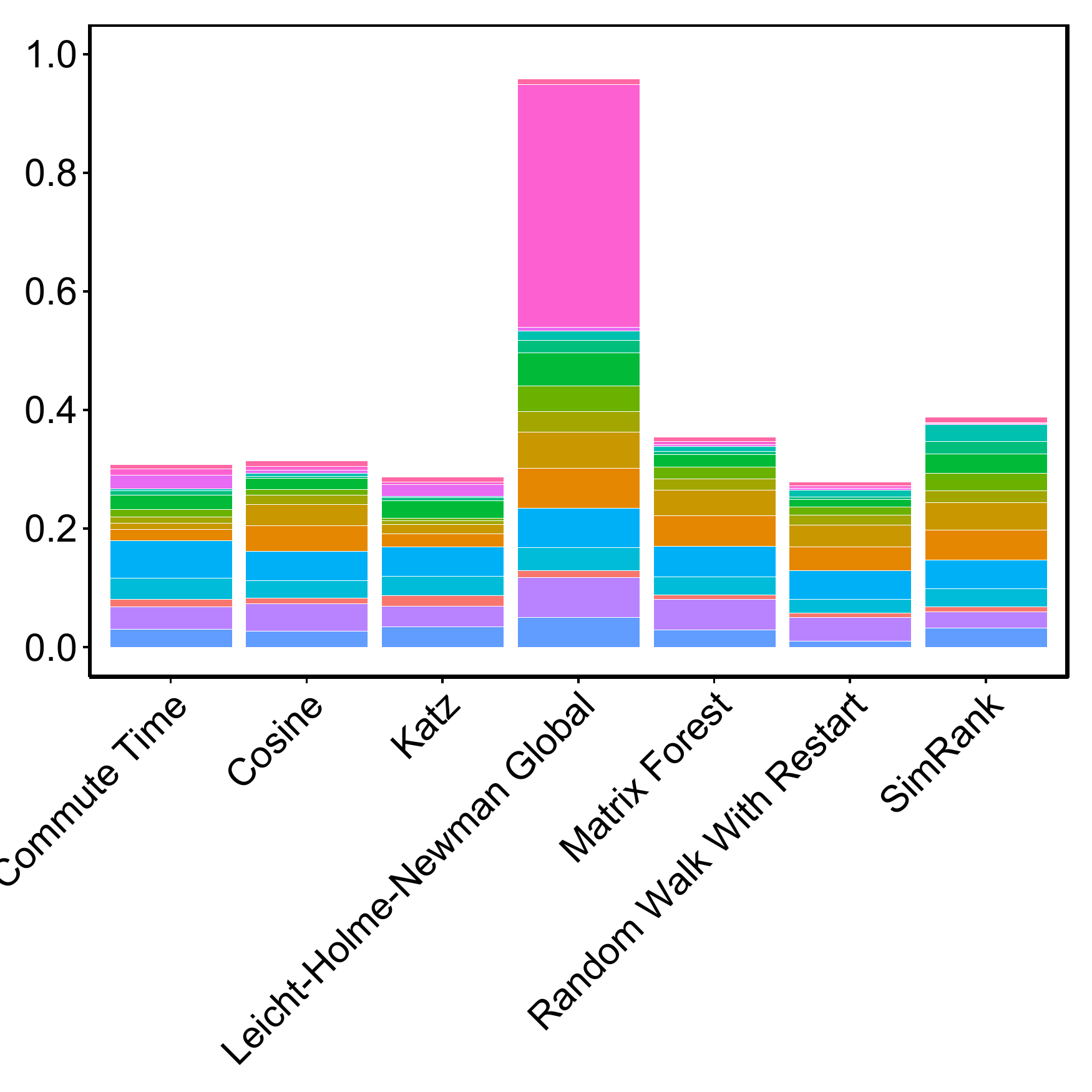} \\
\multicolumn{3}{c}{\includegraphics[width=0.8\linewidth]{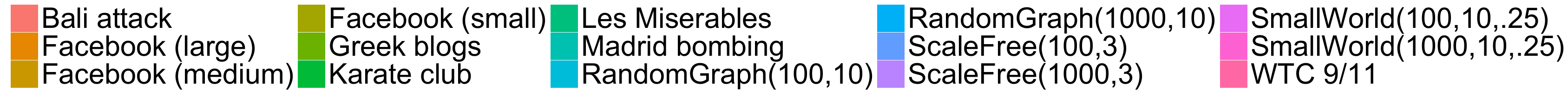}}
\end{tabular}
\caption{Given different \textbf{global similarity} indices, the figure depicts the relative change in $\ROC$ (the area under the ROC curve) and $\AP$ (the average precision) after running OTC and CTR in different networks, where $|\Hide|=\max(10,|E|/100)$ and $b=4|\Hide|$, and the links in $\Hide$ are chosen at random. For each similarity index, the height of the corresponding bar represents the average change taken over all networks, and the height of each segment in that bar is proportional to the change within the corresponding network.}
\label{fig:bars-relative-global-supplementary}
\end{figure*}

\begin{figure*}[ht!]
\centering
\setlength\tabcolsep{1pt}
\renewcommand{\arraystretch}{0.01}
\begin{tabular}{m{.03\textwidth}m{.27\textwidth}m{.27\textwidth}m{.27\textwidth}}
& \multicolumn{1}{c}{WTC 9/11 network}
& \multicolumn{1}{c}{ScaleFree$(100,3)$}
& \multicolumn{1}{c}{Facebook (medium)}\\
\rotatebox{90}{\footnotesize $\ROC$ values for OTC} &
\includegraphics[width=1.03\linewidth]{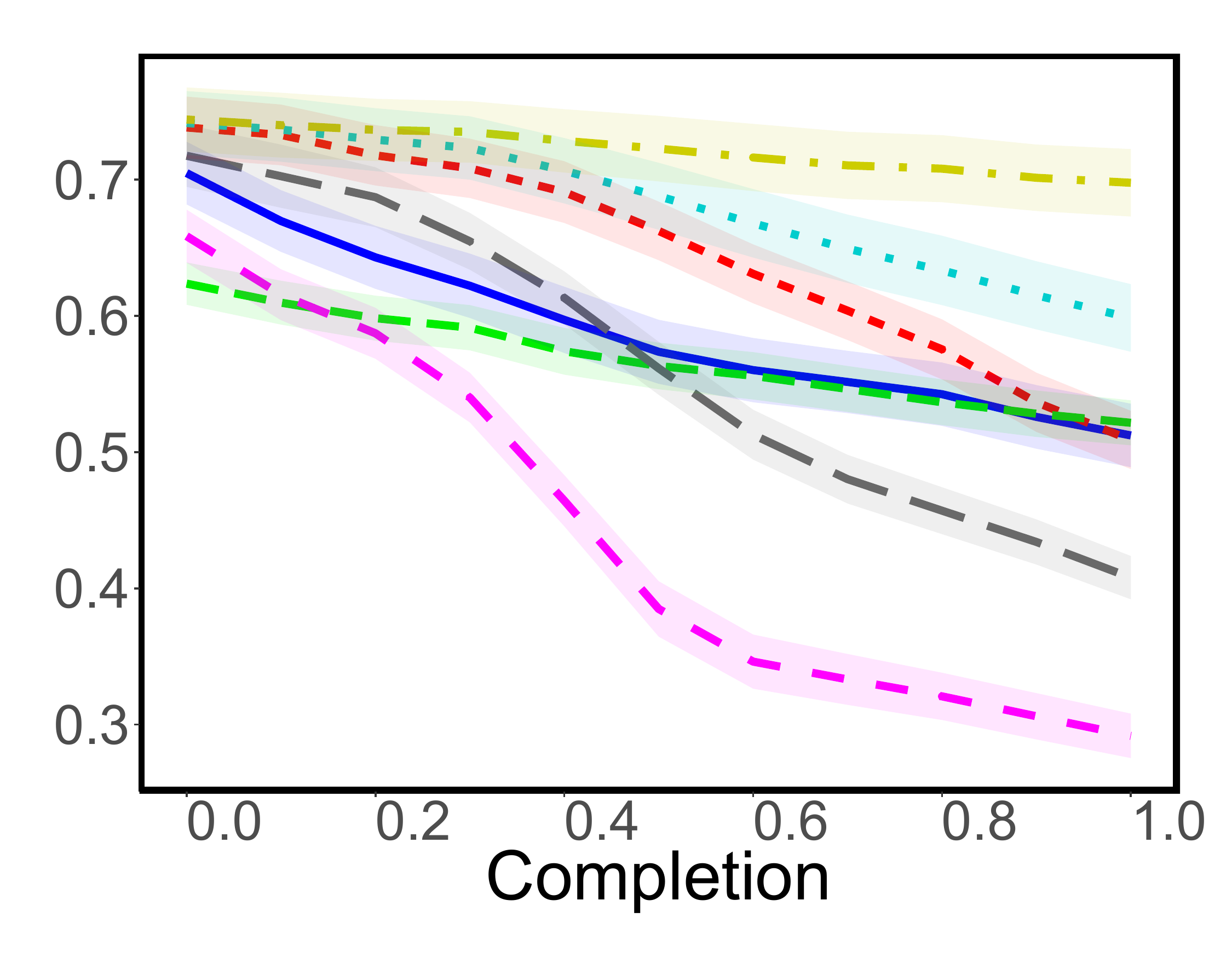} &
\includegraphics[width=1.03\linewidth]{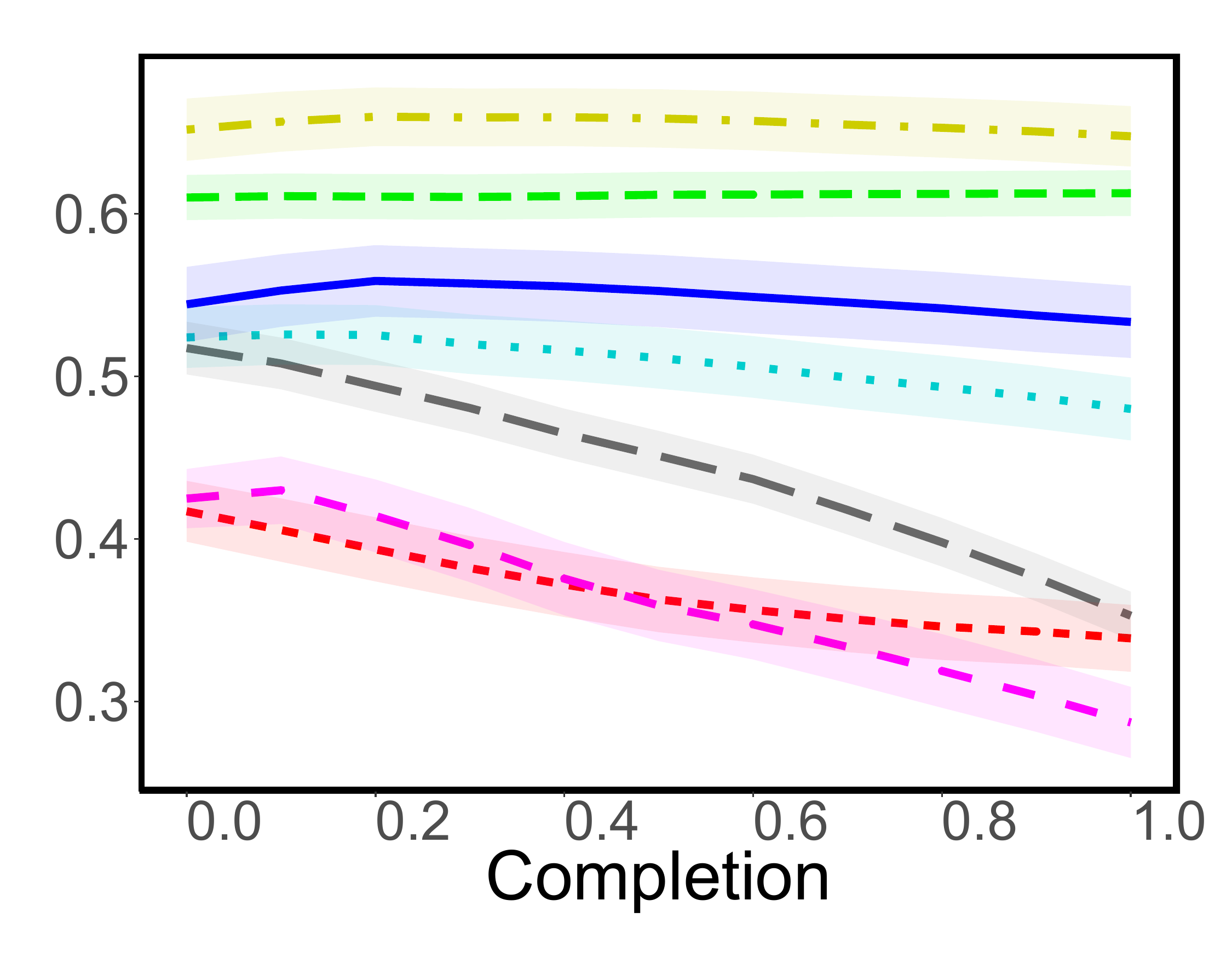} &
\includegraphics[width=1.03\linewidth]{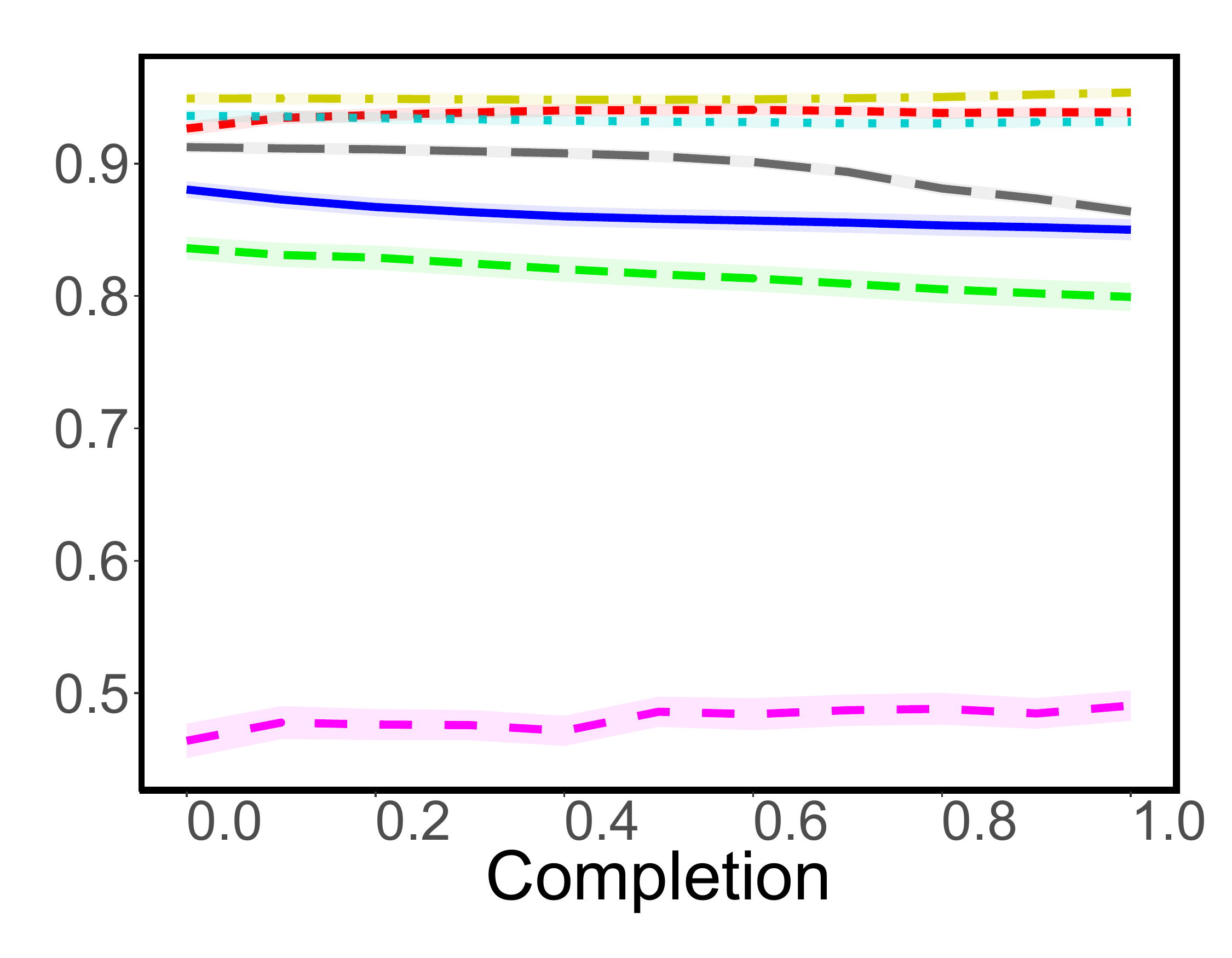}\\
\rotatebox{90}{\footnotesize $\ROC$ values for CTR} &
\includegraphics[width=1.03\linewidth]{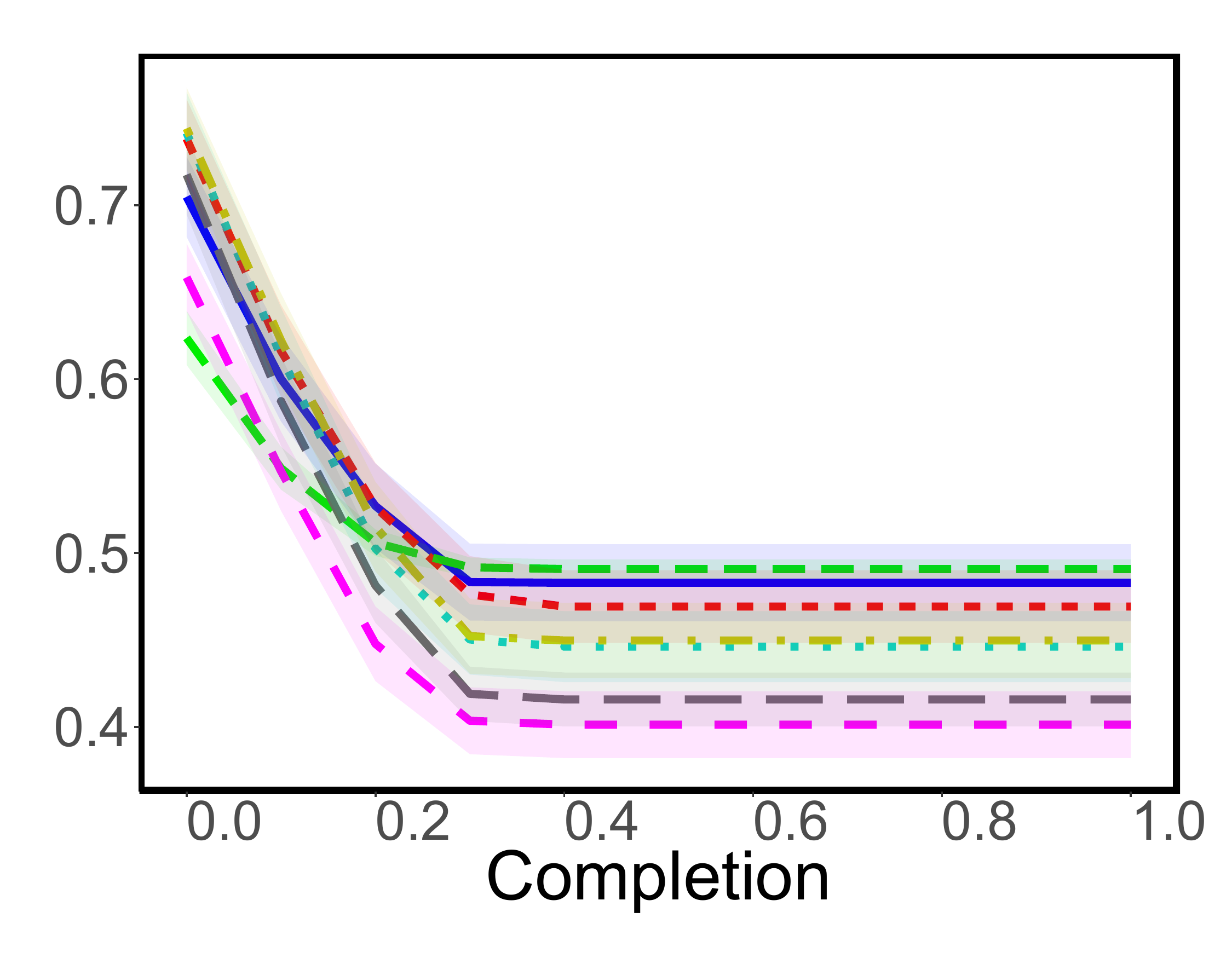} &
\includegraphics[width=1.03\linewidth]{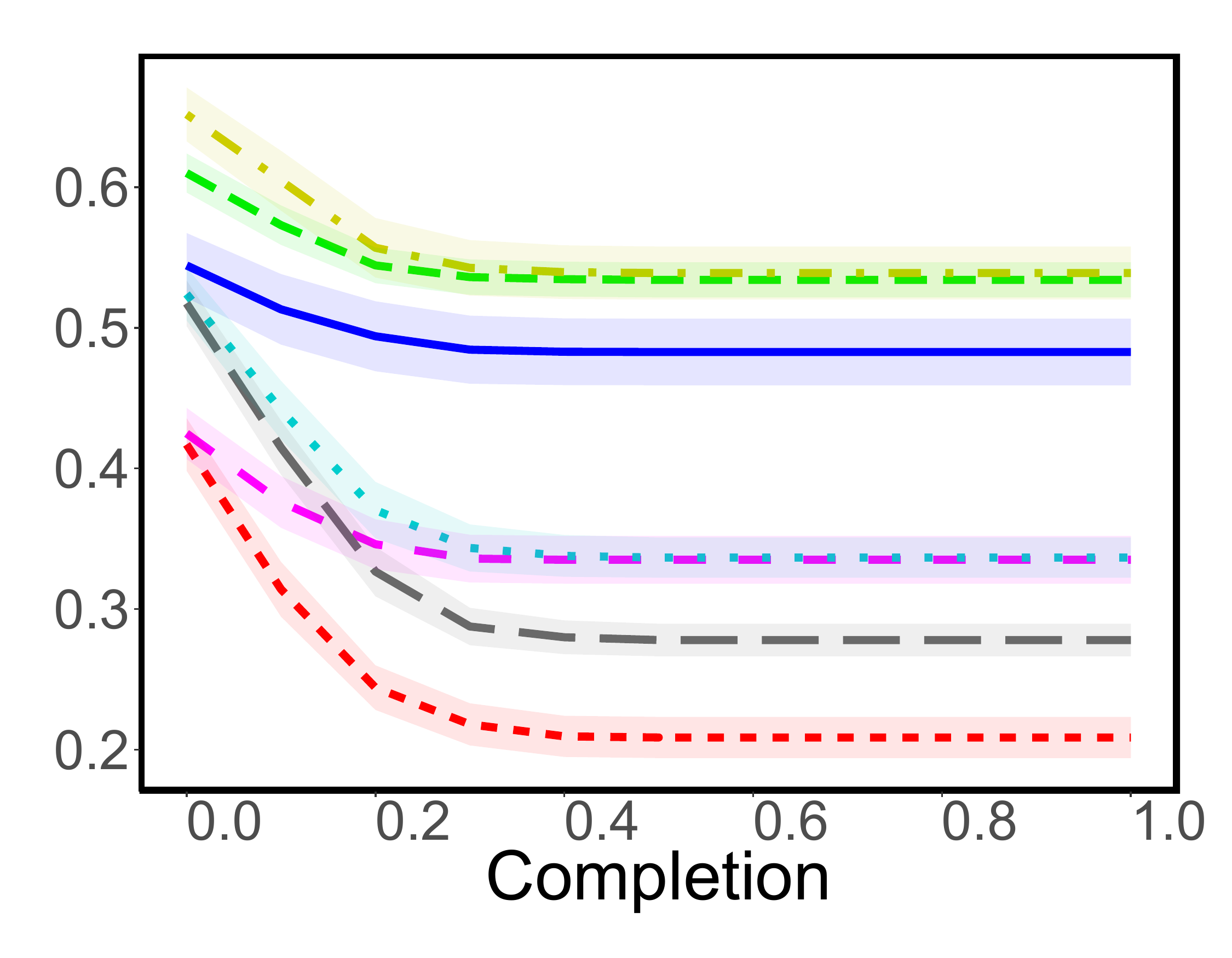} &
\includegraphics[width=1.03\linewidth]{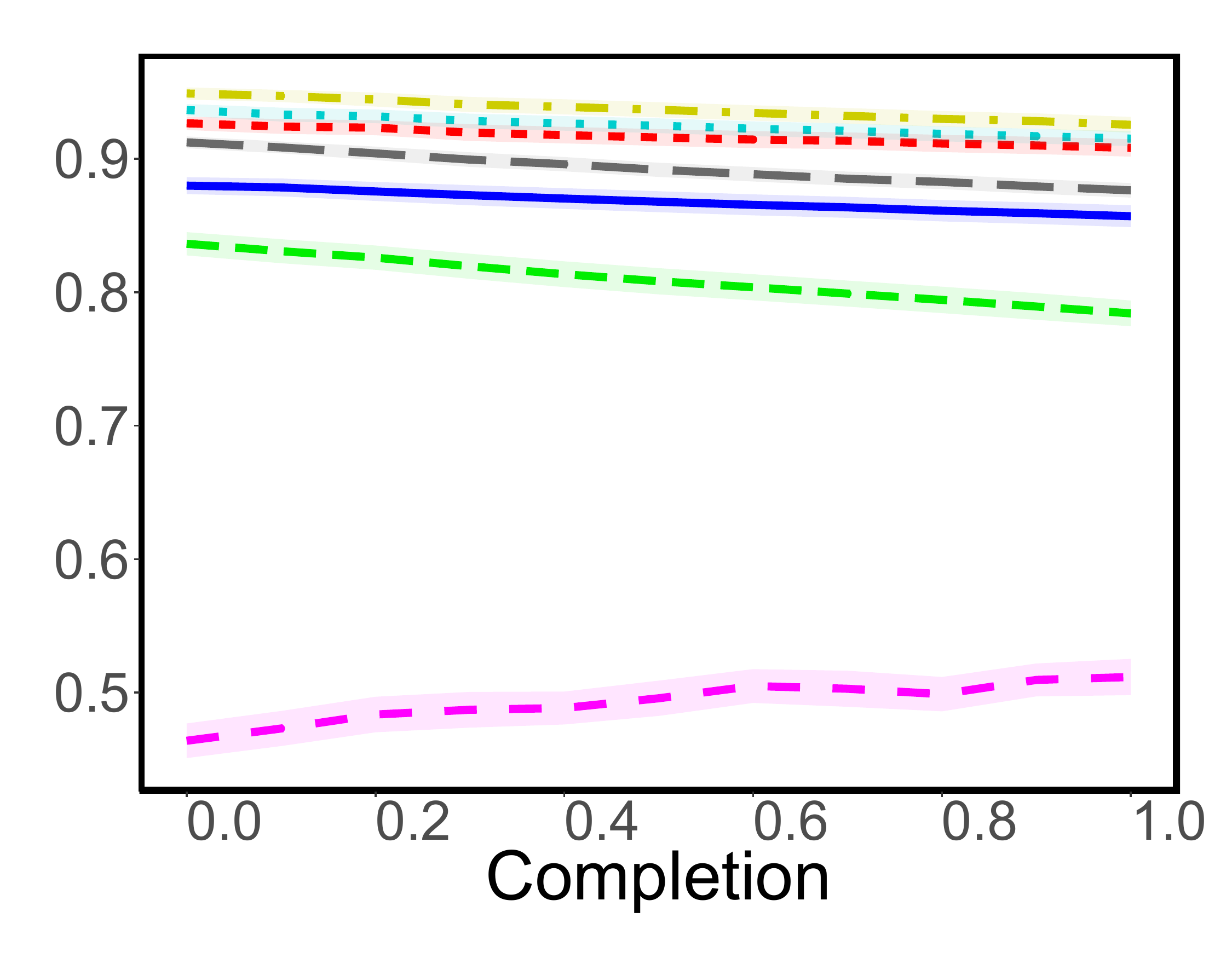} \\
\rotatebox{90}{\footnotesize $\AP$ values for OTC} &
\includegraphics[width=1.03\linewidth]{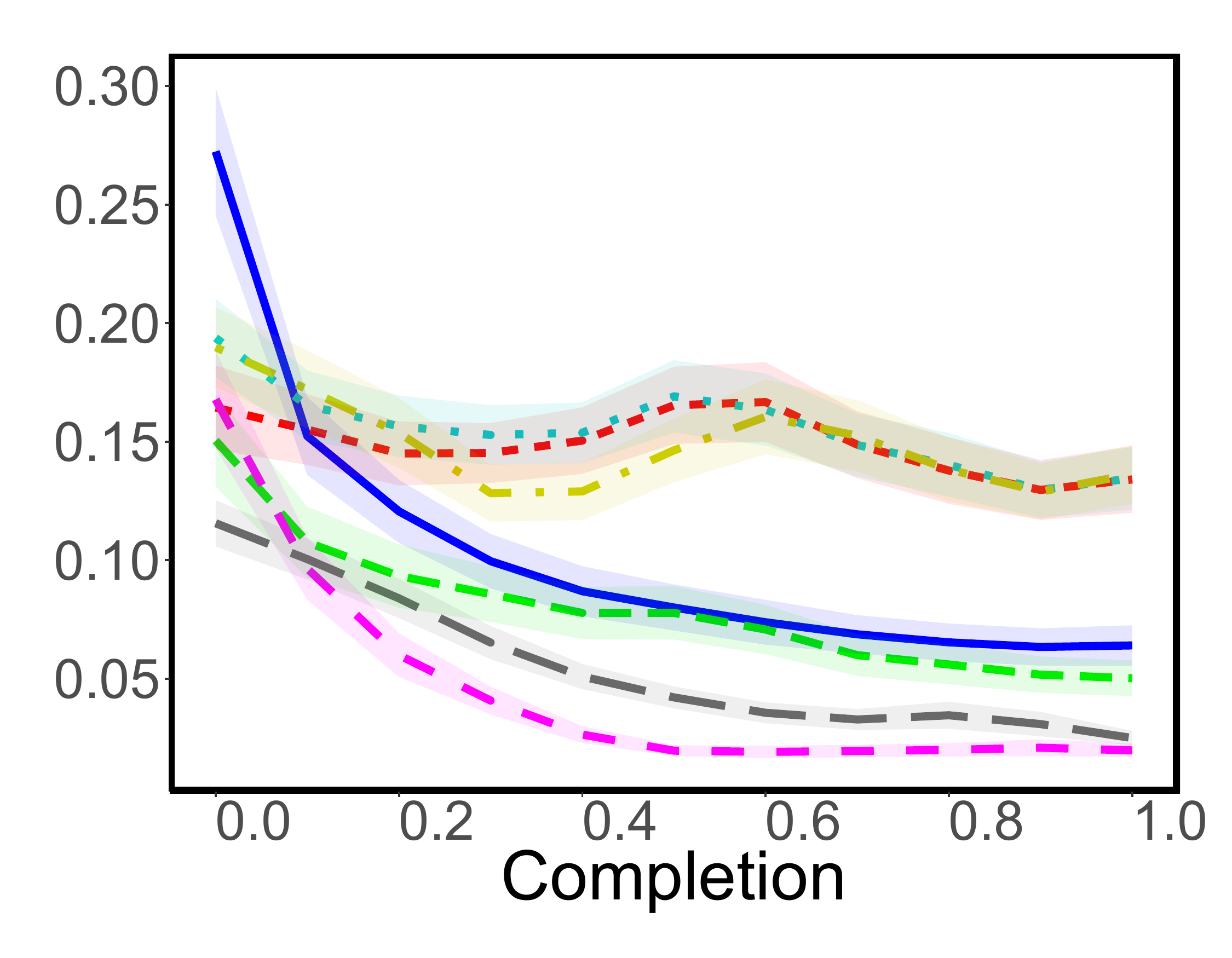} &
\includegraphics[width=1.03\linewidth]{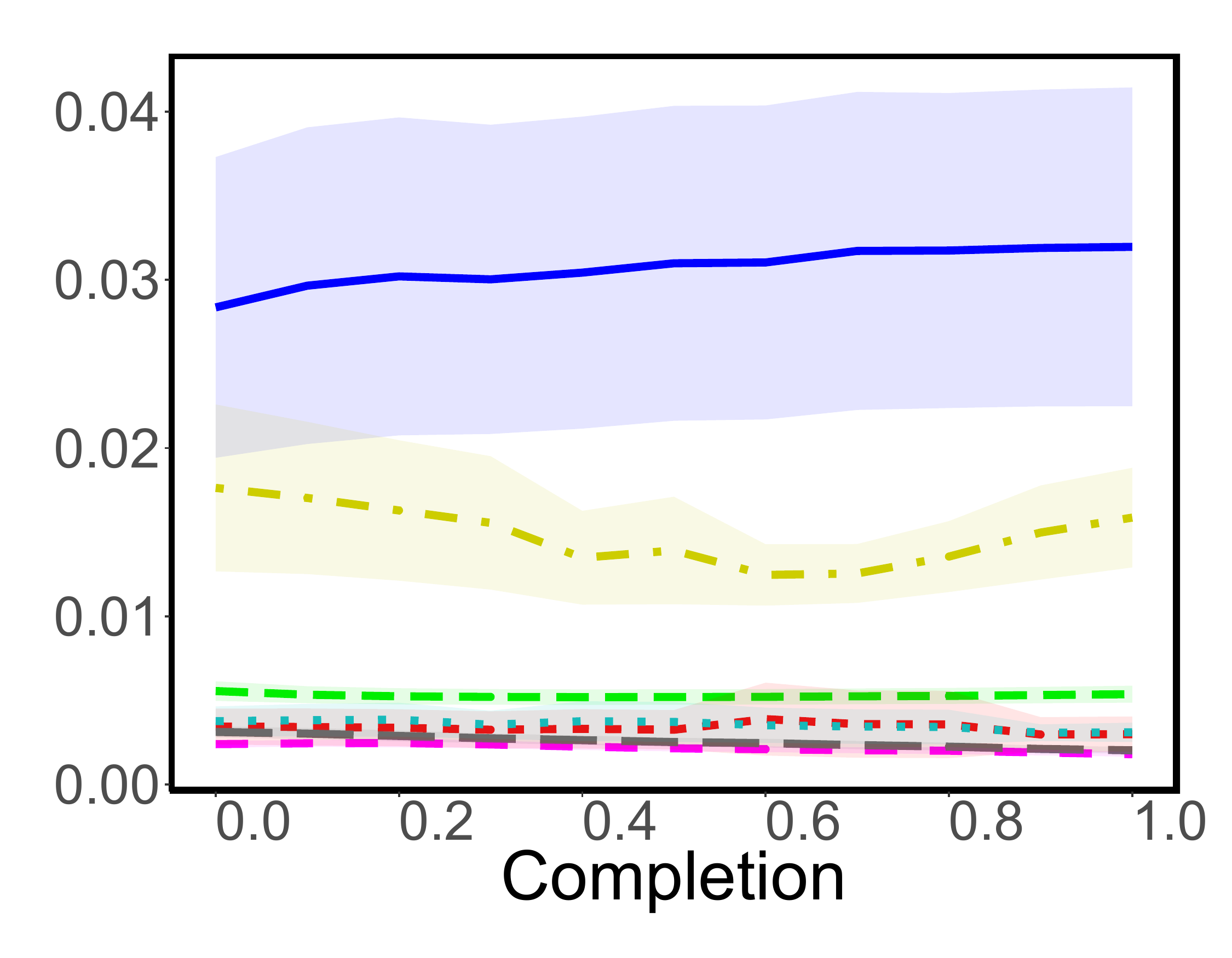} &
\includegraphics[width=1.03\linewidth]{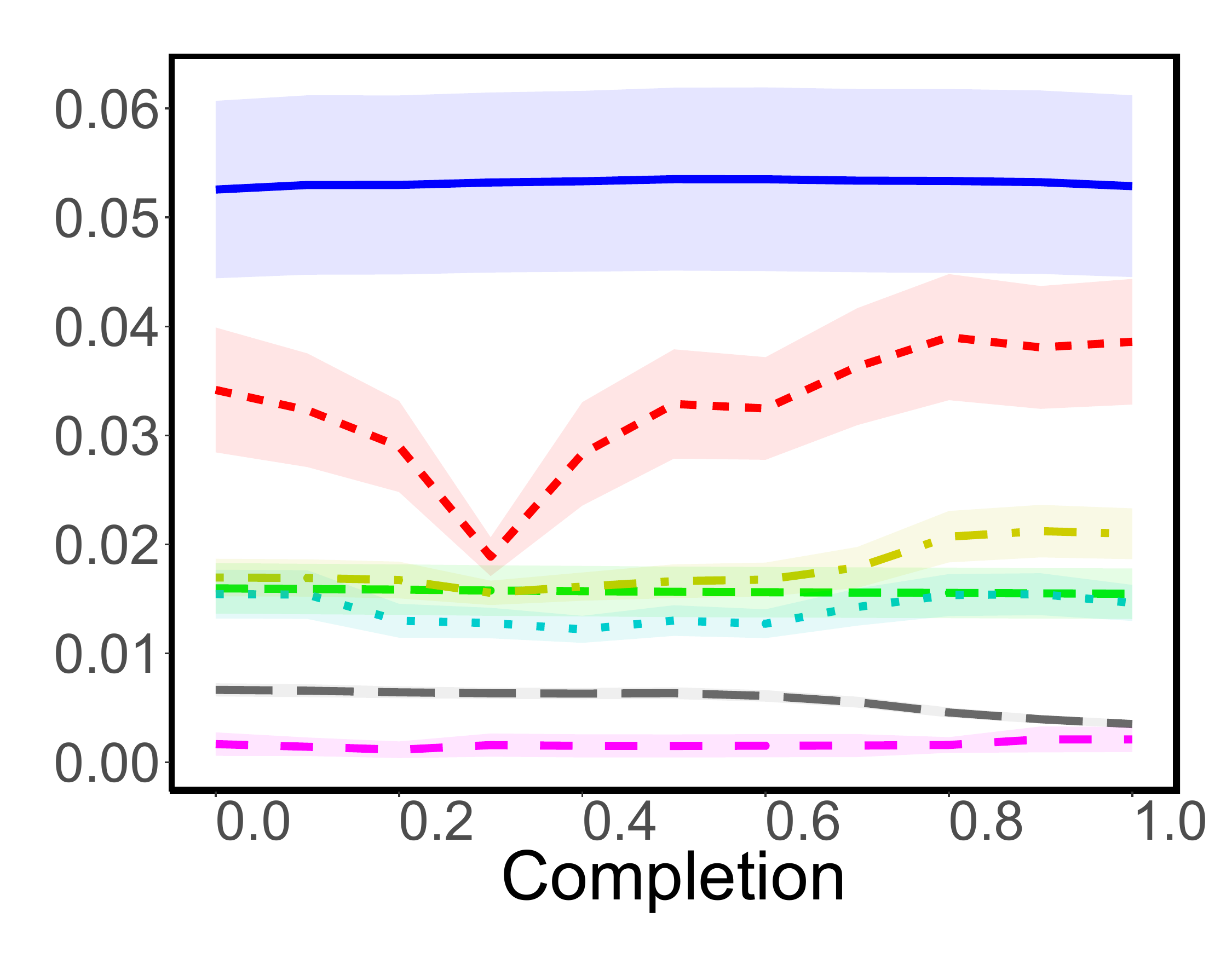} \\
\rotatebox{90}{\footnotesize $\AP$ values for CTR} &
\includegraphics[width=1.03\linewidth]{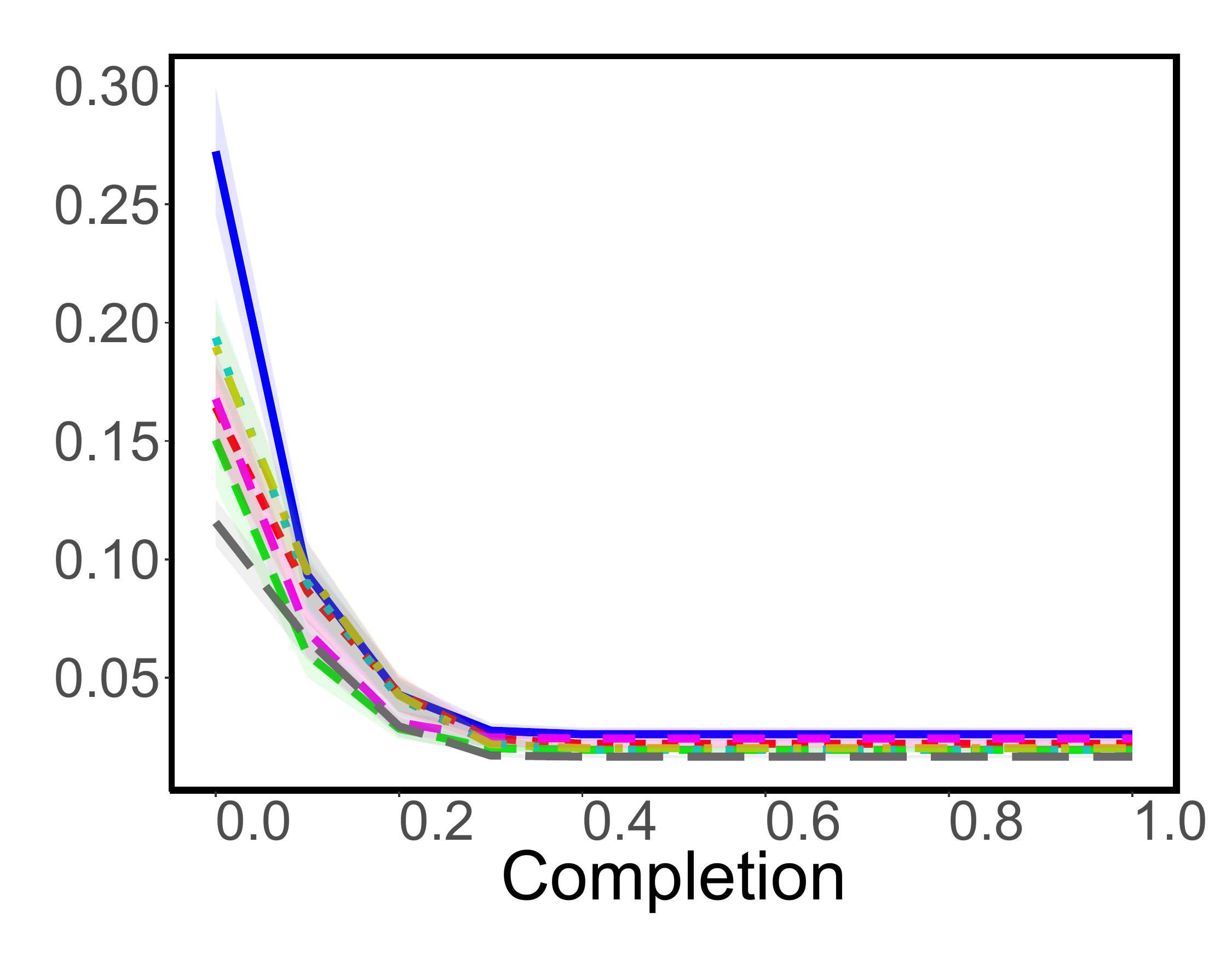} &
\includegraphics[width=1.03\linewidth]{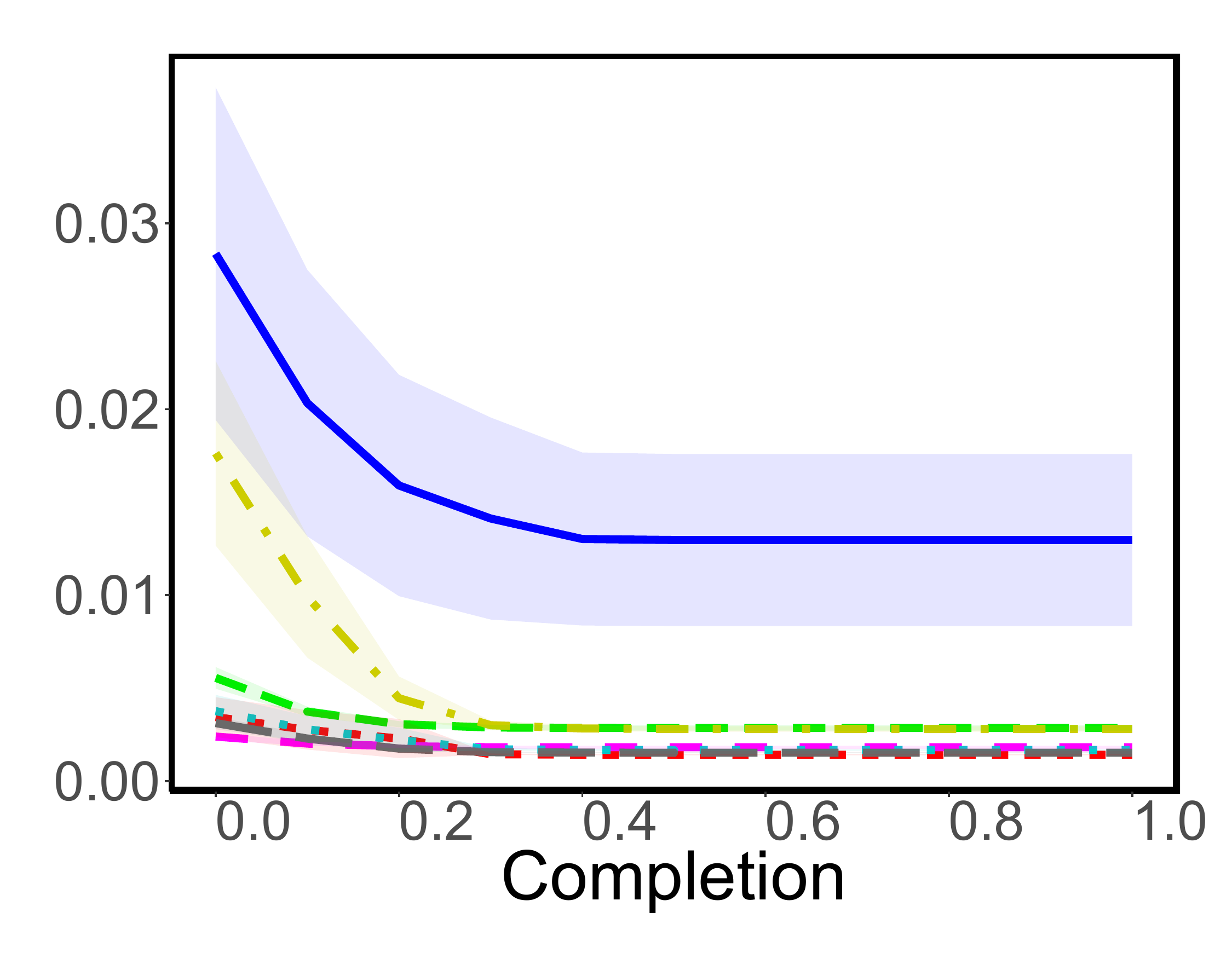} &
\includegraphics[width=1.03\linewidth]{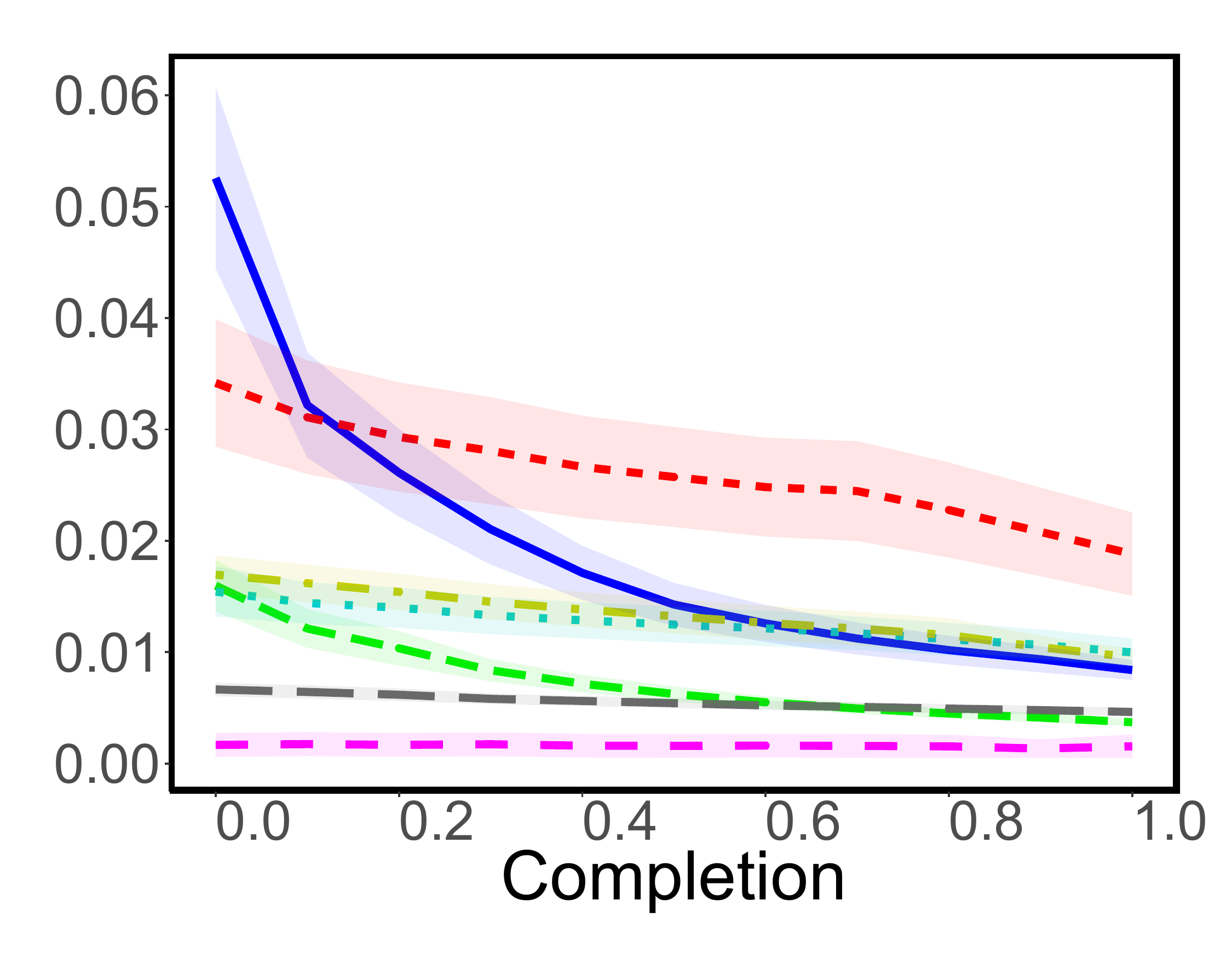} \\
\multicolumn{4}{c}{\includegraphics[width=0.65\linewidth]{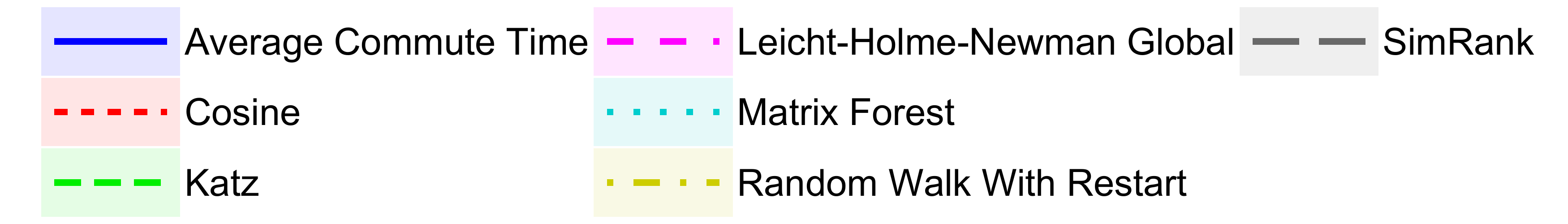}}
\end{tabular}
\caption{Given different \textbf{global similarity} indices, the figure depicts the values of $\ROC$ (the area under the ROC curve) and $\AP$ (the average precision) during the execution of OTC and CTR given $|\Hide|=\max(10,|E|/100)$ and $b=4|\Hide|$ in three networks: (i) \textbf{the WTC 9/11 network}; (ii) \textbf{ScaleFree(100,3)}; and (iii) \textbf{a medium fragment of Facebook}.
In each execution, the links in $\Hide$ are chosen at random. Results are taken as the average over $50$ executions, with coloured areas representing the $95\%$ confidence intervals.}
\label{fig:global-1}
\end{figure*}

\begin{figure*}[tbhp]
\centering
\setlength\tabcolsep{1pt}
\renewcommand{\arraystretch}{0.01}
\begin{tabular}{m{.03\textwidth}m{.27\textwidth}m{.27\textwidth}m{.27\textwidth}}
& \multicolumn{1}{c}{ScaleFree$(1000,3)$}
& \multicolumn{1}{c}{RandomGraph$(100,10)$}
& \multicolumn{1}{c}{RandomGraph$(1000,10)$}\\
\rotatebox{90}{\footnotesize $\ROC$ values for OTC} &
\includegraphics[width=1.03\linewidth]{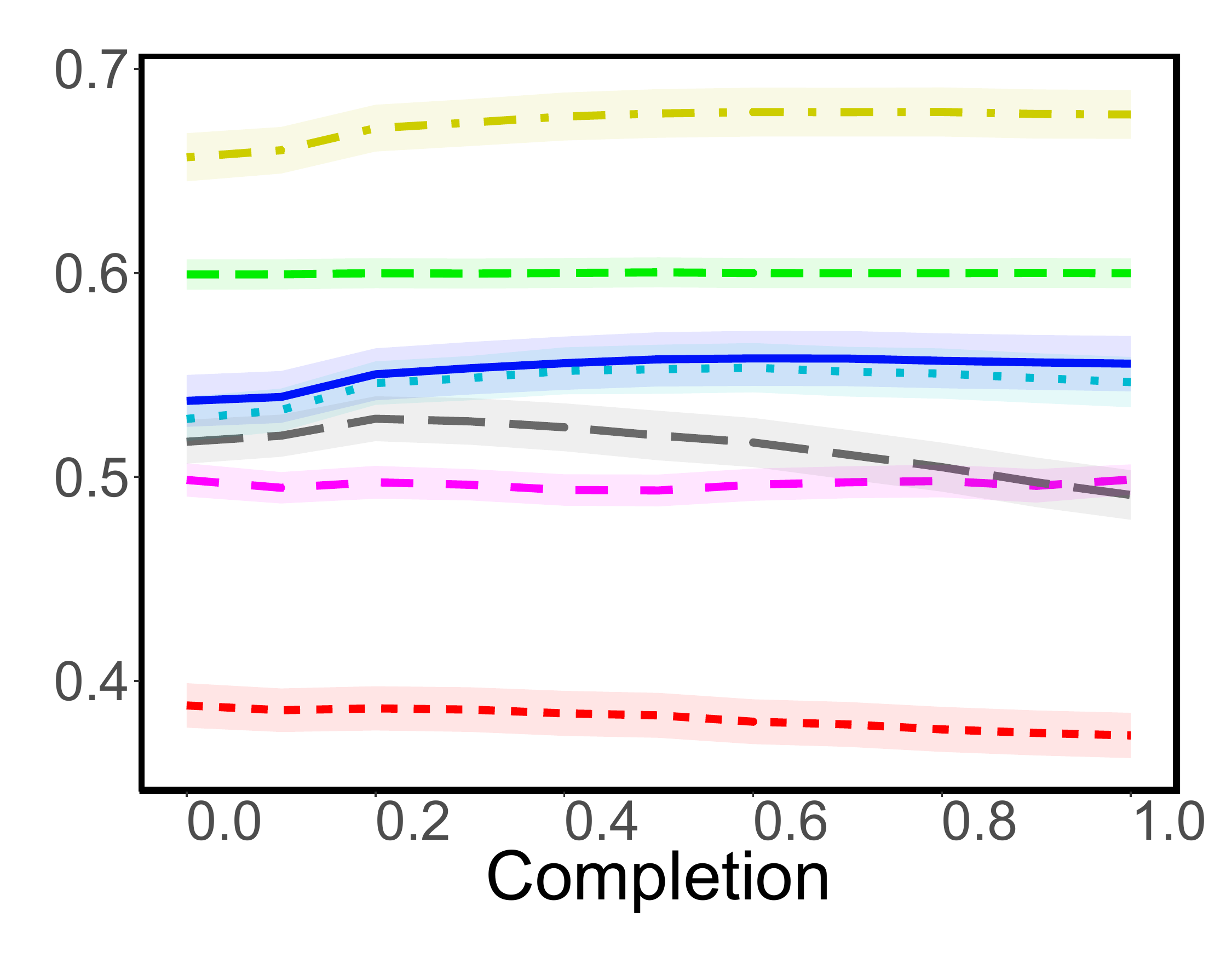} &
\includegraphics[width=1.03\linewidth]{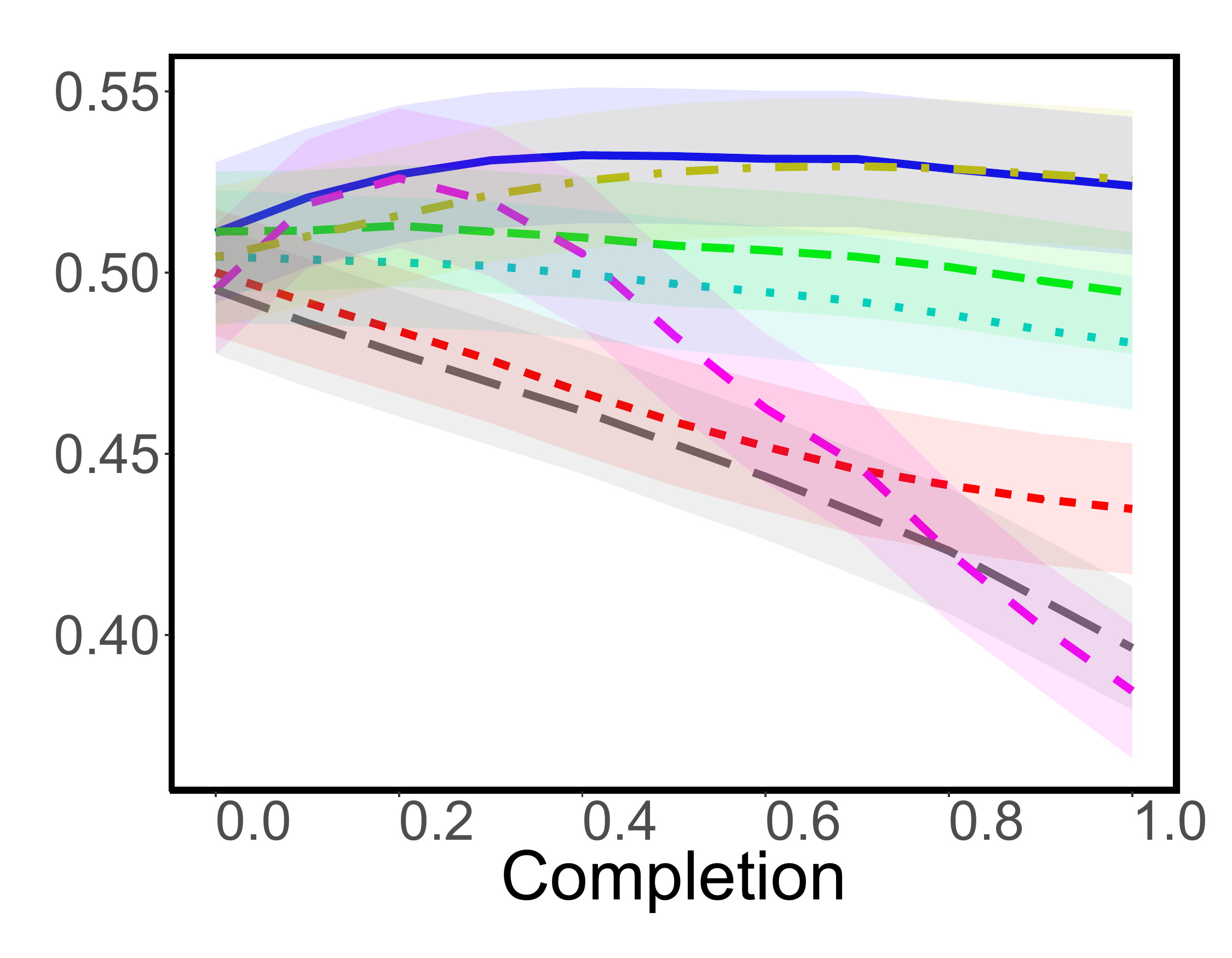} &
\includegraphics[width=1.03\linewidth]{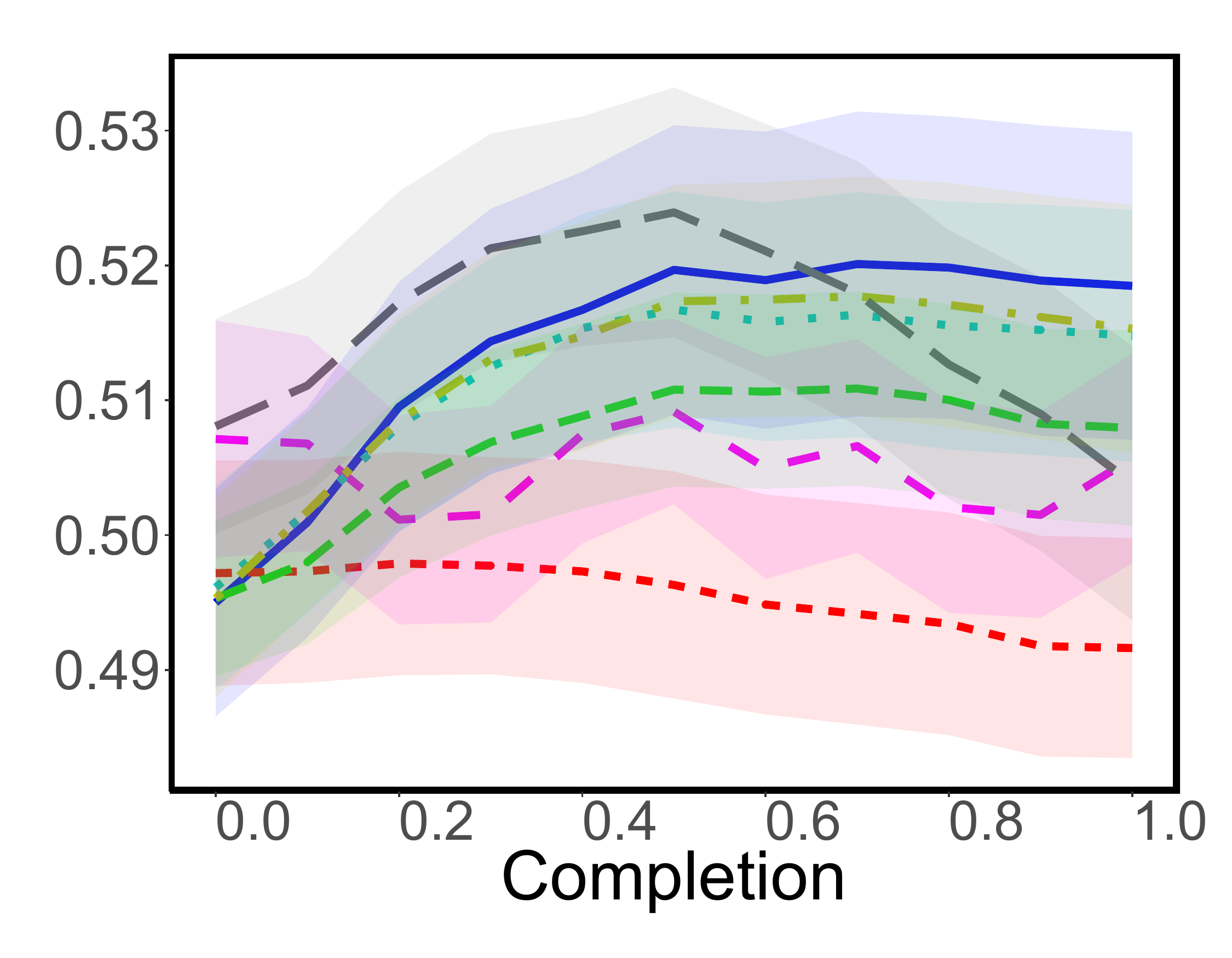}\\
\rotatebox{90}{\footnotesize $\ROC$ values for CTR} &
\includegraphics[width=1.03\linewidth]{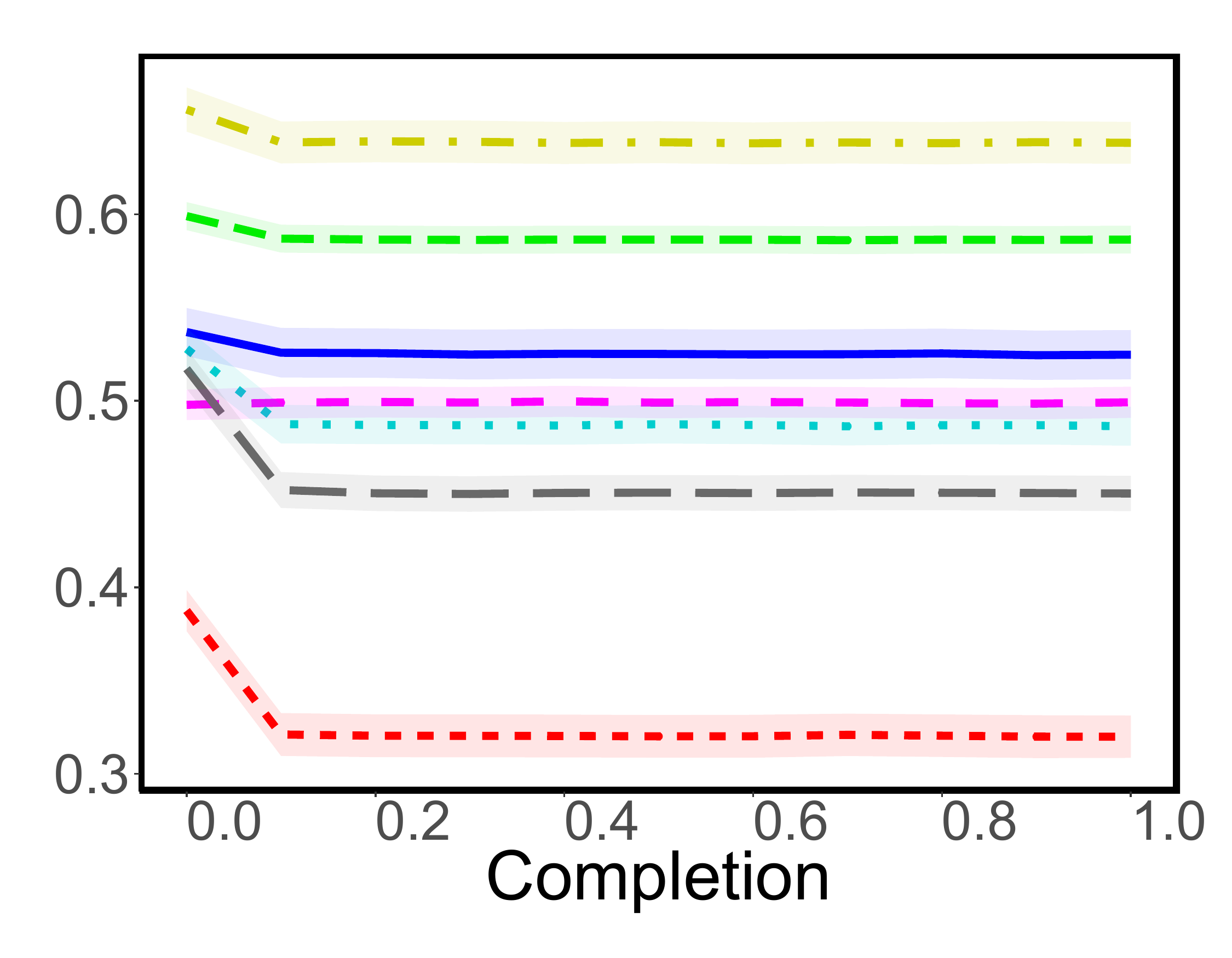} &
\includegraphics[width=1.03\linewidth]{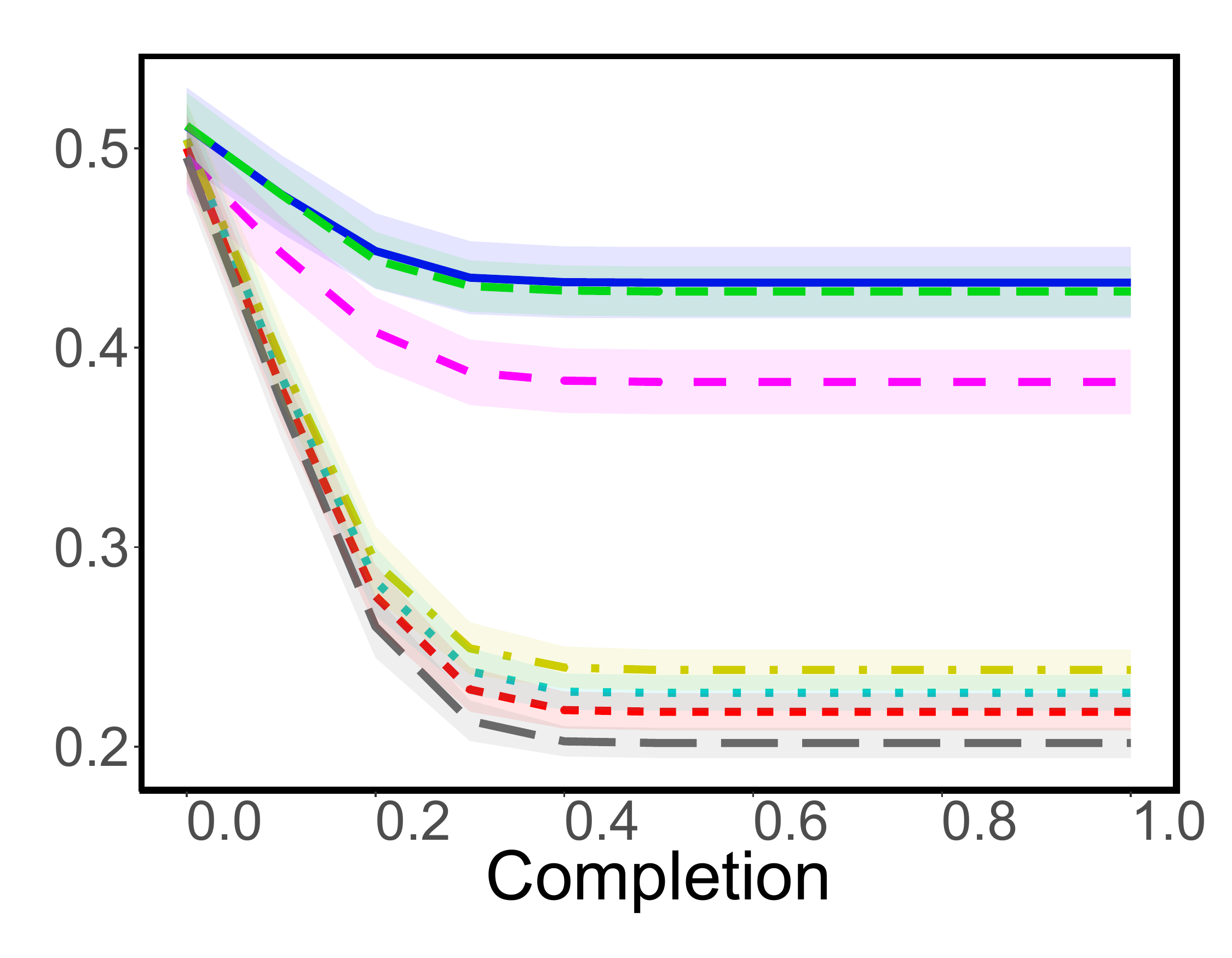} &
\includegraphics[width=1.03\linewidth]{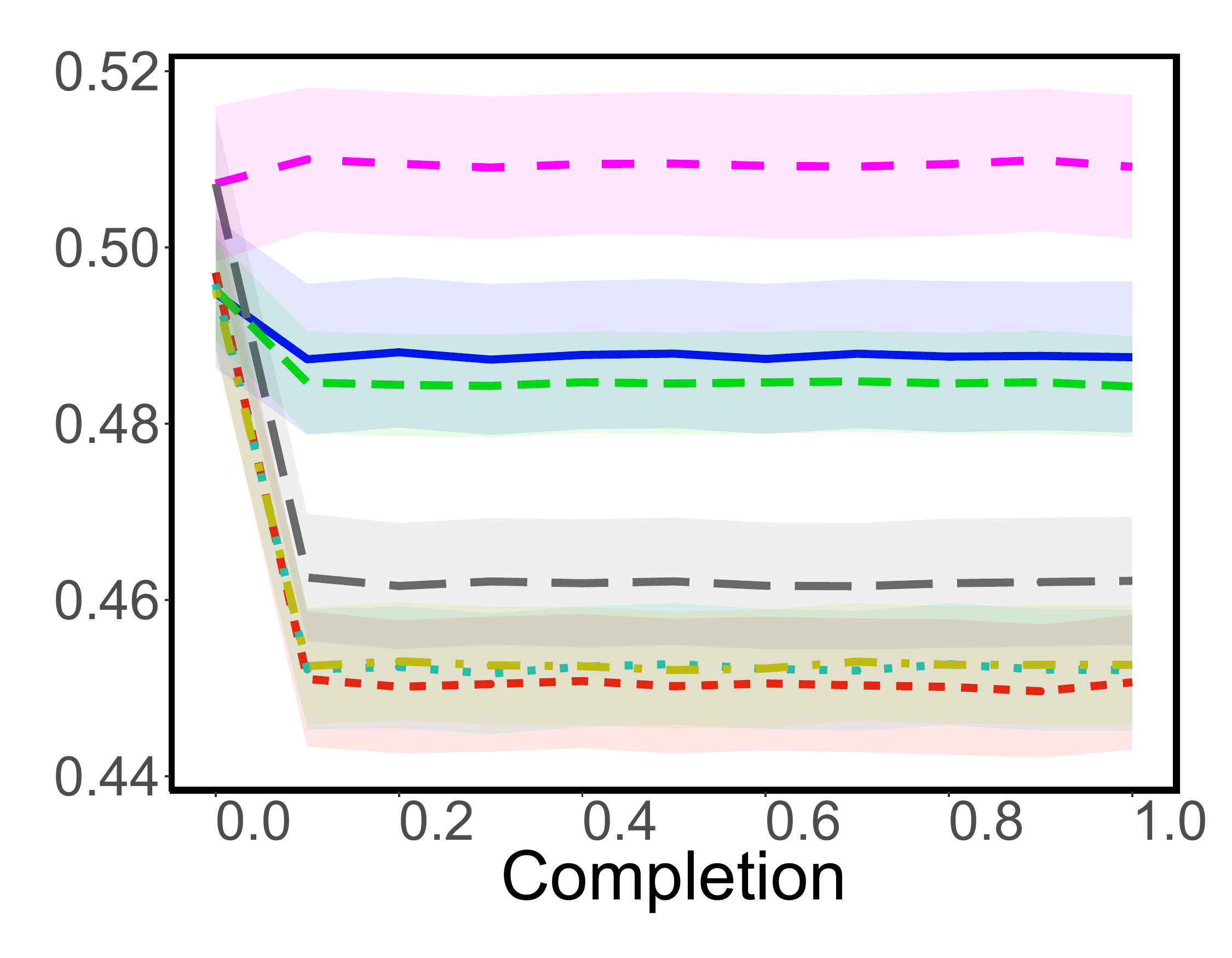} \\
\rotatebox{90}{\footnotesize $\AP$ values for OTC} &
\includegraphics[width=1.03\linewidth]{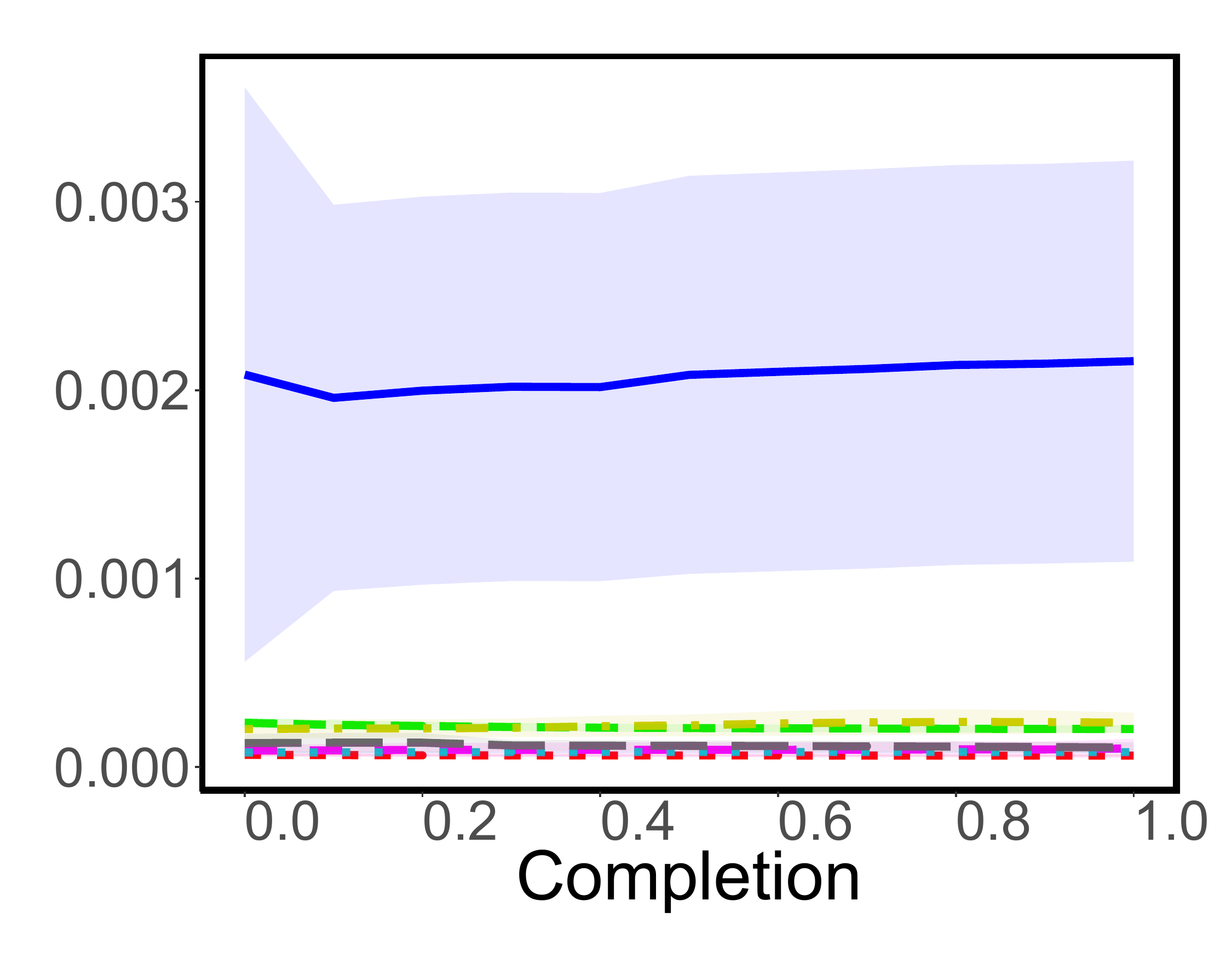} &
\includegraphics[width=1.03\linewidth]{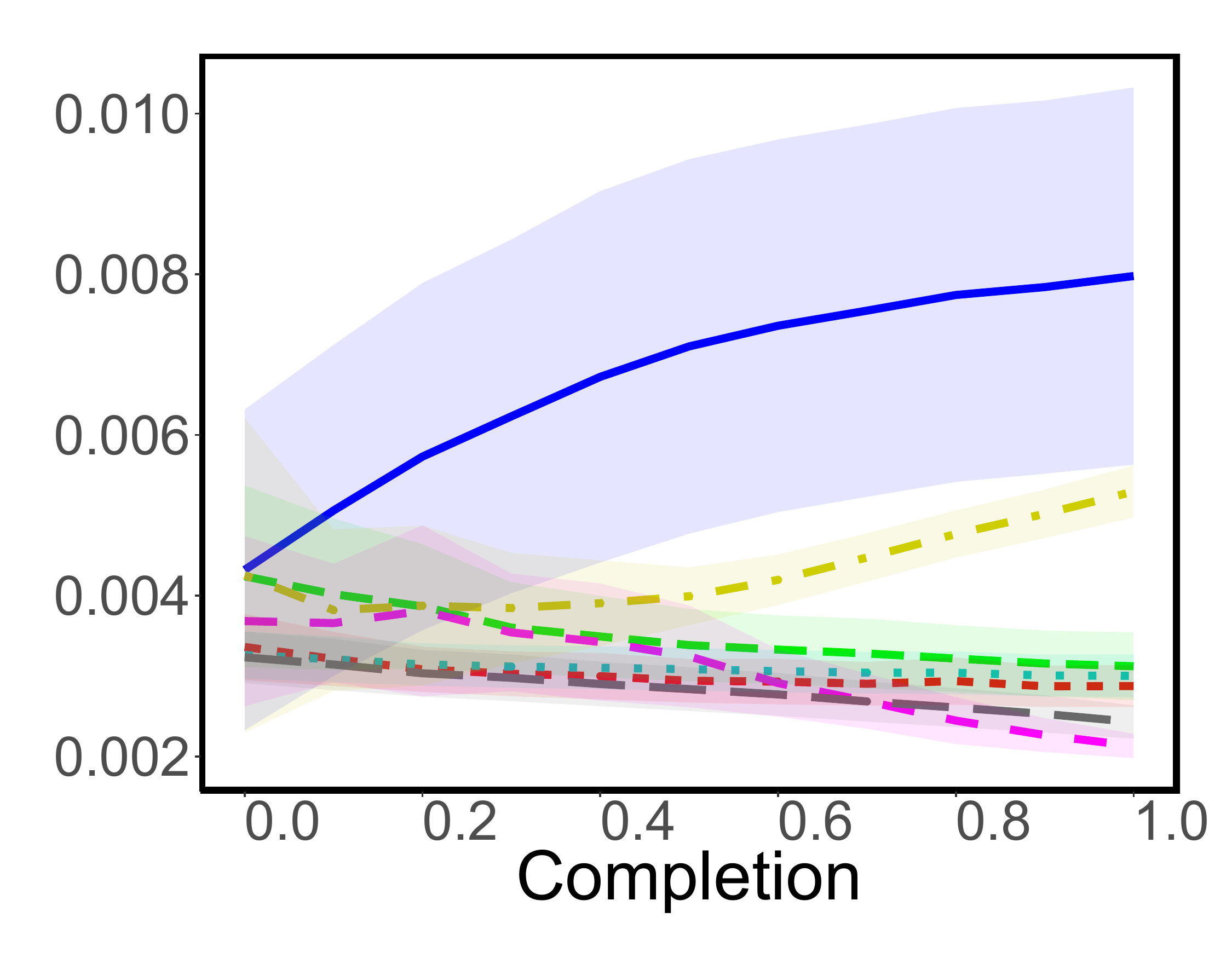} &
\includegraphics[width=1.03\linewidth]{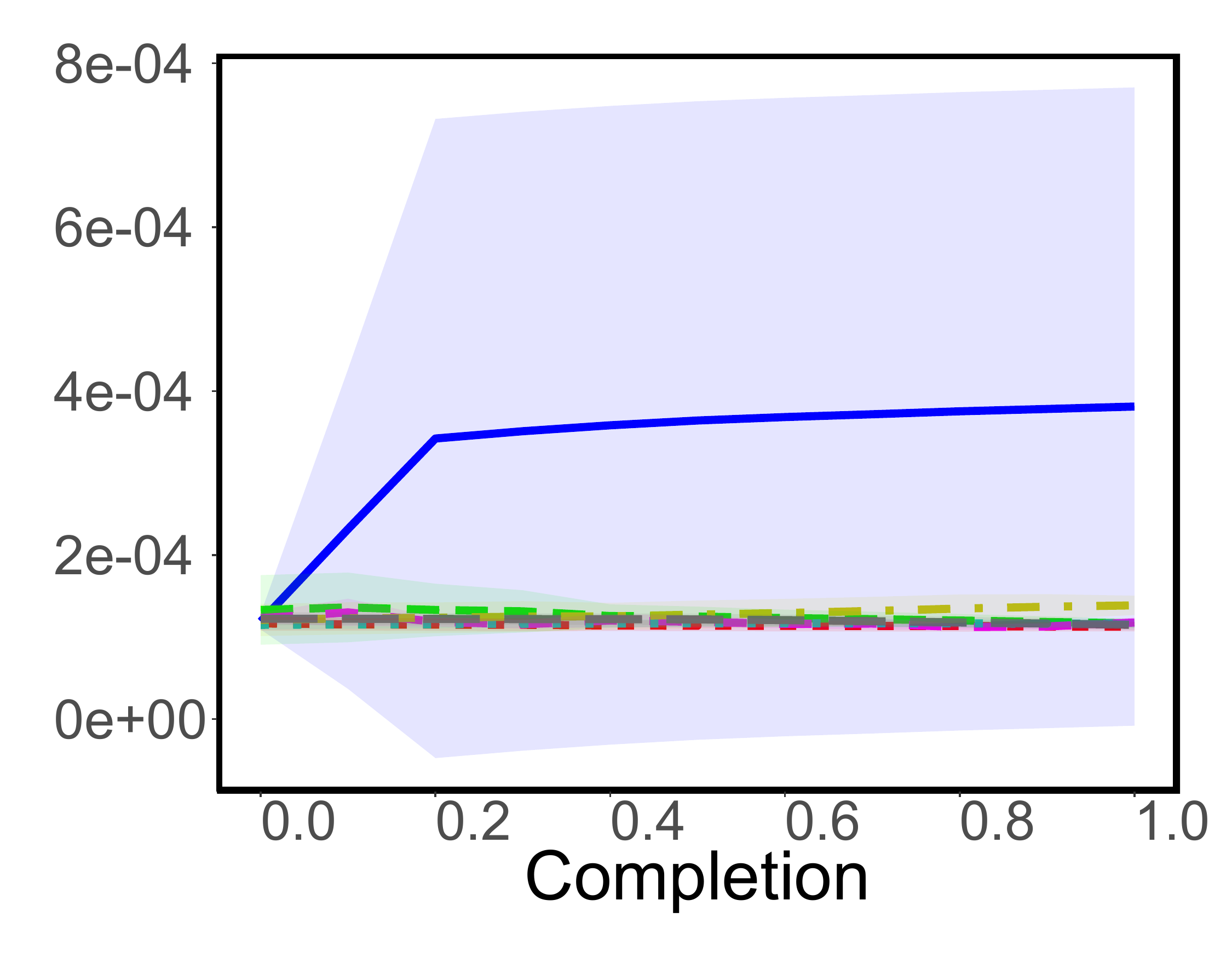} \\
\rotatebox{90}{\footnotesize $\AP$ values for CTR} &
\includegraphics[width=1.03\linewidth]{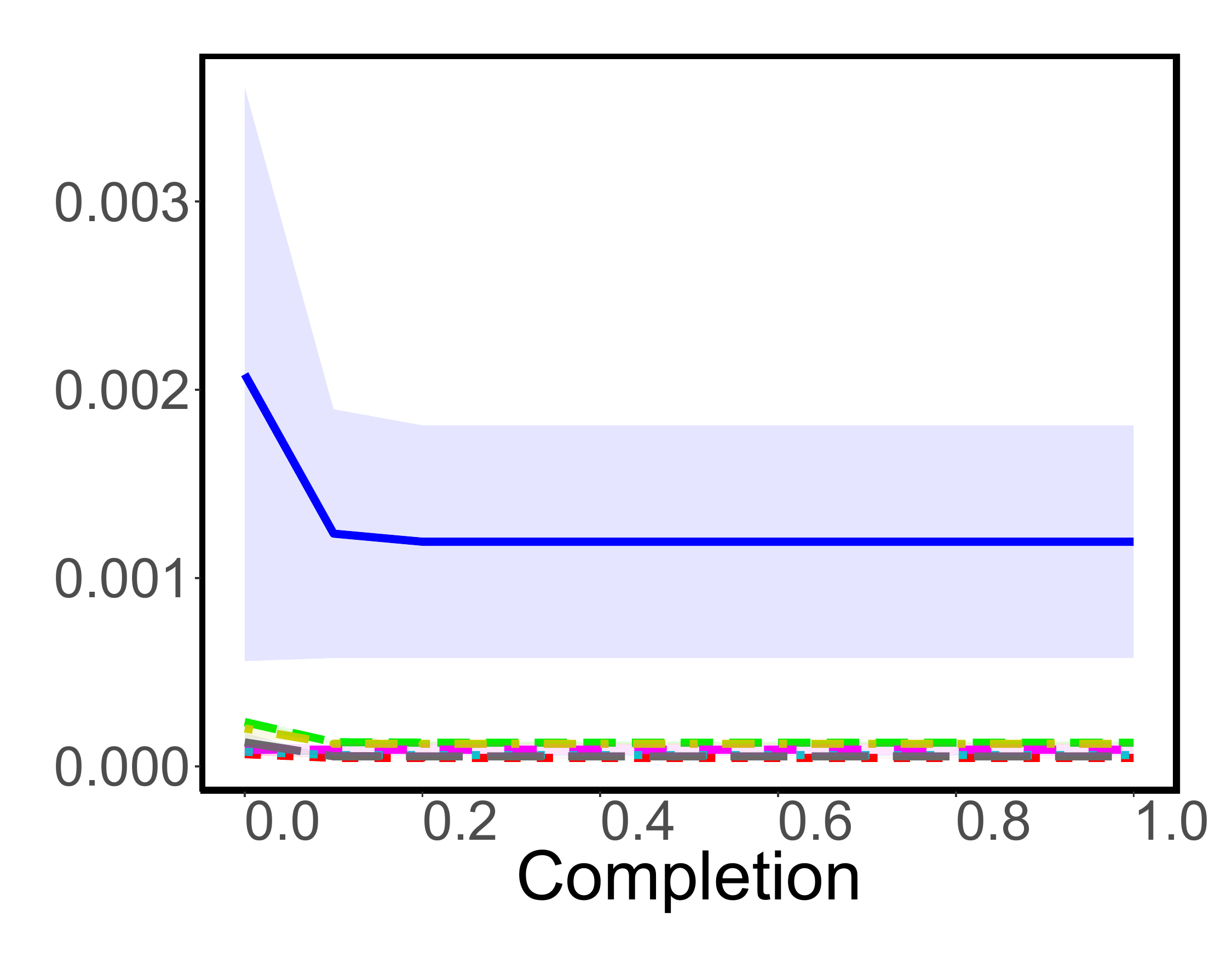} &
\includegraphics[width=1.03\linewidth]{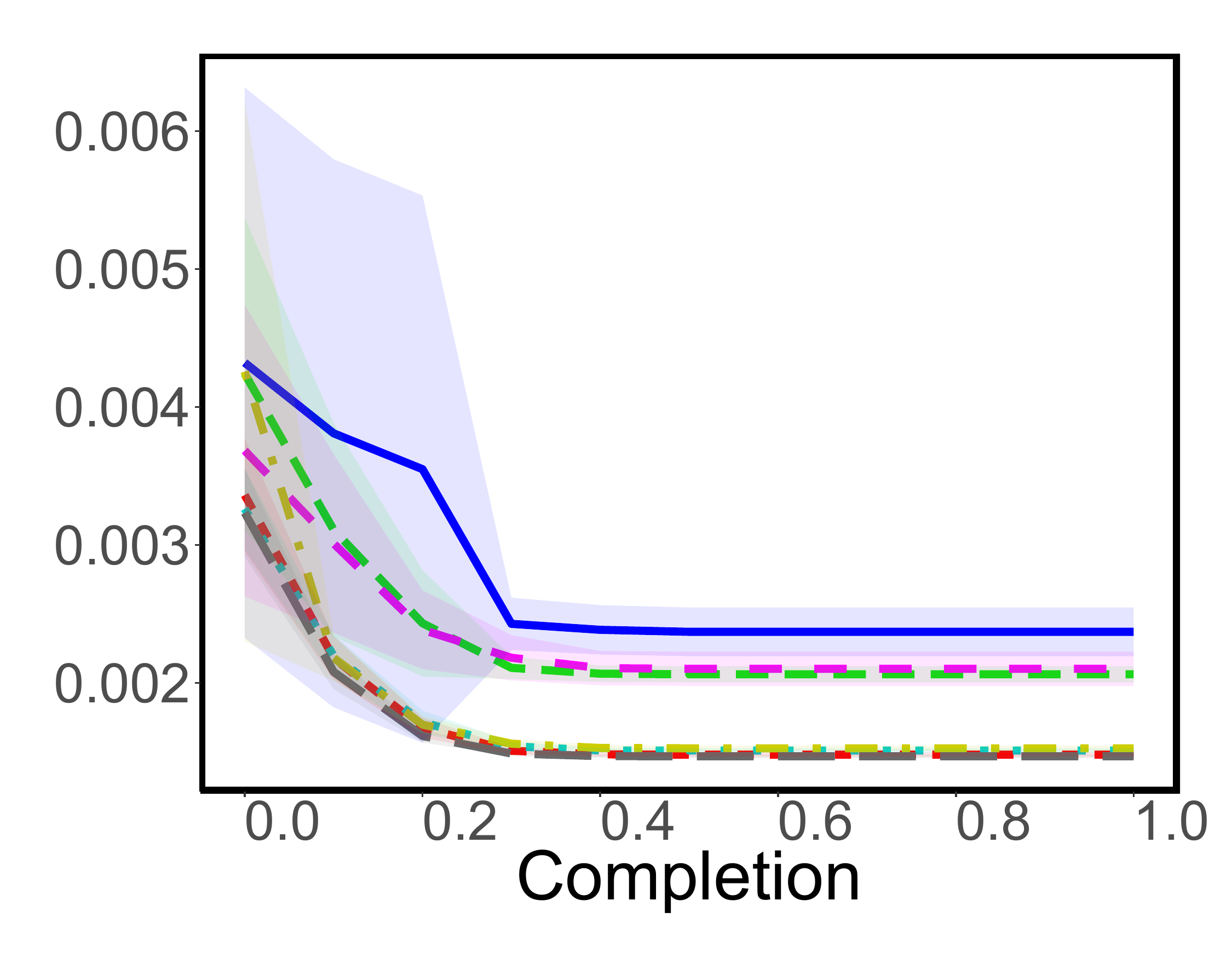} &
\includegraphics[width=1.03\linewidth]{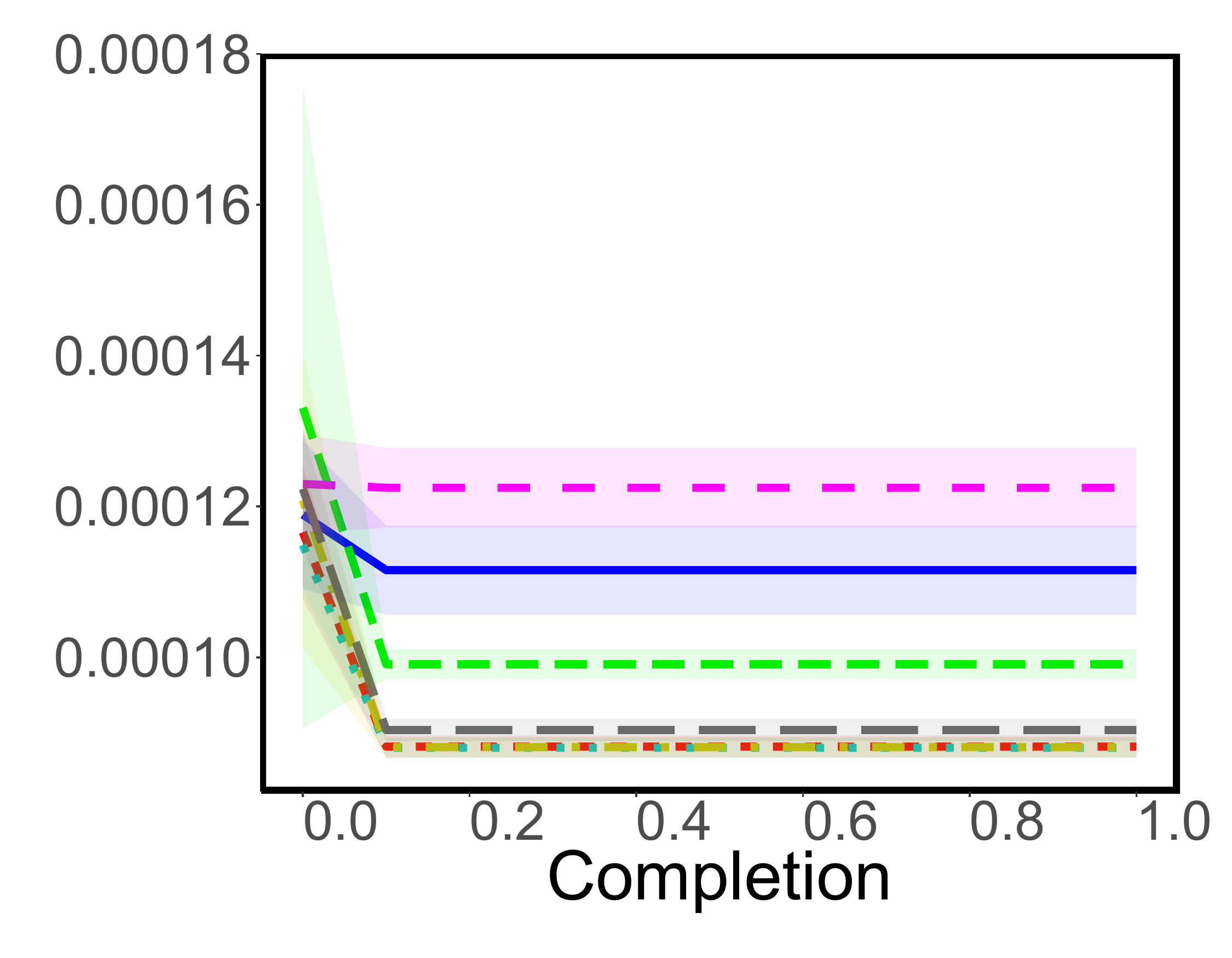} \\
\multicolumn{4}{c}{\includegraphics[width=0.65\linewidth]{figures/plots/global/legend}}
\end{tabular}
\caption{Given different \textbf{global similarity} indices, the figure depicts the values of $\ROC$ (the area under the ROC curve) and $\AP$ (the average precision) during the execution of OTC and CTR given $|\Hide|=\max(10,|E|/100)$ and $b=4|\Hide|$ in three networks: (i) \textbf{ScaleFree(1000,3)}; (ii) \textbf{RandomGraph(100,10)}; and (iii) \textbf{RandomGraph(1000,10)}.
In each execution, the links in $\Hide$ are chosen at random. Results are taken as the average over $50$ executions, with coloured areas representing the $95\%$ confidence intervals.}
\label{fig:global-2}
\end{figure*}

\begin{figure*}[tbhp]
\centering
\setlength\tabcolsep{1pt}
\renewcommand{\arraystretch}{0.01}
\begin{tabular}{m{.03\textwidth}m{.27\textwidth}m{.27\textwidth}m{.27\textwidth}}
& \multicolumn{1}{c}{SmallWorld$(100,10,0.25)$}
& \multicolumn{1}{c}{SmallWorld$(1000,10,0.25)$}
& \multicolumn{1}{c}{Les Mis\'erables network}\\
\rotatebox{90}{\footnotesize $\ROC$ values for OTC} &
\includegraphics[width=1.03\linewidth]{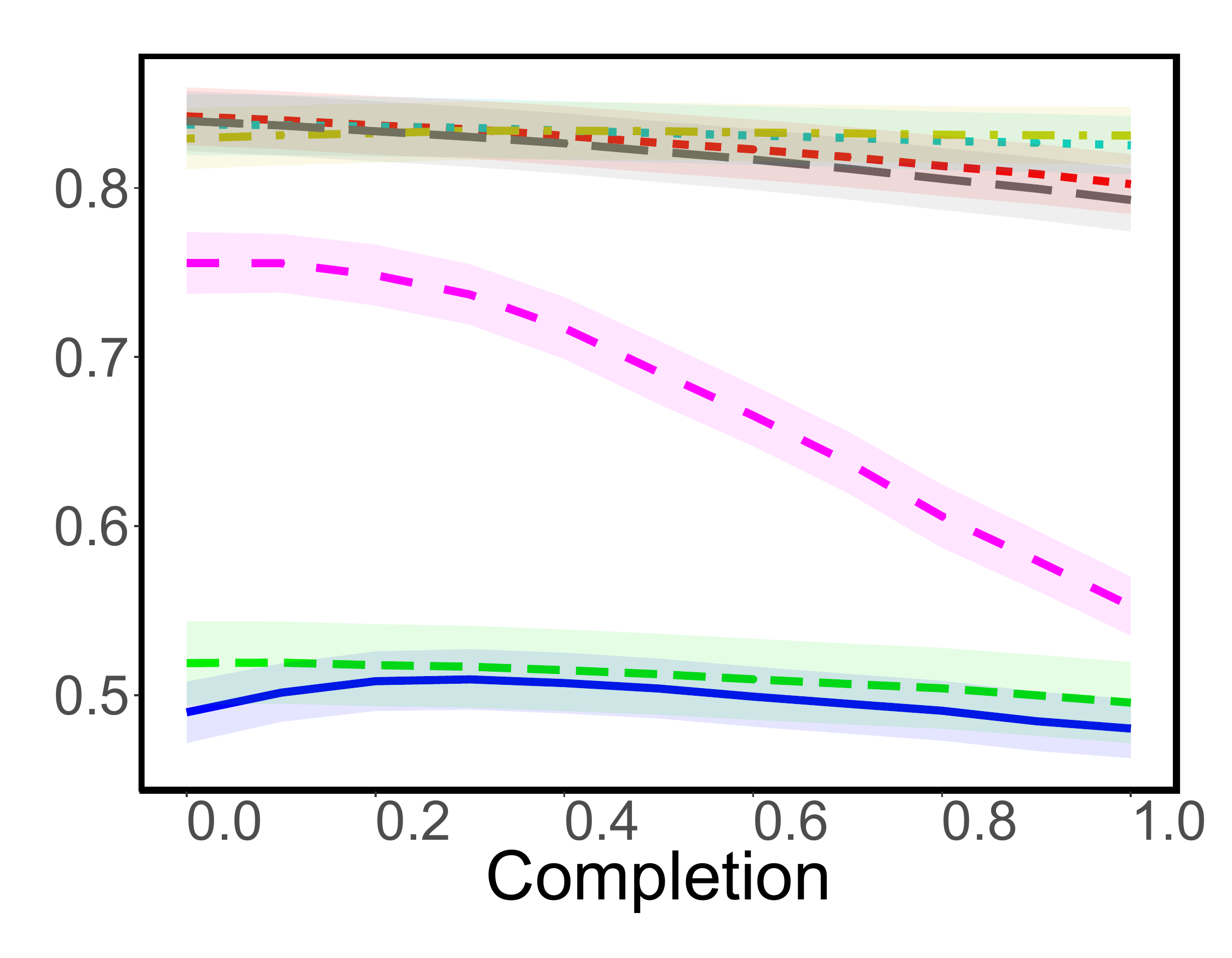} &
\includegraphics[width=1.03\linewidth]{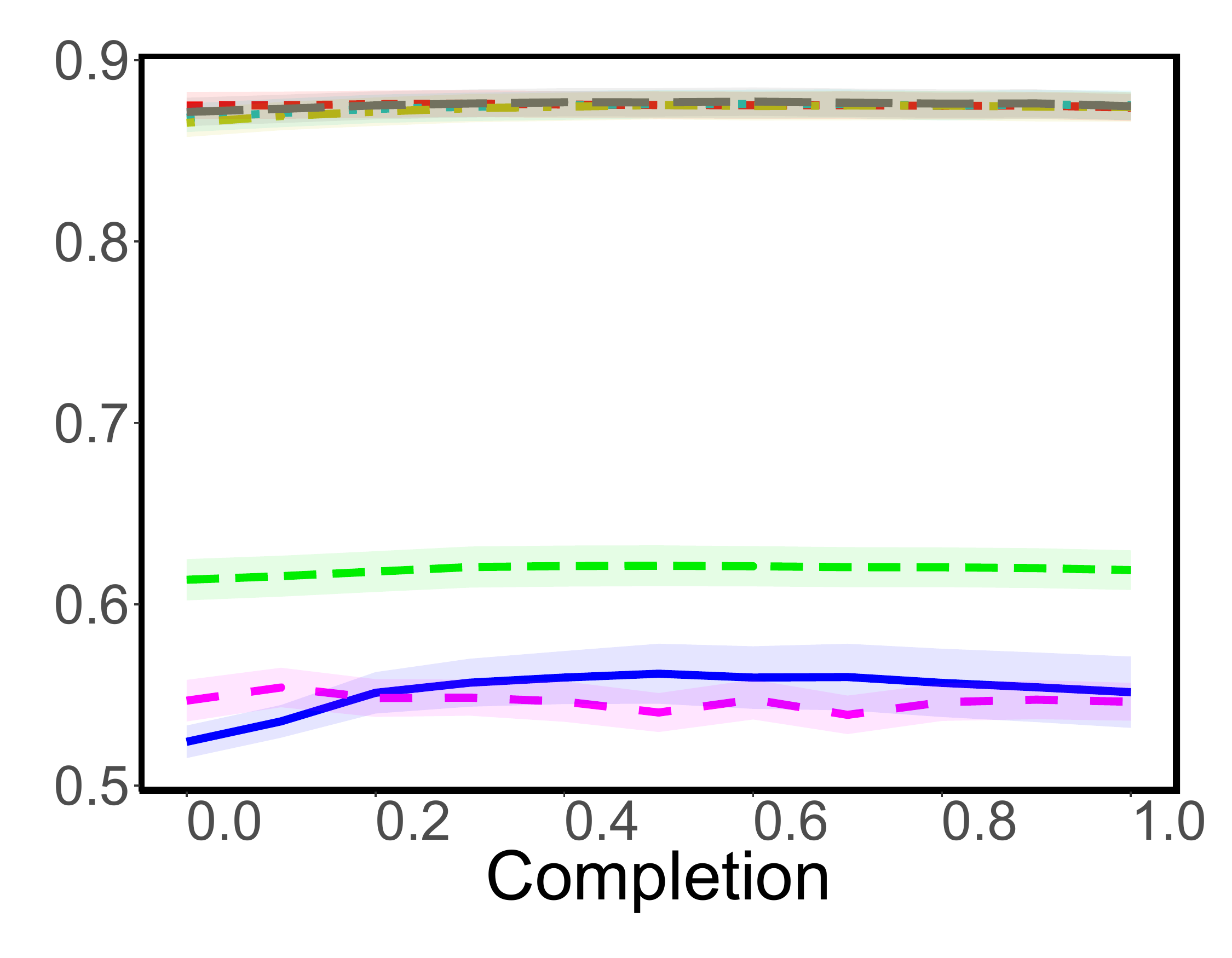} &
\includegraphics[width=1.03\linewidth]{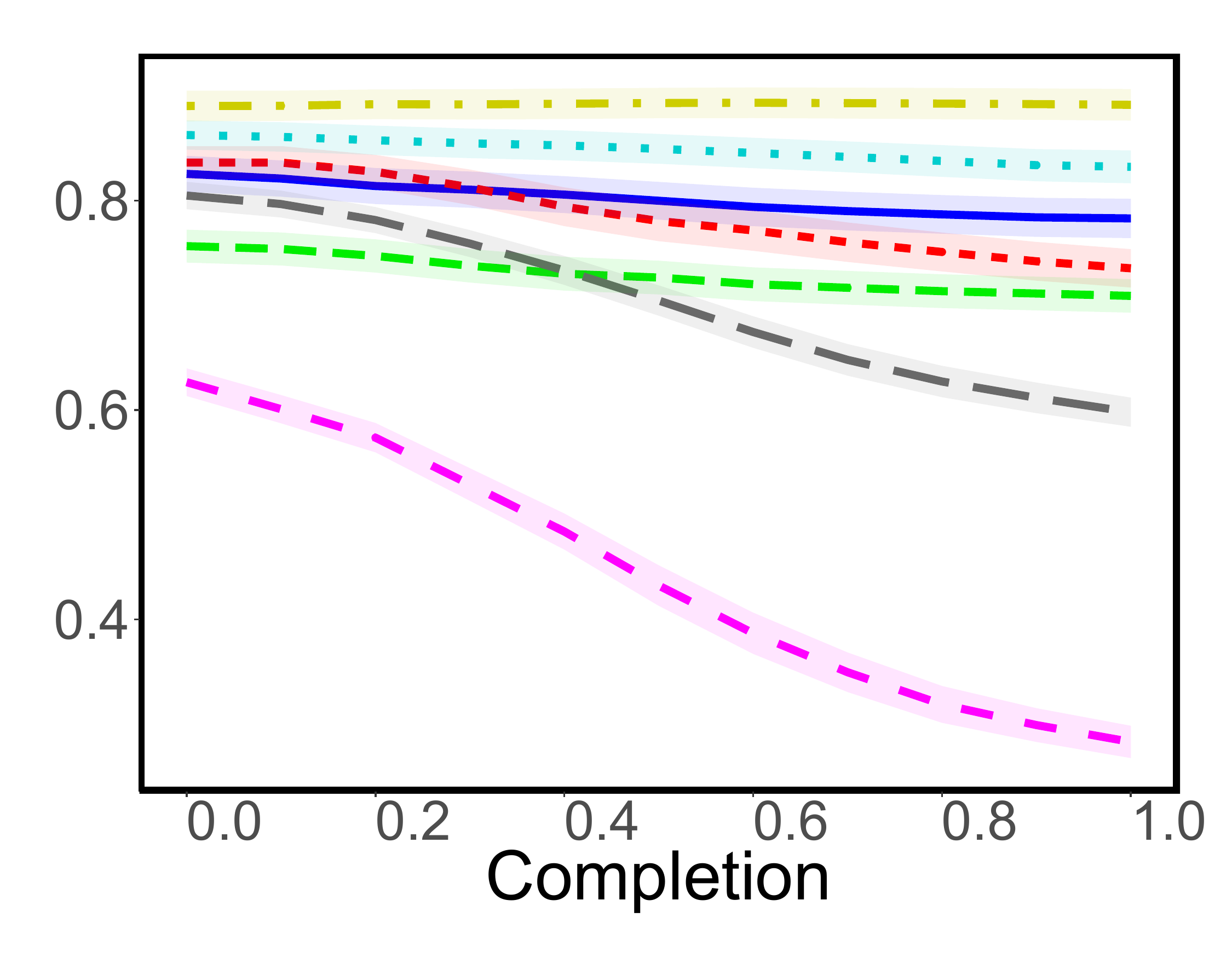}\\
\rotatebox{90}{\footnotesize $\ROC$ values for CTR} &
\includegraphics[width=1.03\linewidth]{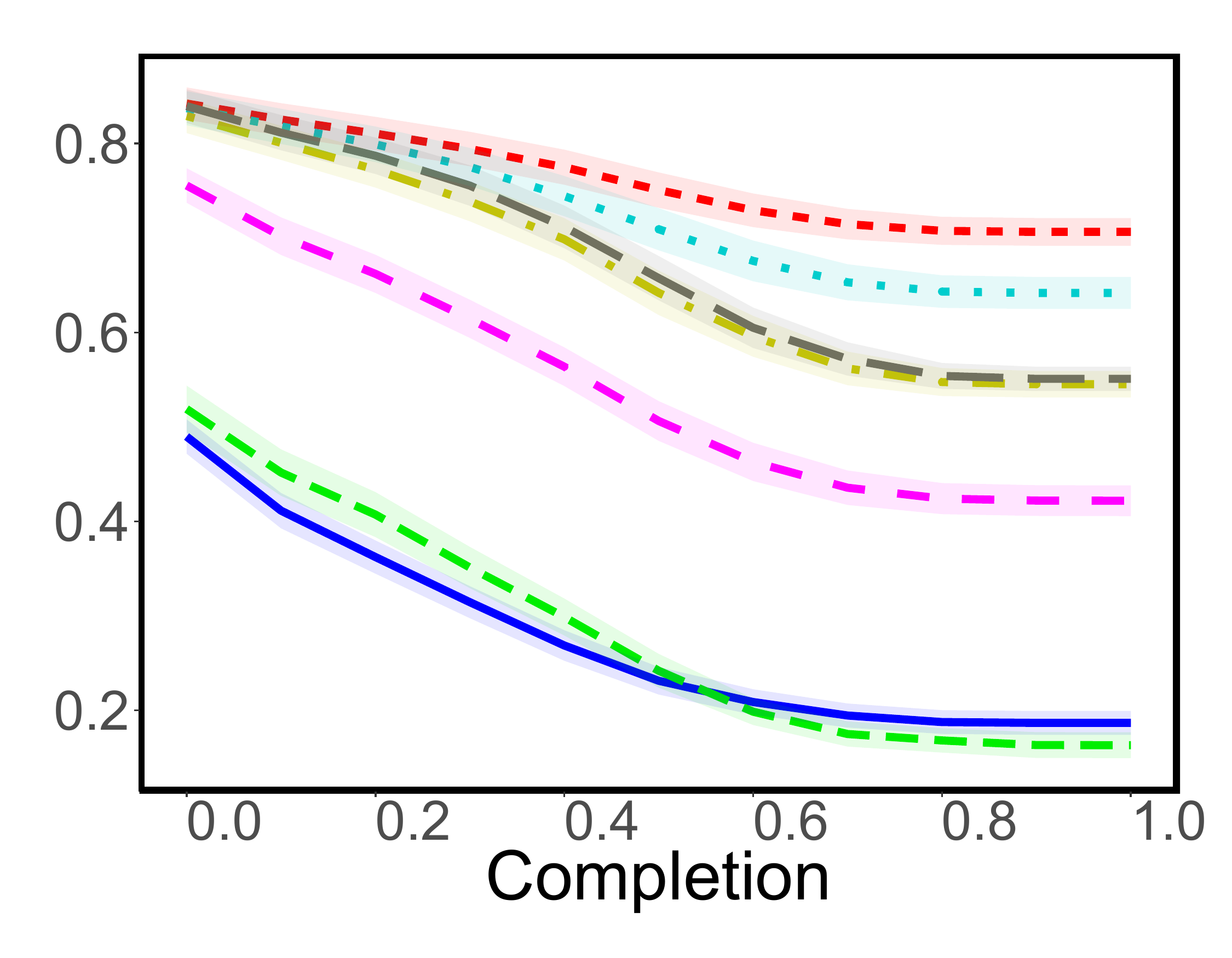} &
\includegraphics[width=1.03\linewidth]{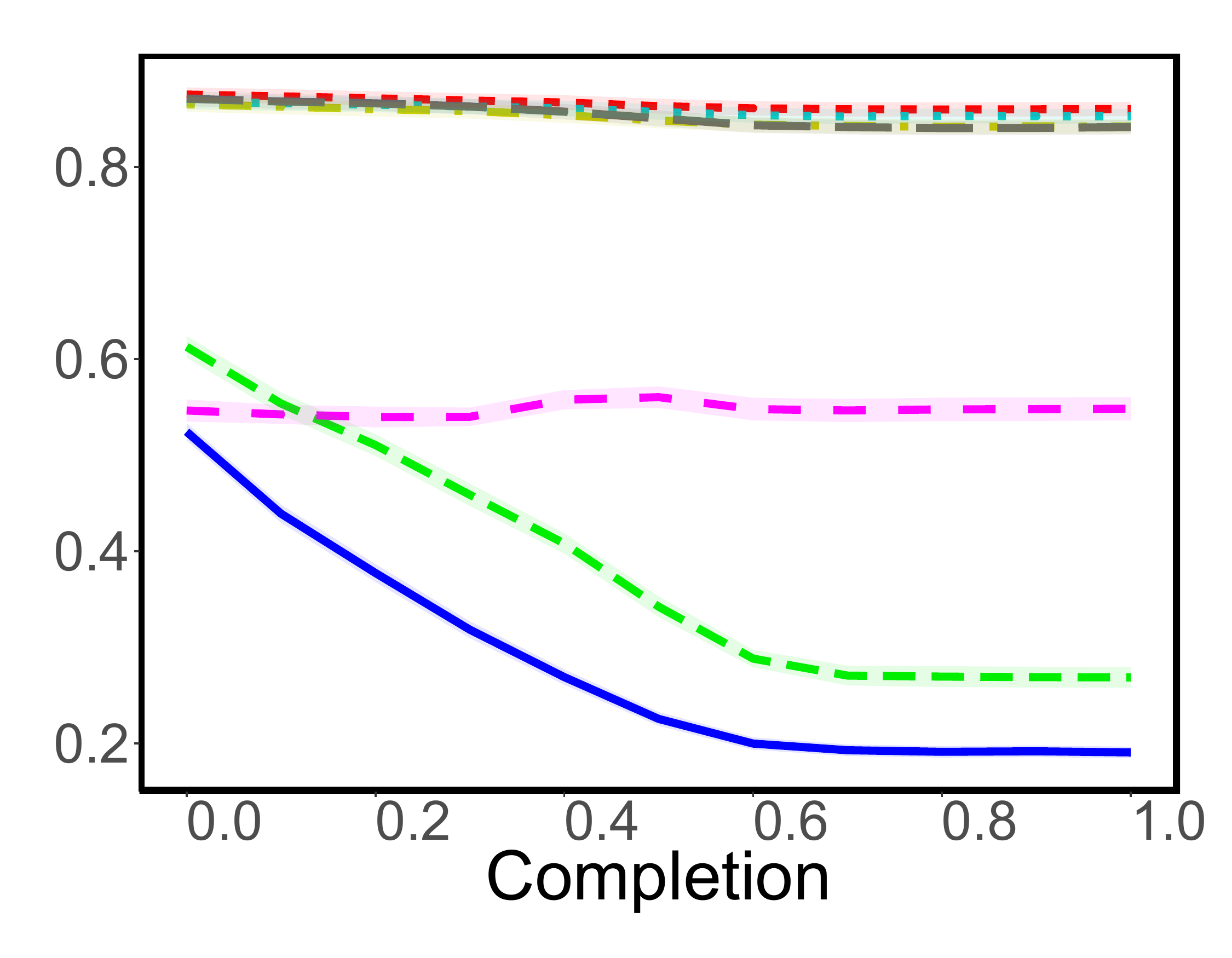} &
\includegraphics[width=1.03\linewidth]{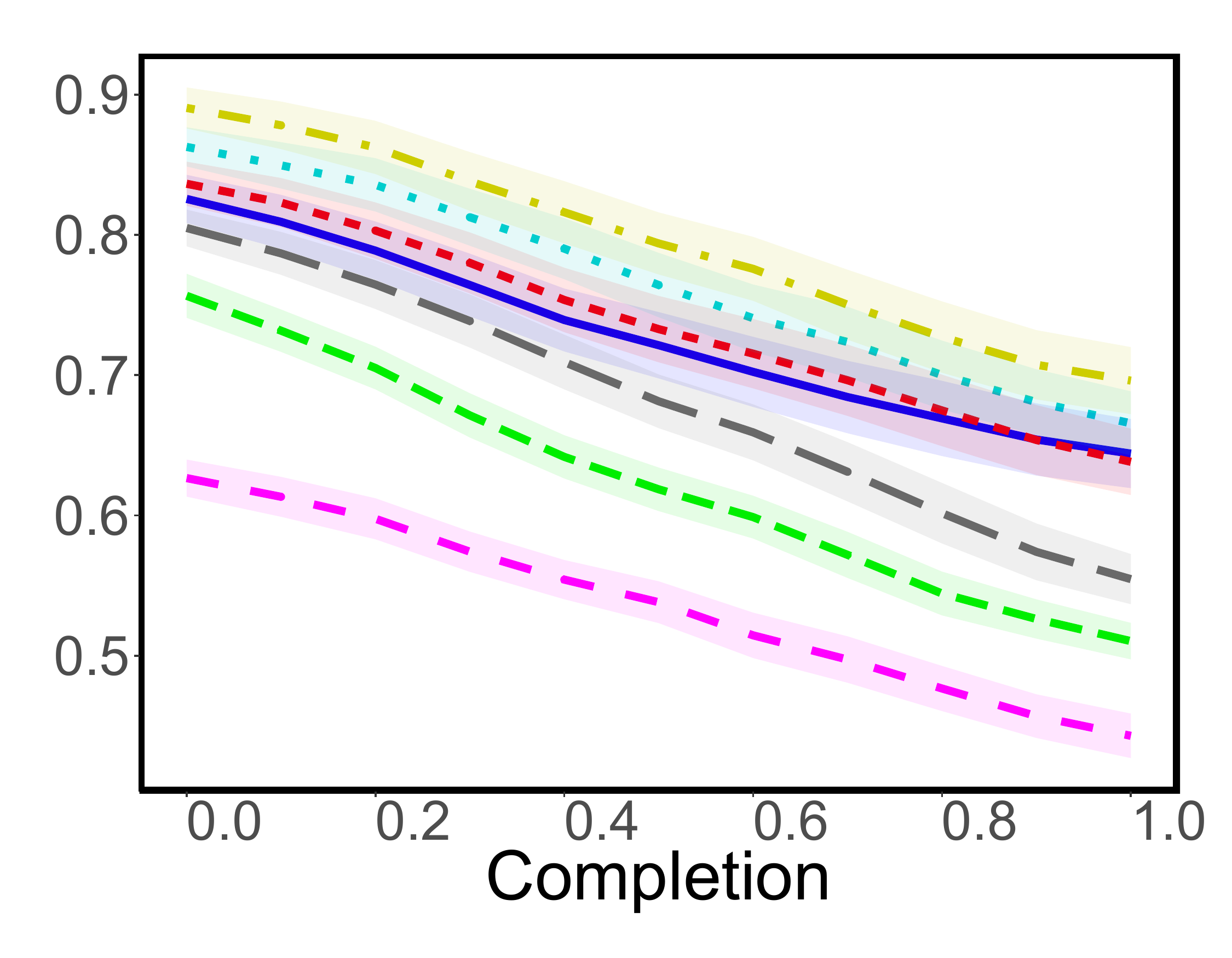} \\
\rotatebox{90}{\footnotesize $\AP$ values for OTC} &
\includegraphics[width=1.03\linewidth]{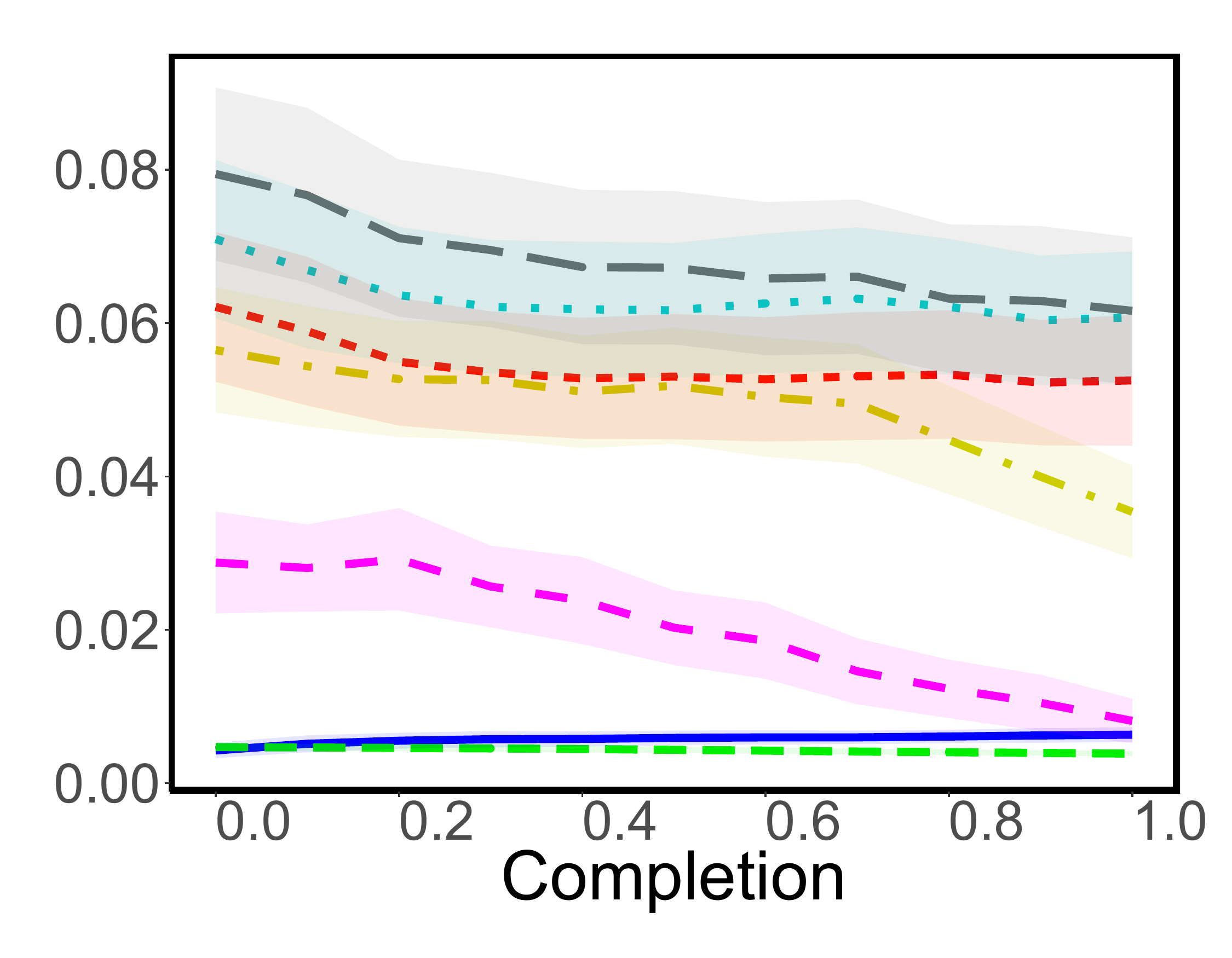} &
\includegraphics[width=1.03\linewidth]{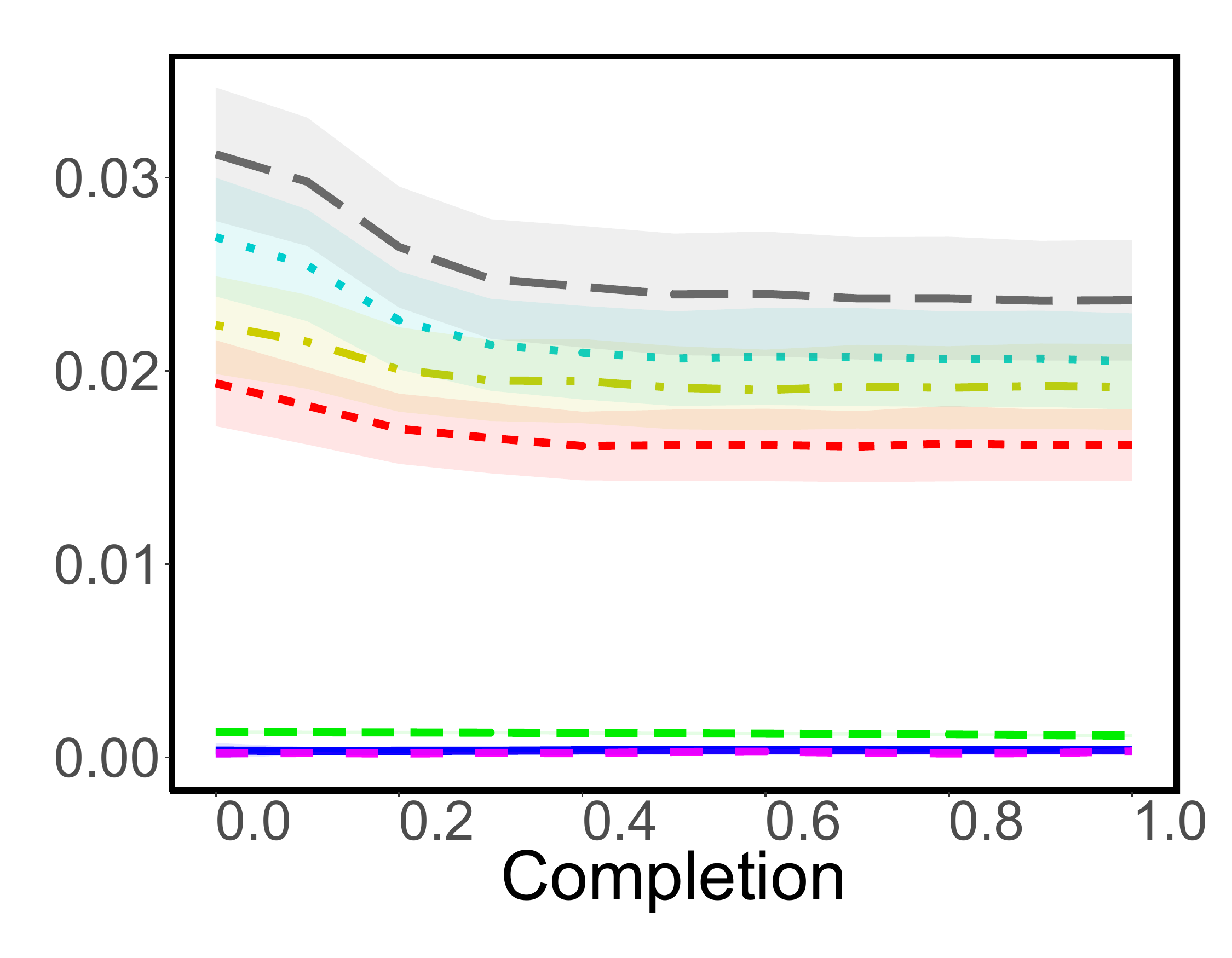} &
\includegraphics[width=1.03\linewidth]{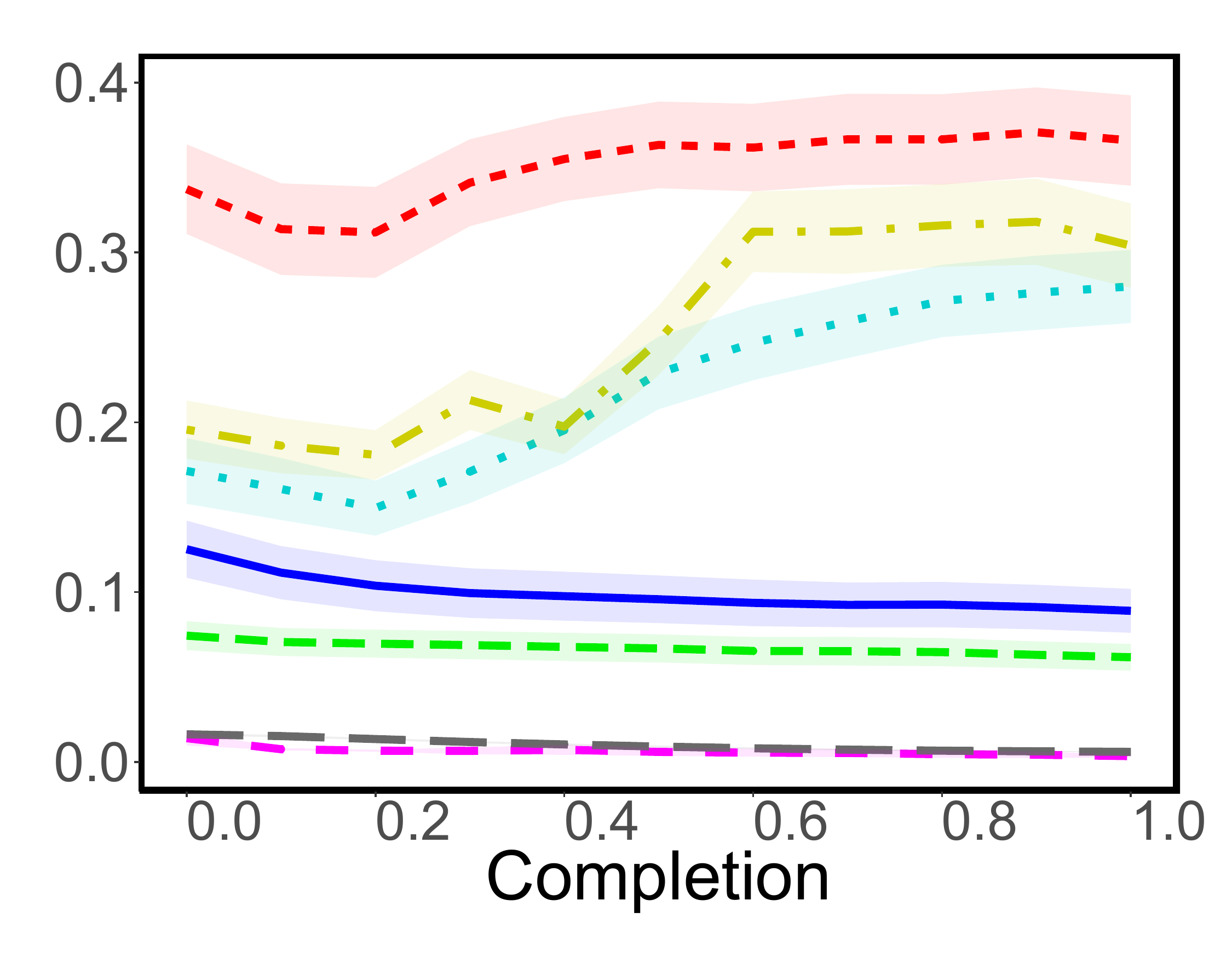} \\
\rotatebox{90}{\footnotesize $\AP$ values for CTR} &
\includegraphics[width=1.03\linewidth]{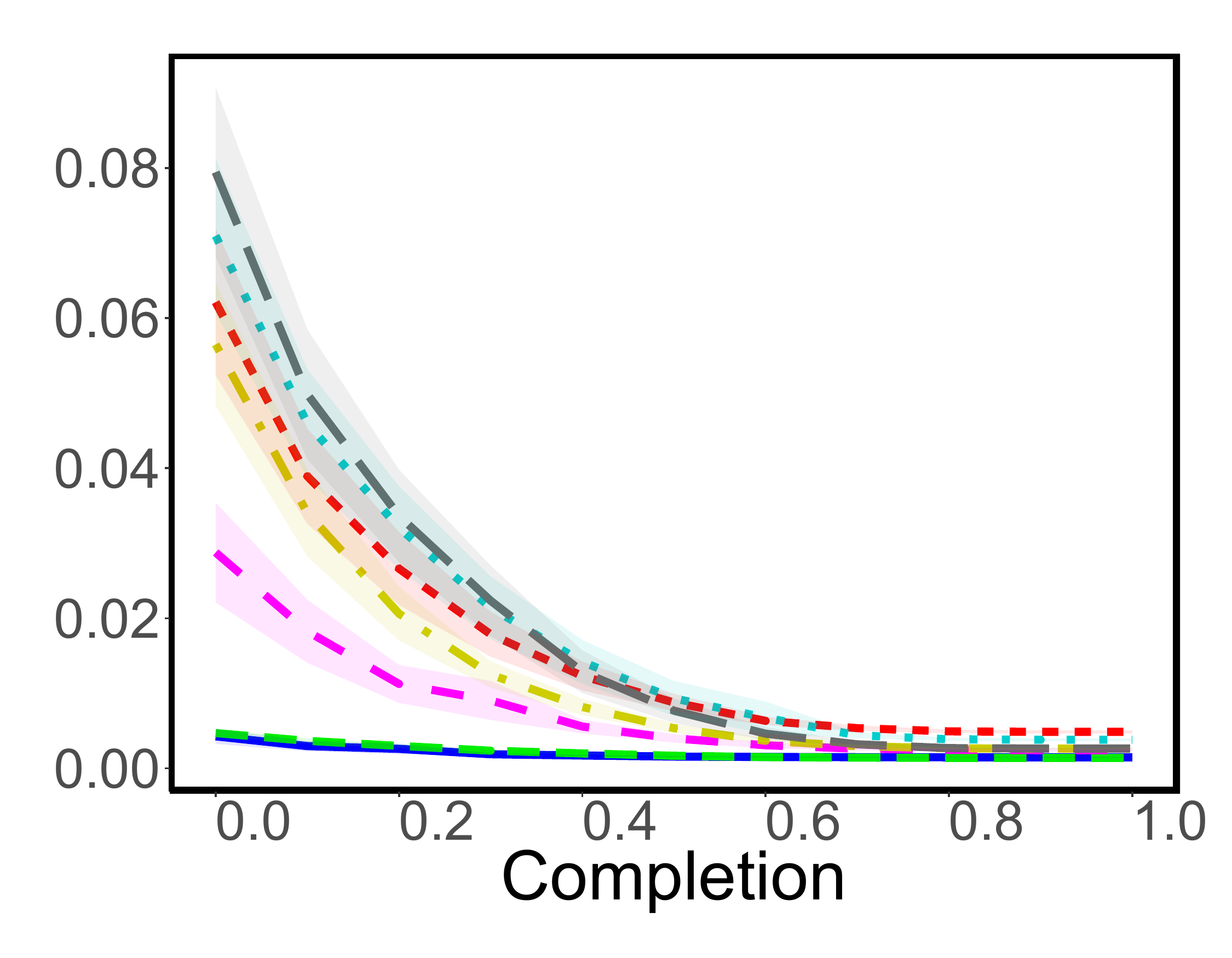} &
\includegraphics[width=1.03\linewidth]{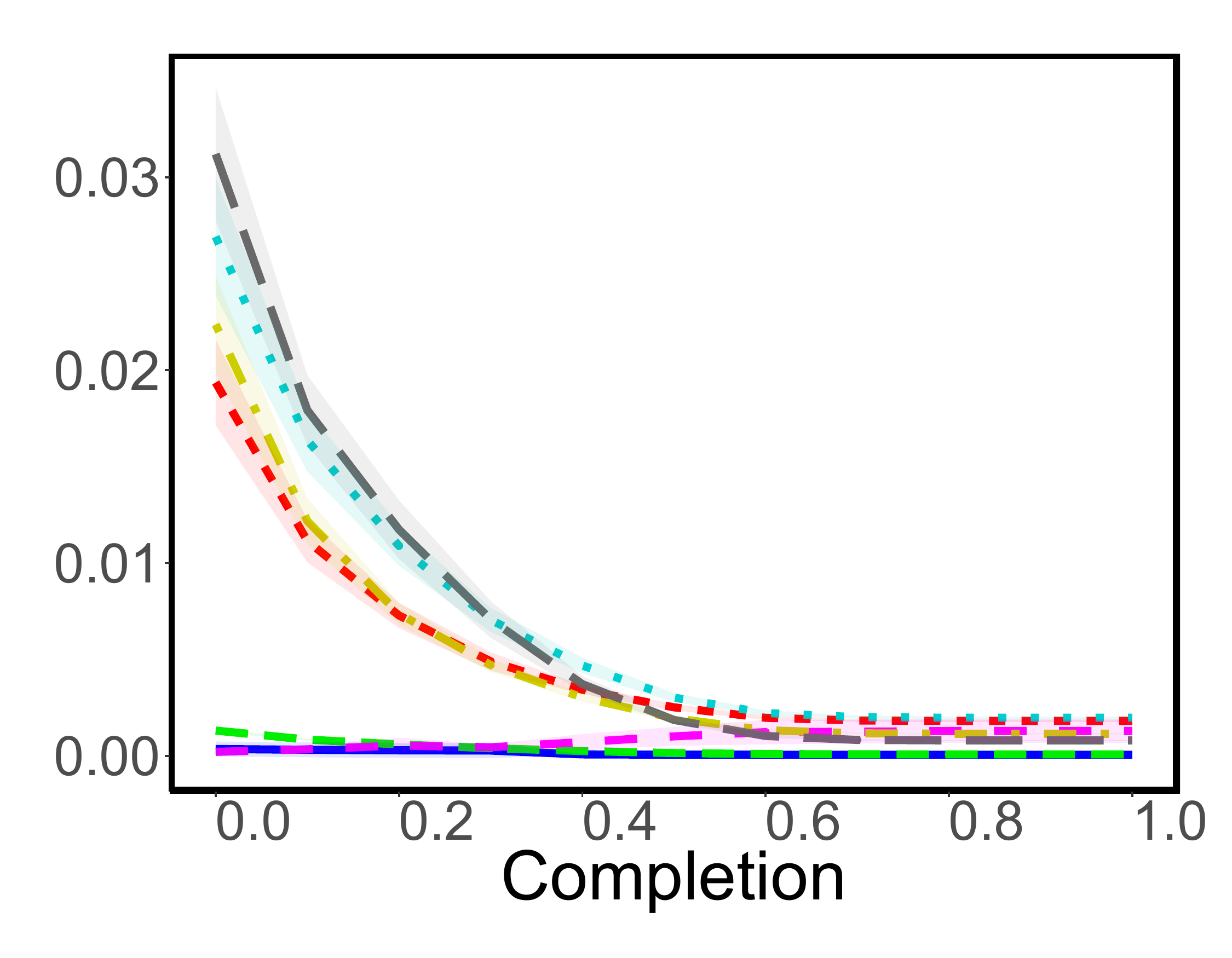} &
\includegraphics[width=1.03\linewidth]{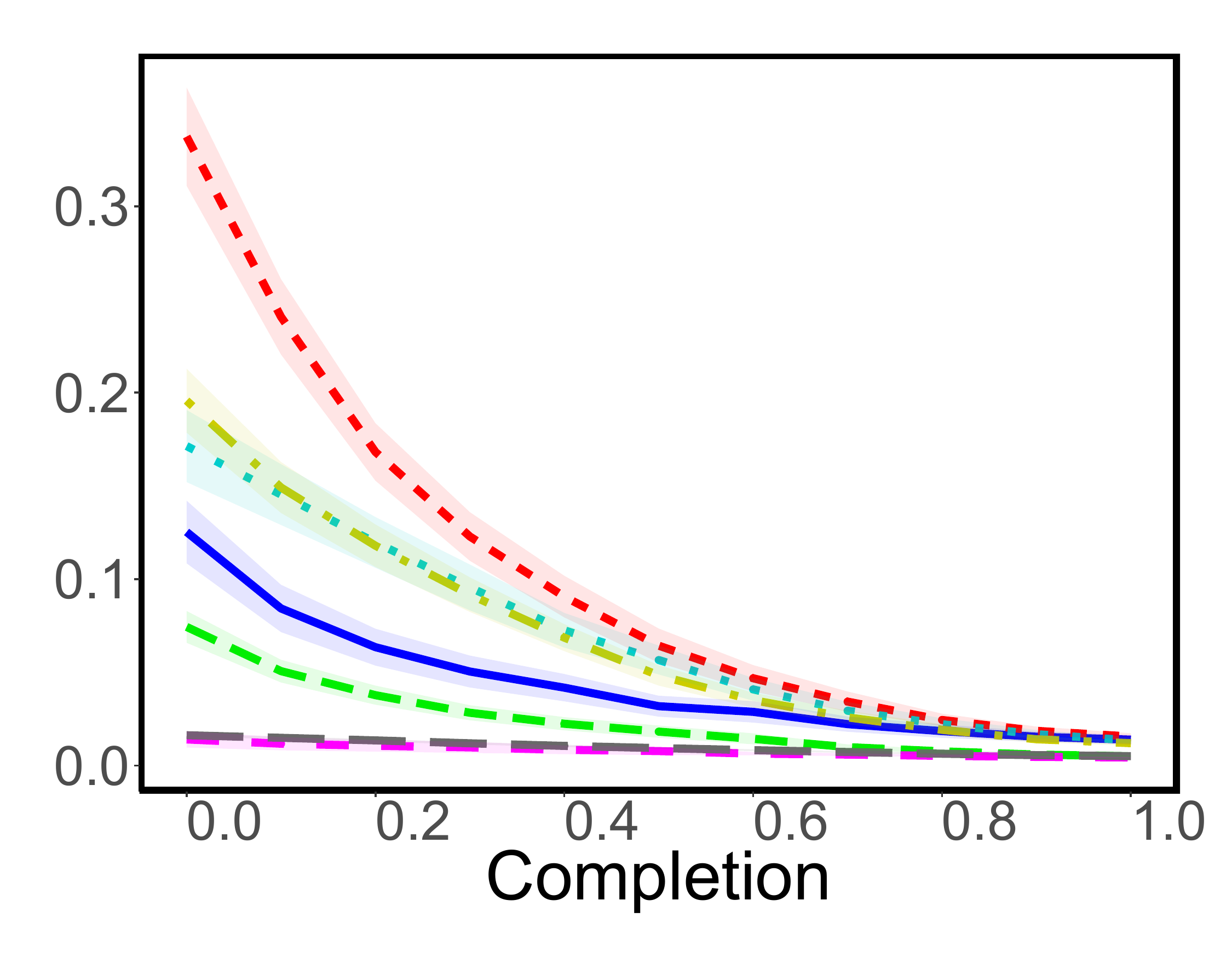} \\
\multicolumn{4}{c}{\includegraphics[width=0.65\linewidth]{figures/plots/global/legend}}
\end{tabular}
\caption{Given different \textbf{global similarity} indices, the figure depicts the values of $\ROC$ (the area under the ROC curve) and $\AP$ (the average precision) during the execution of OTC and CTR given $|\Hide|=\max(10,|E|/100)$ and $b=4|\Hide|$ in three networks: (i) \textbf{SmallWorld(100,10,0.25)}; (ii) \textbf{SmallWorld(1000,10,0.25)}; and (iii) \textbf{Les Mis\'erables network}.
In each execution, the links in $\Hide$ are chosen at random. Results are taken as the average over $50$ executions, with coloured areas representing the $95\%$ confidence intervals.}
\label{fig:global-3}
\end{figure*}

\begin{figure*}[tbhp]
\centering
\setlength\tabcolsep{1pt}
\renewcommand{\arraystretch}{0.01}
\begin{tabular}{m{.03\textwidth}m{.27\textwidth}m{.27\textwidth}m{.27\textwidth}}
& \multicolumn{1}{c}{Facebook fragment (small)}
& \multicolumn{1}{c}{Facebook fragment (large)}
& \multicolumn{1}{c}{Zachary Karate Club}\\
\rotatebox{90}{\footnotesize $\ROC$ values for OTC} &
\includegraphics[width=1.03\linewidth]{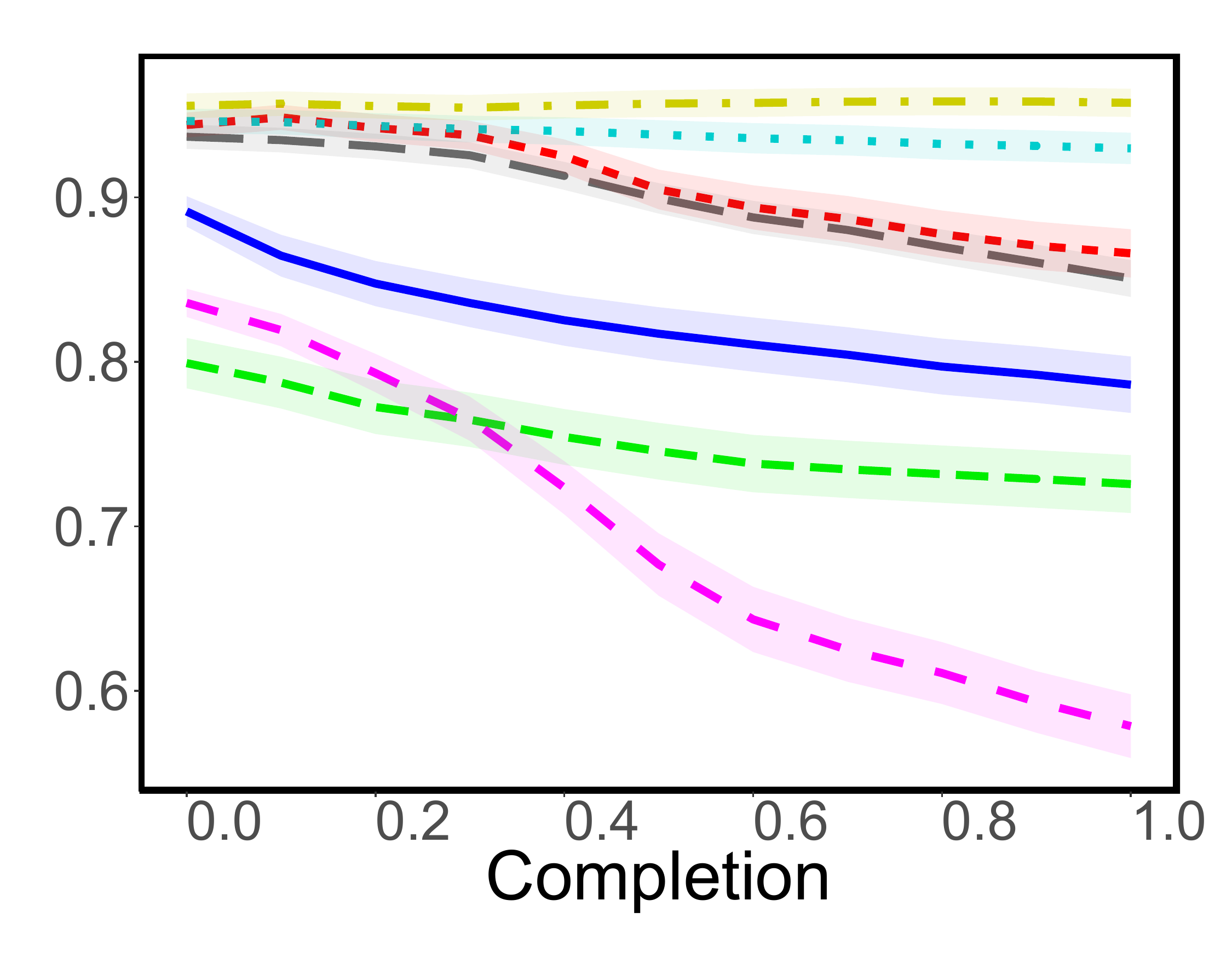} &
\includegraphics[width=1.03\linewidth]{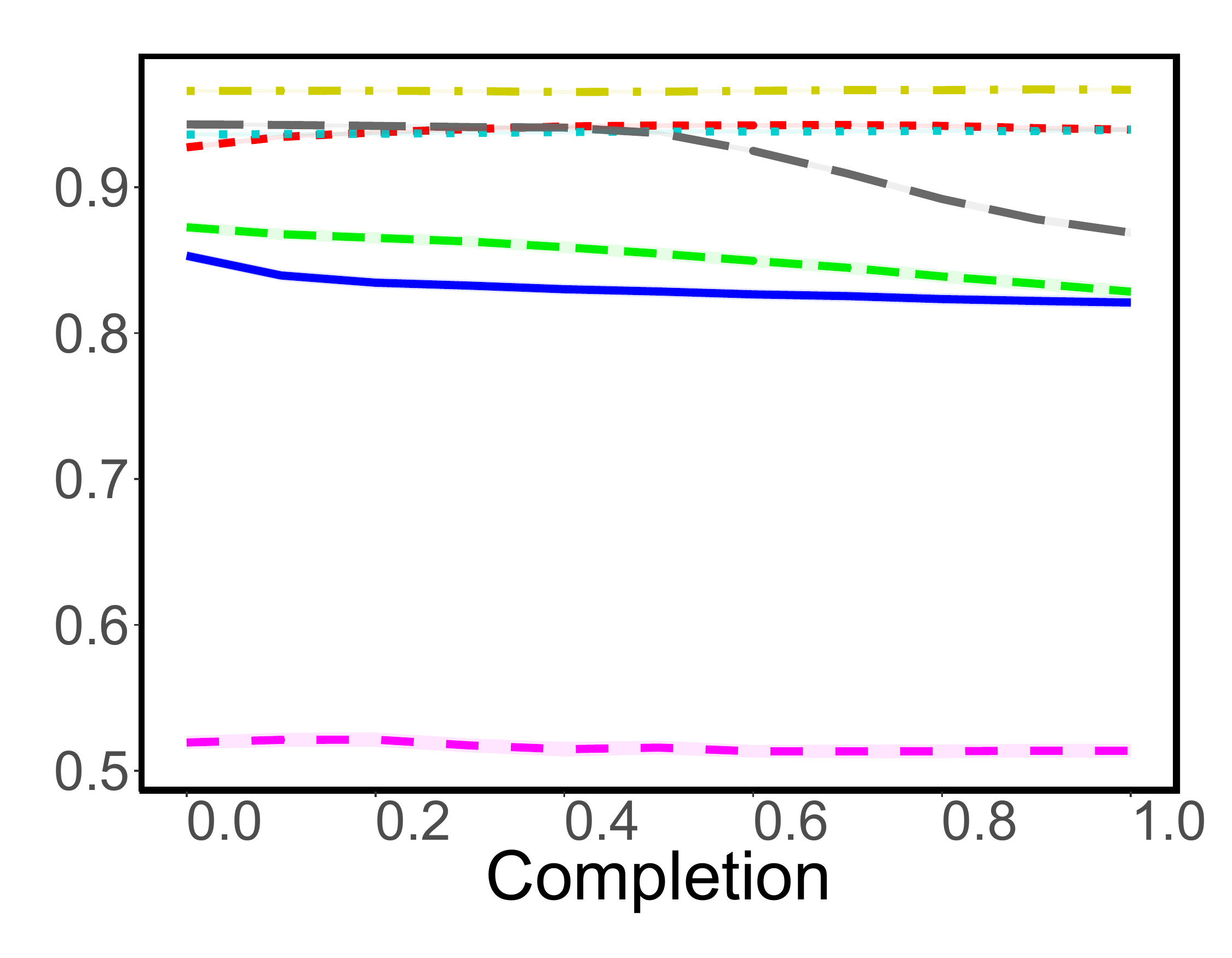} &
\includegraphics[width=1.03\linewidth]{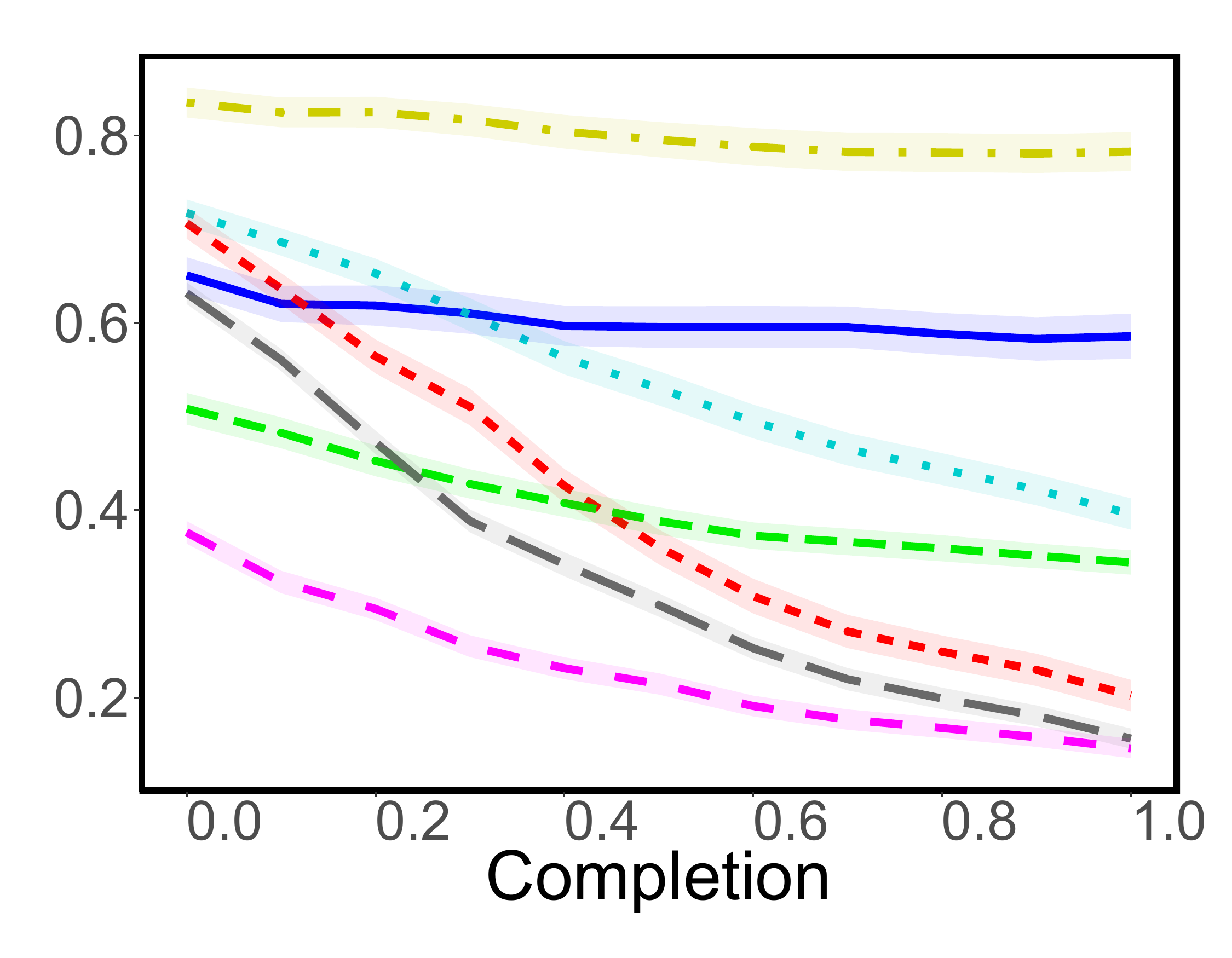}\\
\rotatebox{90}{\footnotesize $\ROC$ values for CTR} &
\includegraphics[width=1.03\linewidth]{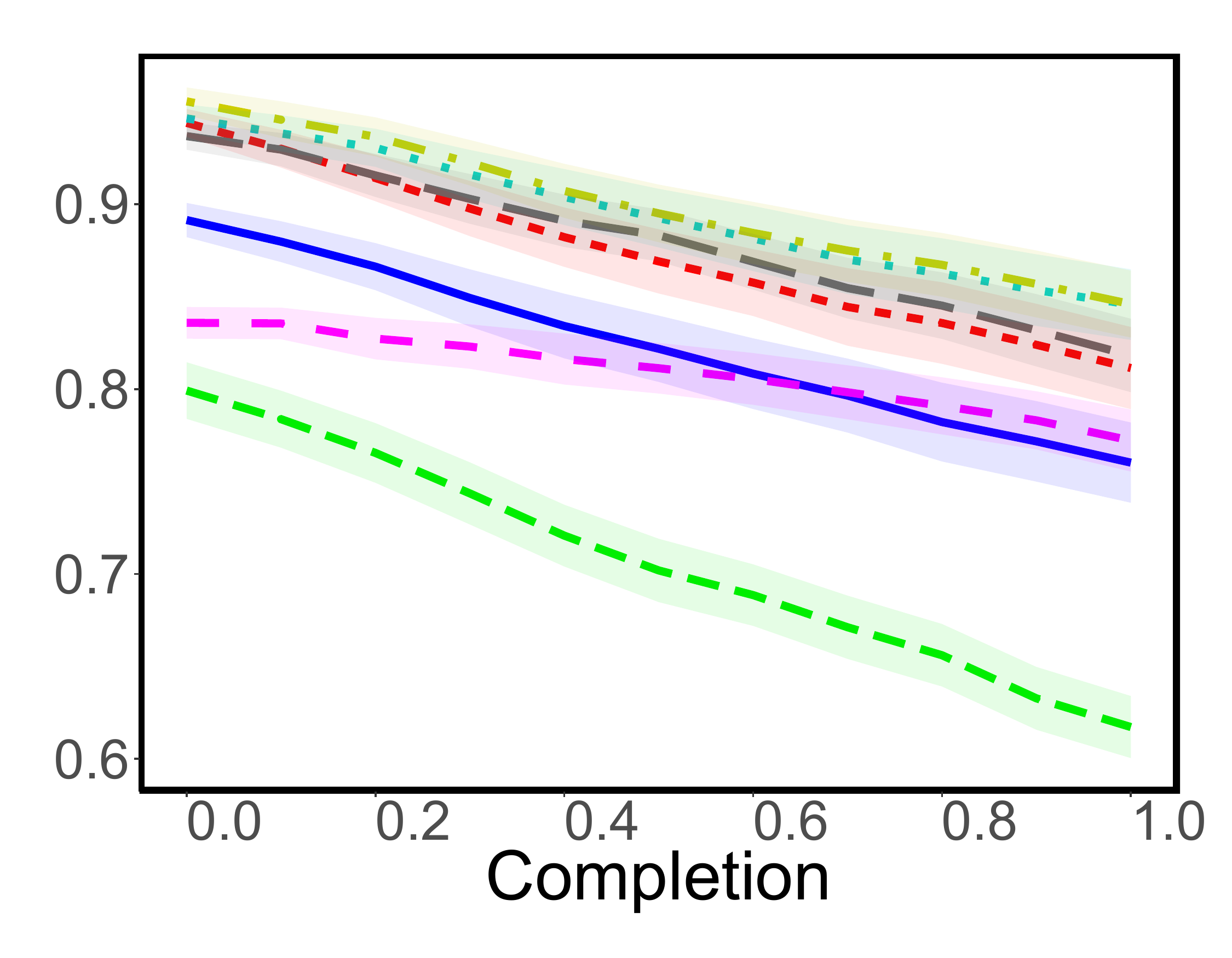} &
\includegraphics[width=1.03\linewidth]{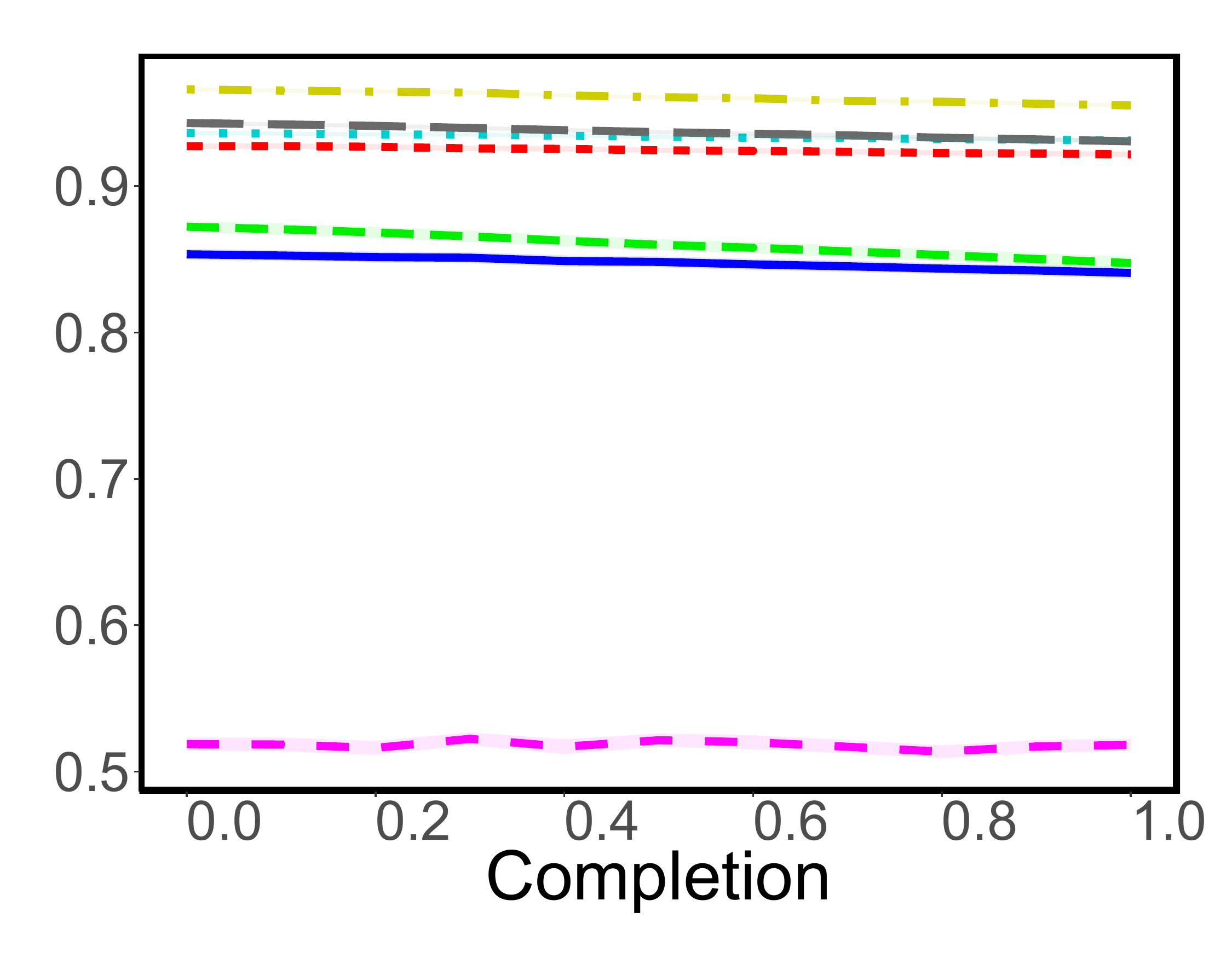} &
\includegraphics[width=1.03\linewidth]{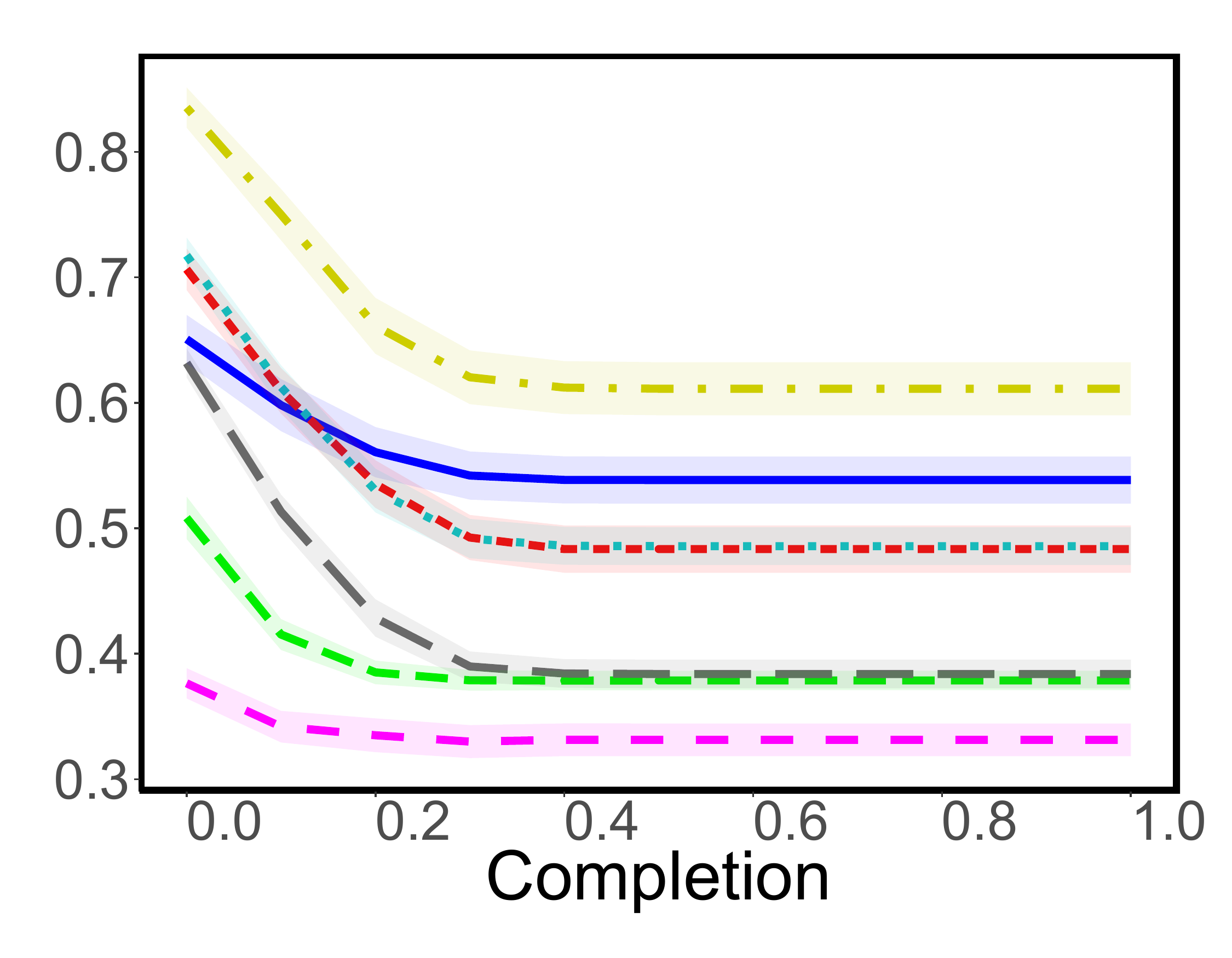} \\
\rotatebox{90}{\footnotesize $\AP$ values for OTC} &
\includegraphics[width=1.03\linewidth]{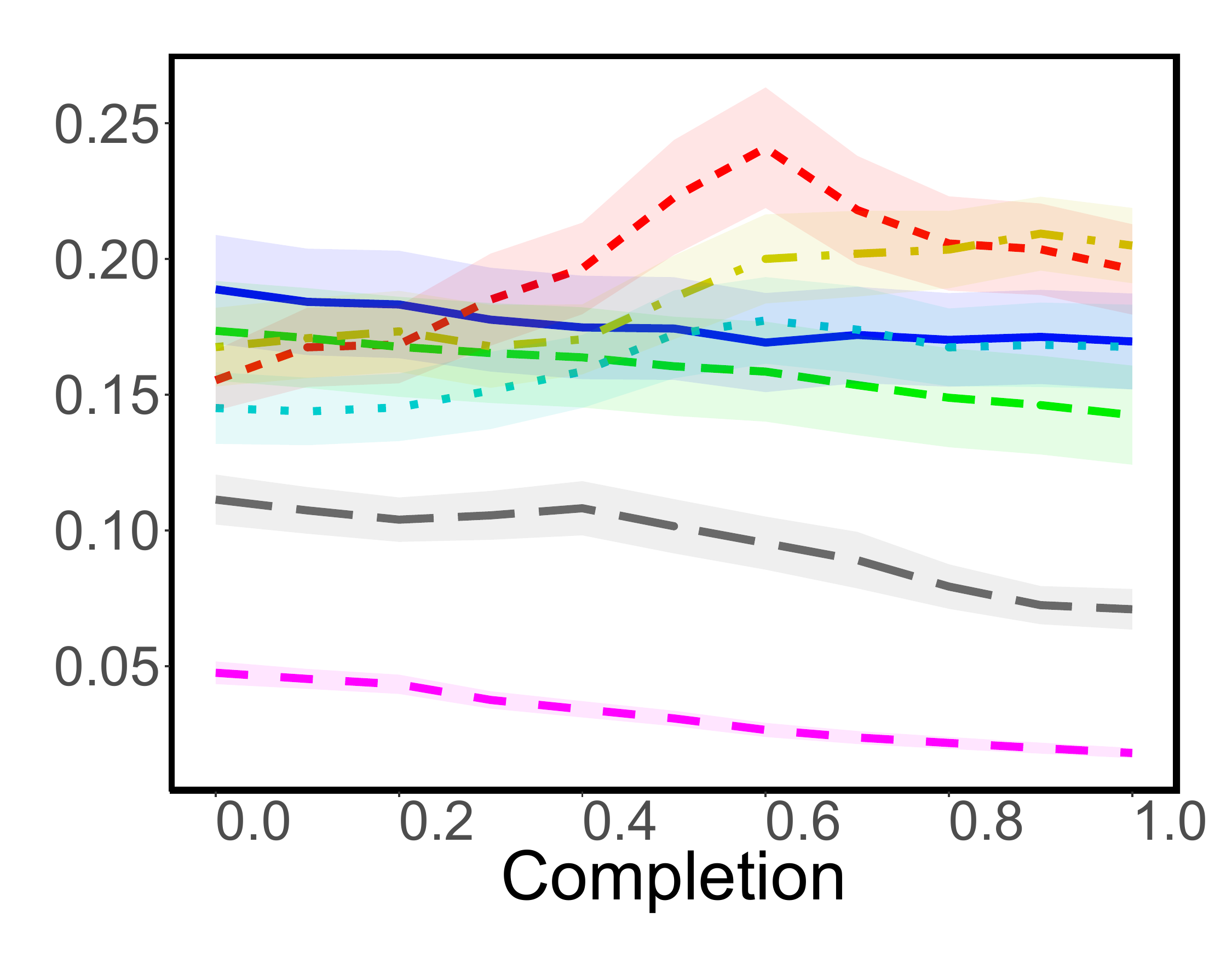} &
\includegraphics[width=1.03\linewidth]{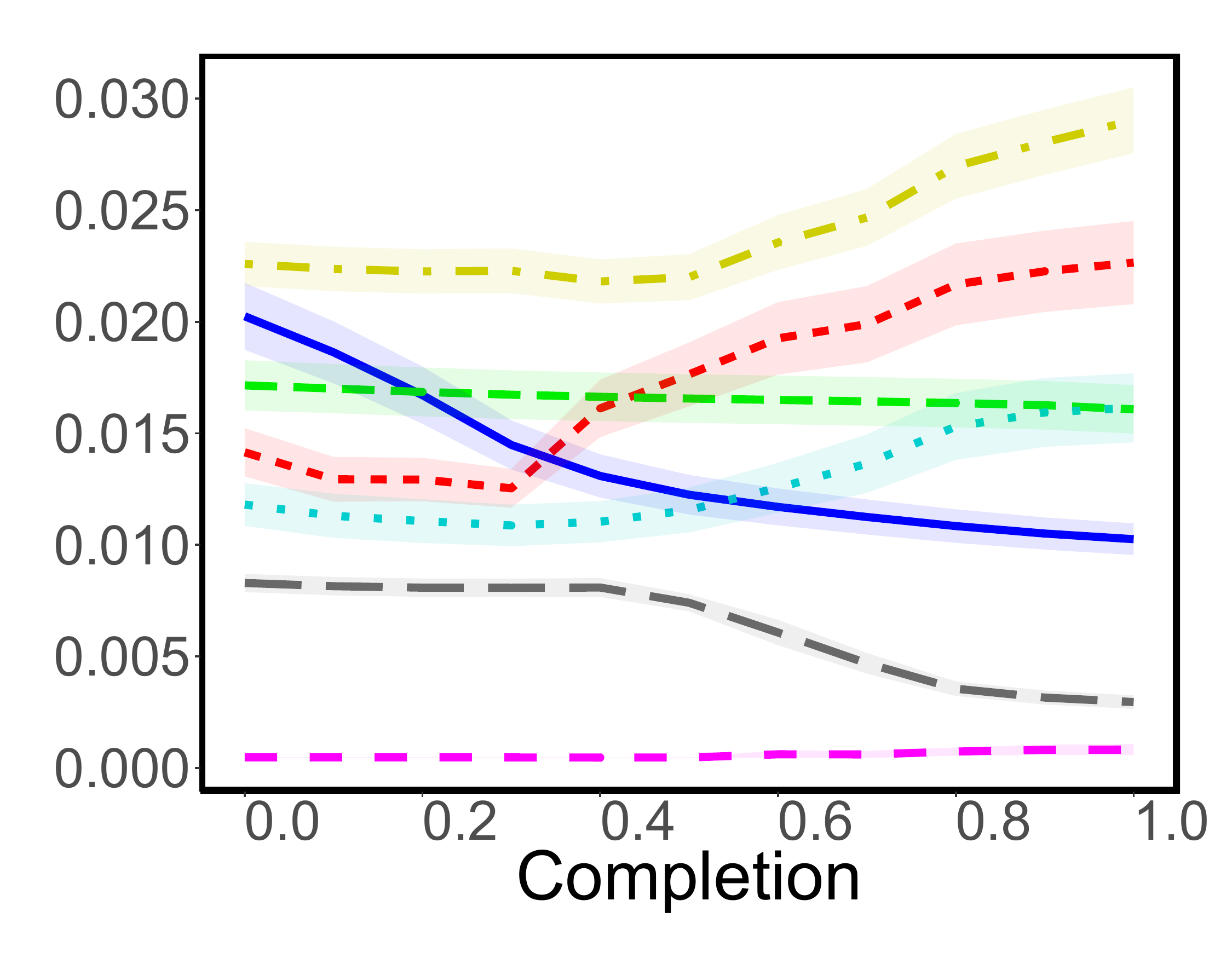} &
\includegraphics[width=1.03\linewidth]{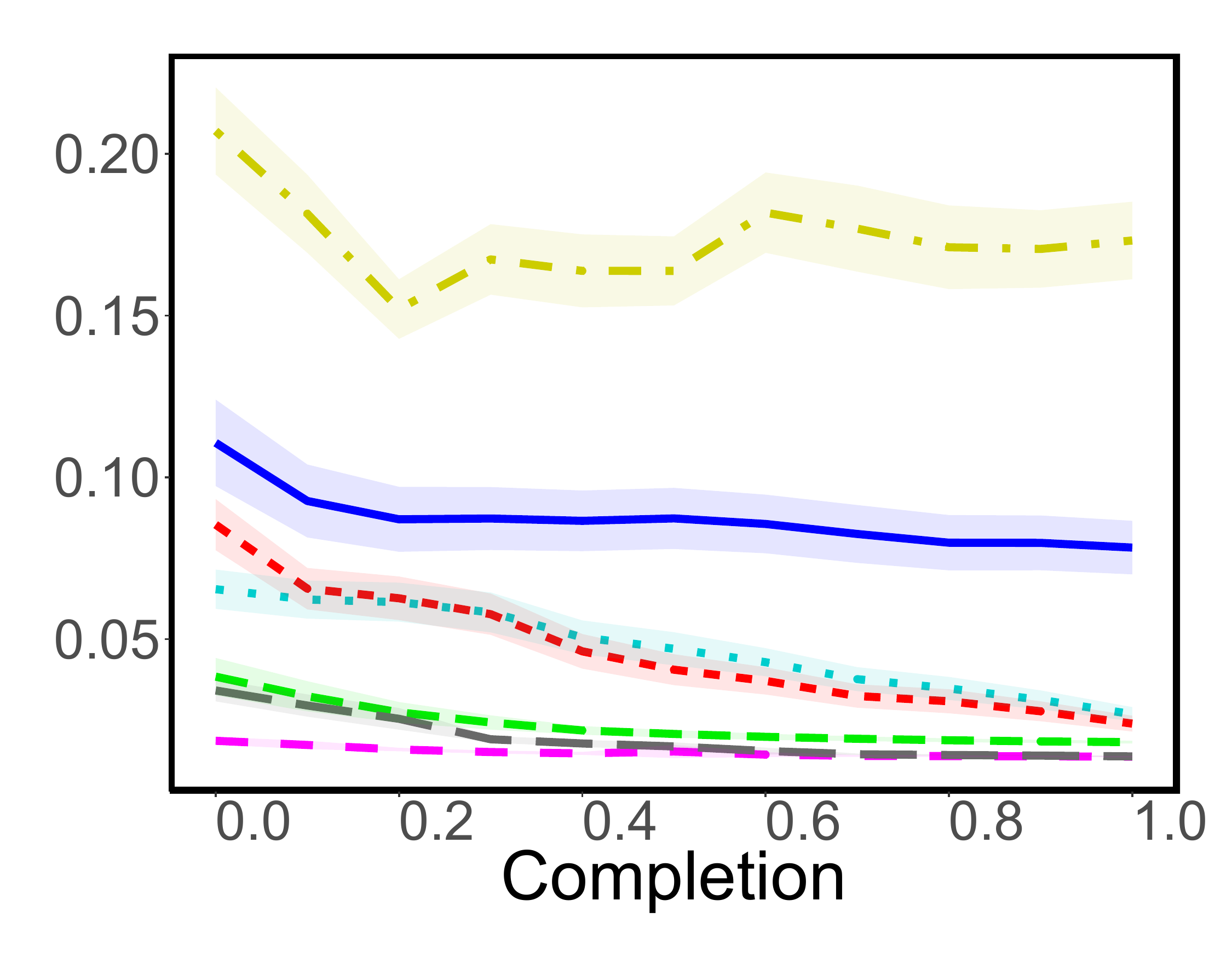} \\
\rotatebox{90}{\footnotesize $\AP$ values for CTR} &
\includegraphics[width=1.03\linewidth]{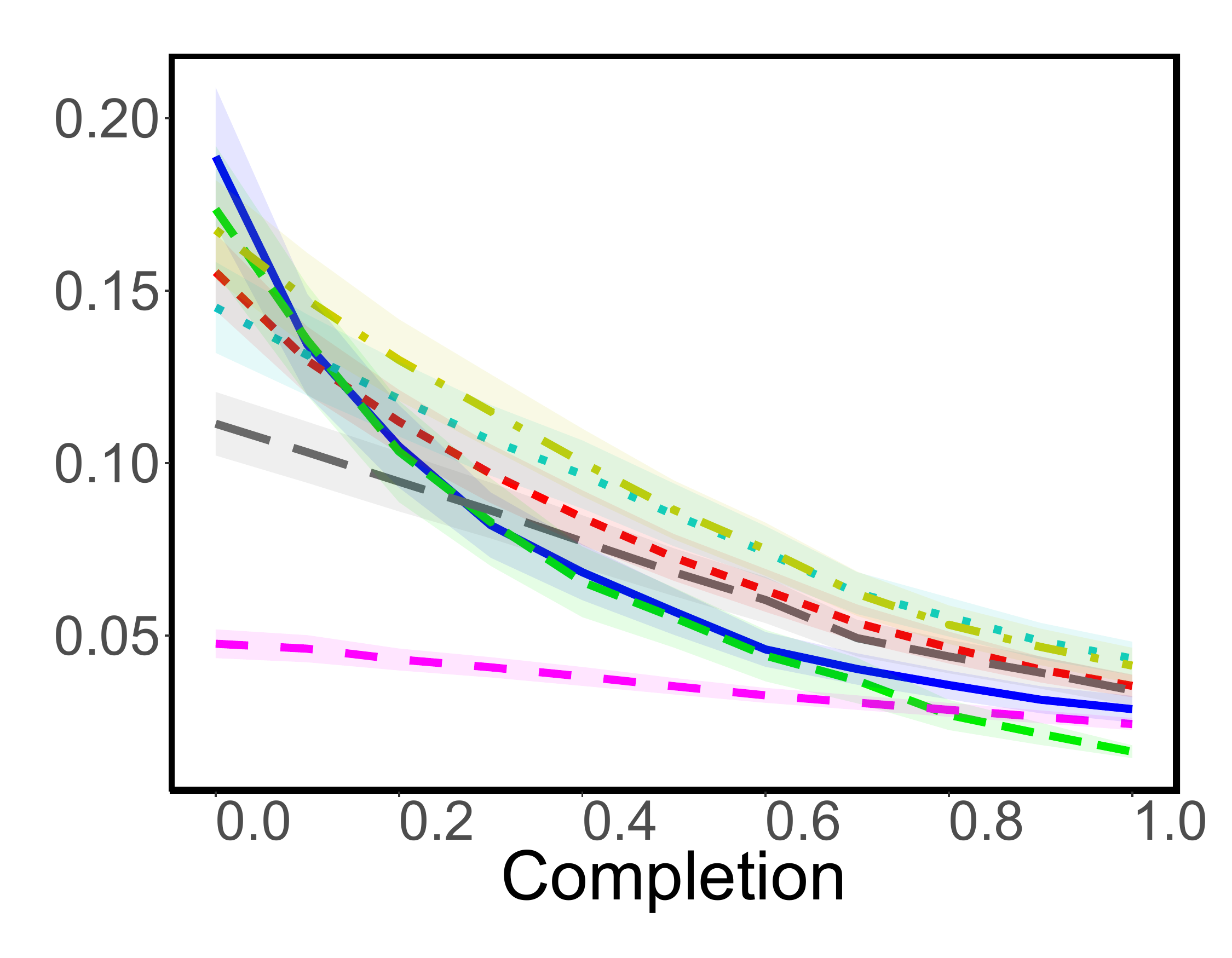} &
\includegraphics[width=1.03\linewidth]{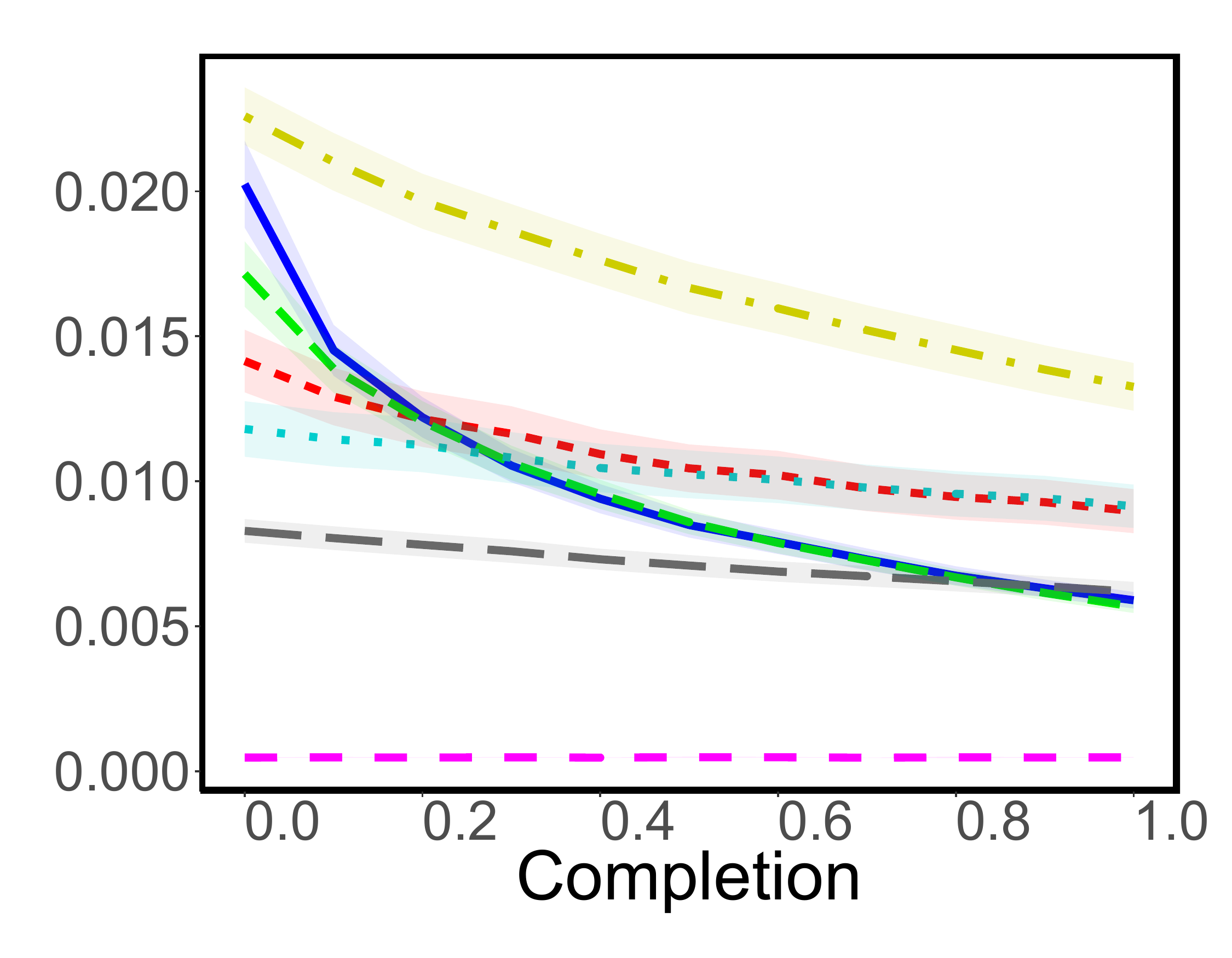} &
\includegraphics[width=1.03\linewidth]{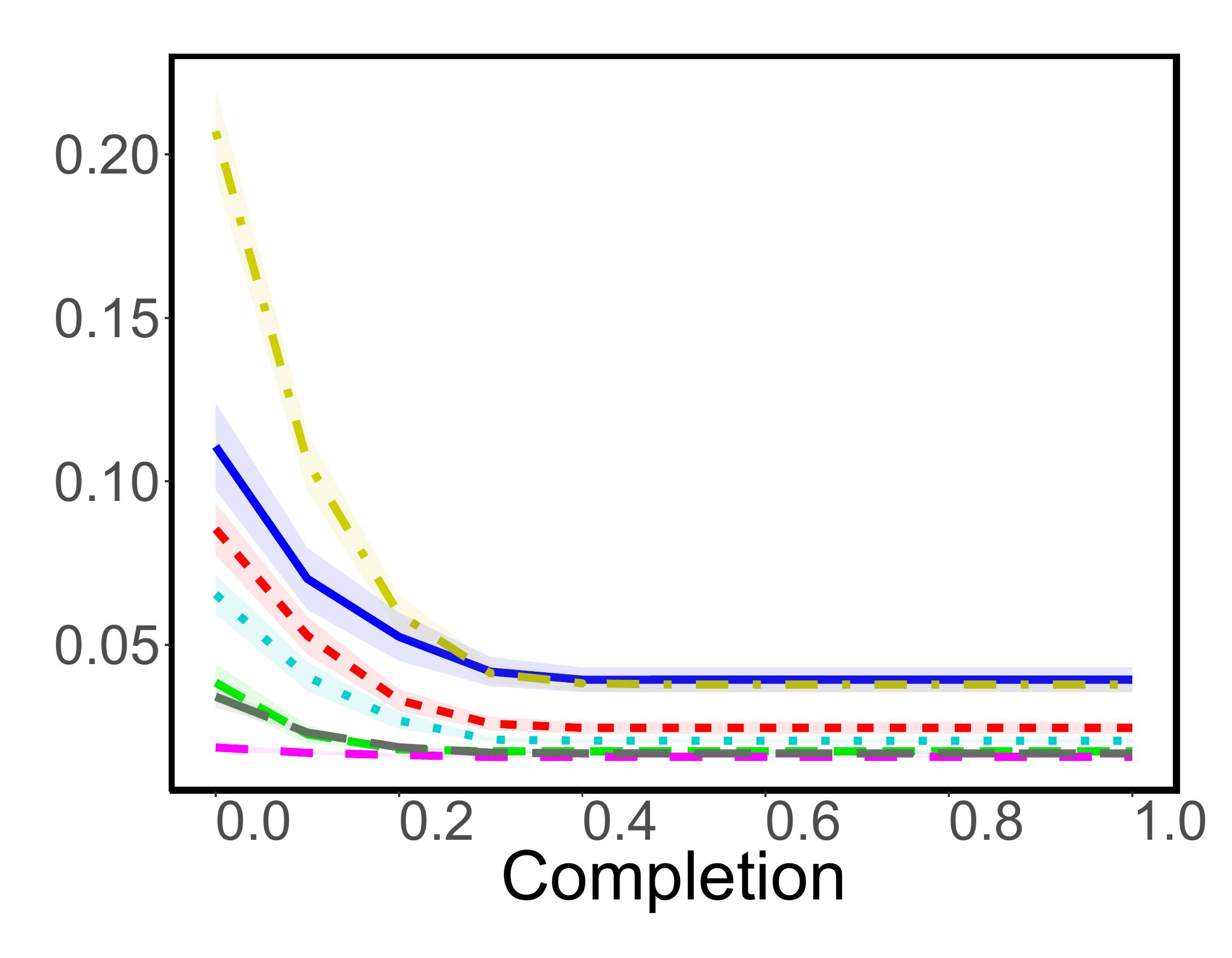} \\
\multicolumn{4}{c}{\includegraphics[width=0.65\linewidth]{figures/plots/global/legend}}
\end{tabular}
\caption{Given different \textbf{global similarity} indices, the figure depicts the values of $\ROC$ (the area under the ROC curve) and $\AP$ (the average precision) during the execution of OTC and CTR given $|\Hide|=\max(10,|E|/100)$ and $b=4|\Hide|$ in three networks: (i) \textbf{A small fragment of Facebook}; (ii) \textbf{a large fragment of Facebook}; and (iii) \textbf{the Zachary Karate club network}.
In each execution, the links in $\Hide$ are chosen at random. Results are taken as the average over $50$ executions, with coloured areas representing the $95\%$ confidence intervals.}
\label{fig:global-4}
\end{figure*}

\begin{figure*}[tbhp]
\centering
\setlength\tabcolsep{1pt}
\renewcommand{\arraystretch}{0.01}
\begin{tabular}{m{.03\textwidth}m{.27\textwidth}m{.27\textwidth}m{.27\textwidth}}
& \multicolumn{1}{c}{Bali-attack network}
& \multicolumn{1}{c}{Madrid-bombing network}
& \multicolumn{1}{c}{Greek political blogs}\\
\rotatebox{90}{\footnotesize $\ROC$ values for OTC} &
\includegraphics[width=1.03\linewidth]{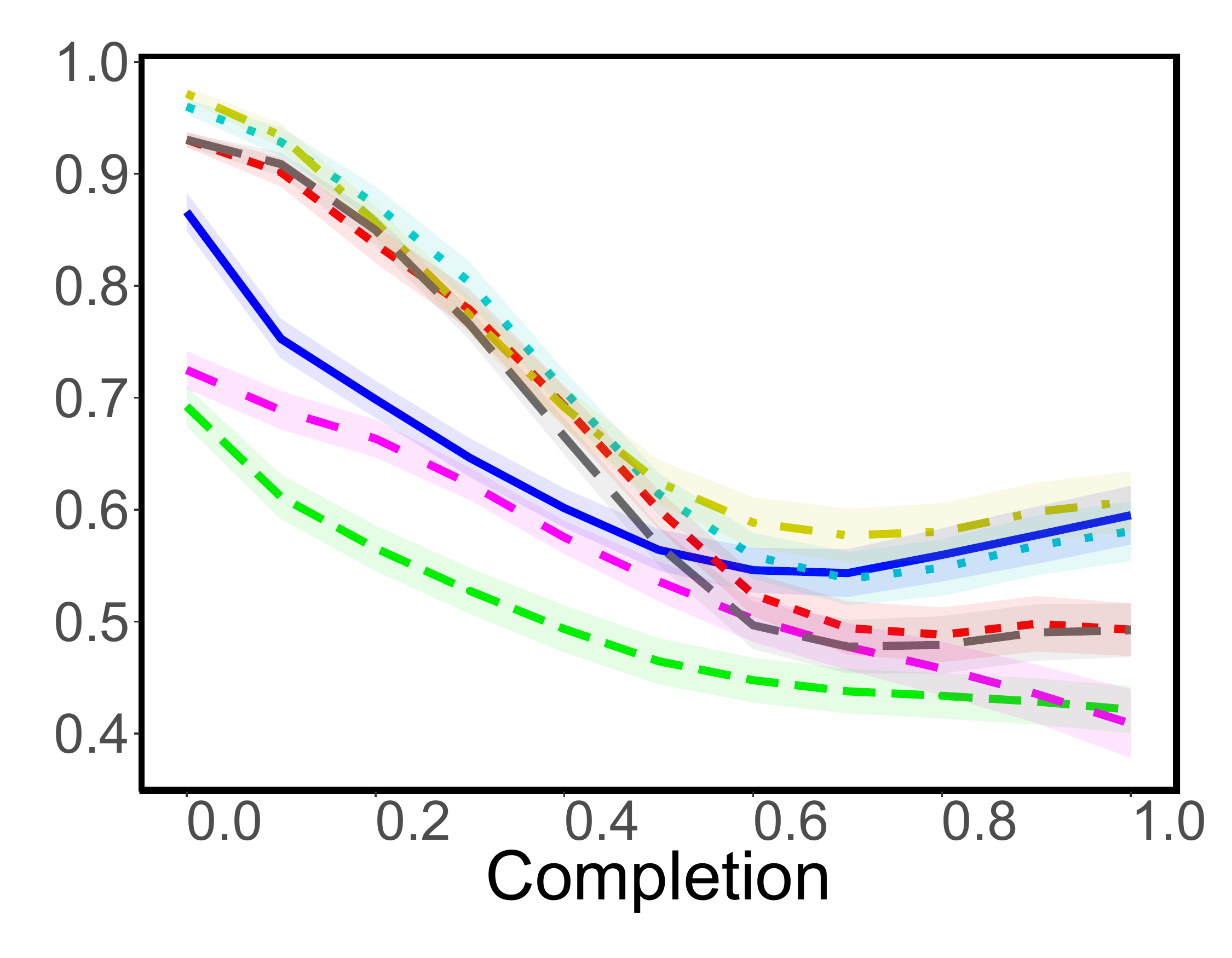} &
\includegraphics[width=1.03\linewidth]{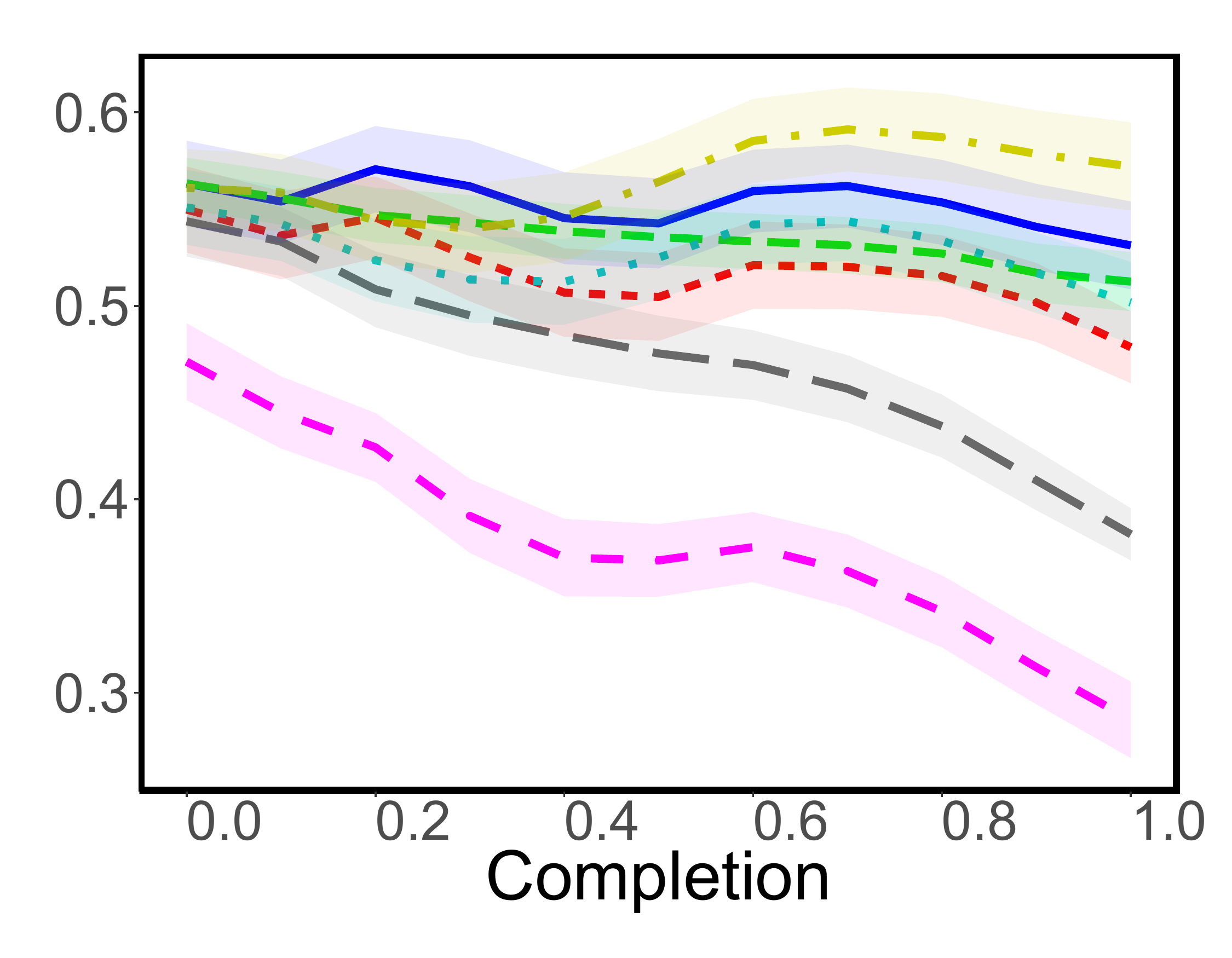} &
\includegraphics[width=1.03\linewidth]{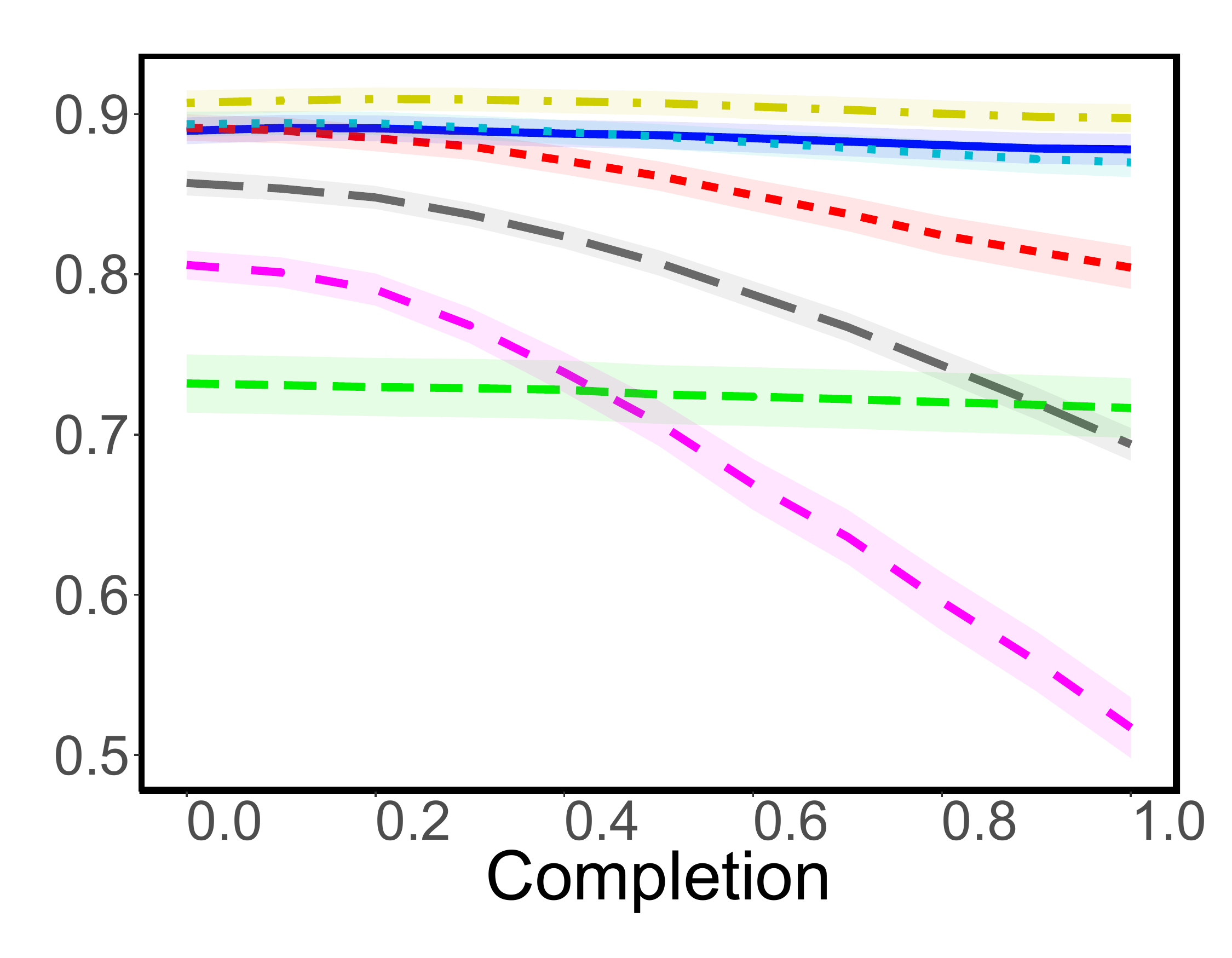}\\
\rotatebox{90}{\footnotesize $\ROC$ values for CTR} &
\includegraphics[width=1.03\linewidth]{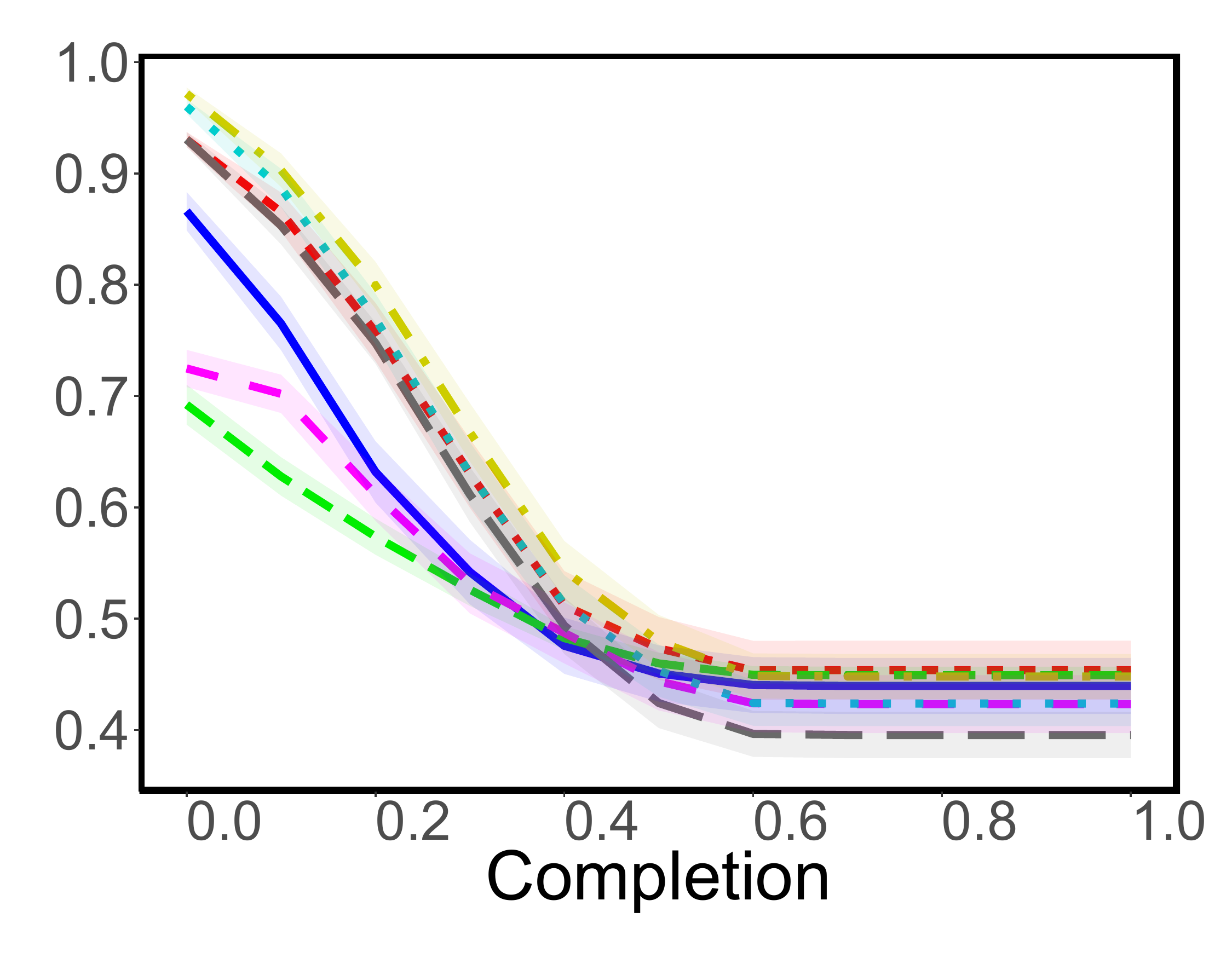} &
\includegraphics[width=1.03\linewidth]{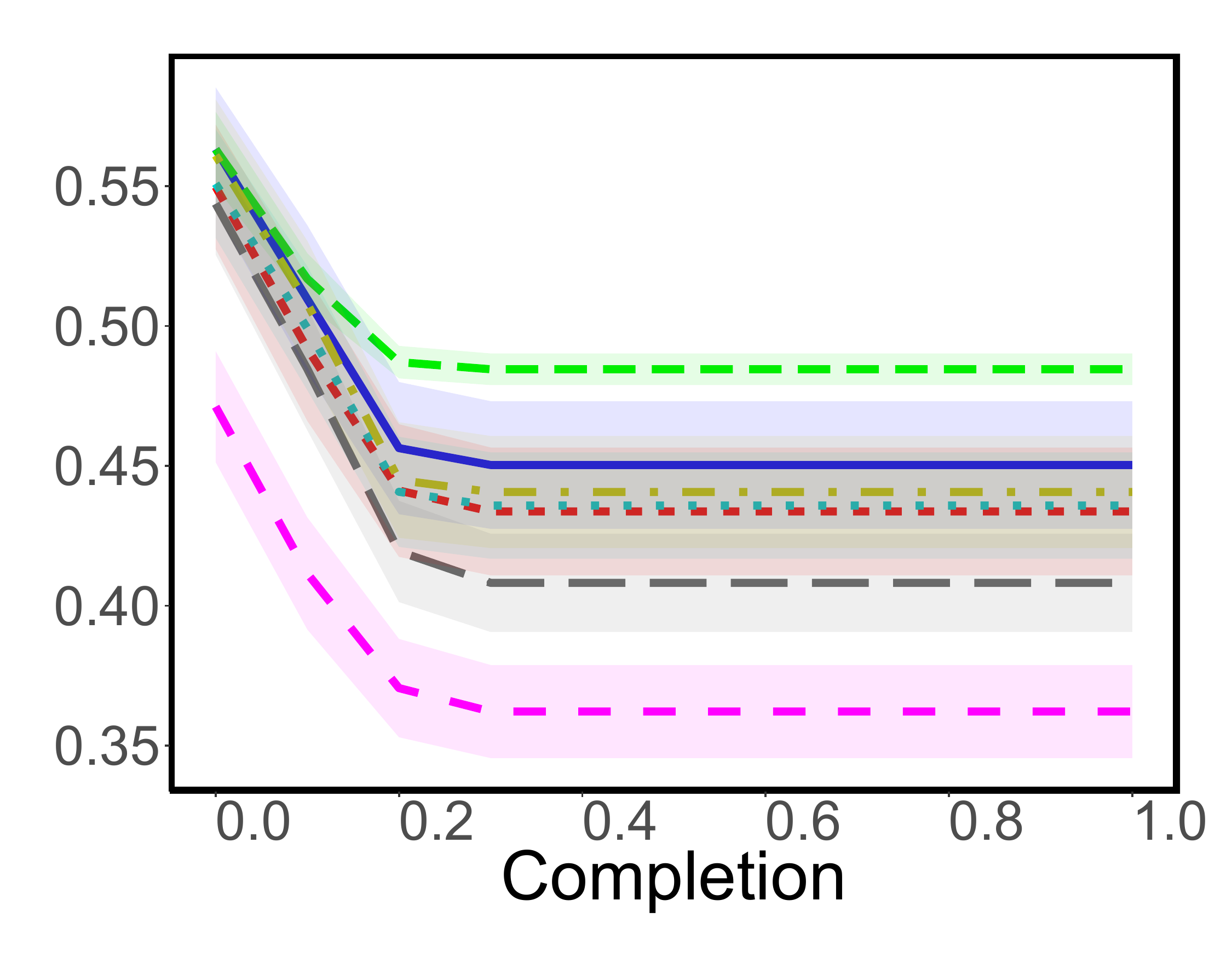} &
\includegraphics[width=1.03\linewidth]{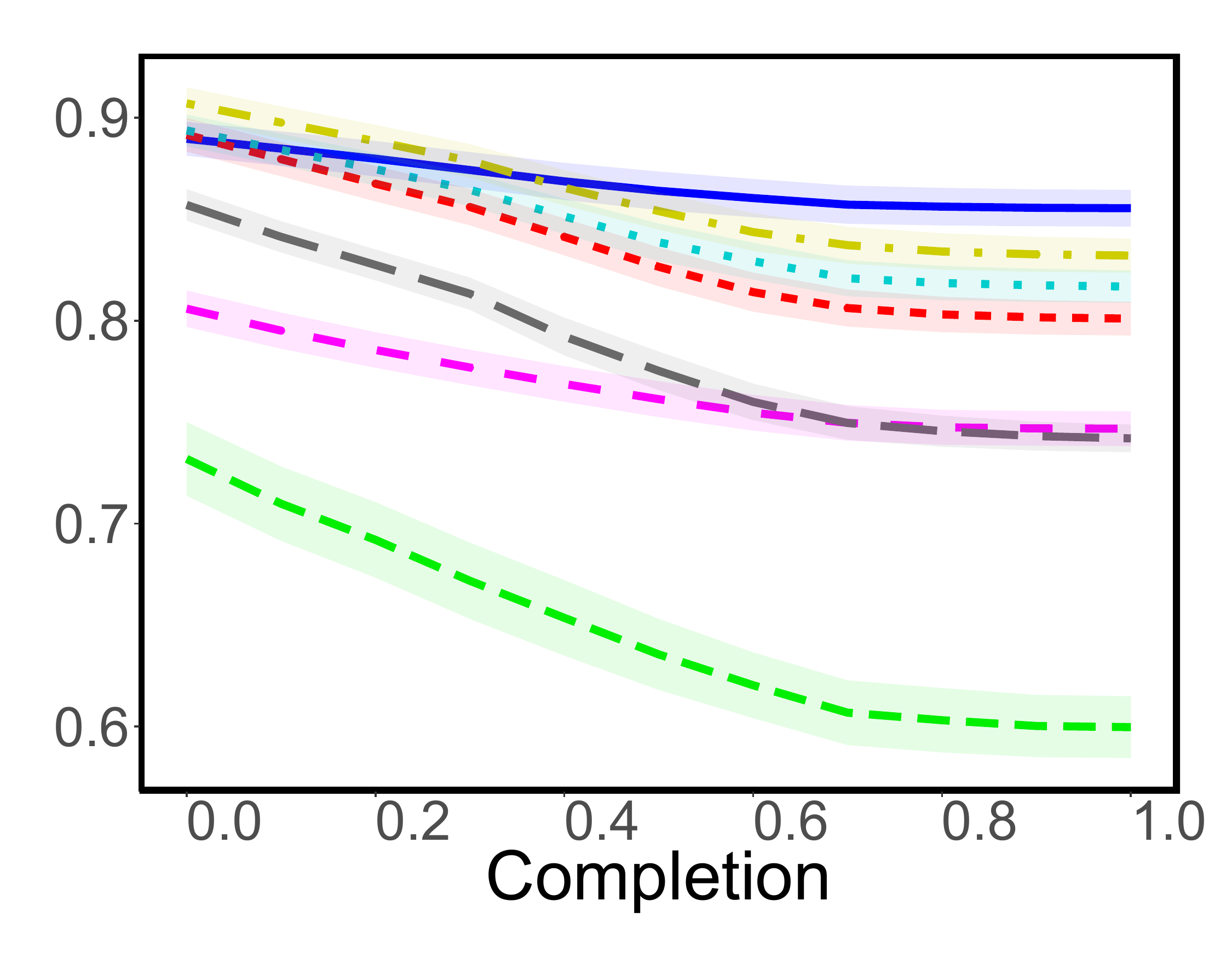} \\
\rotatebox{90}{\footnotesize $\AP$ values for OTC} &
\includegraphics[width=1.03\linewidth]{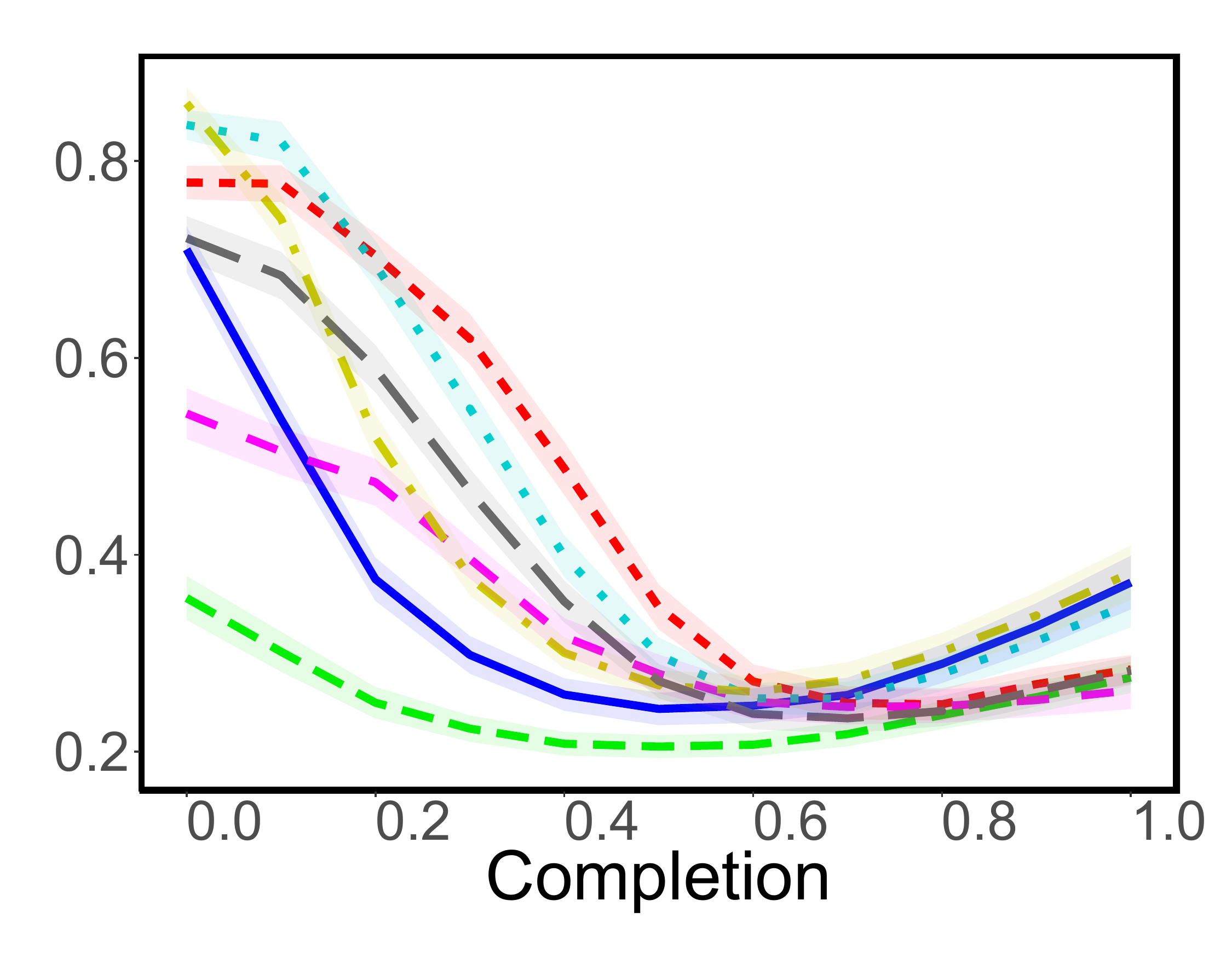} &
\includegraphics[width=1.03\linewidth]{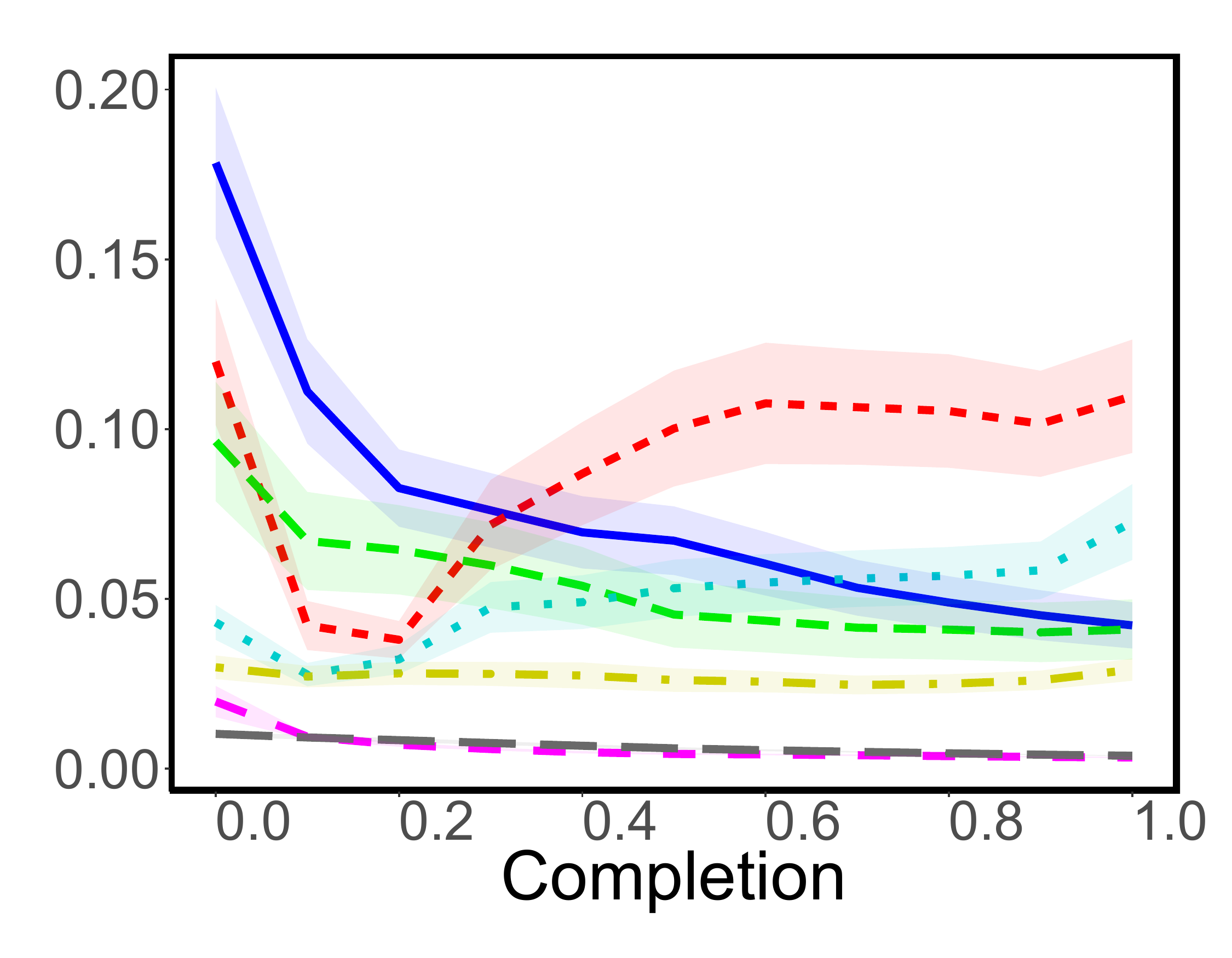} &
\includegraphics[width=1.03\linewidth]{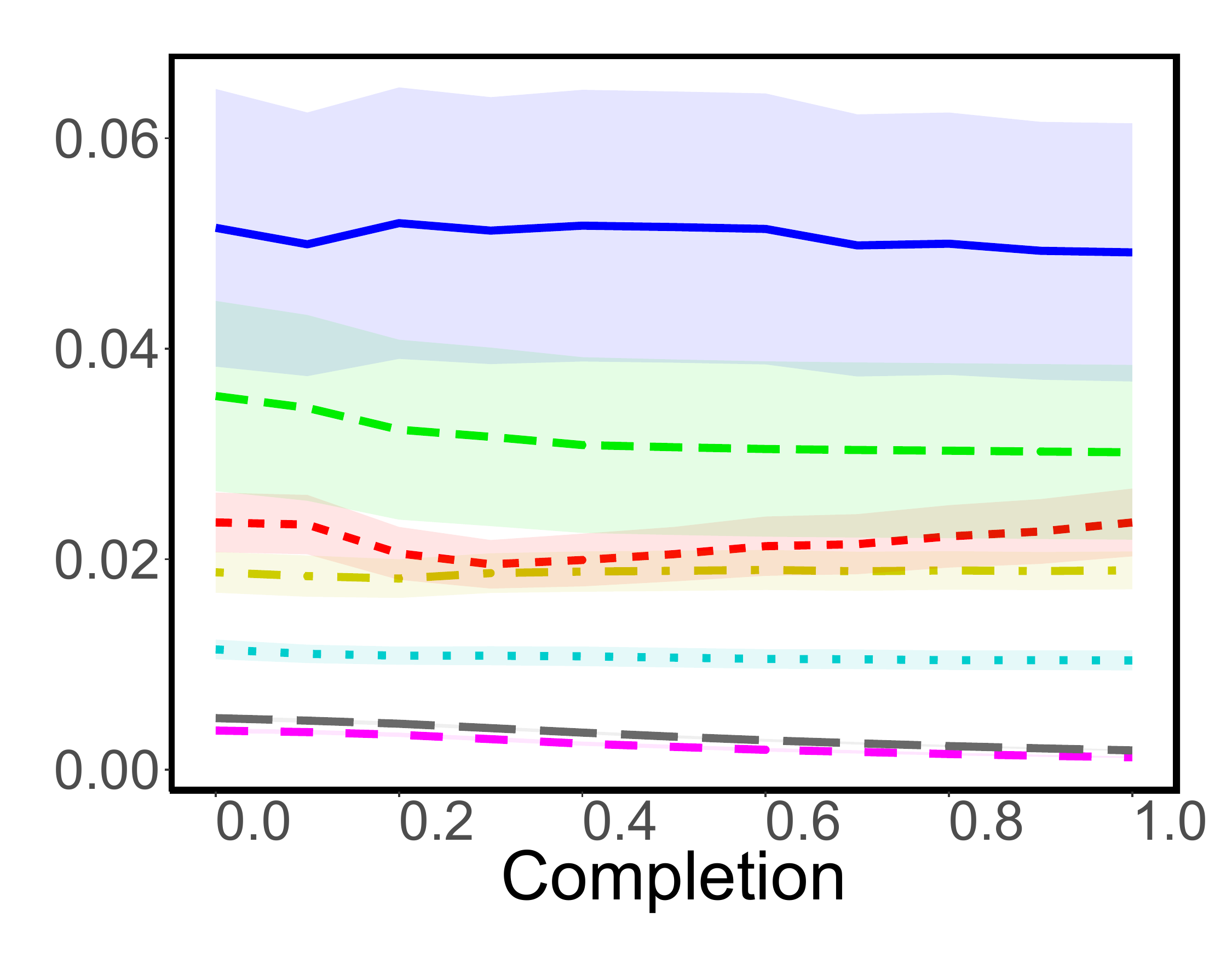} \\
\rotatebox{90}{\footnotesize $\AP$ values for CTR} &
\includegraphics[width=1.03\linewidth]{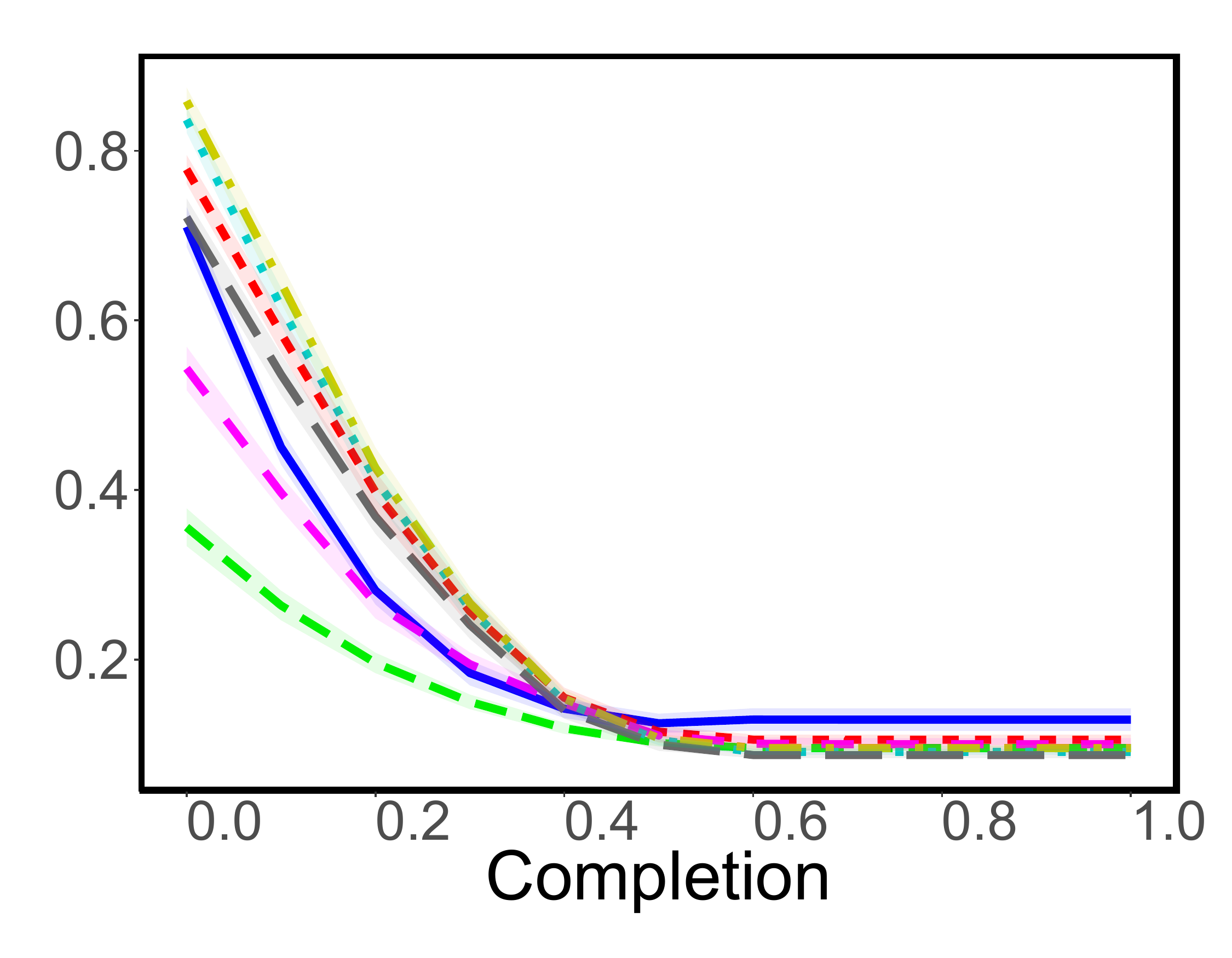} &
\includegraphics[width=1.03\linewidth]{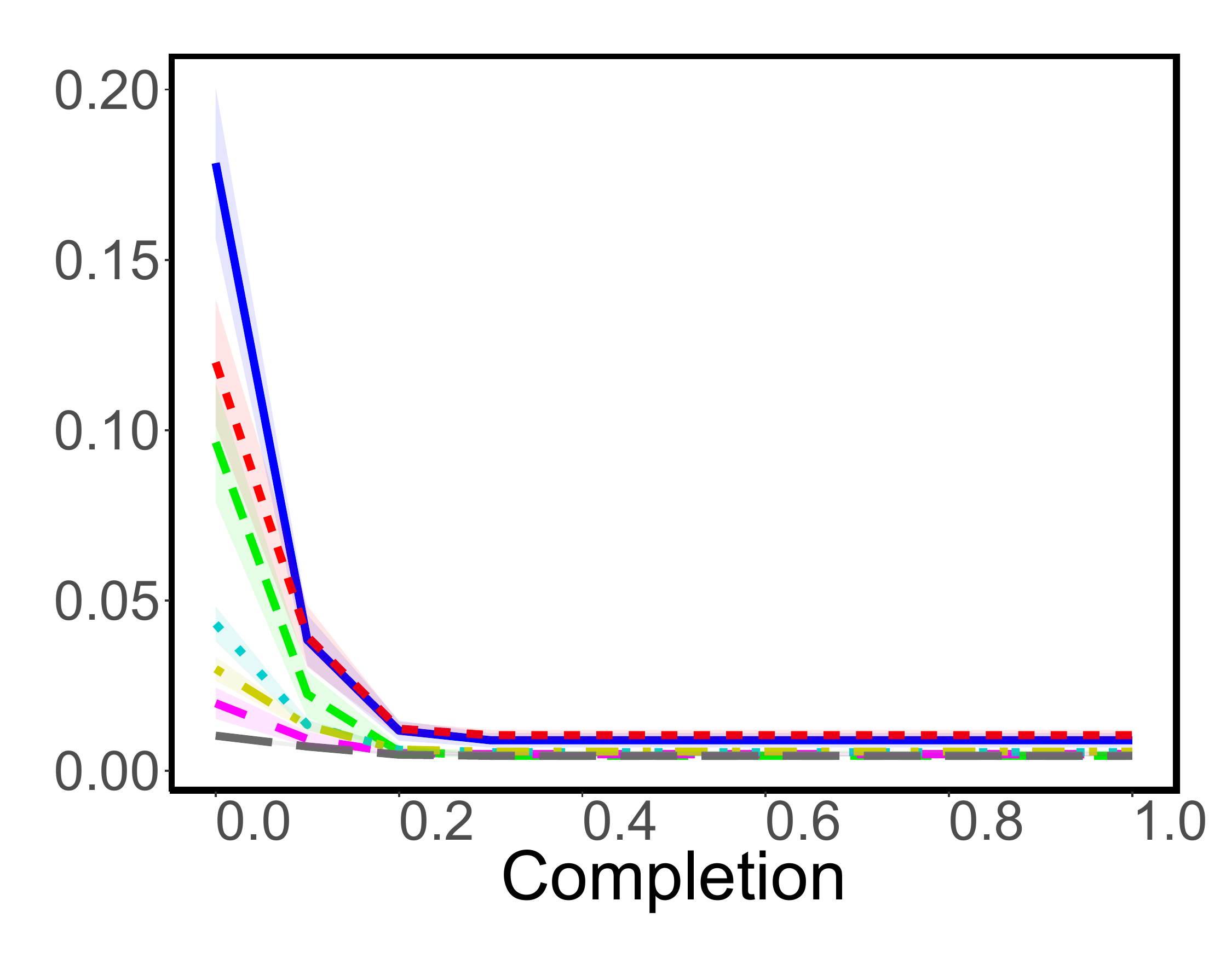} &
\includegraphics[width=1.03\linewidth]{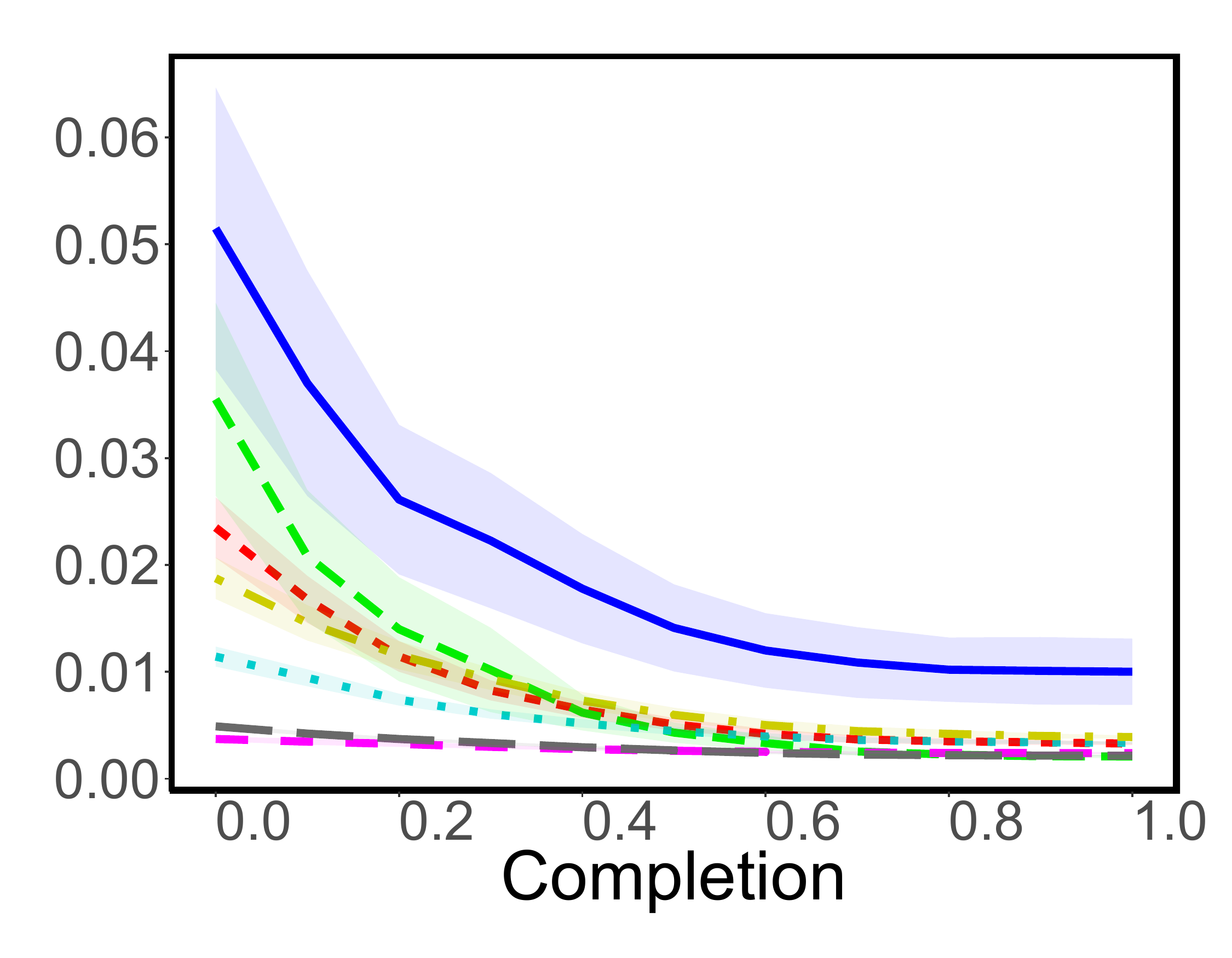} \\
\multicolumn{4}{c}{\includegraphics[width=0.65\linewidth]{figures/plots/global/legend}}
\end{tabular}
\caption{Given different \textbf{global similarity} indices, the figure depicts the values of $\ROC$ (the area under the ROC curve) and $\AP$ (the average precision) during the execution of OTC and CTR given $|\Hide|=\max(10,|E|/100)$ and $b=4|\Hide|$ in three networks: (i) \textbf{the Bali-attack network}; (ii) \textbf{the Madrid-bombing network}; and (iii) \textbf{the Greek political blogs network}.
In each execution, the links in $\Hide$ are chosen at random. Results are taken as the average over $50$ executions, with coloured areas representing the $95\%$ confidence intervals.}
\label{fig:global-5}
\end{figure*}

Next, we summarize the results for \textit{local} similarity indices given all of these networks (Figure~\ref{fig:bars-relative-local-supplementary}), before detailing the result for each network separately (Figures \ref{fig:local-0} to \ref{fig:local-4}). After that, we do the same but for \textit{global} similarity indices, i.e., we start by summarizing the results given all of the above networks (Figure~\ref{fig:bars-relative-global-supplementary}), and then detail the result for each network separately (Figures~\ref{fig:global-1} to \ref{fig:global-5}).  


\clearpage
\subsection{Evaluating CTR and OTC Against Local Link Prediction Algorithms}\label{sec:supplementary:local:evaluation}

\clearpage
\subsection{Evaluating CTR and OTC Against Global Link Prediction Algorithms}\label{sec:supplementary:global:evaluation}

\clearpage
\subsection{A Practical Telecommunication Scenario}\label{sec:supplementary:Telecommunication}

\begin{figure}[ht!]
\centering
\setlength\tabcolsep{1pt}
\renewcommand{\arraystretch}{0.01}
\begin{tabular}{m{.03\linewidth}m{.33\linewidth}m{.33\linewidth}m{.33\linewidth}}
&
\multicolumn{1}{c}{OTC\vspace*{0.1cm}} &
\multicolumn{1}{c}{OTC \& CTR\vspace*{0.1cm}} &
\multicolumn{1}{c}{CTR\vspace*{0.1cm}} \\
\rotatebox{90}{$\ROC$ value} &
\includegraphics[width=0.95\linewidth]{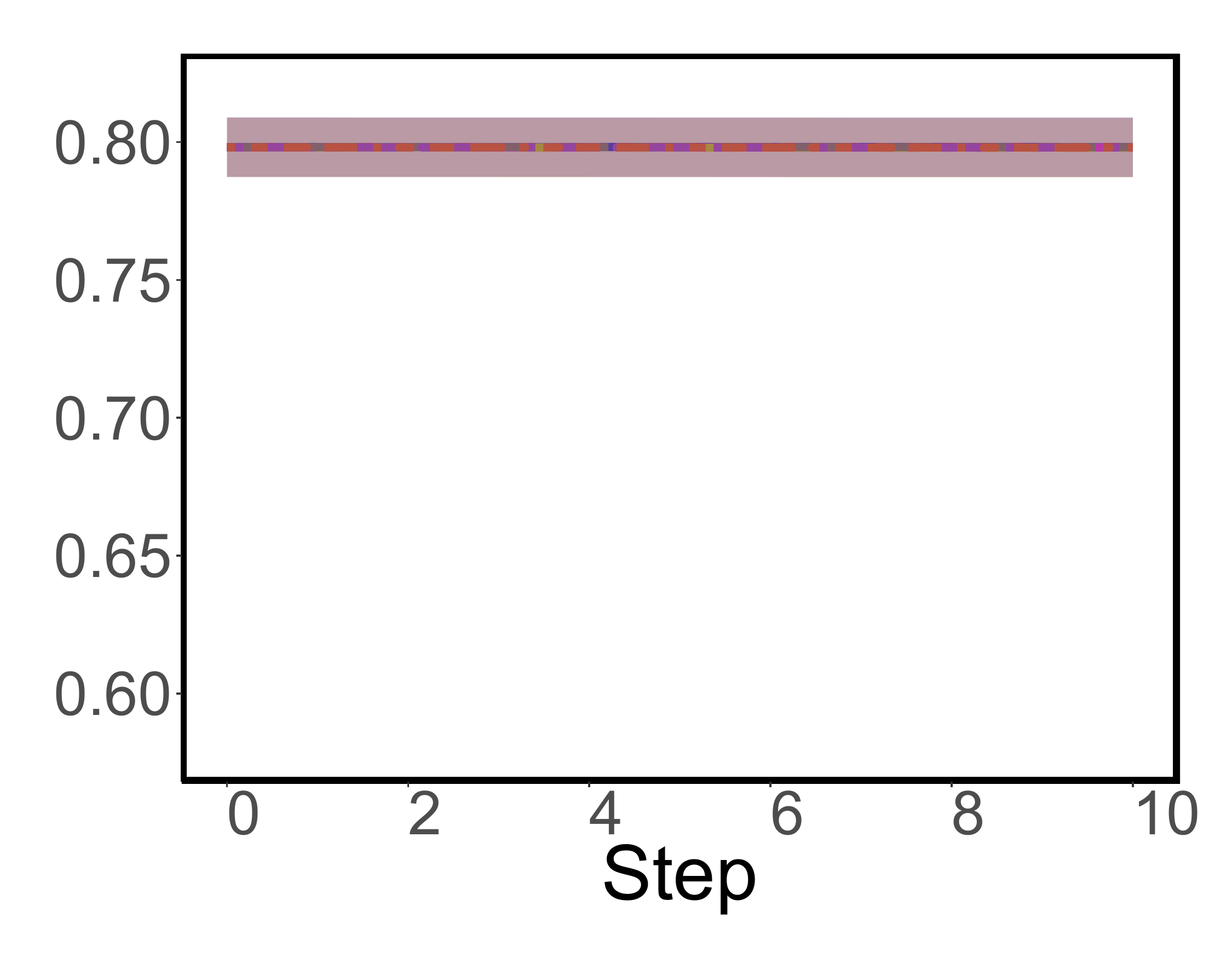} &
\includegraphics[width=0.95\linewidth]{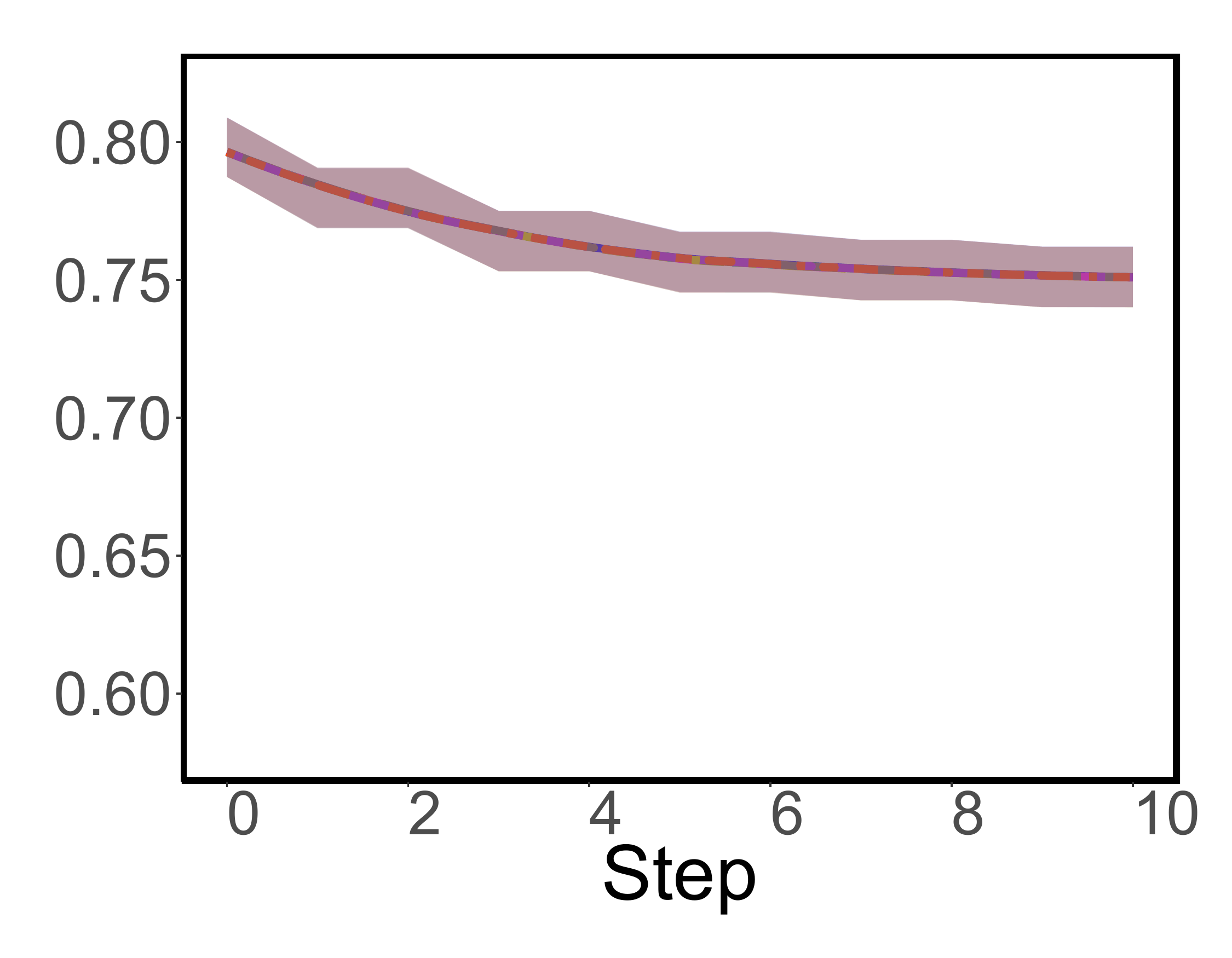} &
\includegraphics[width=0.95\linewidth]{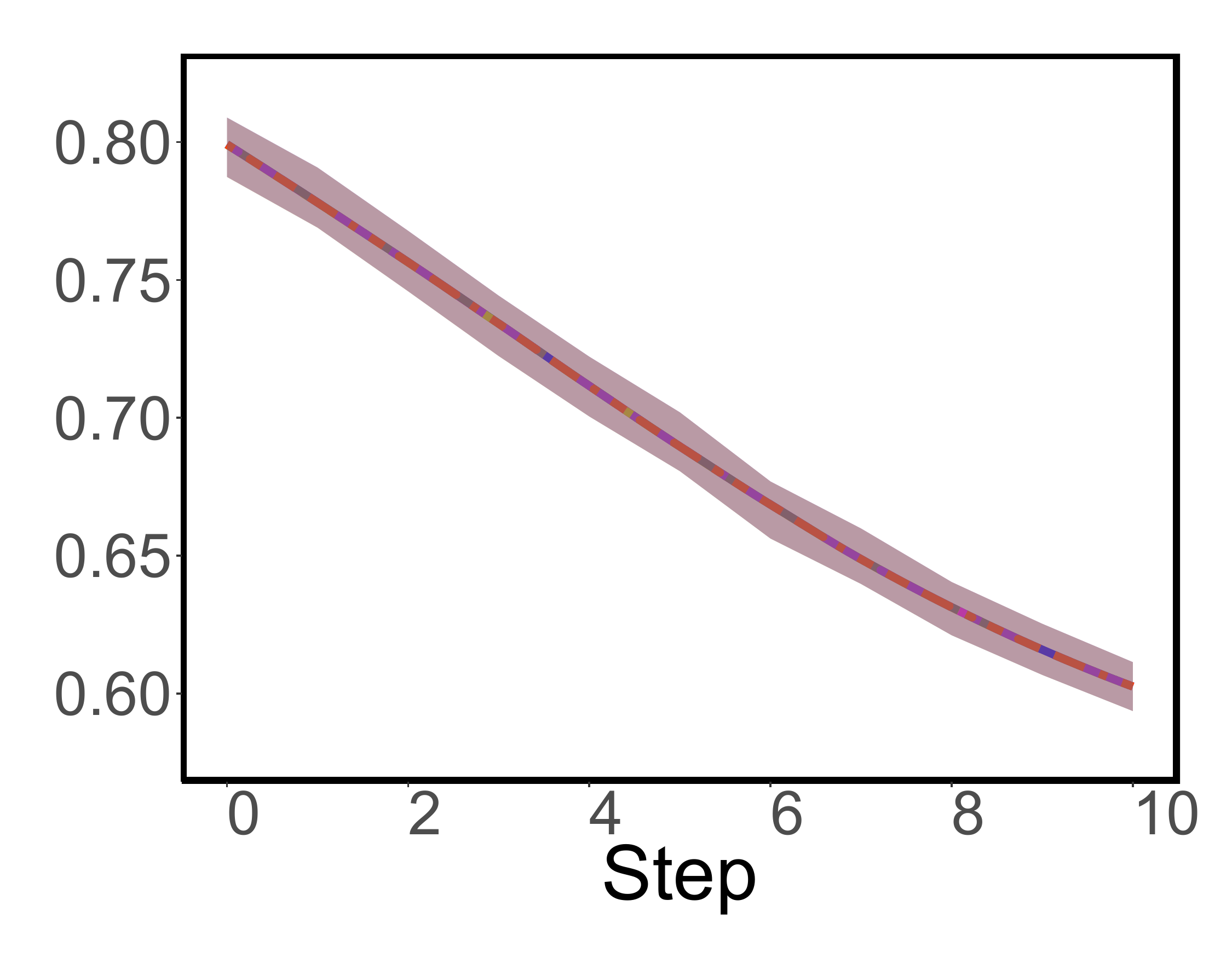} \\
\rotatebox{90}{$\AP$ value} &
\includegraphics[width=0.95\linewidth]{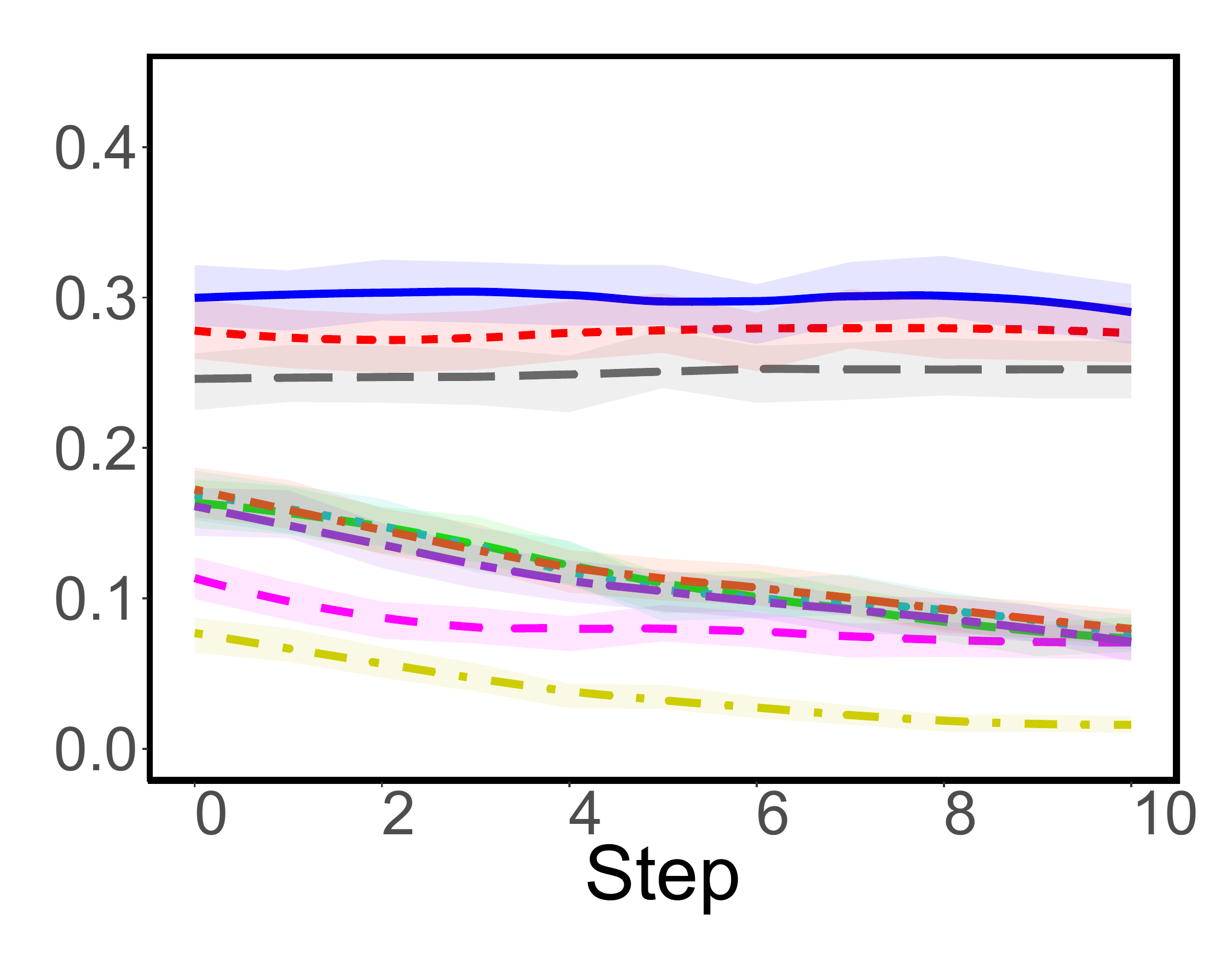} &
\includegraphics[width=0.95\linewidth]{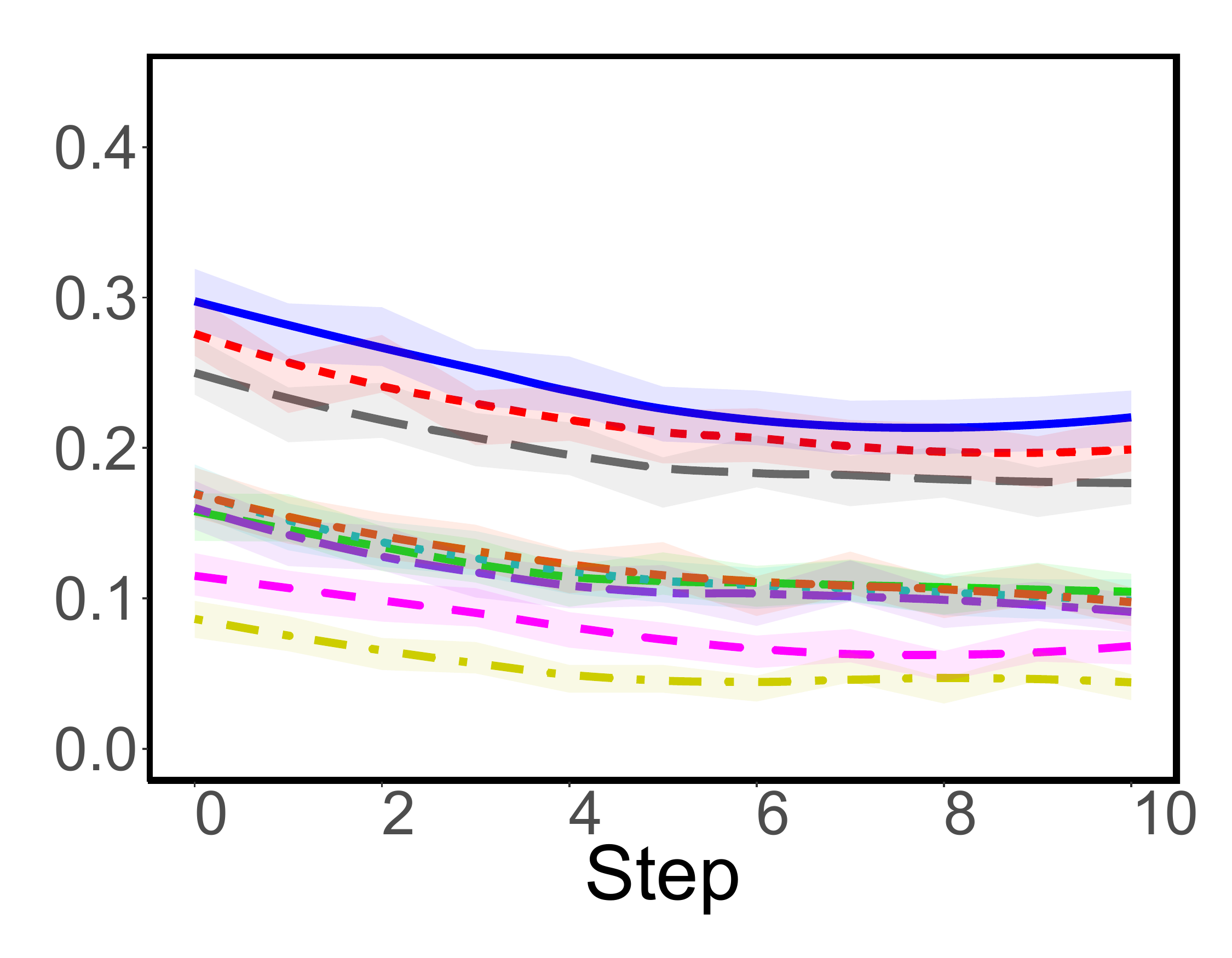} &
\includegraphics[width=0.95\linewidth]{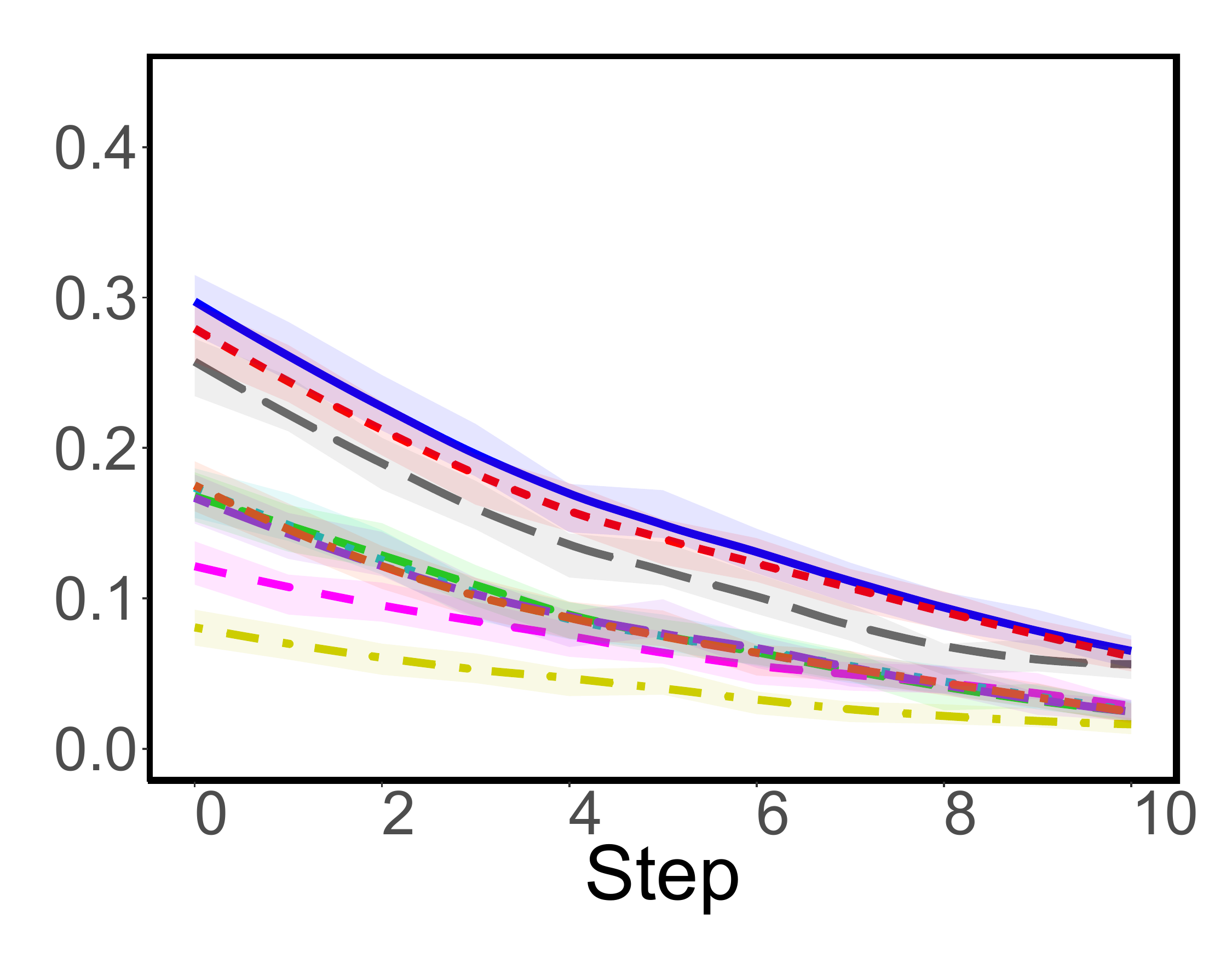} \\
\multicolumn{4}{c}{\includegraphics[width=0.8\linewidth]{figures/plots/single/legend}}
\end{tabular}
\caption{Given different local similarity indices, and a telecommunication network consisting of $56,073$ nodes and $174,608$ edges, the figure depicts the average $\ROC$ and $\AP$ during the execution of OTC and CTR given a budget $b=10$, where $H$ contains just a single link to be hidden. More specifically, for each similarity index, we consider the $1,000$ highest-ranked links, and for each such link, $(v,u)$, we run the heuristic once where the evader is $v$ and another where the evader is $u$. This entire process is repeated 10 times, and the average results are reported with the coloured areas representing the $95\%$ confidence intervals.
}
\label{fig:single-telecommunication-small}
\end{figure}

\noindent In the main article, we evaluated OCT and CTR given a budget of 10 and a single link to hide in a telecommunication network consisting of $248,763$ nodes and $829,725$ edges. The nodes of that network corresponded to all the users of a particular service provider---Telef\'onica Spain---who live in four geographically continuous districts in the UK, and the links corresponded to all the calls between those users (see Figure~5 in the main article). In this section, we consider a smaller telecommunication network consisting of $56,073$ nodes and $174,608$ links, where the nodes correspond to all users living in just a single district in the UK, and the links correspond to all the calls between those users. The results depicted in Figure~\ref{fig:single-telecommunication-small} exhibit similar trends to those presented in Figure~5, i.e., CTR is effective in terms of both $\AP$ and $\ROC$, while OTC is less effective in terms of $\AP$ and not effective at all in terms of $\ROC$; mixing the two heuristics does not seem to produce any synergistic effects.

\clearpage

\section{Evaluating the Runtime of CTR and OTC}\label{sec:runtime}

\begin{figure}[tbht!]
\centering
\setlength\tabcolsep{1pt}
\renewcommand{\arraystretch}{0.01}
\begin{tabular}{m{.5\textwidth}m{.5\textwidth}}
&\\
  \multicolumn{1}{c}{{\fontsize{12}{12}\selectfont{\ \ \ OTC\smallskip}}}
& \multicolumn{1}{c}{{\fontsize{12}{12}\selectfont{\ \ \ \ \ \ CTR\smallskip}}}\\
\includegraphics[width=0.95\linewidth]{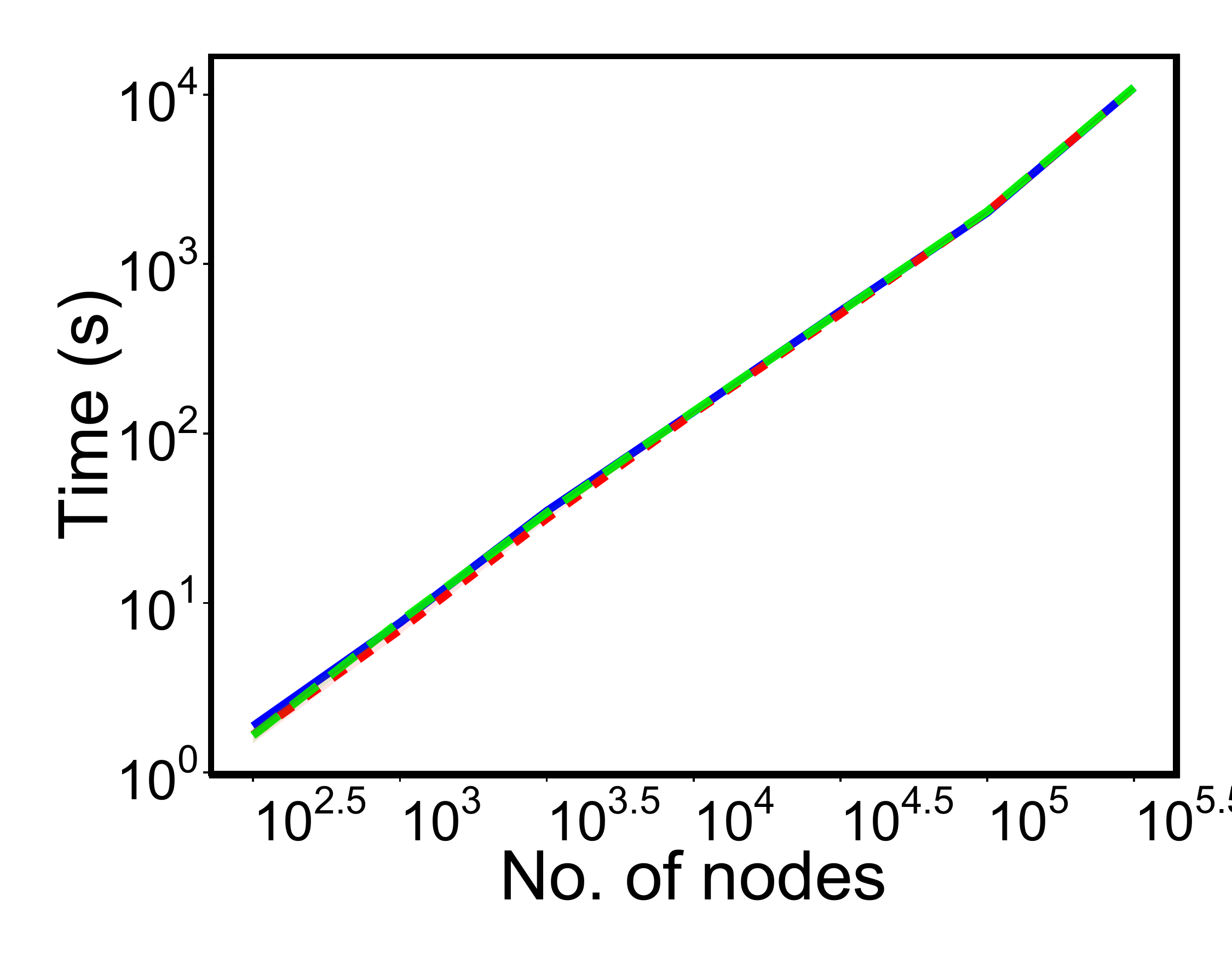} &
\includegraphics[width=0.95\linewidth]{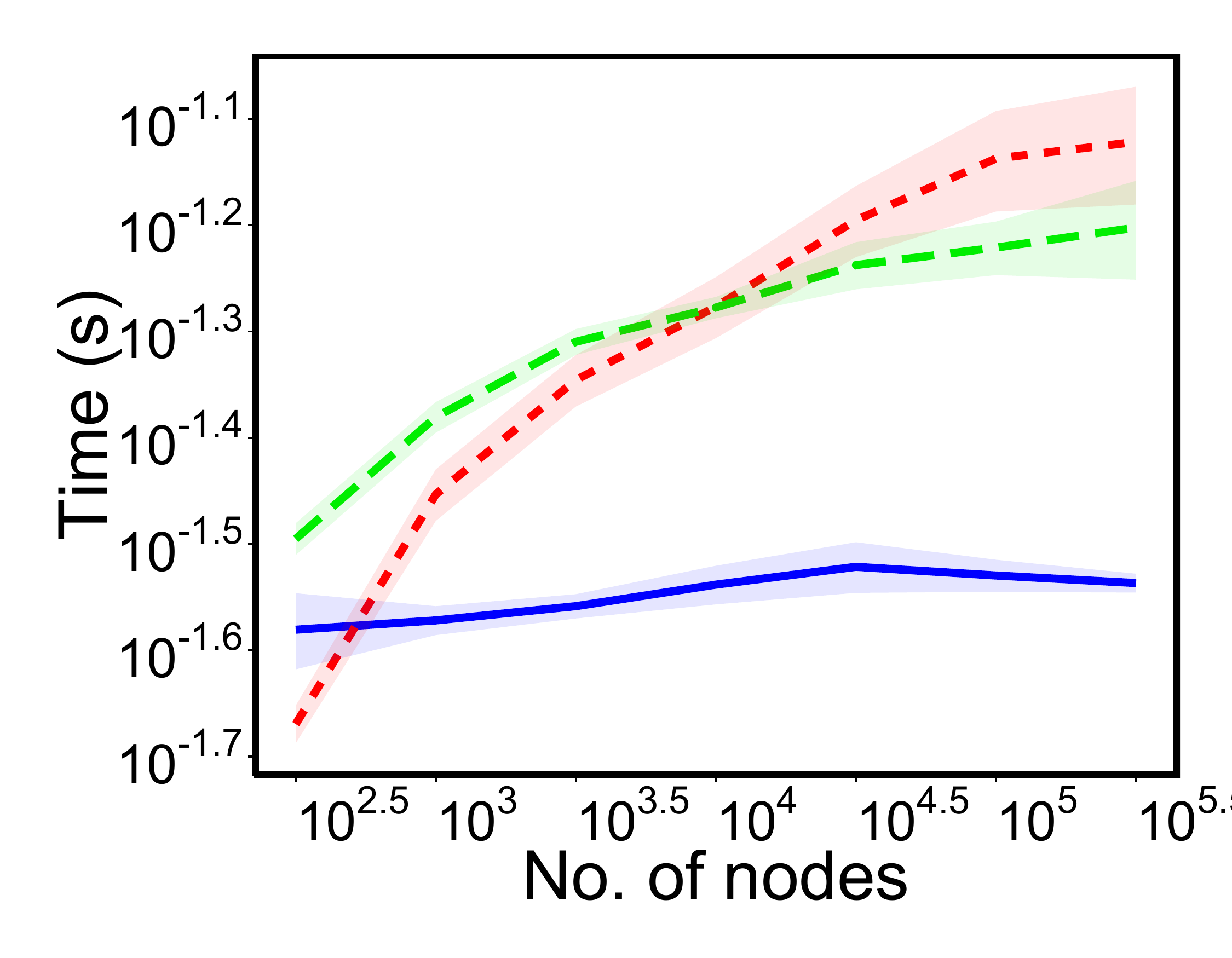} \\
\multicolumn{2}{c}{\includegraphics[width=0.65\linewidth]{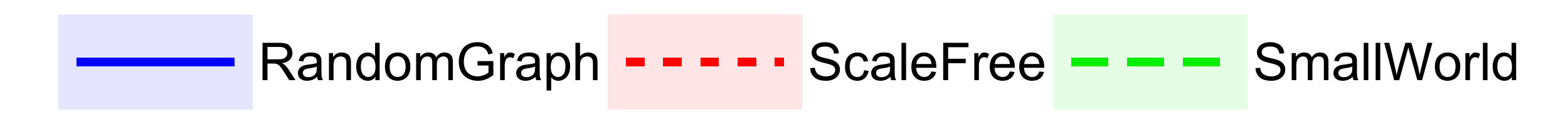}}
\end{tabular}
\caption{The average runtime (in seconds) of OTC and CTR given $|\Hide|=100$ and $b=4|\Hide|$, and given three types of networks: (i) ScaleFree$(n,3)$; (ii) SmallWorld$(n,10,0.25)$; and (iii) RandomGraph$(n,10)$, with $n$ varying from $100$ to $100,000$.}
\label{fig:runtime}
\end{figure}

\noindent In this section, we empirically evaluate how the runtime of OTC and CTR increases with the size of the network.\footnote{\footnotesize Runtime was measured on a modern-day PC, with an Intel Xeon E5-2697 v2 and 16GB DDR3 RAM.} To this end, we considered three standard types of randomly-generated networks, namely (i) Scale-free networks, (ii) Small-world networks, and (iii) Erdos-Renyi networks. For each of these networks, we varied the number of nodes from $100$ to $100,000$, and measured the runtime of OTC and CTR given $100$ edges to hide, and given a budget of $400$. Figure~\ref{fig:runtime} depicts the average runtime taken over $50$ experiments, with the shaded areas representing $95\%$ confidence intervals. As can be seen, CTR is significantly faster than OTC. In fact, the runtime of CTR did not exceed 1 millisecond even when the number of nodes reached $100,000$, regardless of the network-generation model. This shows that CTR is applicable on massive networks. On the other hand, the runtime of OTC increases much more rapidly, and almost reaches 3 hours when the number of nodes reaches $100,000$. The figure also shows that the runtime of OTC is almost independent of the network-generation model, unlike CTR.

\clearpage

\section{Evaluating the Attack Tolerance of Different Link-Prediction Algorithms}\label{sec:evaluatingAttackTolerance}

\begin{figure*}[ht!]
\centering
\setlength\tabcolsep{1pt}
\renewcommand{\arraystretch}{0.01}
\begin{tabular}{m{.03\linewidth}m{.24\linewidth}m{.24\linewidth}m{.24\linewidth}}
& \multicolumn{1}{c}{\footnotesize ScaleFree$(n,d)$}
& \multicolumn{1}{c}{\footnotesize SmallWorld$(n,d,0.25)$}
& \multicolumn{1}{c}{\footnotesize RandomGraph$(n,d)$}\\
\rotatebox{90}{\footnotesize OTC-$\ROC$} &
\includegraphics[width=\linewidth]{figures/plots/params/basic-ScaleFree-otc-roc-bysize} &
\includegraphics[width=\linewidth]{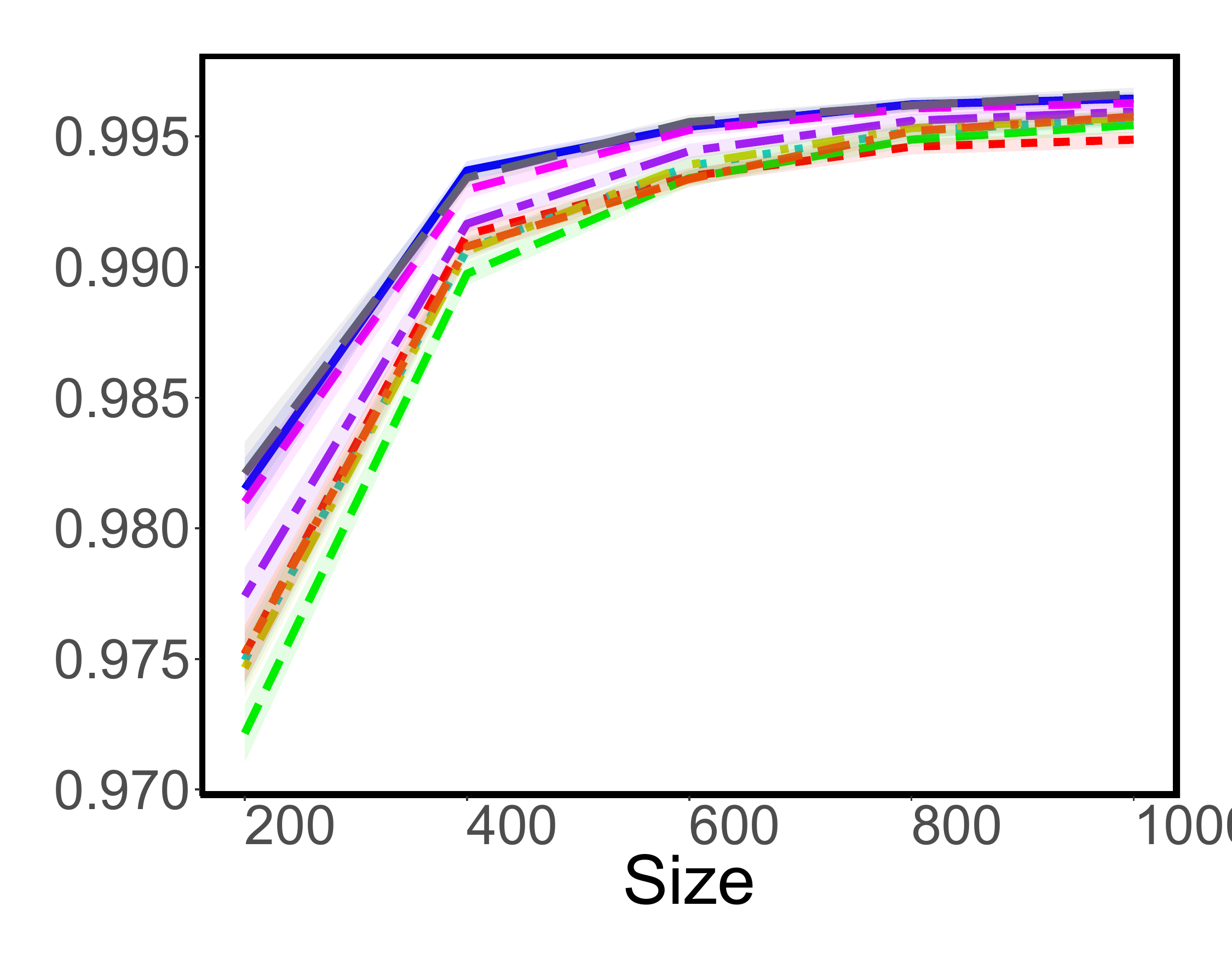} &
\includegraphics[width=\linewidth]{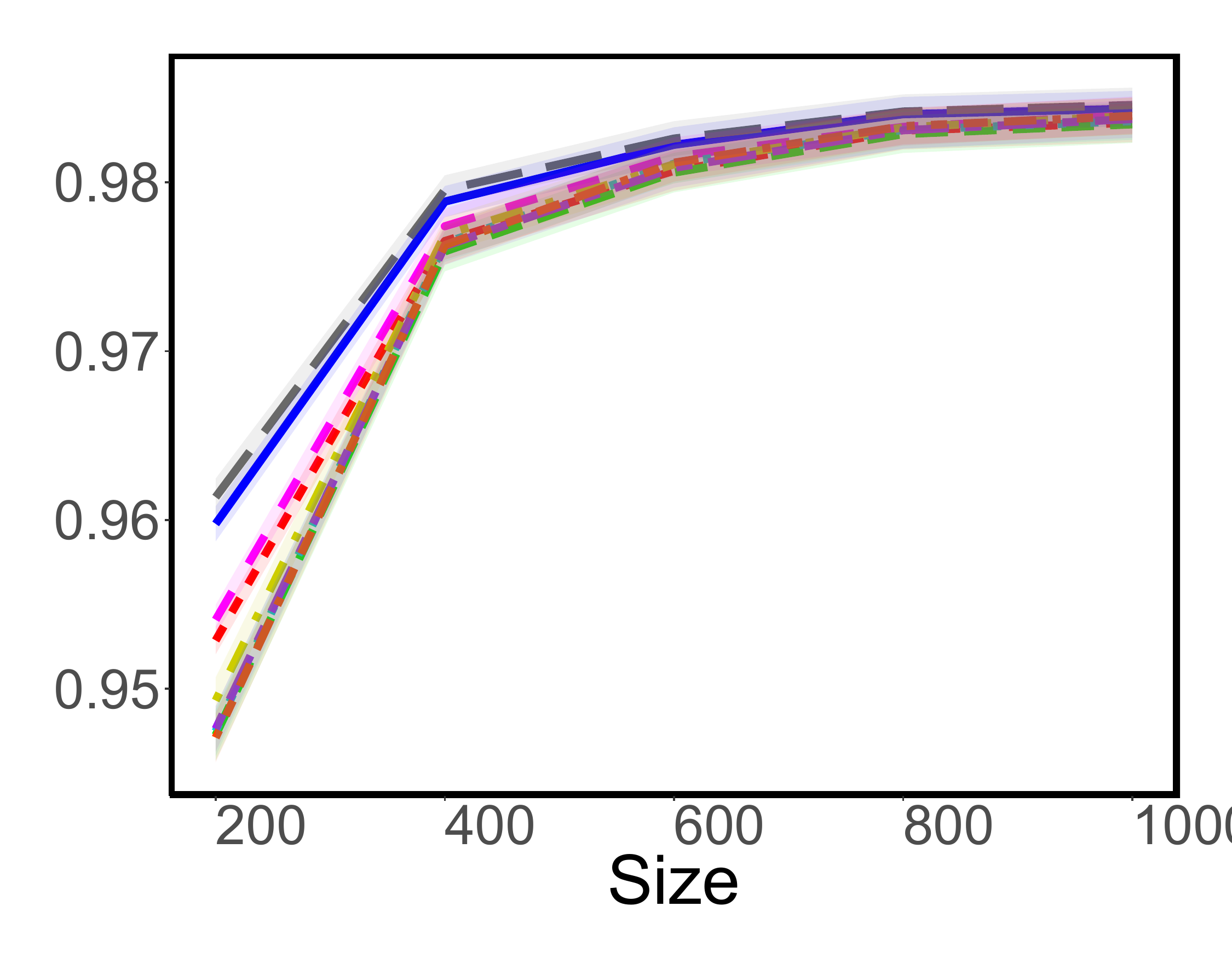} \\
\rotatebox{90}{\footnotesize CTR-$\ROC$} &
\includegraphics[width=\linewidth]{figures/plots/params/basic-ScaleFree-ctr-roc-bysize} &
\includegraphics[width=\linewidth]{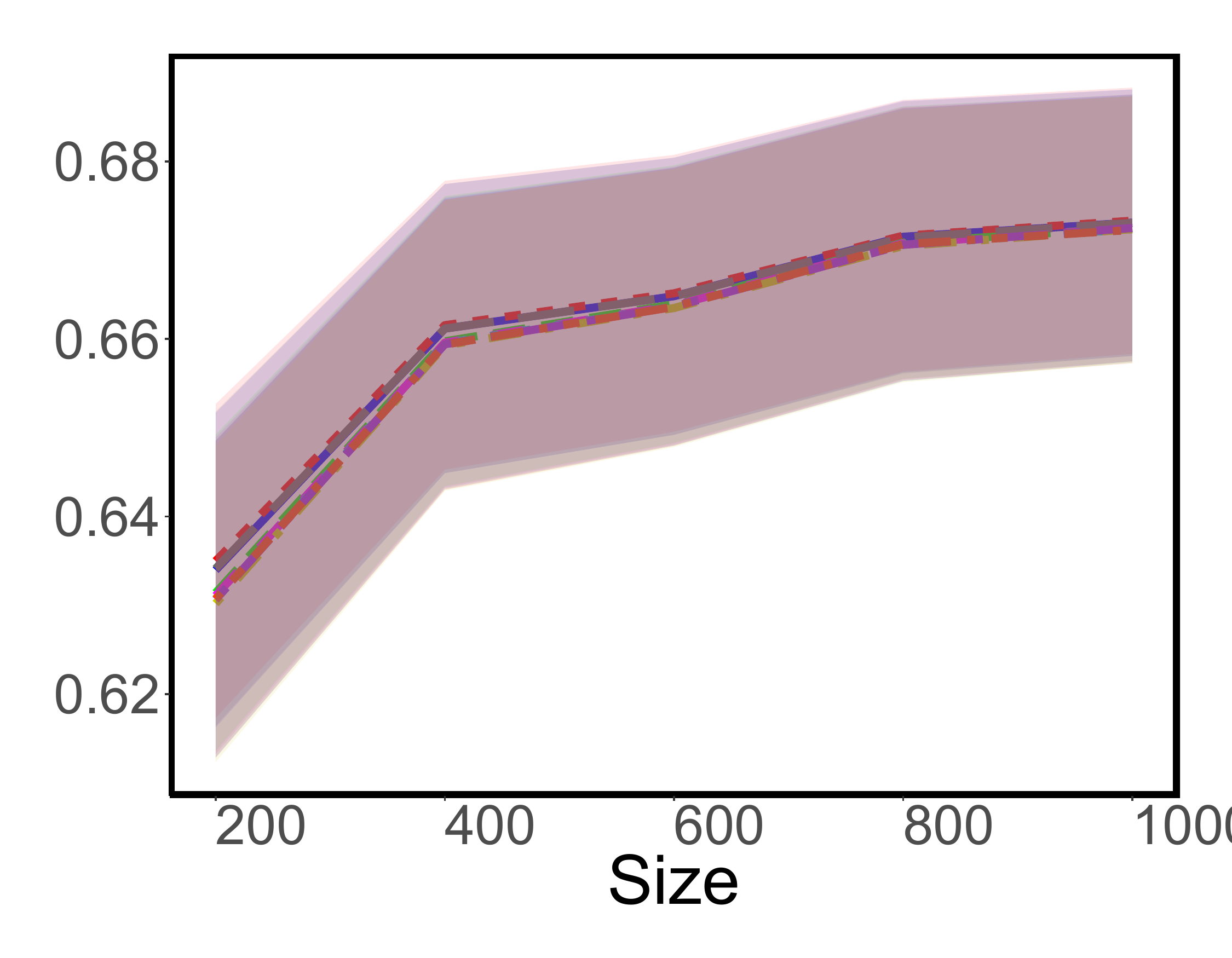} &
\includegraphics[width=\linewidth]{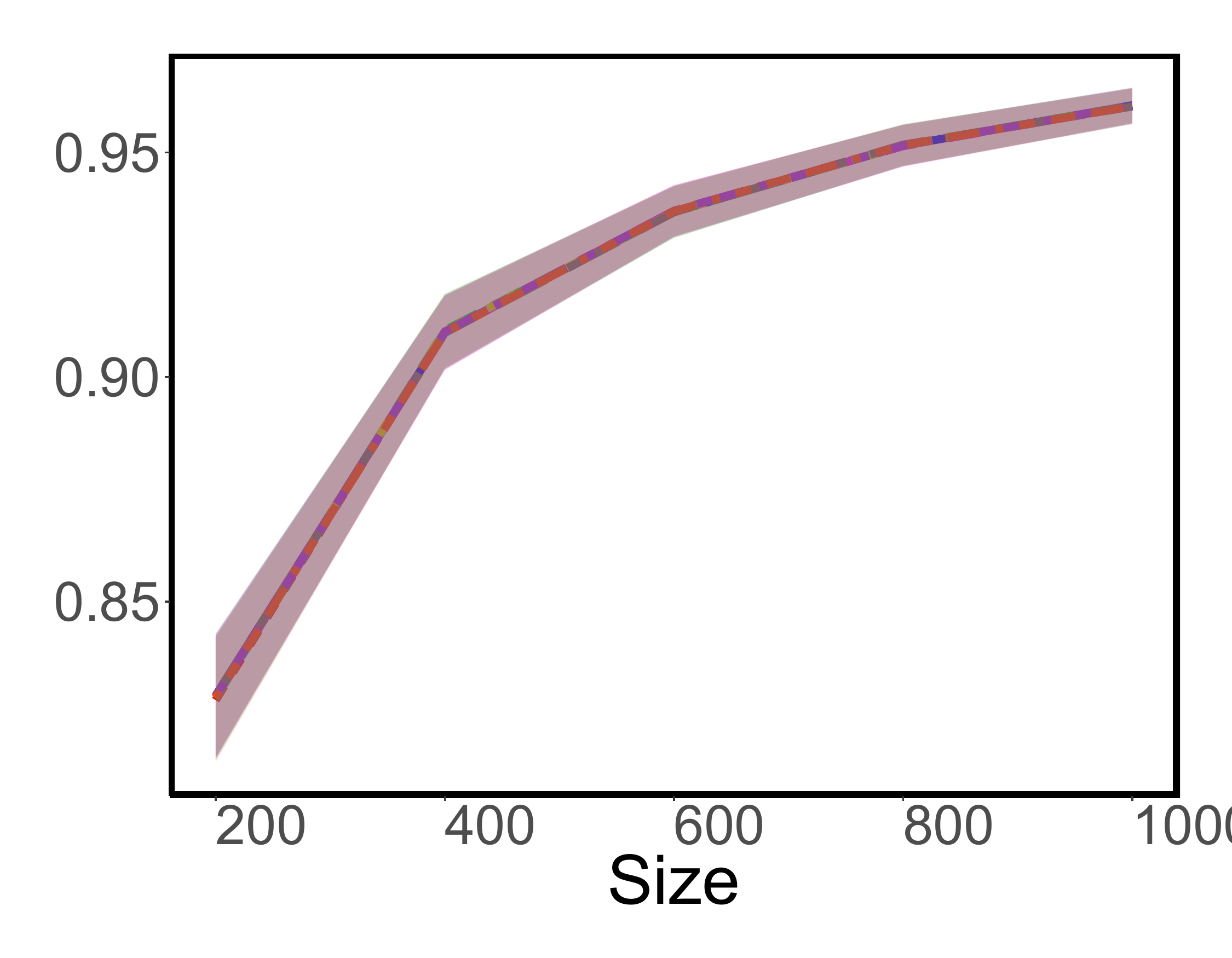} \\
\rotatebox{90}{\footnotesize OTC-$\AP$} &
\includegraphics[width=\linewidth]{figures/plots/params/basic-ScaleFree-otc-ap-bysize} &
\includegraphics[width=\linewidth]{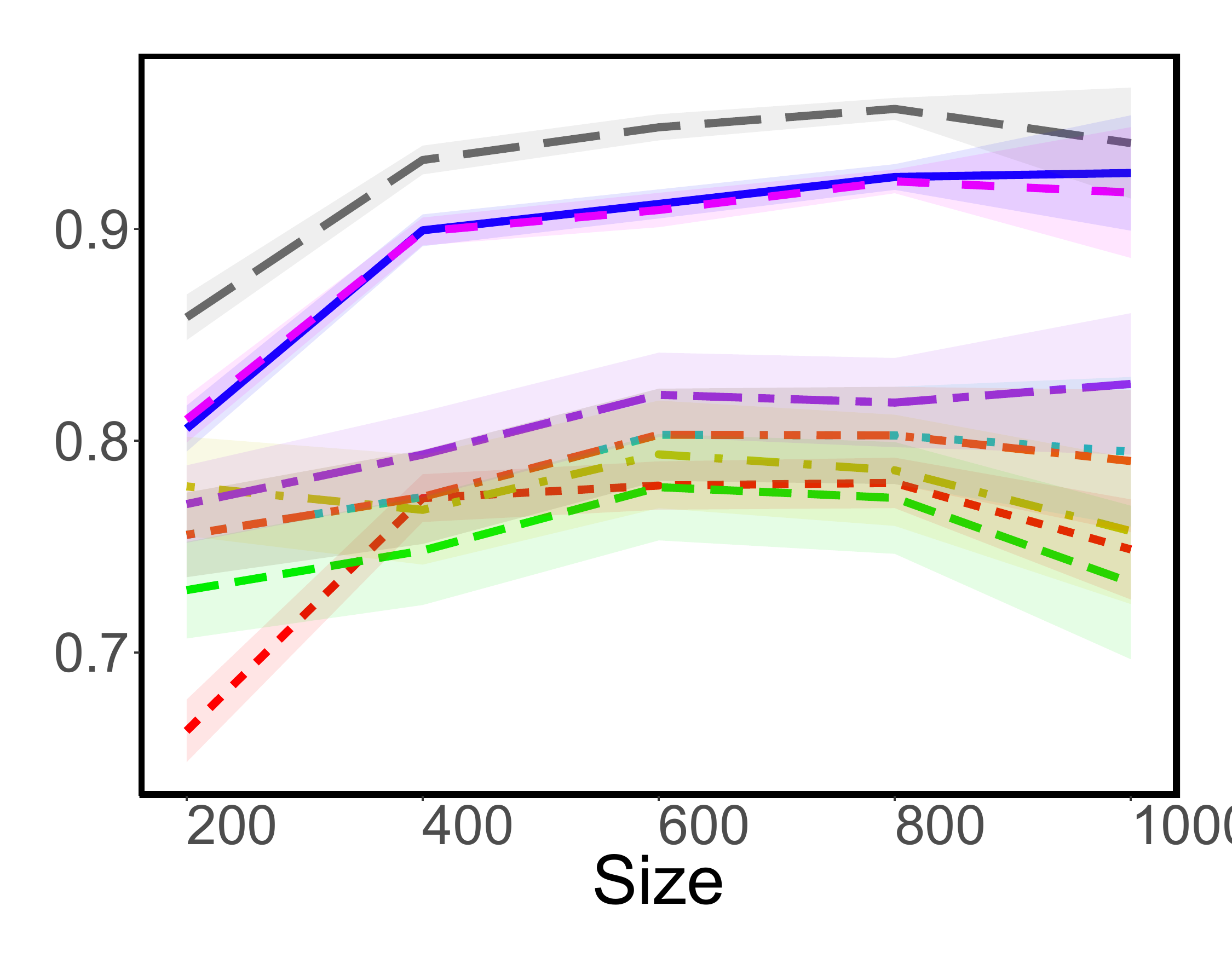} &
\includegraphics[width=\linewidth]{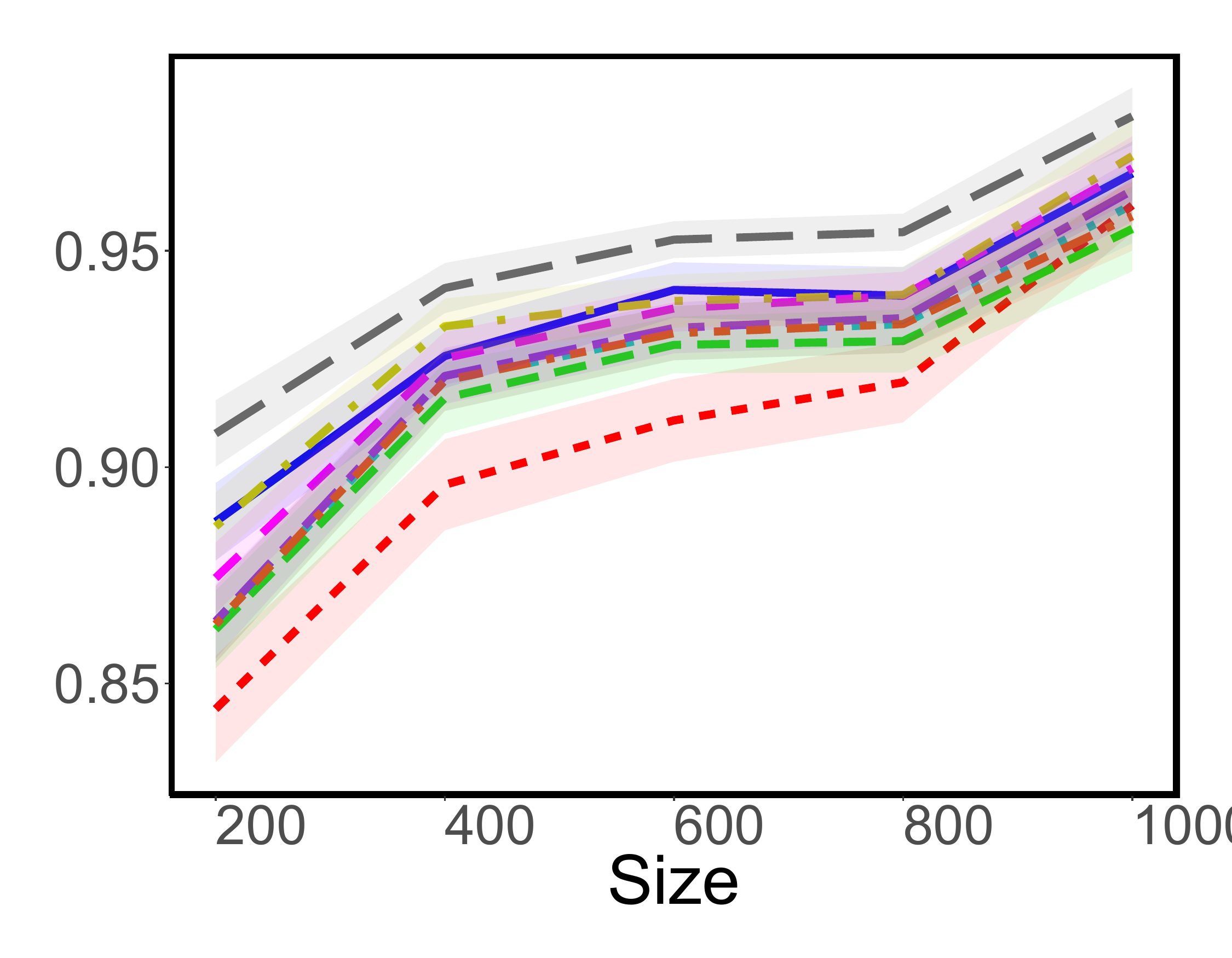} \\
\rotatebox{90}{\footnotesize CTR-$\AP$} &
\includegraphics[width=\linewidth]{figures/plots/params/basic-ScaleFree-ctr-ap-bysize} &
\includegraphics[width=\linewidth]{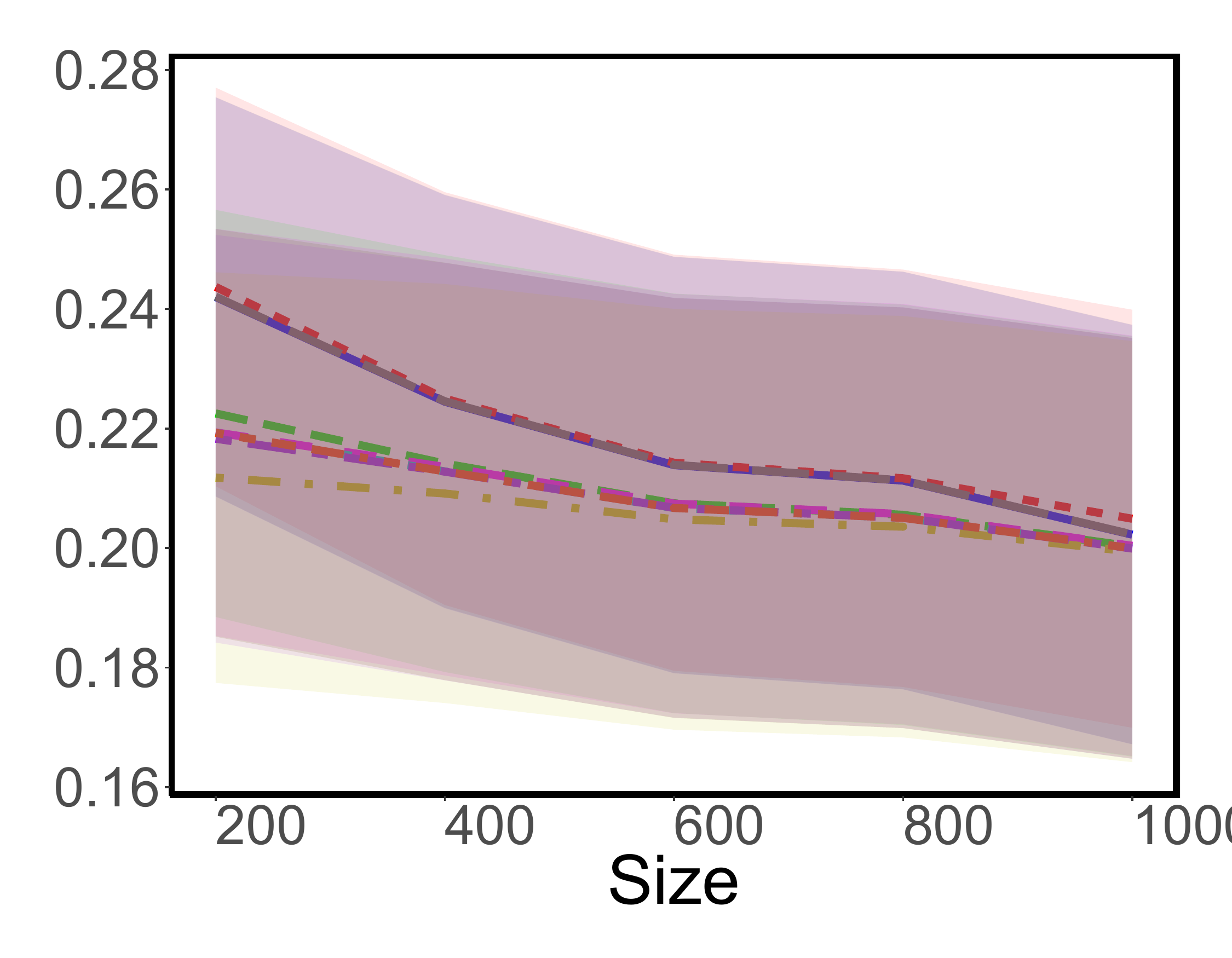} &
\includegraphics[width=\linewidth]{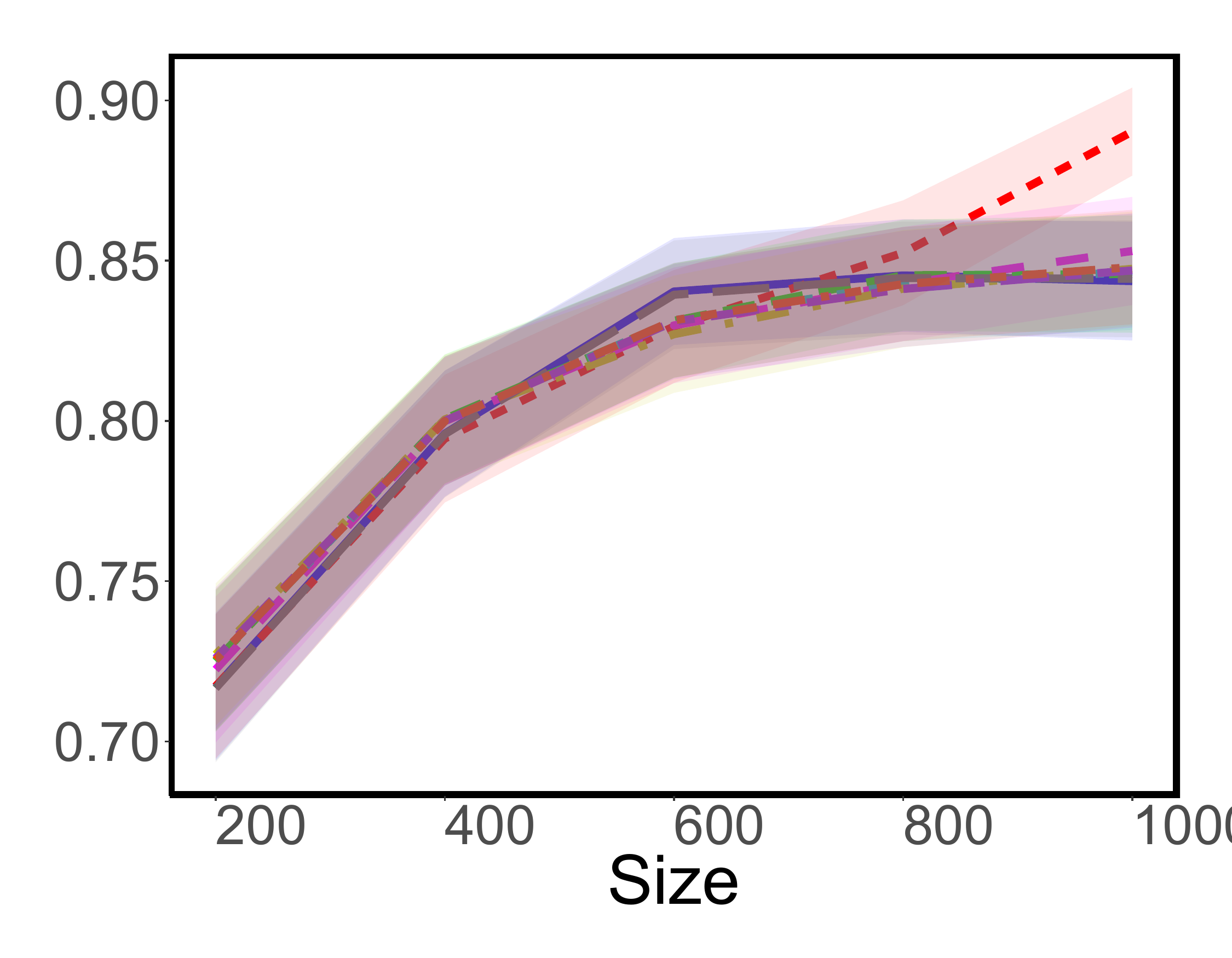} \\
\multicolumn{4}{c}{\includegraphics[width=0.6\linewidth]{figures/plots/params/basic-legend}}
\end{tabular}
\caption{Evaluating the attack tolerance of different \textbf{local similarity} indices against OTC (which adds edges) and CTR (which removes edges) by measuring the relative change in $\ROC$ (the area under the ROC curve) and $\AP$ (the average precision) while varying the \textbf{number of nodes}, $n$, in three types of networks: (i) ScaleFree$(n,d)$; (ii) SmallWorld$(n,d,0.25)$; and (iii) RandomGraph$(n,d)$. For each $n$, we report the average degree over $d=2,4,\ldots,10$. The links in $\Hide$ are chosen at random, where $|\Hide|=100$ and $b=4|\Hide|$. The entire experiment is repeated $50$ times and the average is reported with coloured areas representing the $95\%$ confidence intervals.}
\label{fig:bysize:supplementary}
\end{figure*}

\begin{figure*}[tbhp]
\centering
\setlength\tabcolsep{1pt}
\renewcommand{\arraystretch}{0.01}
\begin{tabular}{m{.03\linewidth}m{.25\linewidth}m{.25\linewidth}m{.25\linewidth}}
& \multicolumn{1}{c}{\footnotesize ScaleFree$(n,d)$}
& \multicolumn{1}{c}{\footnotesize SmallWorld$(n,d,0.25)$}
& \multicolumn{1}{c}{\footnotesize RandomGraph$(n,d)$}\\
\rotatebox{90}{\footnotesize OTC-$\ROC$} &
\includegraphics[width=\linewidth]{figures/plots/params/basic-ScaleFree-otc-roc-bydegree} &
\includegraphics[width=\linewidth]{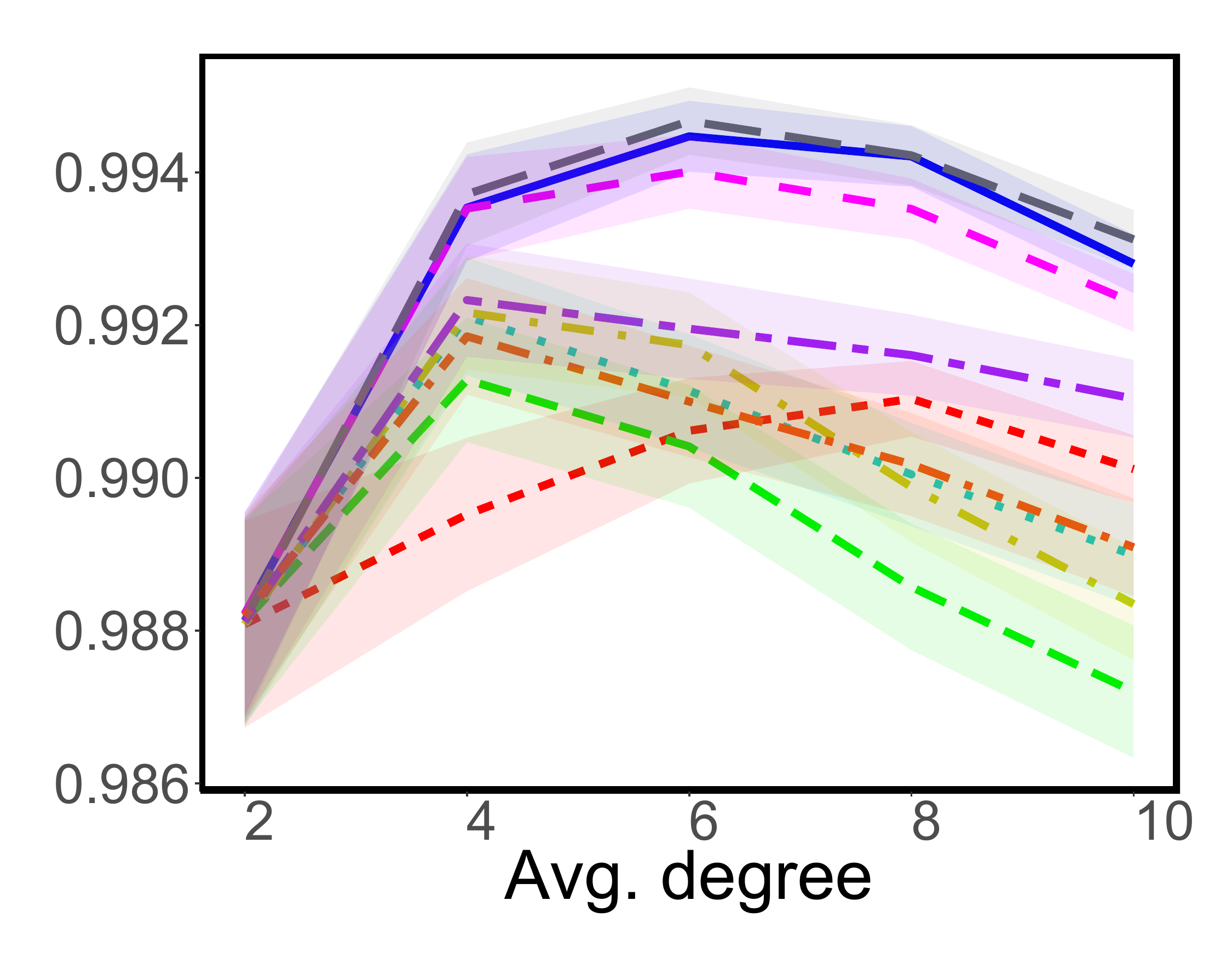} &
\includegraphics[width=\linewidth]{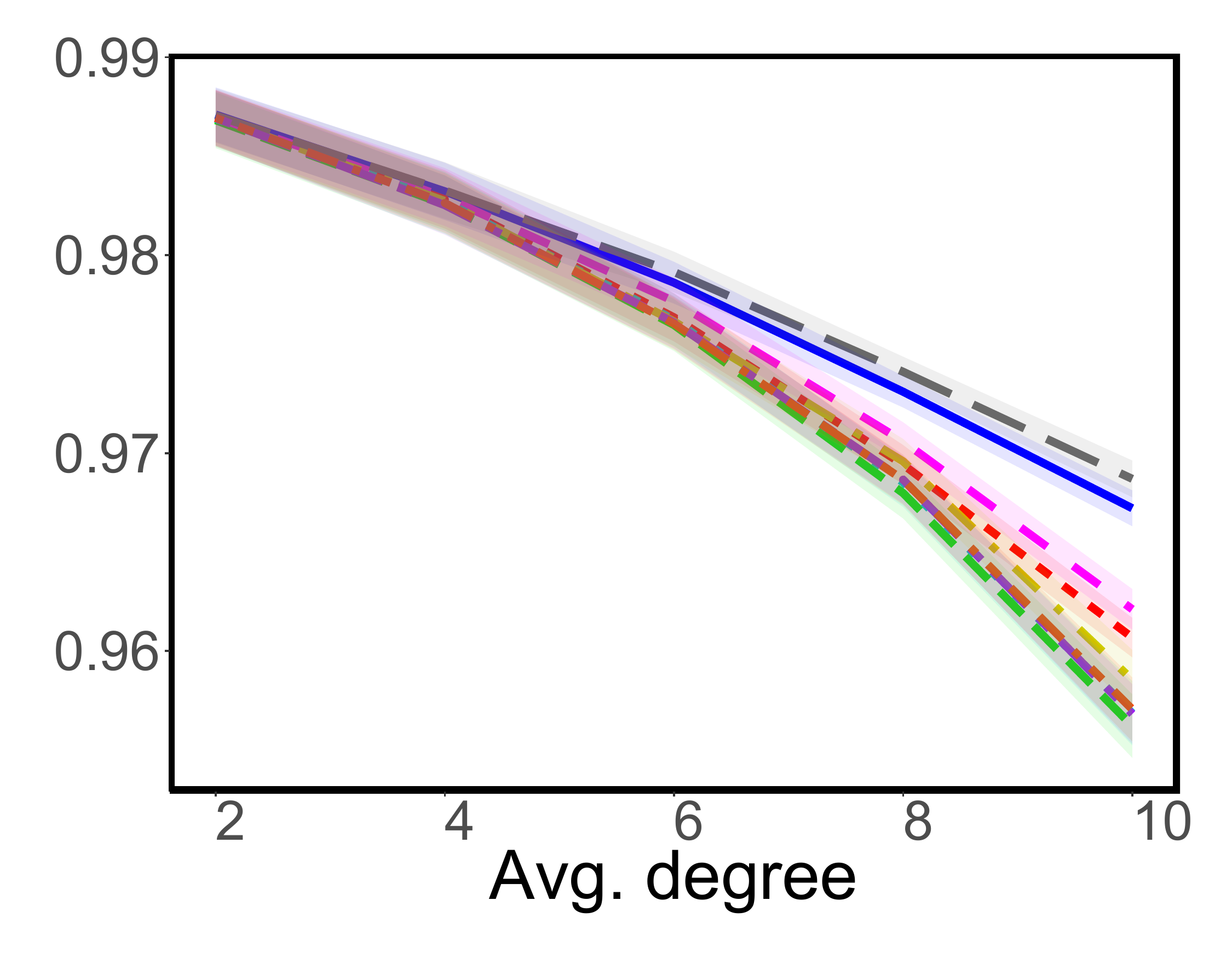} \\
\rotatebox{90}{\footnotesize CTR-$\ROC$} &
\includegraphics[width=\linewidth]{figures/plots/params/basic-ScaleFree-ctr-roc-bydegree} &
\includegraphics[width=\linewidth]{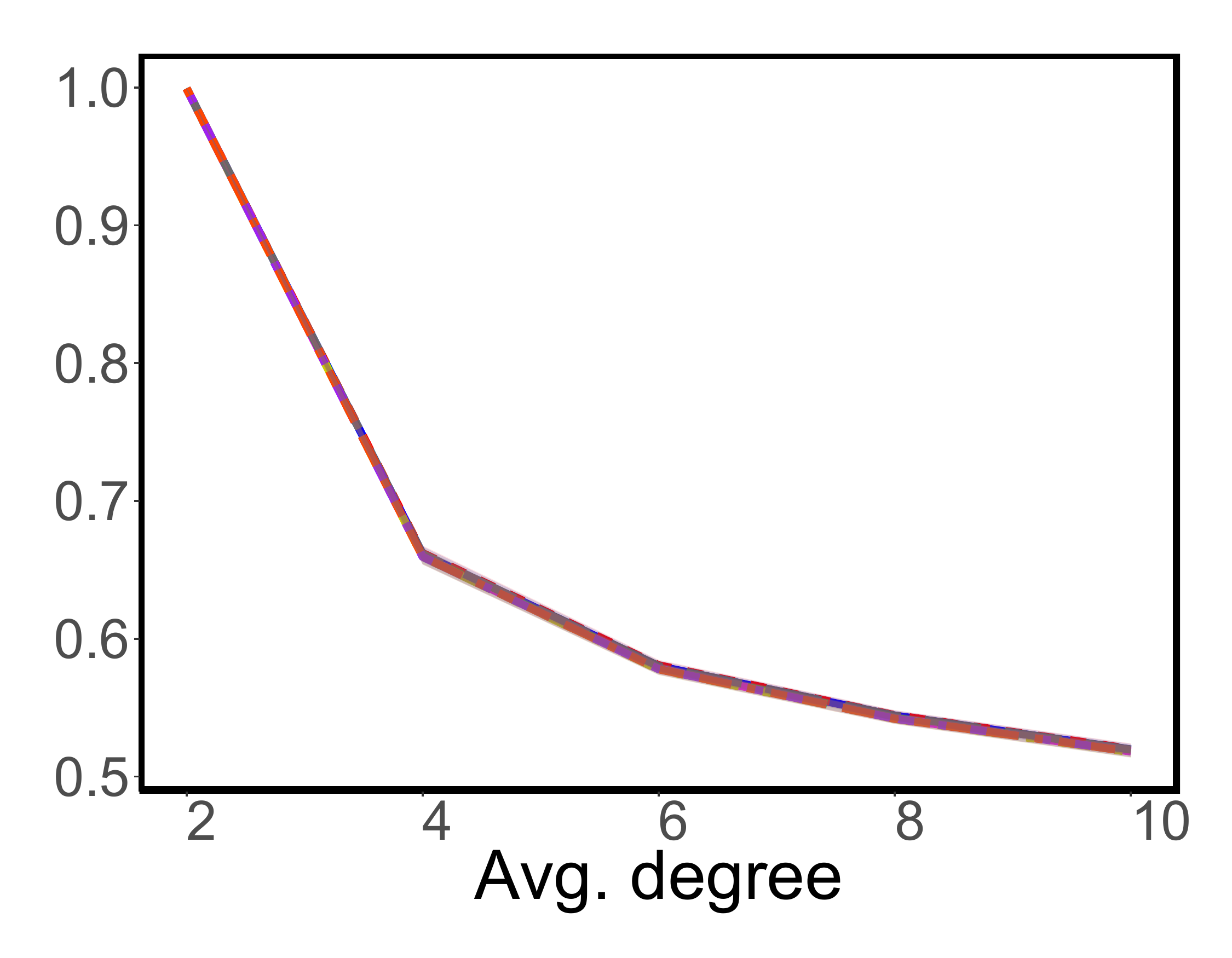} &
\includegraphics[width=\linewidth]{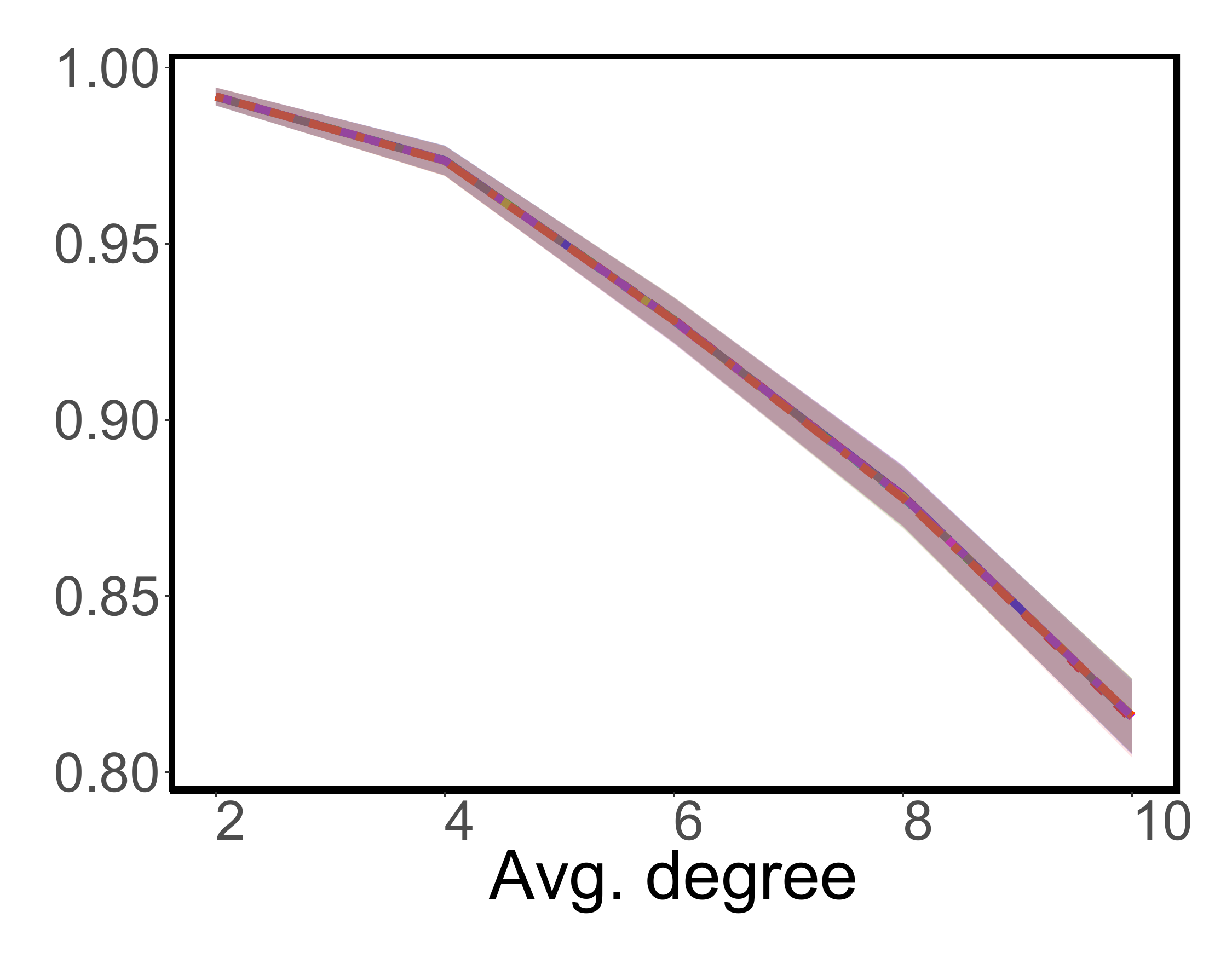} \\
\rotatebox{90}{\footnotesize OTC-$\AP$} &
\includegraphics[width=\linewidth]{figures/plots/params/basic-ScaleFree-otc-ap-bydegree} &
\includegraphics[width=\linewidth]{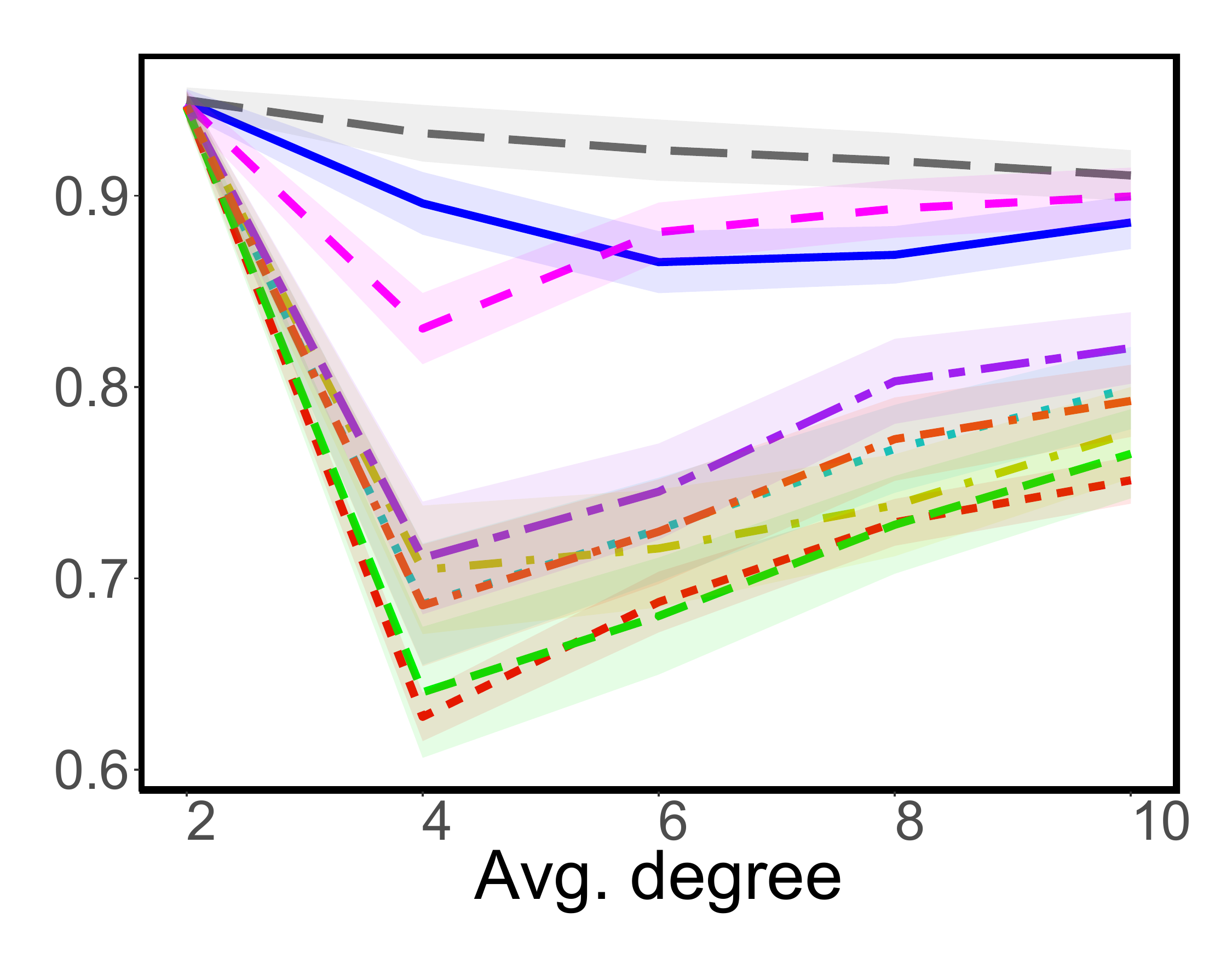} &
\includegraphics[width=\linewidth]{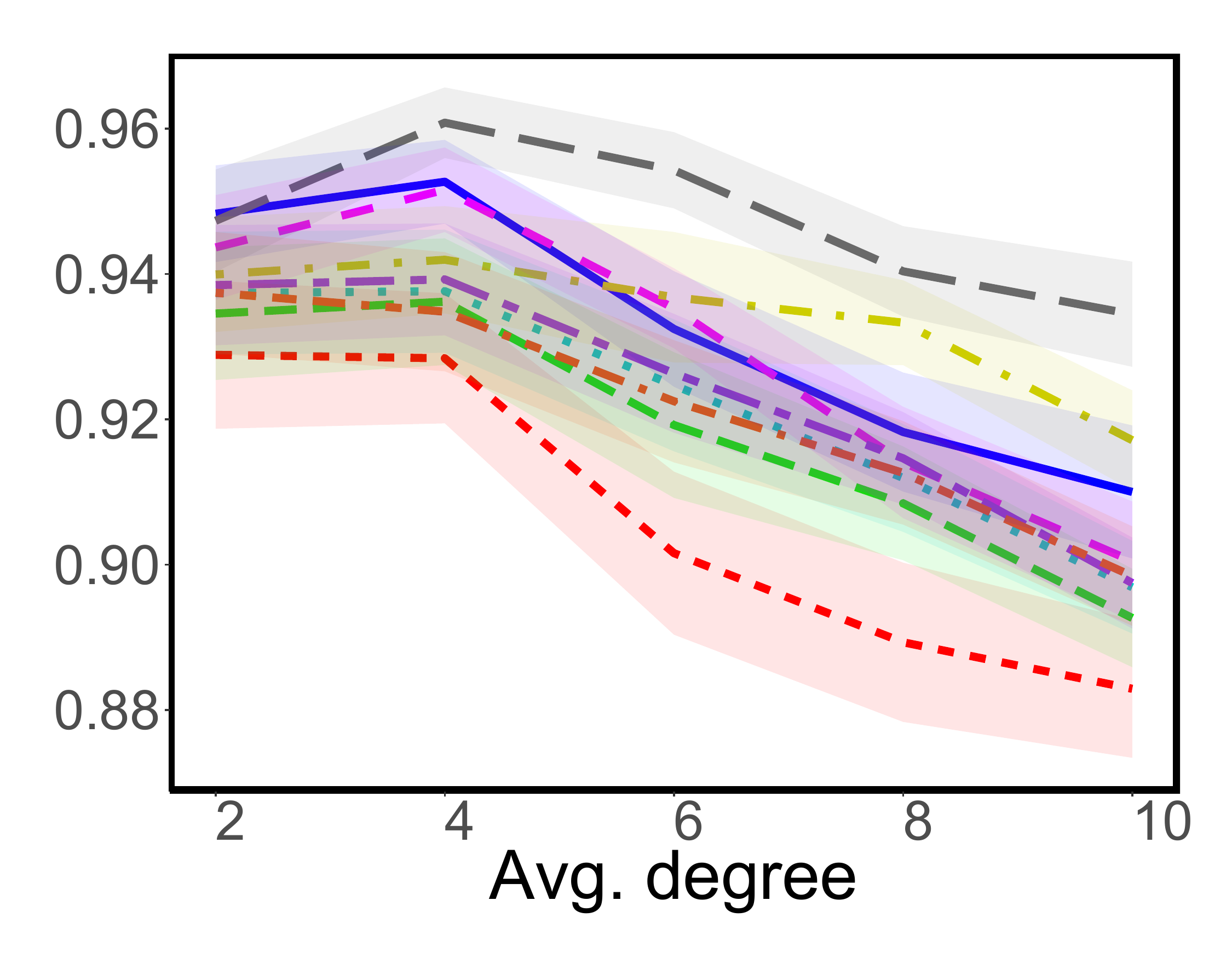} \\
\rotatebox{90}{\footnotesize CTR-$\AP$} &
\includegraphics[width=\linewidth]{figures/plots/params/basic-ScaleFree-ctr-ap-bydegree} &
\includegraphics[width=\linewidth]{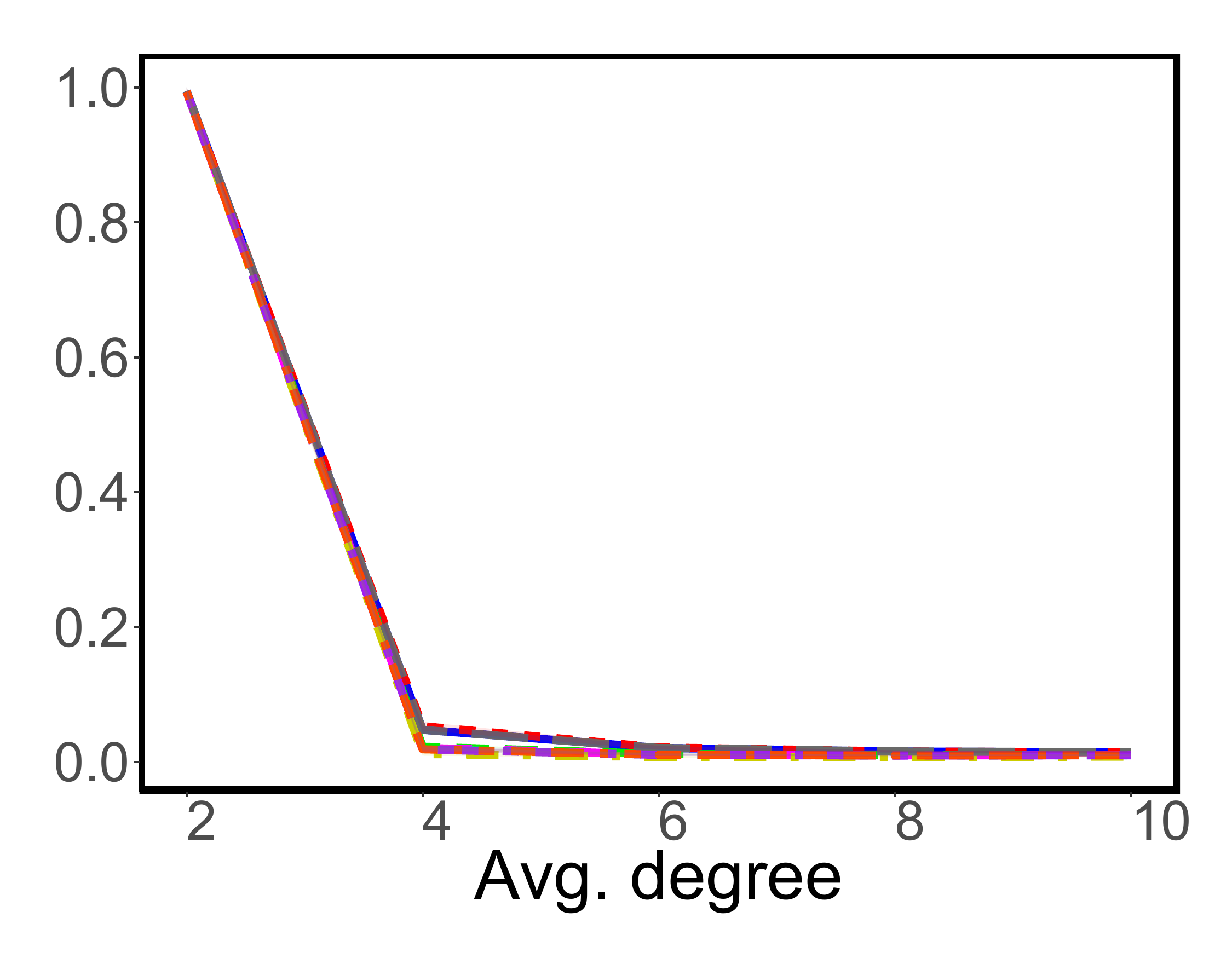} &
\includegraphics[width=\linewidth]{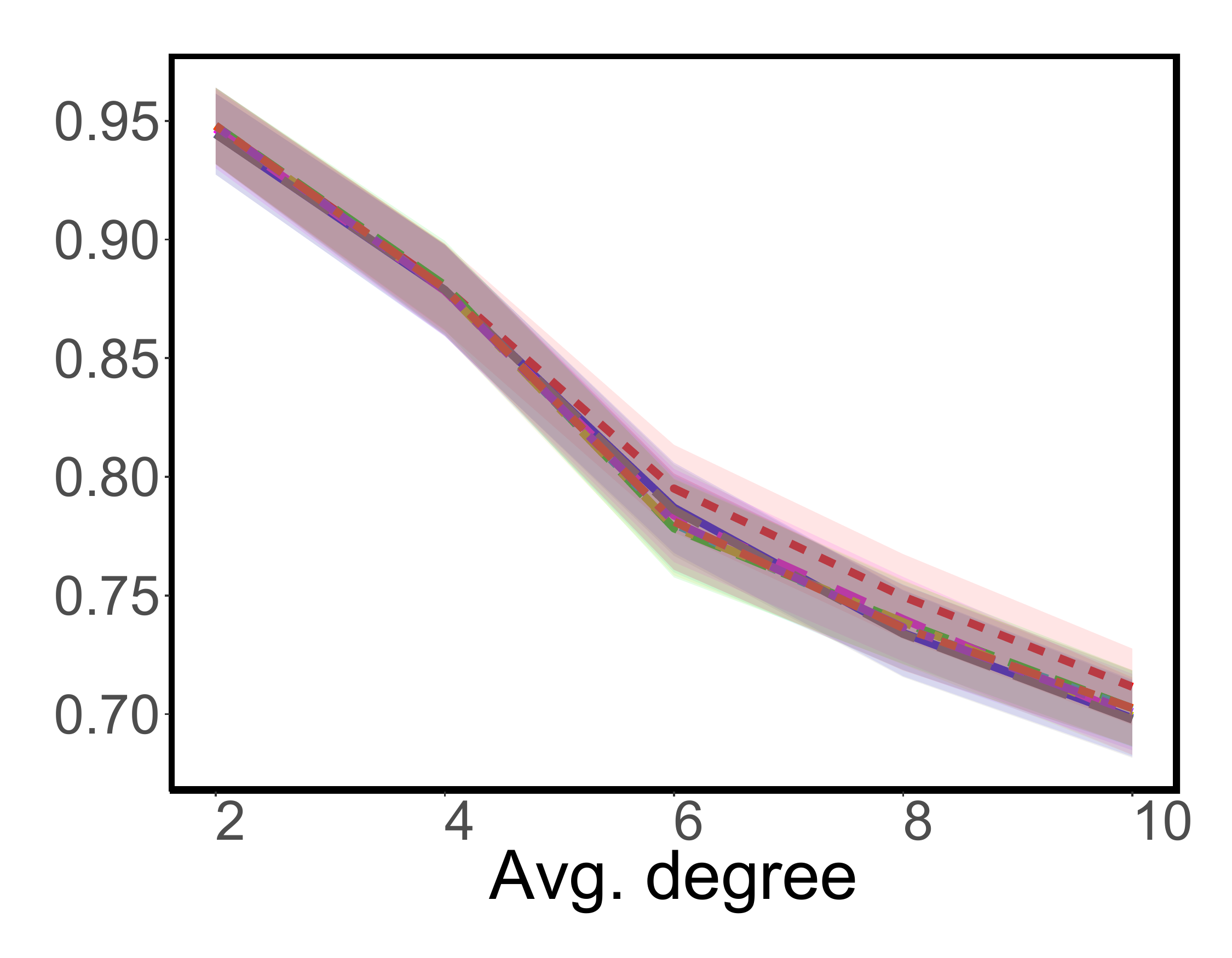} \\
\multicolumn{4}{c}{\includegraphics[width=0.6\linewidth]{figures/plots/params/basic-legend}}
\end{tabular}
\caption{Evaluating the attack tolerance of different \textbf{local similarity} indices against OTC (which adds edges) and CTR (which removes edges) by measuring the relative change in $\ROC$ (the area under the ROC curve) and $\AP$ (the average precision) while varying the \textbf{average degree}, $d$, in three types of networks: (i) ScaleFree$(n,d)$; (ii) SmallWorld$(n,d,0.25)$; and (iii) RandomGraph$(n,d)$. For each $d$, we report the average over $n=200,400,\ldots,1000$. The links in $\Hide$ are chosen at random, where $|\Hide|=100$ and $b=4|\Hide|$. The entire experiment is repeated $50$ times and the average is reported with coloured areas representing the $95\%$ confidence intervals.}

\label{fig:bydegree:supplementary}
\end{figure*}

\end{document}